\documentclass[twocolumn]{aastex631}
\usepackage{graphicx} 
\usepackage{xcolor}
\usepackage{colortbl}
\usepackage{soul}
\usepackage{todonotes}
\usepackage[version=4]{mhchem}
\usepackage{isotope}
\usepackage{amsmath, amsthm, amssymb, amsfonts}
\usepackage[T1]{fontenc}
\usepackage{mathptmx}

\usepackage{multirow}
\usepackage{booktabs}
\usepackage{hyperref}
\usepackage{CJK}

\begin{document} 
\title{Properties and Possible Physical Origins of $\gamma$-ray Emission in Extreme Synchrotron Blazars}

\correspondingauthor{Jin Zhang}
\email{j.zhang@bit.edu.cn}

\author[0000-0003-2547-1469]{Ji-Shun Lian}
\affiliation{School of Physics, Beijing Institute of Technology, Beijing 100081, People's Republic of China; j.zhang@bit.edu.cn}

\author[0009-0003-9471-4724]{Jia-Xuan Li}
\affiliation{School of Physics, Beijing Institute of Technology, Beijing 100081, People's Republic of China; j.zhang@bit.edu.cn}

\author[0000-0002-3883-6669]{Ze-Rui Wang}
\affiliation{College of Physics and Electronic Engineering, Qilu Normal University, Jinan 250200, People’s Republic of China}

\author[0009-0001-4710-5120]{Rui-Qi Huang}
\affiliation{School of Physics, Beijing Institute of Technology, Beijing 100081, People's Republic of China; j.zhang@bit.edu.cn}

\author[0000-0001-6863-5369]{Hai-Ming Zhang\dag}
\affiliation{Guangxi Key Laboratory for Relativistic Astrophysics, School of Physical Science and Technology, Guangxi University, Nanning 530004, People's Republic of China}

\author[0000-0003-3554-2996]{Jin Zhang\dag}
\affiliation{School of Physics, Beijing Institute of Technology, Beijing 100081, People's Republic of China; j.zhang@bit.edu.cn}

\begin{abstract}

Extreme synchrotron blazars, characterized by a first peak in their broadband spectral energy distributions (SEDs) at frequencies exceeding $10^{17}$ Hz, often exhibit a second peak beyond 1~TeV. These sources serve as ideal laboratories for studying particle acceleration and radiation mechanisms in relativistic jets. In this work, we systematically analyze the $\sim$16-year Fermi-LAT observational data for 25 extreme high-synchrotron-peaked BL Lacs (EHBLs). The results indicate that the majority of these sources display stable or low flux levels in the GeV band, with only 6 sources showing significant variability at a confidence level exceeding 5$\sigma$. The time-averaged spectra over the 16-year period for most EHBLs are well described by a hard power-law model, with photon indices predominantly clustered between 1.7 and 1.8. Using Fermi-LAT data in conjunction with multiwavelength observations compiled from the literature, we construct broadband SEDs for these EHBLs and fit them with a one-zone synchrotron + synchrotron-self-Compton (SSC) model. We find that this simplified theoretical framework is sufficient for modeling the observed SEDs of most of these EHBLs, albeit requiring relatively higher electron energies compared to other $\gamma$-ray emitting HBLs, and at times under-representing the UV emission. Based on the SED fitting results, we investigate the physical properties of the emission regions in these EHBLs and compare them with those of other $\gamma$-ray emitting HBLs. Consistent with other GeV--TeV $\gamma$-ray-emitting BL Lacs, the jets in these EHBLs are marked by low radiation efficiency and low magnetization.

\end{abstract}


\section{Introduction}\label{introduction}

Blazars are a subclass of active galactic nuclei (AGNs), characterized by relativistic jets oriented at small viewing angles \citep{1987PhR...154....1B, 1995PASP..107..803U}. They are further classified into Flat-spectrum radio quasars (FSRQs) and BL Lacertae objects (BL Lacs) based on the presence or absence of prominent optical emission lines \citep{1991ApJS...76..813S, 1991ApJ...374..431S, 2003ApJ...585L..23F}. The broadband spectral energy distributions (SEDs) of blazars, spanning from radio to $\gamma$-ray bands, are characterized by two humps and are dominated by the non-thermal emission from their relativistic jets, although thermal emission from the host galaxy, the accretion disk, and the dusty torus can lead to a more complex SED shape. The first hump is attributed to synchrotron radiation emitted by relativistic electrons within the jet, while the second hump is typically explained by inverse Compton scattering of low-energy photons by these same relativistic electrons. The seed photons for this scattering process may originate from the synchrotron radiation itself, a mechanism referred to as synchrotron self-Compton (SSC; e.g., \citealt{1992ApJ...397L...5M, 1996ApJ...461..657B, 1997A&A...320...19M, 2009RAA.....9..777Z, 2012ApJ...752..157Z, 2014ApJ...788..104Z}), or from external fields such as the accretion disk, broad-line region, or dusty torus, in which case it is referred to as external Compton (EC; e.g., \citealt{1992A&A...256L..27D, 1994ApJ...421..153S, 2000ApJ...545..107B, 2009MNRAS.397..985G, 2009ApJ...704...38S, 2014ApJ...788..104Z}). Based on the peak frequency ($\nu_{\rm s, p}$) of the synchrotron emission, blazars are generally classified into three categories: low-synchrotron-peaked (LSP) blazars with $\nu_{\rm s, p} \le 10^{14}$ Hz, intermediate-synchrotron-peaked (ISP) blazars with $10^{14}<\nu_{\rm s, p}<10^{15}$ Hz, and high-synchrotron-peaked (HSP) blazars with $ \nu_{\rm s, p} \ge 10^{15}$ Hz \citep{2010ApJ...716...30A, 2016ApJS..226...20F, 2019NewA...7301278S, 2020ApJS..247...16A}. FSRQs are almost exclusively LSP blazars \citep{2012MNRAS.420.2899G}.

There also exists a small subset of BL Lacs that exhibit extremely high synchrotron-peaked (EHSP) behavior, with $\nu_{\rm s, p} \geq 10^{17}$ Hz \citep{2019MNRAS.486.1741F}. Moreover, the TeV-emitting EHSP BL Lacs (EHBLs) typically display a second spectral component peaking at $\geq 1$ TeV in their SEDs \citep{2019NewA...7301278S, 2020NatAs...4..124B}. These extreme peak characteristics in the broadband SEDs of BL Lacs can be observed both during flaring episodes and in quiescent states. These sources are characterized by hard spectra in both the soft X-ray and 0.1--1.0 TeV energy ranges, with photon spectral indices less than 2. For TeV-emitting BL Lacs, the one-zone synchrotron + SSC model is commonly employed to successfully reproduce their broadband SEDs \citep{2010MNRAS.401.1570T, 2012ApJ...752..157Z}. This model has also been applied to explain the broadband SEDs of EHBLs (e.g., \citealt{2001A&A...371..512C, 2002A&A...386..833G, 2015MNRAS.451..611B, 2018MNRAS.477.4257C, 2020ApJS..247...16A, 2020ApJ...889..149D, 2022MNRAS.517L..16T, 2025A&A...700A.229L}).  However, such modeling often requires extreme physical conditions within the emission region, such as a magnetic field strength significantly lower than the equipartition value \citep{2010MNRAS.401.1570T, 2018MNRAS.477.4257C, 2022Galax..10..105S}, an extremely hard electron spectrum \citep{2020NatAs...4..124B, 2022Galax..10..105S, 2024A&A...682A.134G}, a minimum electron Lorentz factor exceeding $10^3$ \citep{2010MNRAS.401.1570T, 2015MNRAS.451..611B, 2015MNRAS.448..910C, 2022MNRAS.512.1557A}, or a Doppler factor greater than 40 \citep{2006MNRAS.368L..52K, 2015MNRAS.448..910C, 2018MNRAS.477.4257C, 2019NewA...7301278S}. These extreme BL Lacs offer a unique opportunity to investigate various high-energy astrophysical phenomena. They serve as ideal candidates for studying jet physics, particularly particle acceleration and radiation mechanisms, as well as for constraining models of the extragalactic background light (EBL) and the intergalactic magnetic field.

Variability across almost all energy bands, observed over different timescales ranging from a few minutes to several years, has been identified as a common characteristic of blazars (e.g., \citealt{2007ApJ...669..862A, 2008Natur.452..966M, 2017A&ARv..25....2P}). Previous studies have shown that only a small fraction of HSP BL Lacs (HBLs) exhibit significant variability in the GeV $\gamma$-ray band, a proportion that is substantially lower than that observed in LSP BL Lacs (LBLs) and ISP BL Lacs (IBLs, \citealt{2011ApJ...743..171A, 2024ApJS..270...22W}). Most TeV-emitting BL Lacs also appear to remain stable over multi-year timescales in the TeV $\gamma$-ray band, although large statistical uncertainties prevent definitive conclusions regarding flux variations \citep{2020NatAs...4..124B}. To date, Fermi-LAT has accumulated more than 17 years of observational data for these bright blazars, providing an opportunity to investigate their variability and radiation properties in the GeV band, as well as their flux states during TeV emission detections.

In this paper, we assemble a sample of TeV-emitting EHSP blazars, comprising 23 of the 24 sources listed in Table 1 of \citet{2020NatAs...4..124B}; RX J1136.5+6737 is excluded due to insufficient observational data. Additionally, two further sources, MRC 0910--208 and 1RXS J1958, are included based on their classification as EHBLs in the TeVCat catalog\footnote{\url{http://tevcat2.uchicago.edu/}}. The final sample consists of 25 sources, all classified as EHBLs. We conduct a systematic analysis of the 16-year Fermi-LAT observational data for these sources to investigate their GeV $\gamma$-ray emission properties. Furthermore, we construct their broadband SEDs by combining the derived GeV-band spectra with archival data from the TeV and other wavelength bands, aiming to explore the underlying $\gamma$-ray radiation mechanisms. The methodology for Fermi-LAT data analysis is described in Section~\ref{sec_Fermi-LAT}, and the corresponding results are presented in Section~\ref{sec_Fermi-result}. The construction of SEDs and the results of model fitting are detailed in Section~\ref{sec_SED}. Discussion and a summary are provided in Sections~\ref{sec_dis} and \ref{sec_Summary}, respectively. Throughout, $H_0=71$ km s$^{-1}$ Mpc$^{-1}$, $\Omega_{\rm m}=0.27$, and $\Omega_{\Lambda}=0.73$ are adopted.

\section{Fermi-LAT Observations and Data Analysis}\label{sec_Fermi-LAT}

The PASS 8 data from the Fermi-LAT satellite within a 15$^\circ$ circular region of interest (ROI), centered on the radio positions of these EHBLs, were selected and downloaded from the Fermi Science Support Center\footnote{\url{https://fermi.gsfc.nasa.gov/cgi-bin/ssc/LAT/LATDataQuery.cgi}}. The dataset spans from 2008 August 4 to 2024 September 22 (MJD 54682--60575), covering an energy range from 0.1 GeV to 1 TeV. The analysis was conducted using the publicly available software packages \texttt{fermitools} (version 2.2.0) and \texttt{Fermipy} \citep[v1.1;][]{Wood2017}. Following the Fermi-LAT data selection recommendations\footnote{\url{https://fermi.gsfc.nasa.gov/ssc/data/analysis/documentation/Cicerone/Cicerone_Data_Exploration/Data_preparation.html}}, we employed event class ``SOURCE'' (evclass=128) and event type ``FRONT+BACK'' (evtype=3) for the binned likelihood analysis. To eliminate the contamination of $\gamma$-ray emission from the Earth's limb, we applied a zenith angle cut of less than 90$^\circ$. The standard data quality filter ``(DATA\_QUAL>0)\&\&(LAT\_CONFIG==1)'' and the instrument response functions P8R3\_SOURCE\_V3 were used. The source model includes all $\gamma$-ray sources listed in the 4FGL-DR4 catalog \citep{2020ApJS..247...33A} within the ROI, as well as two background models: the isotropic emission model ("$\rm iso\_P8R3\_SOURCE\_V3\_V1.txt$") and the galactic diffuse emission model ("$\rm gll\_iem\_v07.fits$"). During the likelihood fitting process, only the parameters of sources located within 7.0$^\circ$ of the ROI center and the normalization parameters of the two background models were allowed to vary; all other parameters were fixed to the values in the 4FGL-DR4.

For all sources in our sample, the spectral models recommended in the 4FGL-DR4 were consistently applied during data analysis. Specifically, the log-parabola (LP) model is 
\begin{equation}
    \frac{d N}{d E}=N_{\gamma,0}\left(\frac{E}{E_{\mathrm{b}}}\right)^{-\left(\Gamma_\gamma+\beta \log \left(\frac{E}{E_{\mathrm{b}}}\right)\right)},
\end{equation}
where $E_{\rm b}$ is the scale parameter of photon energy, $\Gamma_{\gamma}$ is the photon spectral index, and $\beta$ is the curvature parameter.
If $\beta=0$, the LP model turns into a power-law (PL) model,  
\begin{equation}
    \frac{d N}{d E}=N_{\gamma,0}\left(\frac{E}{E_{\mathrm{b}}}\right)^{-\Gamma_\gamma}.
\end{equation}
For Mrk 421, a special function, PLSuperExpCutoff4, is required to describe its spectrum, which is structured as
{\footnotesize
\begin{equation}
    \frac{d N}{d E}= \begin{cases}N_{\gamma,0}\left(\frac{E}{E_0}\right)^{-\Gamma_\gamma-\frac{\beta}{2} \ln \frac{E}{E_0}-\frac{\beta b}{6} \ln ^2 \frac{E}{E_0}-\frac{\beta b^2}{24} \ln ^3 \frac{E}{E_0}}, & \text { if }\left|b \ln \frac{E}{E_0}\right|<1 e^{-2}, 
    \\ N_{\gamma,0}\left(\frac{E}{E_0}\right)^{-\Gamma_\gamma+\beta / b} \exp \left(\frac{\beta}{b^2}\left(1-\left(\frac{E}{E_0}\right)^b\right)\right) & \text { otherwise },\end{cases} 
\end{equation}}
where $N_{\gamma,0}$ is normalization (flux density) at $E_0$, $E_0$ is the scale parameter and $\beta$ is the local curvature at $E_0$. $E_0$ and \textit{b} are fixed at the value of 4FGL-DR4 as required, and $\Gamma_{\gamma}$ is the photon spectral index.

We investigated potential new sources present in the ROI. The test statistic (TS) was utilized to evaluate the significance of a $\gamma$-ray source signal, where $\rm TS=2(log\mathcal{L}_{\rm src}-log\mathcal{L}_{\rm null})$. Here, $\mathcal{L}_{\rm src}$ and $\mathcal{L}_{\rm null}$ denote the likelihood values of the background with and without the target source, respectively. A maximum TS value of $\geq$25 in the residual TS map indicates the detection of a new $\gamma$-ray source at the 5$\sigma$ confidence level. 
We found a new $\gamma$-ray source located at (R.A.=$166.20^\circ\pm0.19^\circ$, Decl.=$-24.85^\circ\pm0.18^\circ$) with a TS = 34.1, which is 1.4$^\circ$ away from 1ES 1101--232. The flux of this new source is $(1.63\pm0.33)\times10^{-12}~\rm erg\ cm^{-2}\ s^{-1}$ with a photon spectral index of $\Gamma_\gamma=2.49\pm0.16$ in the 0.1--1000 GeV energy band. The presence of this newly identified source has been accounted for in the data analysis of 1ES 1101--232. 

The long-term light curves were generated using the best-fit results, in which the normalization and spectral index of the target source were left free while all other sources within the ROI were kept fixed. The light curves were produced using the binned likelihood method with a time bin size of 90 days. A likelihood-based statistic was used to estimate the variability of the GeV light curves  (e.g., \citealt{2012ApJS..199...31N,2019ApJ...884...91P,2020ApJS..247...33A}). Following the definition provided by \citet{2012ApJS..199...31N}, we calculated the variability index (TS$_{\rm var}$) as 
\begin{equation}
    \label{eqTSvar}	\mathrm{TS}_{\mathrm{var}}=2\sum_{i=0}^{N}\left[\log\left(\mathcal{L}_i\left(F_i\right)\right)-\log\left(\mathcal{L}_i\left(F_{\text{global}}\right)\right)\right],
\end{equation}
where $N$ represents the number of time bins of the light curves, $F_i$ denotes the flux in the $i$-th bin, $F_{\rm glob}$ is the best fit flux over the entire time interval, and $\mathcal{L}_{i}(F_i)$ is the likelihood associated with the $i$-th bin. It should be noted that for a light curve comprising $N=65$ time bins in this analysis, a value of $\rm{TS}_{var}\geq138.8$ is required to identify a source as variable at the 5$\sigma$ confidence level, corresponding to $N - 1 = 64$ degrees of freedom.

\section{Fermi-LAT Data Analysis Results}\label{sec_Fermi-result}

\subsection{Spectral Properties}

The 16-year Fermi-LAT average spectra for the 25 EHBLs in the 0.1--1000 GeV band, along with the corresponding fitting results, are presented in Figure \ref{Fig: LCs}. Among these, 18 spectra are well explained by a PL function, while 6 spectra require a curved LP model for an accurate fit; the only exceptional spectrum of Mrk 421 necessitates a more complex functional form for an appropriate representation. This is presumably because of the averaging over the 16-year observation data of this highly variable source, as Mrk 421 exhibits significant variability in both flux and spectrum. All sources display a hard spectra in the GeV band with $\Gamma_\gamma\le2.0$, as listed in Table \ref{tab_LAT}. The $\Gamma_\gamma$ values of these EHBLs range from 1.51 to 2.01, with a concentration between 1.70 and 1.80. The hardest and most curved spectrum is observed from RGB J0710+591, characterized by $\Gamma_\gamma = 1.51 \pm 0.08$ and $\beta = 0.060 \pm 0.026$. These hard GeV $\gamma$-ray spectra suggest the presence of a high-energy peak in the broadband SEDs of these sources. Additionally, we calculate the 16-year average flux in the 0.1--100 GeV band for the 25 EHBLs, which ranges from $1.20\times 10^{-12}\ \rm erg\ cm^{-2}\ s^{-1}$ to $402.02\times 10^{-12}\ \rm erg\ cm^{-2}\ s^{-1}$, as detailed in Table \ref{tab_LAT}. Mrk 421 displays the highest flux, exceeding the average flux of MRC 0910--208 by more than two orders of magnitude.

\subsection{Long-term Light curves}

Using the 16-year Fermi-LAT observation data, we generate the long-term $\gamma$-ray light curves of the 25 EHBLs in the 0.1--1000 GeV band with time bins of 90 days, as illustrated in Figure \ref{Fig: LCs}. Additionally, we compute TS$_{\rm var}$ values to quantify the variability of these sources, as summarized in Table \ref{tab_LAT}. It is found that the majority of sources in the sample do not exhibit significant flux variability. According to the TS$_{\rm var}$ criterion, only six sources, namely Mrk 421, Mrk 501, 1ES 1218+304, 1ES 1712+502, 1ES 1959+650, and 1ES 2344+514, show variability at a confidence level exceeding $5\sigma$.  For several EHBLs, long-term $\gamma$-ray detections are either absent or sparse, with only upper limits reported in their light curves. In these cases, the number of detection points is substantially lower than the number of upper limits, including SHBL J001355.9--185406, 1ES 0229+200, 1ES 0347--121, PKS 0548--322, PGC 2402248, RBS 0723, 1ES 1440+122, and 1ES 2037+521 (featuring fewer than 20 detection points in their respective light curves). These findings suggest that the emission of these EHBLs remains below the Fermi-LAT detection threshold over a period of three months.

In the long-term Fermi-LAT light curves, the time intervals corresponding to the TeV band observations used for SED modeling in Section \ref{sec_SED} are indicated by pink shading, as illustrated in Figure \ref{Fig: LCs}. For sources whose TeV observations were conducted prior to the launch of the Fermi satellite, no such shaded regions appear in their Fermi-LAT light curves; these sources include 1ES 0347--121, PKS 0548--322, 1ES 1101--232, and 1ES 2356--309. As shown in Figure \ref{Fig: LCs}, the TeV detections of these EHBLs are generally not associated with a high-flux state in the GeV energy band.

\section{SED Construction and Model Fitting}\label{sec_SED}
\subsection{SED Construction}\label{sec_SED-Construction}

The majority of these sources may exhibit EHBL characteristics only during flaring states and do not remain in the EHBL state continuously. For example, Mrk 421, Mrk 501, and 1ES 1959+650 are well-known HBLs that occasionally display extreme behavior \citep{2018A&A...620A.181A, 2019MNRAS.486.1741F}. In this study, we select only one broadband SED per source, with the condition that $\nu_{\rm s, p} \geq 10^{17}$ Hz. Except for the GeV spectra, multi-band data for the 25 EHBLs are compiled from existing literature; the highest flux state in the TeV band, together with the quasi-simultaneously observed multi-band data, is selected for each EHBL in cases where multiple observations are available. The broadband SEDs of the 25 EHBLs are presented in Figure \ref{Fig: SED fitting}, and detailed information regarding the data origins for each source refers to Appendix \ref{appendix: Details}. 

\begin{figure*}
\epsscale{1.18}
\plotone{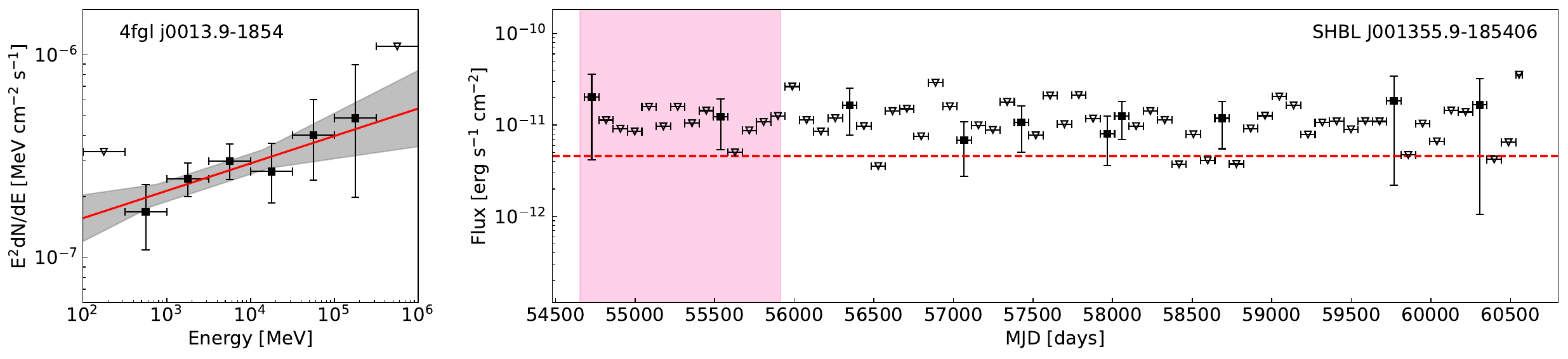}
\caption{Left column: the $\sim$16-yr Fermi-LAT average spectra along with the fitting results, where the open inverted triangles indicate TS<4 for that energy bin. Right column: the long-term Fermi-LAT light curves with time bins of 90 days, where the open inverted triangles indicate the TS value of that time bin less than 9. The horizontal red dash lines represent the $\sim$16 yr average $\gamma$-ray flux of the sources. The pink-shaded regions denote the period covered by the TeV observations, as well as other multiwavelength observations performed during the TeV campaigns. For sources such as 1ES 0347--121, PKS 0548--322, 1ES 1101--232, and 1ES 2356--309, whose TeV observations were conducted prior to the launch of the Fermi satellite, no such shaded regions appear in their Fermi-LAT light curves. For the four BL Lacs (Mrk 421, 1ES 1218+304, 1ES 1727+502 and 1ES 1959+650; marked with a red star in the light-curve figure panels) with significant variability, the contemporaneous Fermi-LAT spectra with the TeV observations are derived to construct their broadband SEDs in Figure \ref{Fig: SED fitting}, that is, their spectra in both the GeV and TeV bands from the observation periods marked as the pink-shaded regions in their light curves. For the other two sources (Mrk 501 and 1ES 2344+514, marked with a red rhombus in the light-curve figure panels) with significant variability, slightly extended observation periods (the green-shaded regions in their light curves) relative to their TeV campaigns (the pink-shaded regions in their light curves) are employed to generate the GeV spectrum for their broadband SED construction. For the remaining 19 objects, the 16-year Fermi-LAT averaged spectra are used to construct their broadband SEDs in Figure \ref{Fig: SED fitting}.}
\label{Fig: LCs} 
\end{figure*}

The GeV spectra of these sources are obtained from the analysis results presented in this work and are selected according to the following criteria: (1) for the 19 sources exhibiting no significant variability, the 16-year averaged spectra are used; (2) for the six BL Lacs (Mrk 421, 1ES 1218+304, 1ES 1727+502, 1ES 1959+650, Mrk 501, and 1ES 2344+514) with significant variability, a reanalysis of Fermi-LAT observational data is conducted. Specifically, the contemporaneous Fermi-LAT spectra with the TeV observations are derived for Mrk 421, 1ES 1218+304, 1ES 1727+502, and 1ES 1959+650, i.e., their spectra in both GeV and TeV bands from the observation periods marked as the pink-shaded regions in their light curves in Figure \ref{Fig: LCs}. For the other two sources, Mrk 501 and 1ES 2344+514, the TeV spectra are from the observations within one day (the pink-shaded regions in their light curves in Figure \ref{Fig: LCs}), and no statistical signal can be obtained from the simultaneous Fermi-LAT observations. Therefore, slightly extended observation periods relative to the TeV campaigns are employed to ensure sufficient statistical significance of the GeV flux measurements, similar to the approach in \citet{2020A&A...637A..86M} and \citet{2020MNRAS.496.3912M}. For Mrk 501, a 16-day period (the green-shaded region in its light curve in Figure \ref{Fig: LCs}) is selected to reanalyze the Fermi-LAT observational data, which corresponds to the MAGIC telescope observations during its very high energy flaring activity in July 2014 \citep{2020A&A...637A..86M}. The obtained GeV spectrum is consistent with the result of the MAGIC Collaboration using a 10-day time window centered on the very high energy observation time. For 1ES 2344+514, the same Fermi-LAT observation period as that in \citet{2020MNRAS.496.3912M} is chosen, that is, one month centered on the very high energy observation (the green-shaded region in its light curve in Figure \ref{Fig: LCs}). Details of the data reanalysis for the six sources are provided in Table \ref{tab_LAT}.

\subsection{SED Fitting and Results}\label{section: SED Fitting}

To investigate the $\gamma$-ray emission mechanisms of these EHBLs, we employ the simplest theoretical framework, the one-zone leptonic synchrotron + SSC (one-zone SSC for short) model, to reproduce the broadband SEDs of the 25 selected EHBLs. This model is commonly used to explain the SEDs of TeV BL Lacs \citep{2010MNRAS.401.1570T, 2012ApJ...752..157Z, 2024ApJS..271...10W, 2025A&A...700A.229L}. The model calculations follow established computational methods (e.g., \citealt{1979rpa..book.....R, 2001A&A...367..809K, 2013LNP...873.....G, 2017ApJ...842..129C}), with detailed formulations provided in Appendix \ref{section: model}. The spectral output of the model is determined by 9 parameters: the radius $R$, magnetic field strength $B$, and Doppler factor $\delta$ of the emission region; the broken power-law indices $p_1$ and $p_2$ of the electron energy distribution; the minimum ($\gamma_{\rm min}$), break ($\gamma_{\rm b}$), and maximum Lorentz factors ($\gamma_{\rm max}$) of the electrons; and the electron density parameter $N_{\rm 0}$. We assume $\delta=\Gamma$, where $\Gamma$ denotes the bulk Lorentz factor of the emitting region. 

Our strategy of SED fitting is to keep parameter values that are as consistent as possible with the parameter space reported for TeV BL Lacs in previous studies (e.g., \citealt{2010MNRAS.401.1570T, 2012ApJ...752..157Z, 2014MNRAS.439.2933Y, 2022MNRAS.512..137N, 2024ApJ...967..104Z}). The constrained parameter ranges are as follows: $B \in [0.01, 1]$ G, $R \in [10^{16}-10^{17}]$ cm, $\delta_{\rm D} \in [5-30]$, $p_1 \in [1.8-3]$, $p_2 \in [3.5-6]$, $\gamma_{\rm min} \in [1-10^4]$, $\gamma_{\rm b} \in [10^4-10^{6.5}]$, $\gamma_{\rm max} = 10 \times \gamma_{\rm b}$, and $N_{\rm 0} \in [10^{-3}-10^5]$ $\rm cm^{-3}$. Given the numerous free parameters, the well-known intrinsic degeneracies among them, and the limited availability of broadband data that may be not strictly simultaneous, an adjustment “by eye” is still the most common approach for SED fitting \citep{2022Galax..10..105S}. We aim to identify a plausible set of model parameters that can reproduce the observed data, with main focus on the X-ray and $\gamma$-ray spectra. Therefore, the modeling results presented here do not represent an exhaustive exploration of all possible parameter combinations or scenarios that could potentially fit the data within the framework of our model.

The fitting results are presented in Figure \ref{Fig: SED fitting}. Overall, the one-zone SSC model can well represent the broadband SEDs of these EHBLs. However, for several sources, the GeV spectra are not adequately described by this model. Specifically, RGB J0152+017, 1ES 0347--121, 1ES 0414+009, and 1ES 1741+196 exhibit relatively soft GeV spectra, leading to an excess at the low-energy end compared to model predictions. Additionally, the TeV spectra of four objects—TXS 0210+515, 1ES 0347--121, 1ES 0229+200, and 1ES 1101--232—show a hard spectrum in the very high energy band. These features suggest that a more complex model may be necessary to accurately describe their broadband SEDs, as the emission in the GeV--TeV energy range cannot be fully explained by a single SSC component.

The fitting parameter values for each SED are summarized in Table \ref{tab:Fit Parameters}, and their distributions are shown in Figure \ref{Fig: distribution}. specifically, the magnetic field strength $B$ ranges from 0.01 G to 0.6 G with a median of 0.07 G, while the Doppler factor varies between 5 and 30, with an average value of 14.6. The radii of the emission region are predominantly clustered within $10^{16}-10^{17}$ cm. Our analysis indicates that the one-zone SSC model remains effective in reproducing the broadband SEDs of most EHBLs across the X-ray to TeV energy bands. However, this comes at the expense of (i) requiring relatively high values of $N_0$, $\gamma_{\min}$, and $\gamma_{\rm b}$ for certain sources, and (ii) failing to adequately account for simultaneous optical-UV data, which may originate from distinct emission components (e.g., \citealt{2018MNRAS.477.4257C, 2020NatAs...4..124B}). A detailed comparison of these parameters with those reported for other HBLs is provided in Section \ref{sec_dis}.

The range of the synchrotron peaked frequency $\nu_{\rm s, p}$ is $1.0\times 10^{17}\ \text{Hz} \leqslant  \nu_{\rm s, p} \leqslant 9.7\times 10^{18}\ \text{Hz}$ with a mean value of $1.0\times 10^{18}\ \text{Hz}$, while the range of the SSC peaked frequency $\nu_{\rm c, p}$ is $9.8\times 10^{24}\ \text{Hz} \leqslant  \nu_{\rm c, p} \leqslant 1.2\times 10^{27}\ \text{Hz}$ with a mean value of $1.7\times 10^{26}\ \text{Hz}$.

\subsection{Jet Powers} 

\begin{figure*}[htbp]
\centering
\includegraphics[width=0.8\linewidth]{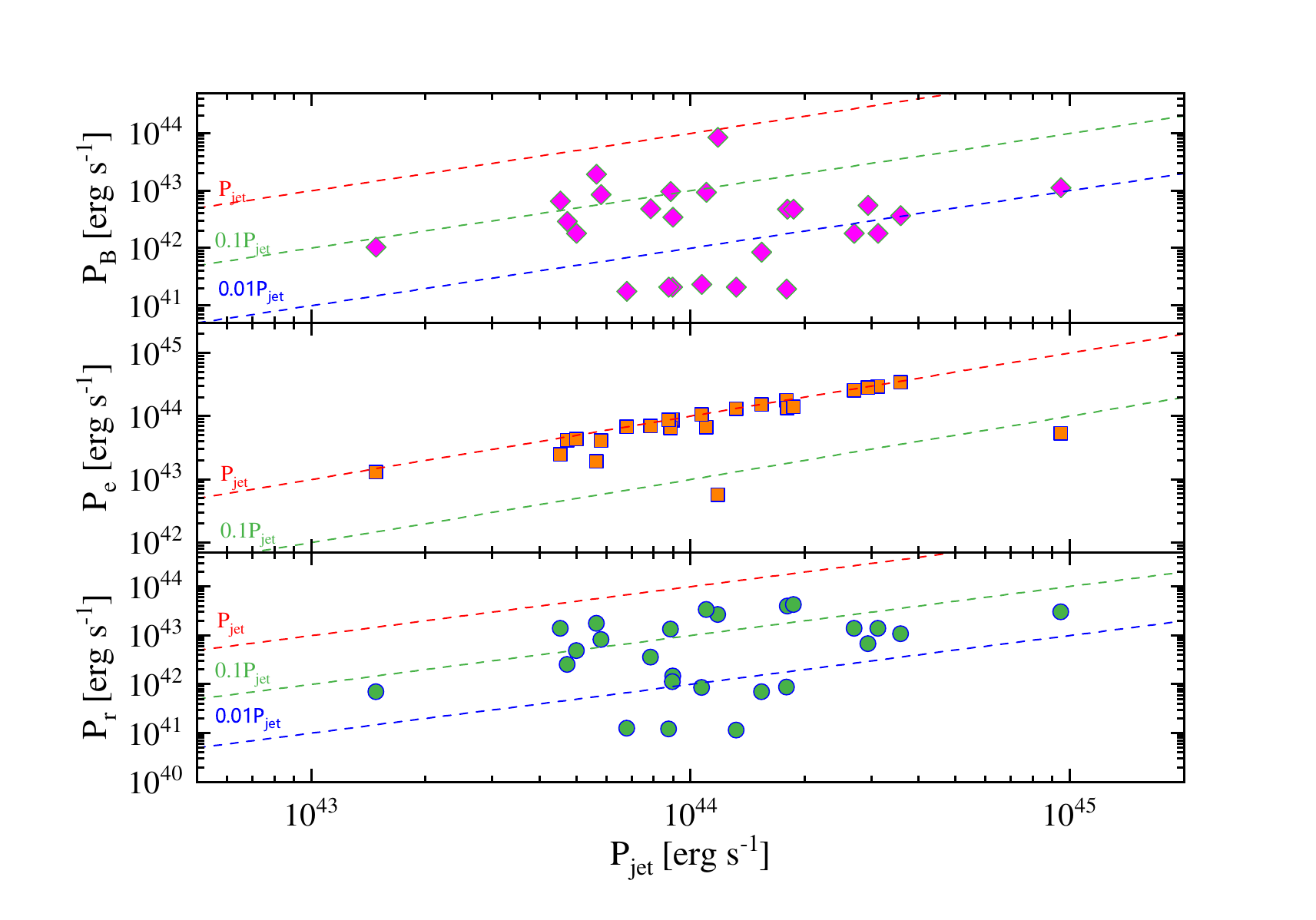}
\caption{$P_{\rm r}$, $P_{\rm e}$, and $P_B$ as functions of $P_{\rm jet}$ for the 25 EHBLs.}
\label{jetpower}
\end{figure*}

Based on the SED fitting parameters and following the method of previous works (e.g., \citealt{2010MNRAS.402..497G, 2012ApJ...752..157Z}), we calculate the jet powers of the 25 EHBLs by assuming that the jets consist of electrons ($P_{\rm e}$), magnetic fields ($P_B$), and radiation ($P_{\rm r}$), i.e., 
\begin{align}
&P_{\rm jet}=\pi R^2 \Gamma^2 c\left(U_{\mathrm{e}}+U_B+U_{\rm r}\right),\\
&U_{\mathrm{e}} = m_{\mathrm{e}} c^2 \int N(\gamma) \gamma \, d \gamma ,\\
&U_{B} = \frac{B^2}{8 \pi}, \\
&U_{\mathrm{r}} = \frac{L_{\mathrm{bol}}}{ 4 \pi R^2 c \delta^4},
\end{align}
where $L_{\rm bol}$ is the bolometric luminosity. The derived jet power and the powers of each jet component for these EHBLs are given in Table \ref{tab:Jet properties}. We plot $P_{\rm r}$, $P_{\rm e}$, and $P_B$ as a function of $P_{\rm jet}$ in Figure \ref{jetpower}. For the majority of the sources, the jets exhibit low radiation efficiency and low magnetization, i.e., $P_{\rm r}/P_{\rm jet} < 0.1$ and $P_{B}/P_{\rm jet} < 0.1$.

\section{Discussion}\label{sec_dis}
\begin{figure*}[htbp!]
        \centering
        \includegraphics[angle=0,width=0.9\linewidth]{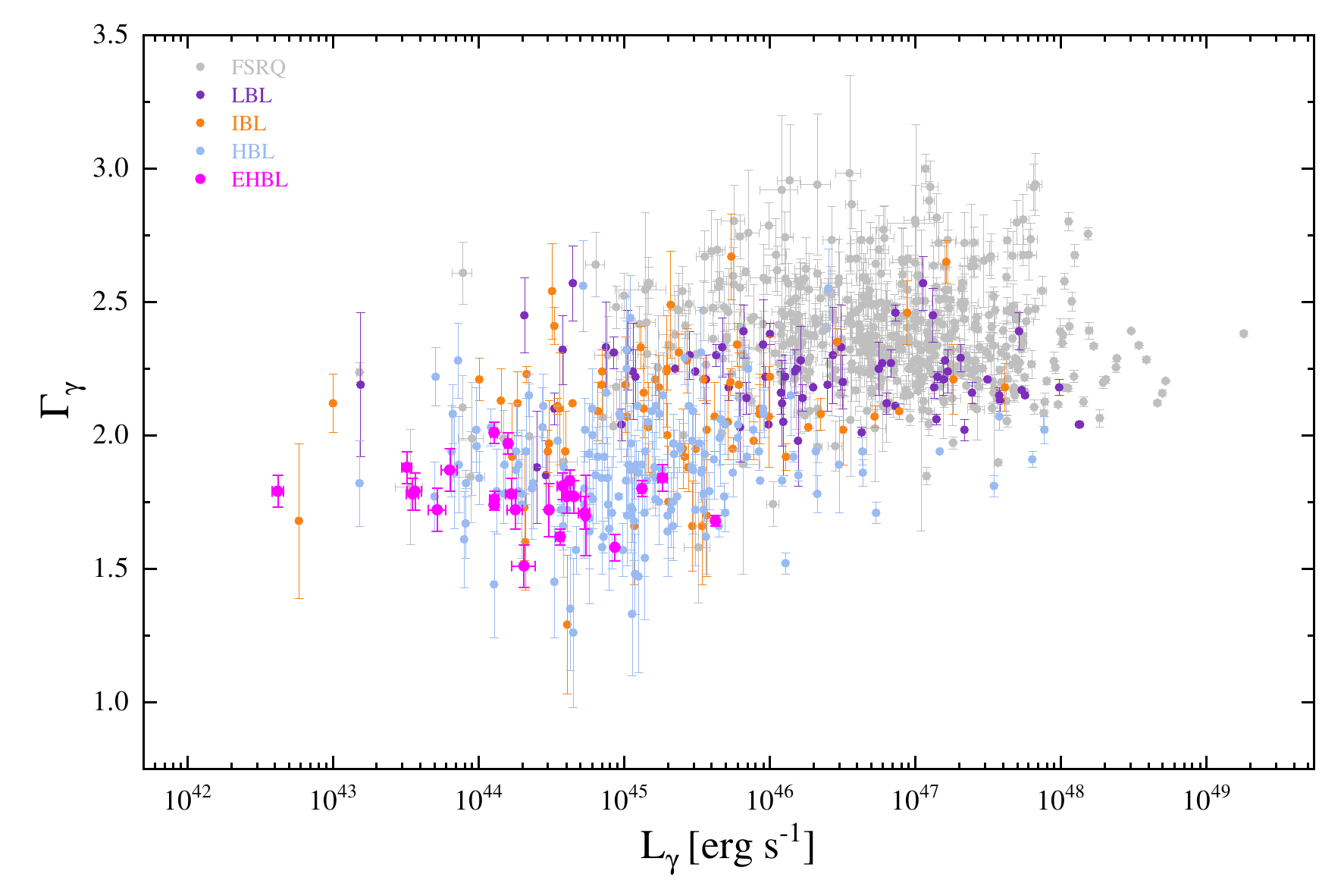}    
    \caption{The photon spectral index ($\Gamma_{\gamma}$) versus the $\gamma$-ray average luminosity ($L_{\gamma}$). Data for FSRQs, LBLs, IBLs, HBLs are taken from the 4FGL-DR3 \citep{2022ApJS..260...53A}.}
     \label{Fig: Lum-index}
\end{figure*}

To investigate the properties of these EHBL with respect to other HBLs, we perform a comparative analysis between the sample of 25 EHBLs and broader blazar populations. Based on Fermi-LAT data analysis results for these 25 EHBLs, we present a plot of $\Gamma_{\gamma}$ versus $L_{\gamma}$ in Figure \ref{Fig: Lum-index}, comparing them with a large sample of blazars from the 4FGL-DR3 catalog \citep{2022ApJS..260...53A}. Here, $L_{\gamma}$ refers to the luminosity in the 0.1--100 GeV energy band for both the 4FGL-DR3 catalog sources and the 25 EHBLs. The results show that these EHBLs are located within the HBL region of the $\Gamma_{\gamma}-L_{\gamma}$ parameter space and exhibit, on average, the hardest $\gamma$-ray spectra among all blazar subclasses. To assess whether the EHBLs differ statistically from typical HBLs in terms of $L_{\gamma}$, $\Gamma_{\gamma}$, and their joint distribution, we apply the Kolmogorov--Smirnov (K--S) test. This K--S test yields a probability value $p_{\rm KS}$; a two-dimensional K--S test result above 0.2 (or above 0.1 for one-dimensional tests) suggests strong evidence for no significant difference between the samples, whereas $p_{\rm KS} < 10^{-4}$ indicates strong evidence of a statistically significant difference. The computed $p_{\rm KS}$ values are approximately $1.3\times10^{-6}$ for the $L_{\gamma}$ distribution, $9.4\times 10^{-4}$ for the $\Gamma_{\gamma}$ distribution, and $4.3\times 10^{-5}$ for the joint $L_{\gamma}-\Gamma_{\gamma}$ distribution. These results indicate that the distributions of $L_{\gamma}$ and $\Gamma_{\gamma}$ exhibit distinct differences between EHBLs and HBLs.  

\begin{figure*}[htbp]
\centering
\includegraphics[width=0.33\linewidth]{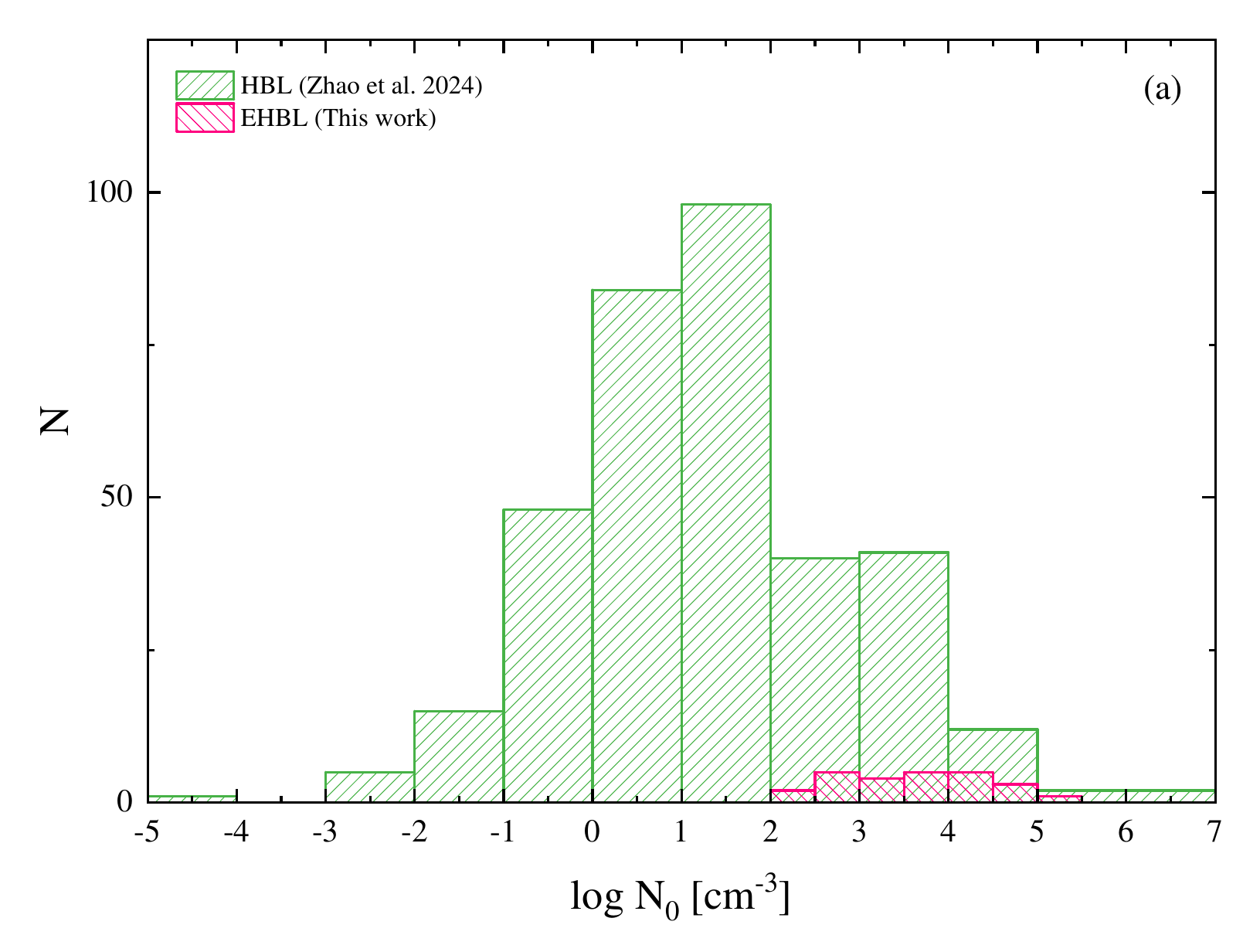}
\includegraphics[width=0.33\linewidth]{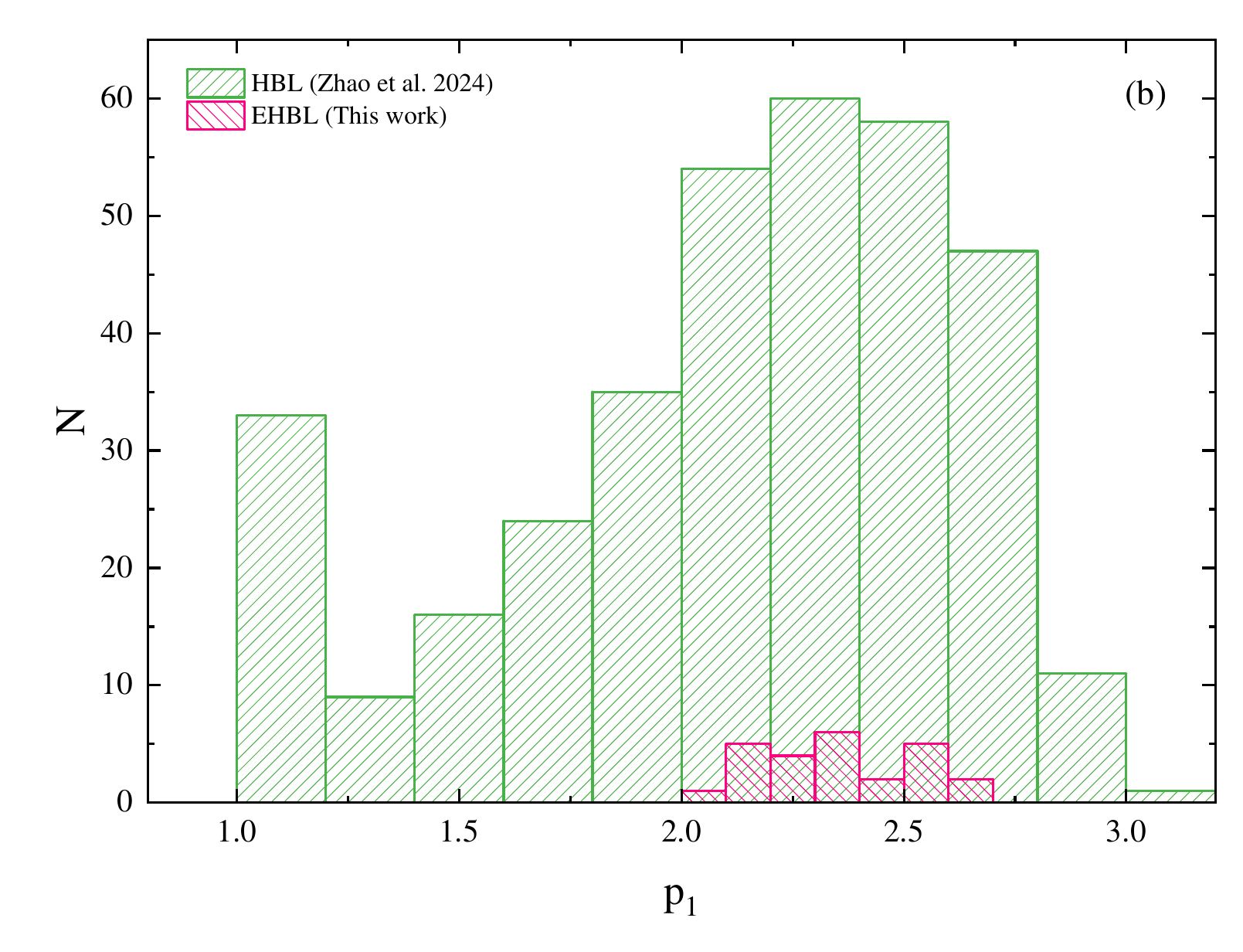}
\includegraphics[width=0.33\linewidth]{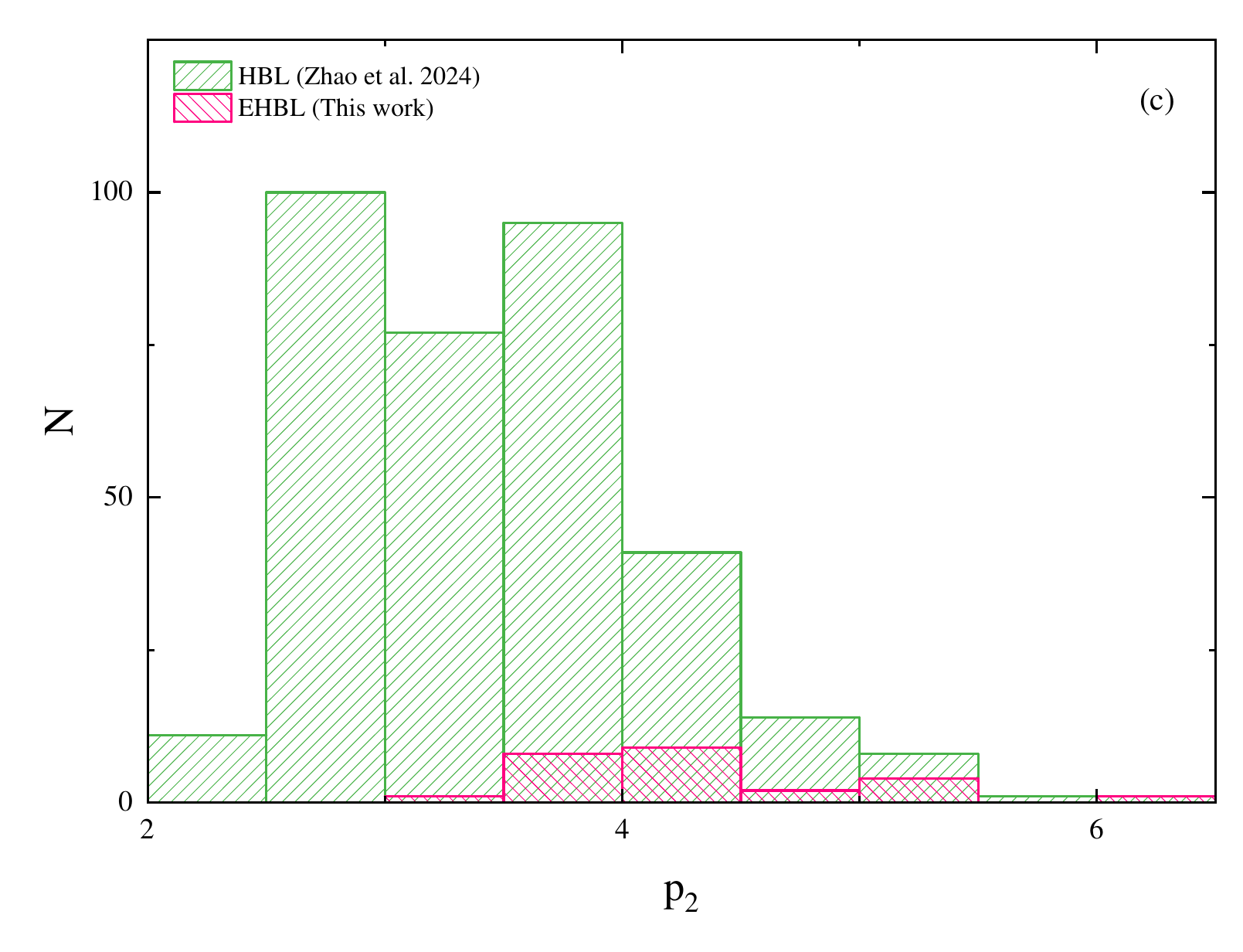} \\
\includegraphics[width=0.33\linewidth]{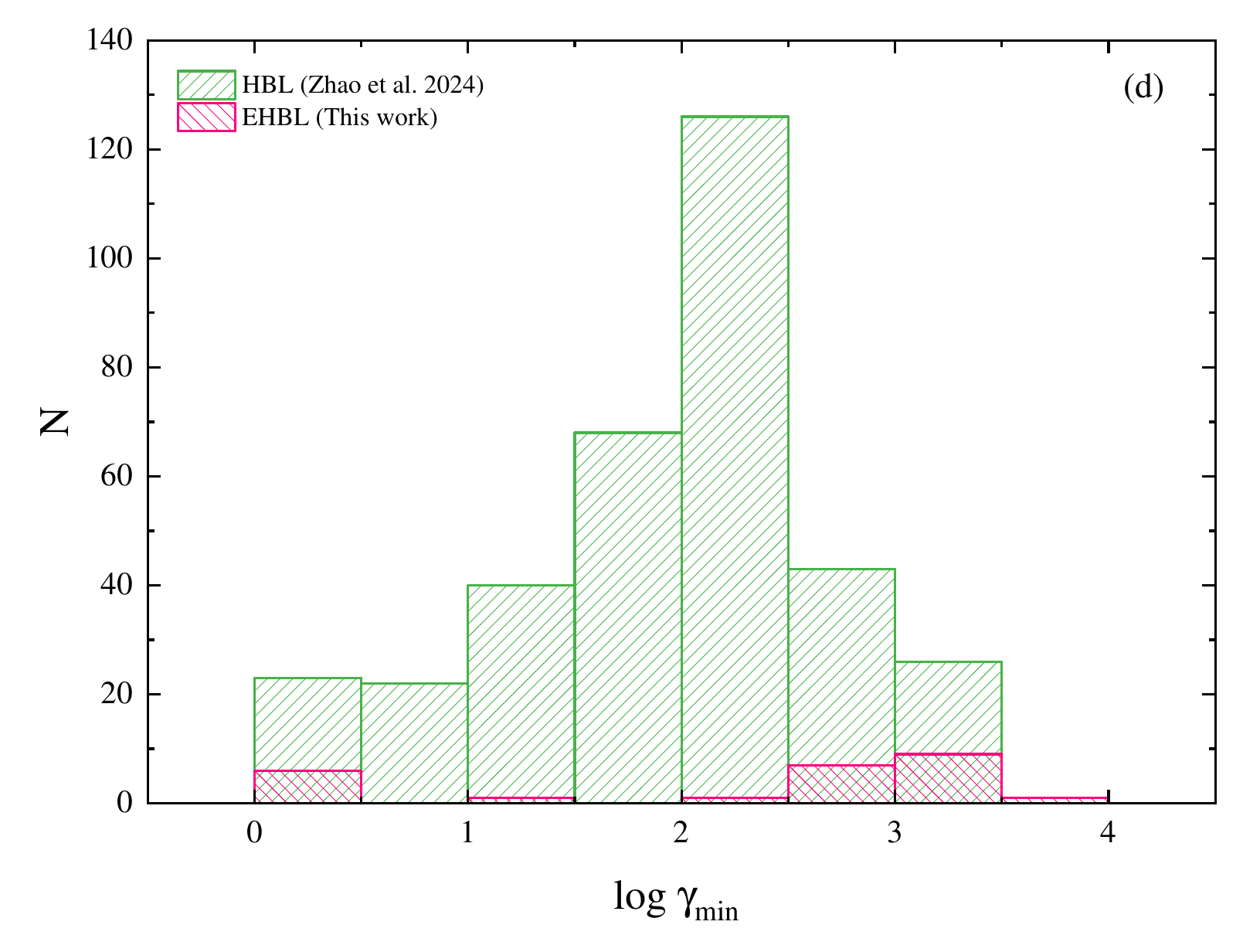} 
\includegraphics[width=0.33\linewidth]{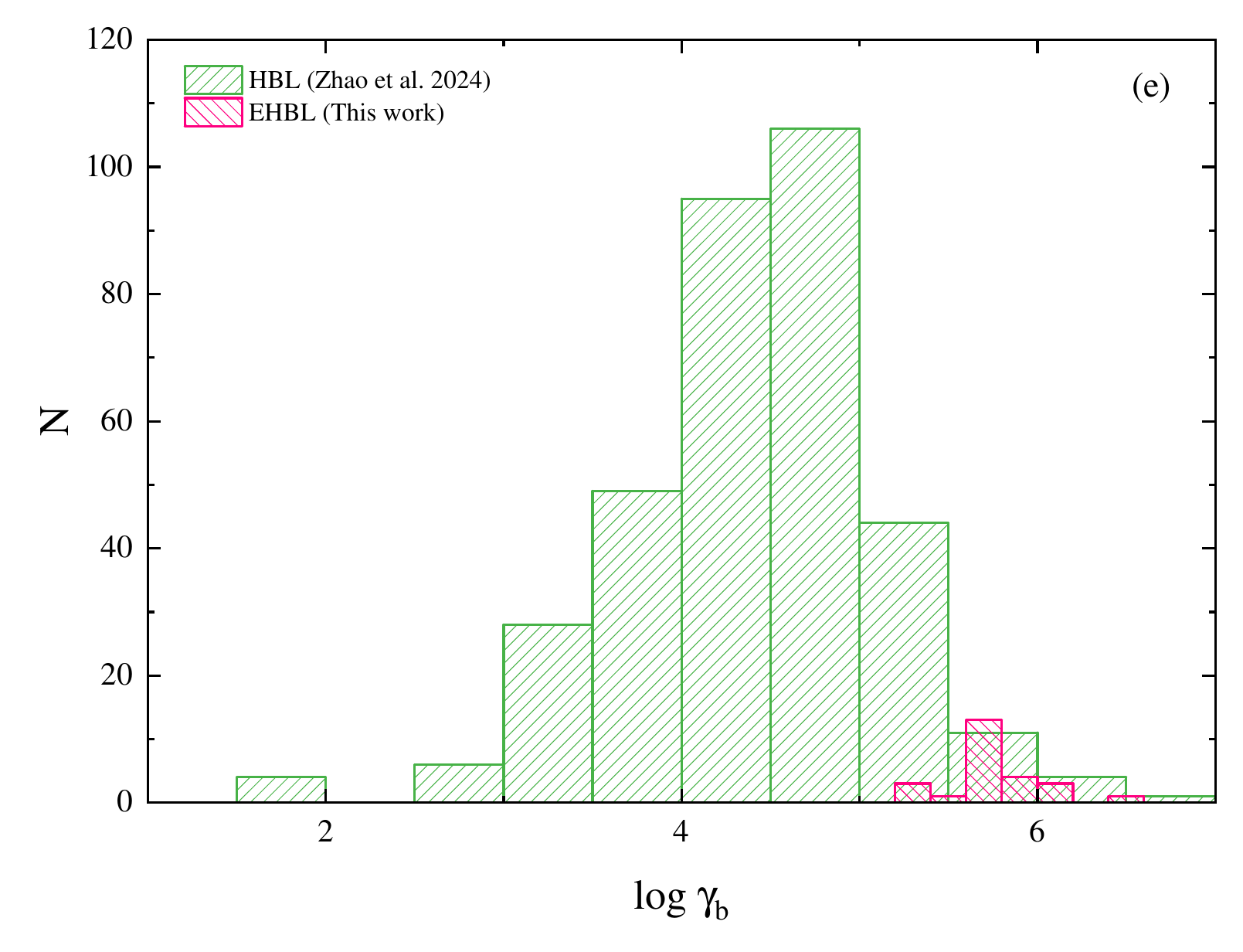}
\includegraphics[width=0.33\linewidth]{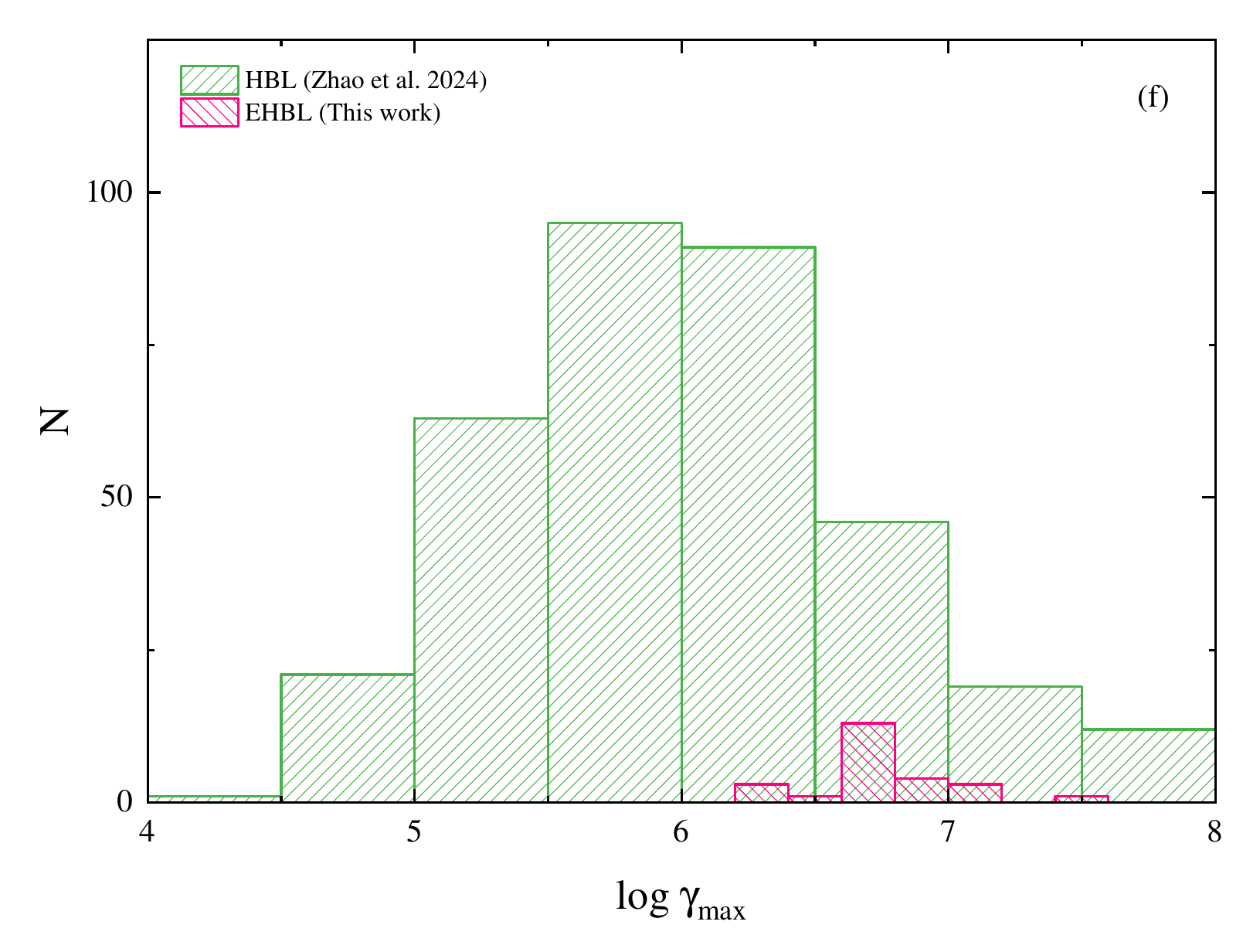}\\
\includegraphics[width=0.33\linewidth]{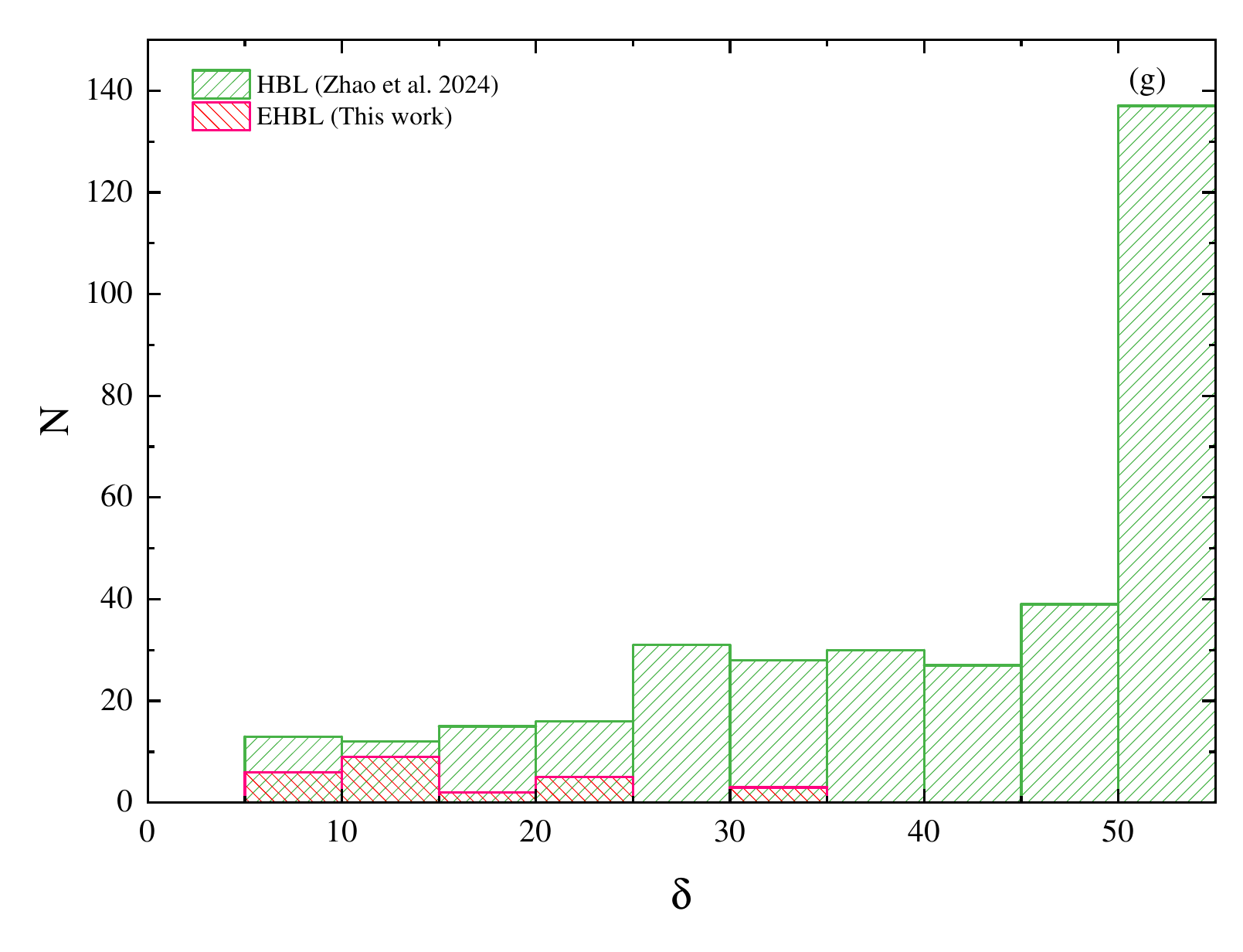}
\includegraphics[width=0.33\linewidth]{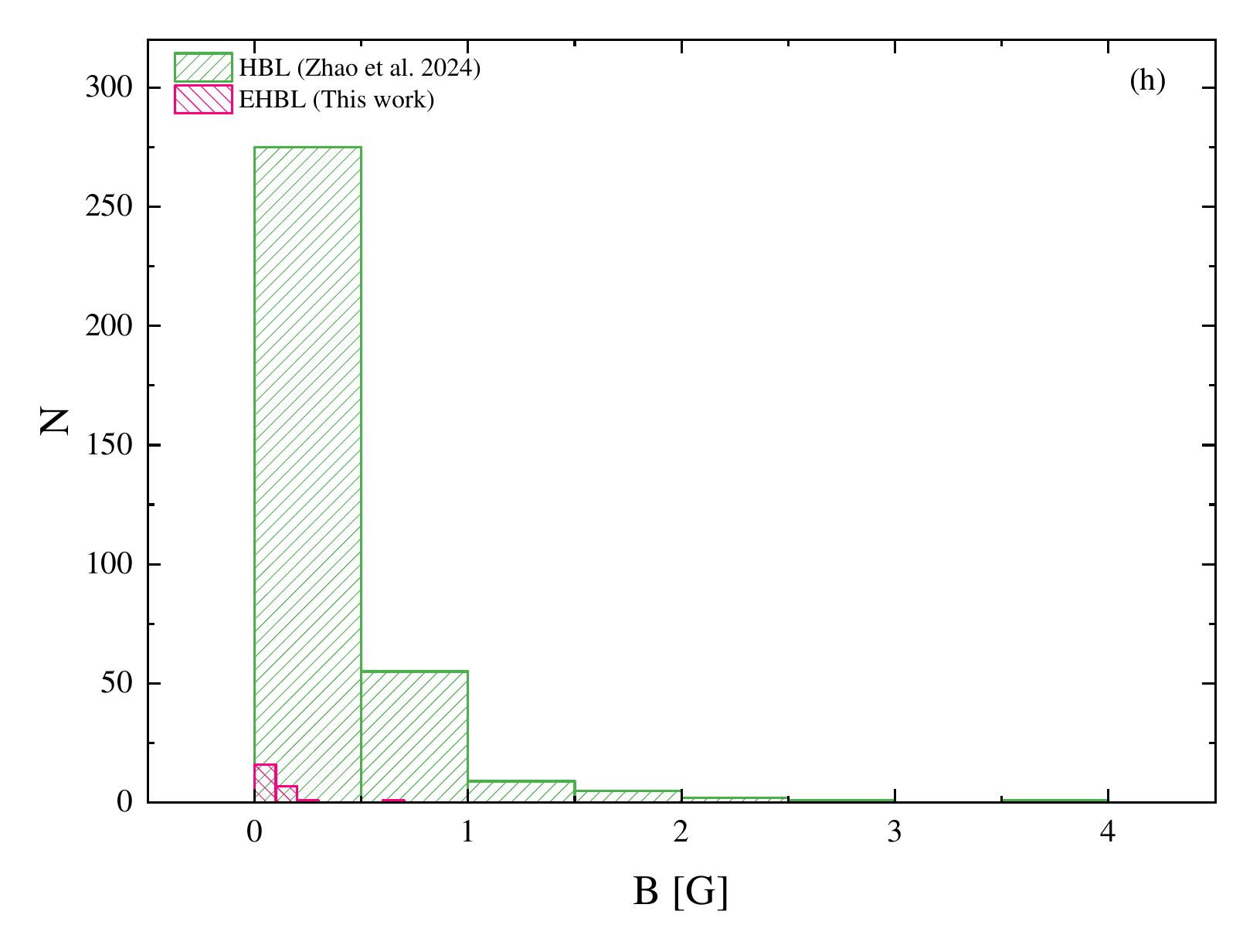}
\includegraphics[width=0.33\linewidth]{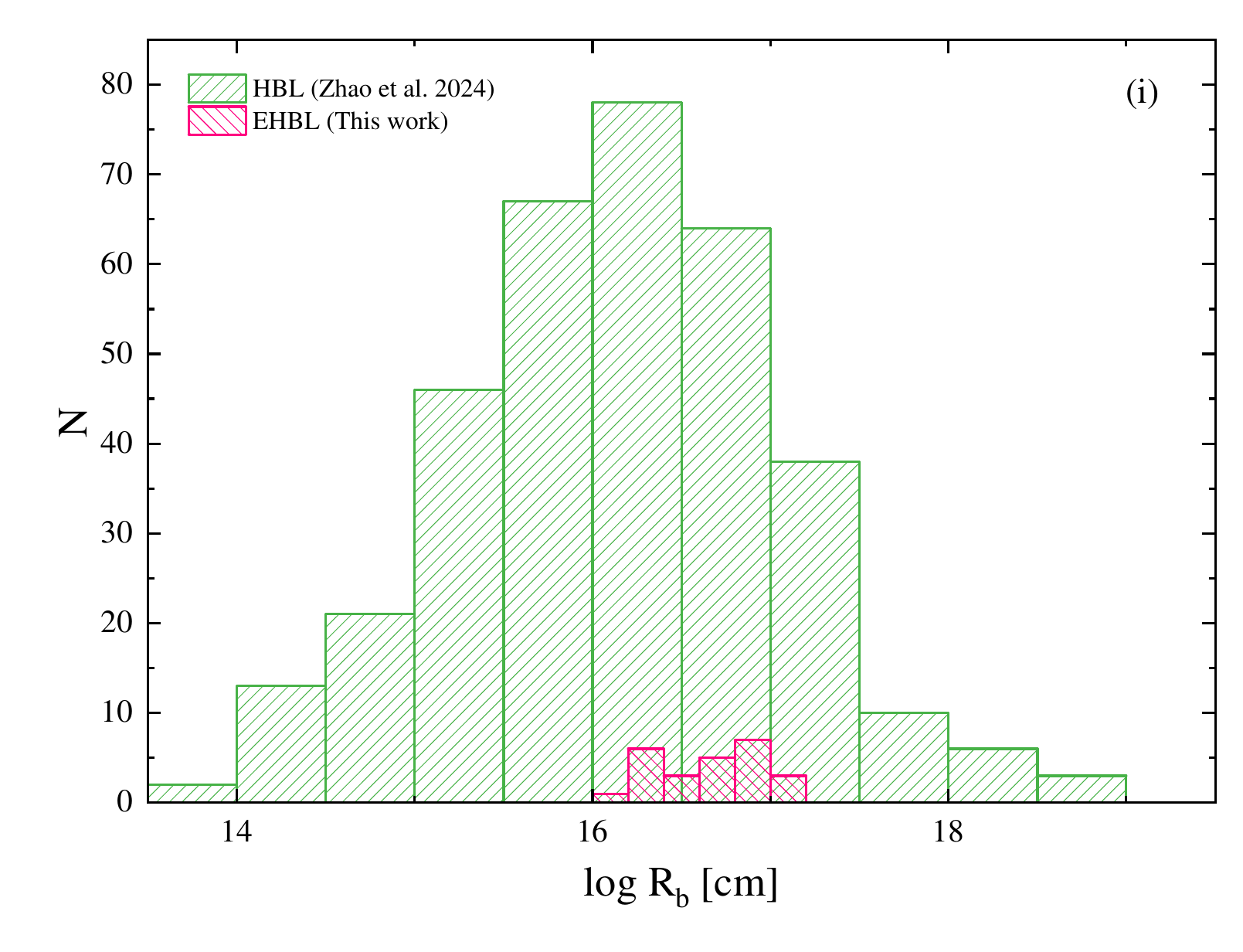}
\caption{The distributions of the SED fitting parameters for the 25 EHBLs are represented by red shaded regions. For comparison, the corresponding parameter data for a HBL sample from \citet{2024ApJ...967..104Z} are also shown, indicated by green shaded regions. 
\label{Fig: distribution}}
\end{figure*}

The distributions of the derived parameters by the SED fitting for these EHBLs are shown in Figure \ref{Fig: distribution}. For comparison, the parameter distributions of a HBL sample from \citet{2024ApJ...967..104Z} are also presented in Figure \ref{Fig: distribution}, where the broadband SEDs of HBLs are also modeled using the one-zone SSC framework. Although the parameter distributions of these EHBLs fall within the overall range observed for HBLs, we observe that, on average, EHBLs exhibit higher values of $N_0$, $\gamma_{\rm b}$, and $\gamma_{\rm max}$, and lower values of $B$ and $\delta$ compared to HBLs. The elevated $\gamma_{\rm b}$ and $\gamma_{\rm max}$ values are necessary to account for the higher synchrotron peak energies characteristic of EHBLs. The median values of $B$ and $\delta$ for these EHBLs are 0.06 G and 12, respectively, while the corresponding medians for HBLs are 0.06 G and 45. We further employ the K--S test to assess the statistical differences between the parameter distributions of EHBLs and HBLs; the resulting $p_{\rm KS}$ values are summarized in Table \ref{Tab: KS test}. The distributions of $N_0$, $\gamma_{\rm b}$, $\gamma_{\rm max}$, and $\delta$ are found to be significantly different between the two samples, while the $B$ distribution shows a marginal deviation. These results indicate that the one-zone SSC model remains applicable for describing the broadband SEDs of most these EHBLs, consistent with previous studies of some EHBLs by \citet{2010MNRAS.401.1570T} and \citet{2012ApJ...752..157Z}. Furthermore, extreme synchrotron sources may represent the high-energy tail of the blazar population (e.g., \citealt{1998MNRAS.301..451G, 2001A&A...371..512C, 2002A&A...386..833G, 2020NatAs...4..124B}).

\begin{figure*}[htbp!]
    \centering
    \includegraphics[angle=0,width=1.0\linewidth]{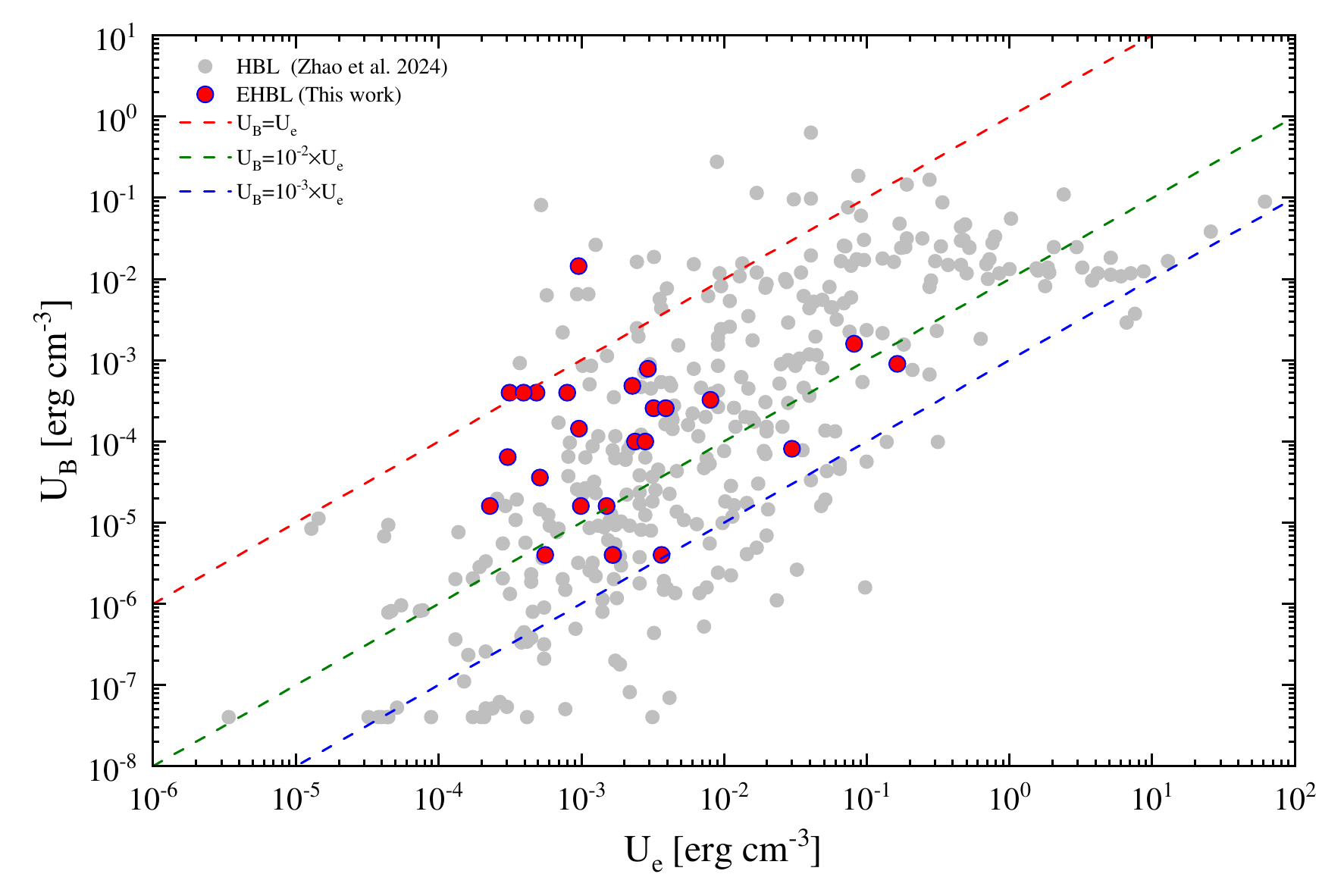}    
    \caption{$U_B$ as a function of $U_{\rm e}$ for the 25 EHBLs (red circles). For comparison, the corresponding parameter data for a HBL sample from \citet{2024ApJ...967..104Z} are also included (gray circles).}
     \label{UB-Ue}
\end{figure*}

Based on the SED fitting parameters, we obtained the values of $U_B$ and $U_{\rm e}$ for the emission regions of these EHBLs and plotted $U_B$ against $U_{\rm e}$ in Figure \ref{UB-Ue}. The corresponding data for HBLs from \citet{2024ApJ...967..104Z} are also included in the same figure. The 25 EHBLs lie within the distribution range of the HBL population. The K--S test comparing the $U_B-U_{\rm e}$ relation between the EHBL and HBL samples yields $p_{\rm KS}\sim3.7\times10^{-4}$, suggesting statistical differences between the two populations. Consistent with the trend observed in HBLs, the majority of EHBLs exhibit $U_B < U_{\rm e}$, indicating a deviation from the equipartition condition. However, all EHBLs have $U_B/U_{\rm e} > 10^{-3}$, and on average, they display higher $U_B/U_{\rm e}$ ratios compared to HBLs. Using the Fermi-LAT observation data and the one-zone SSC model, \citet{2010MNRAS.401.1570T} and \citet{2012ApJ...752..157Z} independently modeled the broadband SEDs of a TeV-emitting BL Lac sample, finding that most sources in their samples also show $U_B < U_{\rm e}$. Furthermore, \citet{2022MNRAS.512..137N} suggested that weakly magnetized jets are likely required for sources exhibiting strong TeV $\gamma$-ray emission, which is consistent with the characteristics of magnetically dominated FSRQ jets and particle-dominated GeV--TeV BL Lac jets \citep{2014ApJ...788..104Z}. Additionally, jets may not always be in equipartition between particles and magnetic fields (e.g., \citealp{2018ApJ...853....6L}).

\begin{figure*}[htbp!]
    \centering
    \includegraphics[angle=0,width=1.0\linewidth]{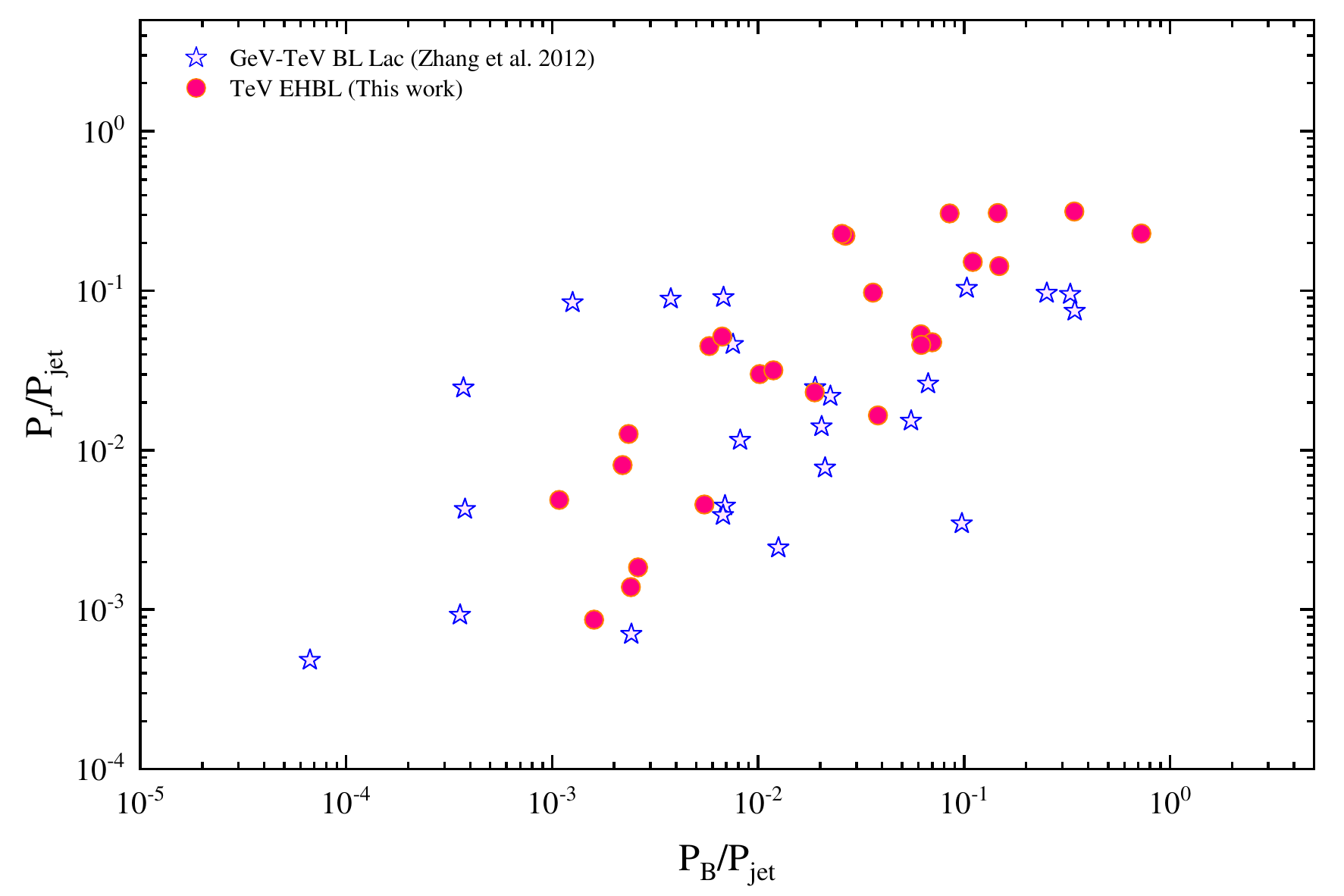}    
    \caption{$P_{\rm r}/P_{\rm jet}$ as a function of $P_B/P_{\rm jet}$ for the 25 EHBLs (red circles). For comparison, the data for a GeV-TeV BL Lac sample from \citet{2012ApJ...752..157Z} are also included (blue stars). }
     \label{Fig: efficiency}
\end{figure*}

We further plotted $P_{\rm r}/P_{\rm jet}$ versus $P_{B}/P_{\rm jet}$ in Figure \ref{Fig: efficiency}, where $P_{\rm r}/P_{\rm jet}$ and $P_{B}/P_{\rm jet}$ represent the radiation efficiency and the magnetization of the jets, respectively. The data for a GeV-TeV emitting BL Lac sample from \citet{2012ApJ...752..157Z} are also shown in Figure \ref{Fig: efficiency}. It should be noted that approximately half of the sources overlap between the two samples; however, different SED data were used. for instance, for Mrk 421 and Mrk 501, data from the high-flux states were selected. On average, these EHBLs exhibit higher values of $P_{\rm r}/P_{\rm jet}$ and $P_{B}/P_{\rm jet}$ compared to the GeV-TeV emitting BL Lacs in \citet{2012ApJ...752..157Z}. Nevertheless, the majority of EHBL jets still display low radiation efficiency and magnetization, with both $P_{\rm r}/P_{\rm jet}$ and $P_{B}/P_{\rm jet}$ below 0.1. The K--S test yields $p_{\rm KS}\sim0.21$ when comparing the $P_{\rm r}/P_{\rm jet}-P_{B}/P_{\rm jet}$ distributions between the two samples, indicating no statistically significant difference.

\section{summary} \label{sec_Summary}

In this work, we systematically analyzed the 16-year Fermi-LAT observation data for a sample of 25 EHBLs. We constructed the broadband SEDs by combining the Fermi-LAT spectra with the TeV data, as well as other multiwavelength observations performed during the TeV campaigns from the literature. For the 19 objects that exhibited no significant variability in the GeV band, we utilized the Fermi-LAT spectra integrated over 16 years, while for the 6 sources with detected significant variability, we employed the Fermi-LAT spectra integrated over the same epoch as the TeV observations. With the caveat of data taken over very different timescales in different energy bands (from years to hours), we performed SED modeling using a one-zone SSC model. Based on the results of the Fermi-LAT data analysis and SED modeling, we conducted a comparative analysis with other $\gamma$-ray emitting HBL samples. The main findings can be summarized as follows.

\begin{itemize}

  \item Most EHBLs exhibit a stable, or even low, flux in the GeV band, with only 6 out of 25 sources showing significant variability at a confidence level exceeding 5$\sigma$.

  \item The majority of the 25 EHBLs show a hard PL spectrum in the 0.1--100 GeV band, with photon spectral indices concentrated between 1.7 and 1.8. 

  \item The one-zone SSC model remains adequate for describing the broadband SEDs of most EHBLs, requiring only relatively higher values of $\gamma_{\rm b}$ and $\gamma_{\max}$ compared to other $\gamma$-ray emitting HBLs.

  \item The derived emission region parameters from SED modeling are consistent with those of other $\gamma$-ray emitting HBLs, but sometimes fail to fully account for the UV flux, and instead indicate a deviation from the equipartition condition, with $U_B/U_{\rm e}<1$.

  \item Similar to other GeV-TeV $\gamma$-ray emitting BL Lacs, the jets in these EHBLs exhibit low radiation efficiency and magnetization.
   
\end{itemize}

\section*{Acknowledgements}
We sincerely appreciate the anonymous referee for the helpful suggestions and valuable comments. This work is supported by the National Key R\&D Program of China (grant 2023YFE0117200) and the National Natural Science Foundation of China (grants 12203024, 12203022, 12022305, 11973050, 12473042, and 12373109).

\begin{deluxetable*}{lcccc cccr}
    \tabletypesize{\footnotesize}
    \tablewidth{5pt}
    \renewcommand{\arraystretch}{1.1} 
    \tablecaption{\label{tab_LAT} Results from Fermi-LAT Observations of 25 EHBLs}
    \tablehead{\rowcolor{gray!25}  \multicolumn{9}{c}{Average Information}\\  \hline  
    \colhead{Source} & \colhead{4FGL Name} & \colhead{z} & \colhead{Energy Flux$^{\clubsuit}$} & \colhead{Spectral Model}& \colhead{$\Gamma_{\gamma}$} & \colhead{$\beta$} &  \colhead{TS}&  \colhead{$\rm TS_{\rm var}$} \\
  ~ &(4FGL)& ~  &(\scriptsize{$\times10^{-12}\ \rm erg\ cm^{-2}\ s^{-1}$})  &   ~    &   ~ & ~ & }
  \startdata  \noalign{\smallskip}       
\scriptsize{SHBL J001355.9-185406}    &  J0013.9-1854  & 0.095  & $2.87\pm0.34 $& PL & $1.87\pm0.08$  & - & 208 &59.4    \\
RGB J0152+017 &  J0152.6+0147    & 0.080   & $8.29\pm0.46 $ & PL      & $2.01\pm0.04$   & - & 949&106.9   \\
TXS 0210+515   &  J0214.3+5145    & 0.049  & $5.85\pm0.47$   & PL  & $1.88\pm0.06$  & - &415 &49.0   \\
1ES 0229+200   & J0232.8+2018    & 0.139    & $3.49\pm0.41$   & PL  & $1.72\pm0.07$& - & 234&31.0   \\
1ES 0347-121   &  J0349.4-1159    & 0.188  & $4.79\pm0.47$   & PL  & $1.77\pm0.06$  & - & 432 &48.1  \\
1ES 0414+009   & J0416.9+0105    & 0.287   & $7.14\pm0.54$    & PL   &$1.84\pm0.05$  & - & 561 &68.4    \\
PKS 0548-322   & J0550.5-3216    & 0.069   & $3.24\pm0.36$  & PL   & $1.79\pm0.07$  & - & 264&43.3 \\
RGB J0710+591  &  J0710.4+5908    & 0.125  & $5.62\pm1.02$ & LP & $1.51\pm0.08$   & $0.060\pm0.026$ & 670 & 66.3  \\
PGC 2402248   &  J0733.4-5152   & 0.065   & $2.63\pm0.35$  & PL  & $1.72\pm0.08$  & - & 183 &45.6     \\
RBS 0723      & J0847.2+1134   & 0.198 & $4.81\pm0.44$   & PL  & $1.79\pm0.06$   & - & 414 & 64.2     \\
MRC 0910-208  &  J0912.9-2102  &  0.198  & $1.20\pm0.64$ & PL  & $1.80\pm0.03$   & - & 1298& 44.6     \\
1ES 1101-232  &  J1103.6-2329  & 0.186  & $5.59\pm0.49$ & PL  & $1.71\pm0.06$   & - & 483 & 38.9     \\
Mrk 421       &  J1104.4+3812  & 0.031    & $402.02\pm4.37$  & \scriptsize{PLSuperExpCutoff4}    & $1.74\pm0.003$   & $0.009\pm0.001$ & 227937 & 3476.4 \\
1ES 1218+304  & J1221.3+3010  & 0.182  & $46.49\pm2.19$ & LP    & $1.68\pm0.02$ & $0.023\pm0.006$ & 8627 & 286.9      \\
1ES 1312-423  & J1315.0-4236  & 0.105  & $6.08\pm0.52$  & PL  & $1.78\pm0.06$    & - & 412&66.8  \\
1ES 1426+428 & J1428.5+4240  & 0.129 & $8.41\pm0.55$  & PL & $1.62\pm0.03$    & - & 1325 &81.3  \\
1ES 1440+122  & J1442.7+1200 & 0.163   & $5.65\pm0.45$ & PL  & $1.78\pm0.05$  & - & 512 & 73.5       \\ 
Mrk 501      &  J1653.8+3945     & 0.034   & $117.30\pm2.92$   & LP    & $1.75\pm0.01$ & $0.021\pm0.003$ & 48673 &959.5       \\
1ES 1727+502  &  J1728.3+5013    & 0.055   & $18.29\pm1.33$      & LP   & $1.76\pm0.03$  & $0.016\pm0.010$ & 4006 & 312.2 \\
1ES 1741+196   &  J1744.0+1935   & 0.084   & $9.37\pm0.50 $  & PL  & $1.97\pm0.04$    & - &897&86.5      \\
\scriptsize{1RXS J195815.6-301119} &  J1958.3-3010  & 0.119  & $11.72\pm0.68$  & PL & $1.83\pm0.04$  & - & 854 & 69.7  \\
1ES 1959+650   & J2000.0+6508    & 0.047     & $106.94\pm2.49$  & LP  &   $1.76\pm0.01$& $0.023\pm0.003$ & 38216 & 2223.4 \\
1ES 2037+521  &  J2039.5+5218  & 0.053 & $5.42\pm0.51$  & PL & $1.78\pm0.06$ & - & 298 & 58.8     \\
1ES 2344+514  & J2347.0+5141   & 0.044    & $28.99\pm1.69$   & LP & $1.74\pm0.02$  & $0.042\pm0.009$ & 5155 & 217.7  \\
H 2356-309   &  J2359.0-3038   & 0.165  & $5.14\pm0.43$ & PL & $1.81\pm0.05$   & - &529 & 60.1   \\  \noalign{\smallskip}  \hline 
\rowcolor{gray!25} \multicolumn{9}{c}{Results from Fermi-LAT Observations during TeV Observations  $^{\blacklozenge}$} \\ \hline
 Source & 4FGL Name & Time & Energy Flux  & Spectral Model &$\Gamma_{\gamma}$ & TS  \\
~ & ~ & (MJD) & ($\times10^{-12}\ \rm erg\ cm^{-2}\ s^{-1}$) & ~ & ~& ~ \\
\noalign{\smallskip}  \hline \noalign{\smallskip} 
Mrk 421 & 4FGL J1104.4+3812 & 55237.0--55250.0 &$773.96\pm124.70$  & PL & $1.58\pm0.05$ & 905.7 \\ 
Mrk 501 & 4FGL J1653.8+3945 & 56854.0--56870.0 & $187.84.40\pm52.09$ & PL & $1.74\pm0.10$ & 160.7 \\ 
1ES 1218+304 & 4FGL J1221.3+3010 & 54801.0--54983.0 & $46.29\pm10.03$ & PL & $1.69\pm0.07$ & 310.1 \\
1ES 1727+502 & 4FGL J1728.3+5013 & 57306.0--57328.0 &$41.67\pm17.65$  &PL & $1.91\pm0.21$ & 40.3 \\
1ES 1959+650 & 4FGL J2000.0+6508 & 57552.0--57554.0 & $605.33\pm253.89$ & PL & $1.70\pm0.15$ & 110.8 \\ 
1ES 2344+514 & 4FGL J2347.0+5141 & 57596.5--57626.5 & $46.88\pm9.52$& PL & $2.02\pm0.19$& 46.4 \\
\noalign{\smallskip}
  \enddata 
  \tablenotetext{\clubsuit}{Within the 0.1--100 GeV band.}
  \tablenotetext{\blacklozenge}{For the six BL Lacs exhibiting significant variability, a reanalysis of Fermi-LAT observational data was conducted. The contemporaneous Fermi-LAT spectra with the TeV observations have been derived for Mrk 421, 1ES 1218+304, 1ES 1727+502, and 1ES 1959+650, which are associated with the observation periods marked as the pink-shaded regions in their light curves in Figure \ref{Fig: LCs}. For Mrk 501, a 16-day period (the green-shaded region in its light curve in Figure \ref{Fig: LCs}) was selected for the reanalysis of the Fermi-LAT observational data, and its TeV spectrum (presented in Figure \ref{Fig: SED fitting}) was obtained from the observation within MJD 56857 (the pink-shaded region in its light curve in Figure \ref{Fig: LCs}). For 1ES 2344+514, a one-month Fermi-LAT observation period (the green-shaded region in its light curve in Figure \ref{Fig: LCs}) centered on the very high energy observation (MJD 57611, the pink-shaded region in its light curve in Figure \ref{Fig: LCs}) was chosen. More details can be found in Section \ref{sec_SED-Construction}.}
 \end{deluxetable*}

\begin{table*}
\setlength{\tabcolsep}{8.5pt}  
\renewcommand{\arraystretch}{1.1} 
\begin{center}
\caption{SED Fitting Results for the 25 EHBLs}
\label{tab:Fit Parameters}
{\footnotesize 
\begin{tabular}{lcccc ccccc cr} 
\hline \hline \noalign{\smallskip}
source & $N_0 [\rm cm^{-3}]$ & $p_1$ & $p_2$ &$\gamma_{\rm min}$&$\gamma_{\rm b}$&$\gamma_{\rm max}$  & $\delta$ &$B$ [G] &$R$ [cm] & $U_B/U_{\rm e}$ & Reference$^{\blacklozenge}$\\ \noalign{\smallskip} \hline \noalign{\smallskip} 
SHBL J001355.9-185406  & 3E3   & 2.3  & 3.8 & 1   & 2.5E5  & 2.5E6  & 12   & 0.09  & 2.8E16  & 4.0E-2 & (1)\\
RGB J0152+017         & 9.5E3 & 2.25 & 5   & 1    & 4E5    & 4E6    & 10   & 0.045 & 2.4E16  & 2.7E-3 & (2--3) \\
TXS 0210+515          & 1E5   & 2.5  & 4   & 1    & 6E5    & 6E6    & 10   & 0.15  & 1E16    & 5.5E-3 & (4)\\
1ES 0229+200          & 7.5E2   & 2.1  & 4 & 1E3  & 1.5E6  & 1.5E7  & 30   & 0.01  & 2.5E16  & 2.4E-3 & (5) \\
1ES 0347-121          & 4E3   & 2.35 & 3.7 & 9E3  & 4E5    & 4E6    & 20   & 0.04  & 6E16    & 2.1E-1 & (6--8) \\ 
1ES 0414+009          & 3E4   & 2.6  & 4.5 & 3E3  & 2E5    & 2E6    & 20   & 0.1   & 6.5E16  & 1.3  & (9) \\
PKS 0548-322          & 2E2   & 2.1  & 4.5 & 2E1  & 6E5    & 6E6    & 7    & 0.1   & 5E16    & 5.0E-1 & (10) \\
RGB J0710+591         & 8E2   & 2.2  & 3.5 & 2E3  & 5.7E5  & 5.7E6  & 30   & 0.03  & 3.1E16  & 6.9E-2 & (11)\\
PGC 2402248           & 5E3   & 2.3  & 4.1 & 1E2  & 6E5    & 6E6    & 13   & 0.08  & 1.6E16  & 7.9E-2 & (12--13)\\ 
RBS 0723              & 1E4   & 2.3  & 3.5 & 2E3  & 6E5    & 6E6    & 20   & 0.05  & 2.2E16  & 4.2E-2 & (4)\\ 
MRC 0910-208          & 7E4   & 2.6  & 6   & 2E3  & 4.5E5  & 4.5E6  & 10   & 0.06  & 8.5E16  & 1.5E-1 & (14)\\ 
1ES 1101-232          & 5E2   & 2.1  & 5   & 7E2  & 2E5    & 2E6    & 12   & 0.6   & 2.1E16  & 1.5E-1 &  (15) \\ 
Mrk 421               & 1E2   & 2    & 4   & 1.5E3& 8.5E5  & 8.5E6  & 22   & 0.01  & 1E17    & 7.1E-3  & (16--19)\\
1ES 1218+304          & 3E4   & 2.5  & 4.2 & 1E3  &   7E5  & 7E6    & 20   & 0.02  & 7.8E16  & 1.1E-2 & (20--26)\\ 
1ES 1312-423          & 2E2   & 2.1  & 4   & 4E2  & 5.5E5  & 5.5E6  & 5    & 0.1   & 9.8E16  & 5.2E-1  &(27)\\
1ES 1426+428          & 5E2   & 2.2  & 3.2 & 9E2  &   4E5  & 4E6    & 12   & 0.1   & 6E16    & 1.0  & (4)\\ 
1ES 1440+122          & 4.8E2 & 2.15 & 4   & 1    & 5E5    & 5E6    & 5    & 0.11   & 1E17   & 2.1E-1 & (28)\\
Mrk 501               & 1E3   & 2.2  & 5   & 1    & 3E6    & 3E7    & 6    & 0.08  & 9E16    & 6.5E-2  & (29)\\
1ES 1727+502          & 7E3   & 2.4  & 5   & 7E2  & 1.2E6  & 1.2E7  & 12   & 0.02  & 8.5E16  & 1.6E-2  & (30)\\
1ES 1741+196          & 5E4   & 2.5  & 4   & 1    & 4E5    & 4E6    & 6    & 0.2   & 3.2E16  & 1.9E-2 & (31--32)\\
1RXS J195815.6-301119 & 4.5E4 & 2.5  & 3.5 & 6E2  &   4E5  & 4E6    & 6    & 0.14  & 5E16    & 2.7E-1  & (14)\\
1ES 1959+650          & 6.8E3 & 2.3  & 3.5 & 4E2  & 7.8E5  & 7.8E6  & 11   & 0.05  & 6.5E16  & 3.5E-2  & (30)\\
1ES 2037+521          & 2.3E4 & 2.5  & 4   & 5E2  & 9E5    & 9E6    & 30   & 0.01  & 2.3E16  & 2.4E-3  & (33) \\ 
1ES 2344+514          & 1.8E4 & 2.4  & 3.5 & 3E2  & 1E6    & 1E7    & 18   & 0.01  & 4E16  & 1.1E-3  & (34)\\
H 2356-309            & 1E3   & 2.3  & 3.5 & 2E3  & 3.2E5  & 3.2E6  & 18   & 0.02  & 1E17  & 6.9E-2  & (35--36)\\ \noalign{\smallskip} 
\hline
\end{tabular}
}   
\end{center}
\tablenotetext{\blacklozenge}{The references for the broadband SED data in Figure \ref{Fig: SED fitting}: (1) \citet{2013A&A...554A..72H}; (2) \citet{2008A&A...481L.103A}; (3) \citet{2003A&A...400...95N}; (4) \citet{2020ApJS..247...16A}; (5) \citet{2014ApJ...782...13A}; (6) \citet{2014ApJ...787..155T}; (7) \citet{2007A&A...473L..25A}; (8) \citet{2013ApJS..207...19B}; (9) \citet{2012A&A...538A.103H}; (10) \citet{2010A&A...521A..69A}; (11) \citet{2010ApJ...715L..49A}; (12) \citet{2023MNRAS.519..854M}; (13) \citet{2019MNRAS.490.2284M}; (14) \citet{2022icrc.confE.823B}; (15) \citet{2007A&A...470..475A}; (16) \citet{2021MNRAS.505.2712D}; (17) \citet{2019MNRAS.487..845B}; (18) \citet{2012A&A...541A.140S}; (19) \citet{2012AIPC.1505..514F}; (20) \citet{2010ApJ...709L.163A}; (21) \citet{2010MNRAS.401..973R}; (22) \citet{2009ApJ...695.1370A}; (23) \citet{2006ApJ...642L.119A}; (24) \citet{2007A&A...467..501T}; (25) \citet{2005A&A...433.1163D}; (26) \citet{2005NewA...11...27C}; (27) \citet{2013MNRAS.434.1889H}; (28) \citet{2016MNRAS.461..202A}; (29) \citet{2020A&A...637A..86M}; (30) \citet{2020A&A...640A.132M}; (31) \citet{2016MNRAS.459.2550A}; (32) \citet{2017MNRAS.468.1534A}; (33) \citet{2019MmSAI..90..164P}; (34) \citet{2020MNRAS.496.3912M}; (35) \citet{2010A&A...516A..56H}; (36) \citet{2001A&A...371..512C}.}
\end{table*}

\begin{deluxetable*}{ccc}
    \renewcommand{\arraystretch}{1.1} 
    \setlength{\tabcolsep}{10pt}  
    \label{Tab: KS test}
    \tablecaption{Results of the K--S Test}
    \tablehead{\colhead{Figure}&\colhead{Parameters}&
    \colhead{ $p_{\rm KS}$} }
  \startdata        
\multirow{3}{*}{Fig.\ref{Fig: Lum-index}}&$L_{\gamma}$  & $1.3\times 10^{-6}$  \\
&$\Gamma_{\gamma}$  & $9.4\times 10^{-4}$  \\
&$L_{\gamma}-\Gamma_{\gamma}$ & $4.3\times 10^{-5}$  \\ \noalign{\smallskip}  \hline \noalign{\smallskip}  
\multirow{9}{*}{Fig.\ref{Fig: distribution}} 
&$N_0$   & $8.8\times 10^{-12}$   \\
&$p_1$   & $1.3\times 10^{-3}$    \\
&$p_2$   & $8.5\times 10^{-6}$   \\
&$\gamma_{\rm min}$   & $2.3\times 10^{-4}$   \\
&$\gamma_{\rm b}$   & $3.7\times 10^{-19}$  \\
&$\gamma_{\rm max}$   & $3.0\times 10^{-11}$   \\
&$B$      & $2.8\times 10^{-2}$    \\
&$\delta$ & $1.6\times 10^{-12}$   \\
&$R$ & $1.3\times 10^{-5}$   \\ \noalign{\smallskip}  \hline \noalign{\smallskip} 
Fig.\ref{UB-Ue} &$U_{B}-U_{\rm e}$ & $3.7\times 10^{-4}$  \\ \noalign{\smallskip}  \hline \noalign{\smallskip} 
Fig.\ref{Fig: efficiency}& $P_{\rm r}/P_{\rm jet}-P_{B}/P_{\rm jet}$ & $2.1\times 10^{-1}$ \\  \noalign{\smallskip}
 \enddata

\end{deluxetable*}

\begin{deluxetable*}{lccccccccccr}
    \tabletypesize{\footnotesize}
    \setlength{\tabcolsep}{10pt}  
    \renewcommand{\arraystretch}{1.1} 
    \tablecaption{\label{tab:Jet properties} Derived Parameters from SED Fitting for the 25 EHBLs}
    \tablehead{
    \colhead{Object} & \colhead{$P_{\rm jet}$} & \colhead{$P_{\rm r}$} & \colhead{$P_{\rm e}$} & \colhead{$P_B$} & \colhead{$\nu_{\rm s,p}$}& \colhead{$L_{\rm s,p}$} &\colhead{$\nu_{\rm c,p}$}&\colhead{$L_{\rm c,p}$} \\
  ~ &[$\rm erg\ s^{-1}$] & [$\rm erg\ s^{-1}$]  & [$\rm erg\ s^{-1}$]  &   [$\rm erg\ s^{-1}$]    &   [$\rm Hz$]    &[$\rm erg\ s^{-1}$]   &[$\rm Hz$]   &[$\rm erg\ s^{-1}$]      }
  \startdata  \noalign{\smallskip} 
SHBL J001355.9-185406     & 8.98E43    &  1.49E42   &   8.49E43   & 3.43E42    &     2.00E17  & 1.10E44      & 4.36E25    & 1.74E43   \\
RGB J0152+017            & 1.64E44    &  1.68E42   &   1.62E44   & 4.37E41    &     1.48E17  & 6.09E43      & 5.88E25    & 4.78E43   \\
TXS 0210+515             & 1.54E44    &  7.02E41   &   1.53E44   & 8.43E41    &     1.20E18  & 2.94E43      & 3.24E25    & 6.10E42   \\
1ES 0229+200             & 8.96E43    &  1.13E42   &   8.83E43   & 2.11E41    &     2.17E18  & 5.35E44      & 1.16E27    & 1.65E44   \\
1ES 0347-121             & 5.81E43    &  8.27E42   &   4.12E43   & 8.63E42    &     3.63E17  & 1.65E45      & 1.53E26    & 4.28E44   \\
1ES 0414+009             & 1.29E44    &  1.54E43   &   4.99E43   & 6.33E43    &     1.04E17  & 2.90E45      & 1.78E25    & 6.18E44   \\
PKS 0548-322             & 1.80E43    &  4.18E42   &   9.19E42   & 4.59E42    &     6.60E17  & 1.39E44      & 1.07E26    & 1.24E43   \\
RGB J0710+591            & 4.73E43    &  2.52E42   &   4.18E43   & 2.92E42    &     1.20E18  & 1.22E45      & 3.73E26    & 1.66E44   \\
PGC 2402248              & 1.48E43    &  7.01E41   &   1.30E43   & 1.04E42    &     1.13E18  & 6.58E43      & 1.07E26    & 7.83E42   \\
RBS 0723                 & 5.00E43    &  4.86E42   &   4.33E43   & 1.81E42    &     1.52E18  & 8.34E44      & 2.06E26    & 3.03E44   \\
MRC 0910-208             & 8.85E43    &  1.34E43   &   6.54E43   & 9.75E42    &     1.40E17  & 4.47E44      & 9.82E24    & 3.01E44   \\
1ES 1101-232             & 1.18E44    &  2.69E43   &   5.72E42   & 8.57E43    &     6.21E17  & 2.82E45      & 4.36E25    & 1.90E44   \\
Mrk 421                  & 2.70E44    &  1.39E43   &   2.55E44   & 1.81E42    &     6.21E17  & 3.05E45      & 6.38E26    & 1.84E45   \\
1ES 1218+304             & 3.58E44    &  1.07E43   &   3.44E44   & 3.65E42    &     3.63E17  & 1.26E45      & 6.24E25    & 9.73E44   \\
1ES 1312-423             & 3.82E43    &  1.83E43   &   1.09E43   & 9.00E42    &     4.90E17  & 2.69E44      & 6.24E25    & 4.82E43   \\
1ES 1426+428             & 5.64E43    &  1.77E43   &   1.92E43   & 1.94E43    &     1.20E18  & 1.32E45      & 1.44E26    & 1.48E44   \\
1ES 1440+122             & 9.49E44    &  3.00E43   &   5.36E43   & 1.13E43    &     4.61E17  & 3.96E44      & 4.36E25    & 1.02E44   \\
Mrk 501                  & 1.74E44    &  6.07E43   &   1.07E44   & 6.99E42    &     9.65E18  & 1.30E45      & 2.77E26    & 2.40E44   \\
1ES 1727+502             & 1.03E44    &  4.03E42   &   9.73E43   & 1.56E42    &     6.60E17  & 2.39E44      & 1.07E26    & 1.06E44   \\
1ES 1741+196             & 2.94E44    &  6.79E42   &   2.82E44   & 5.53E42    &     4.61E17  & 9.13E43      & 1.32E25    & 3.09E43   \\
1RXS J195815.6-301119    & 4.53E43    &  1.39E43   &   2.47E43   & 6.61E42    &     4.61E17  & 1.64E44      & 1.78E25    & 8.42E43   \\
1ES 1959+650             & 1.80E44    &  3.98E43   &   1.35E44   & 4.79E42    &     1.52E18  & 1.61E45      & 1.44E26    & 1.09E45   \\
1ES 2037+521             & 8.75E43    &  1.21E41   &   8.72E43   & 2.11E41    &     6.21E17  & 4.65E43      & 1.44E26    & 9.90E42   \\
1ES 2344+514             & 1.79E44    &  8.73E41   &   1.78E44   & 1.94E41    &     6.60E17  & 8.63E43      & 2.06E26    & 5.81E43   \\
H 2356-309               & 7.84E43    &  3.58E42   &   7.00E43   & 4.86E42    &     1.40E17  & 5.31E44      & 1.07E26    & 1.51E44   \\ \noalign{\smallskip} 
\enddata
\end{deluxetable*}
\clearpage

\figsetstart
\figsetnum{1}
\figsettitle{The $\sim$16-year Fermi-LAT average spectra and the long-term light curves in the 0.1--1000 GeV band for the 25 EHBLs.}

\begin{figure*}
       \centering
        \includegraphics[angle=0,width=0.33\linewidth]{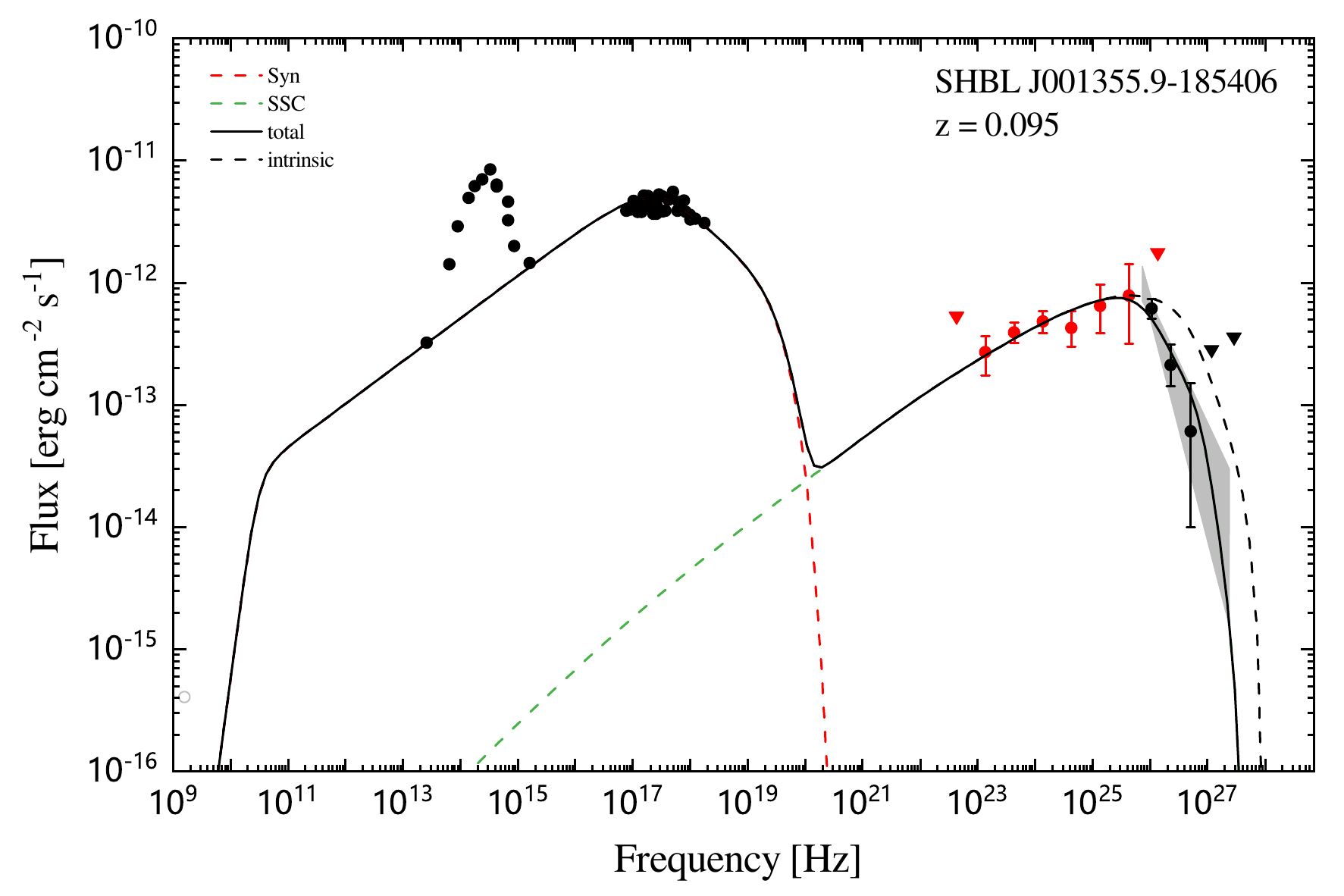}
        \includegraphics[angle=0,width=0.33\linewidth]{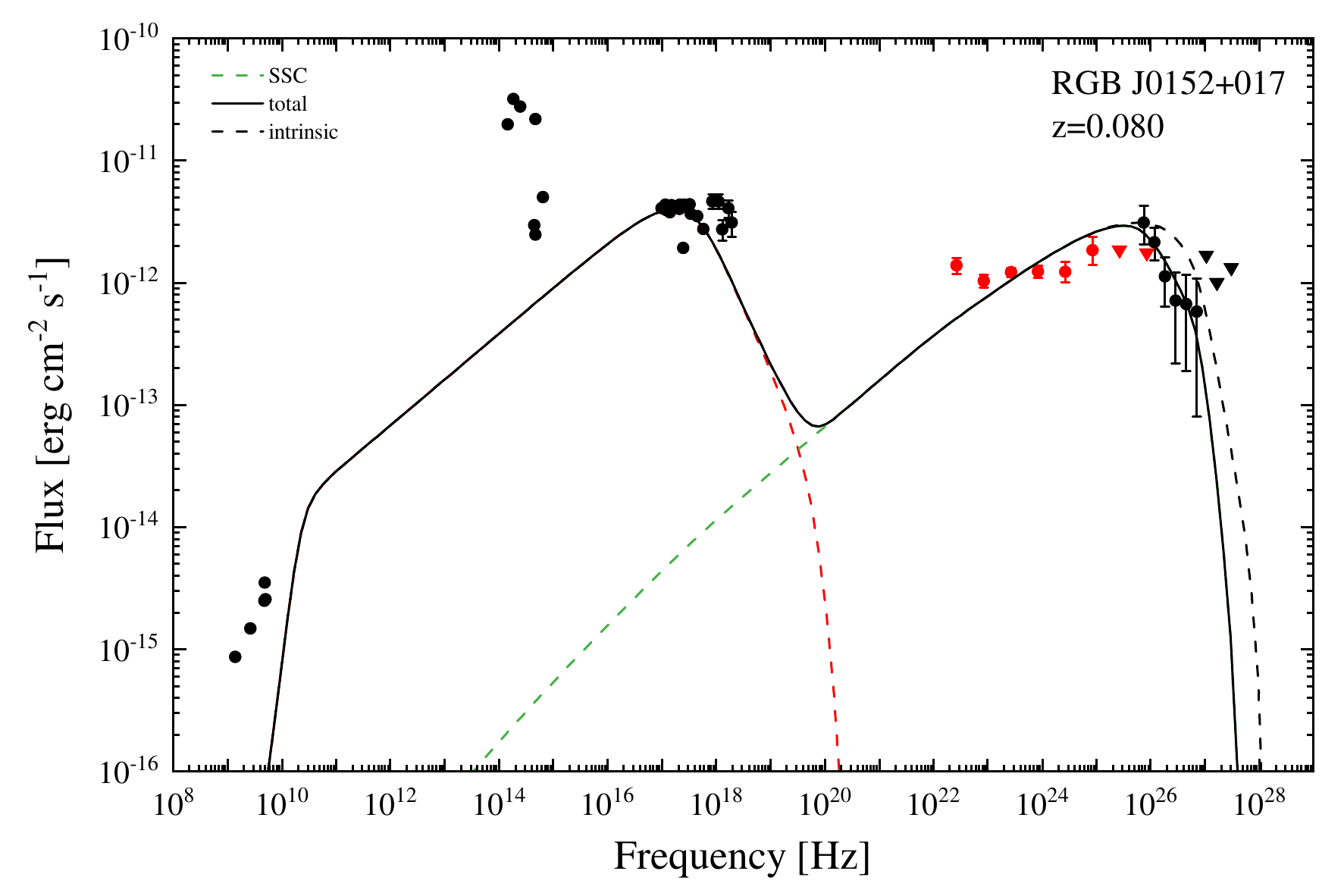}
       \includegraphics[angle=0,width=0.33\linewidth]{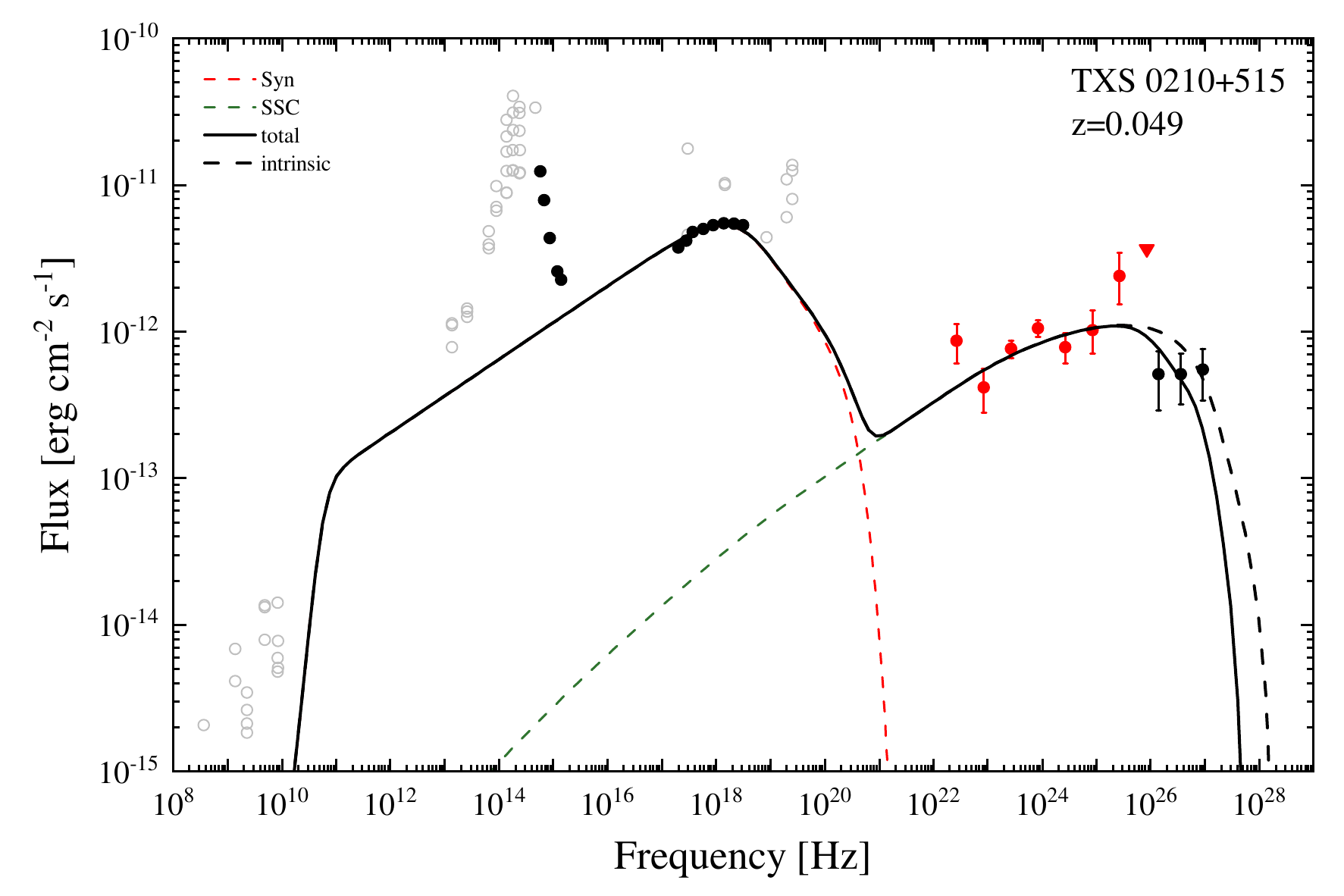} \\  
       \includegraphics[angle=0,width=0.33\linewidth]{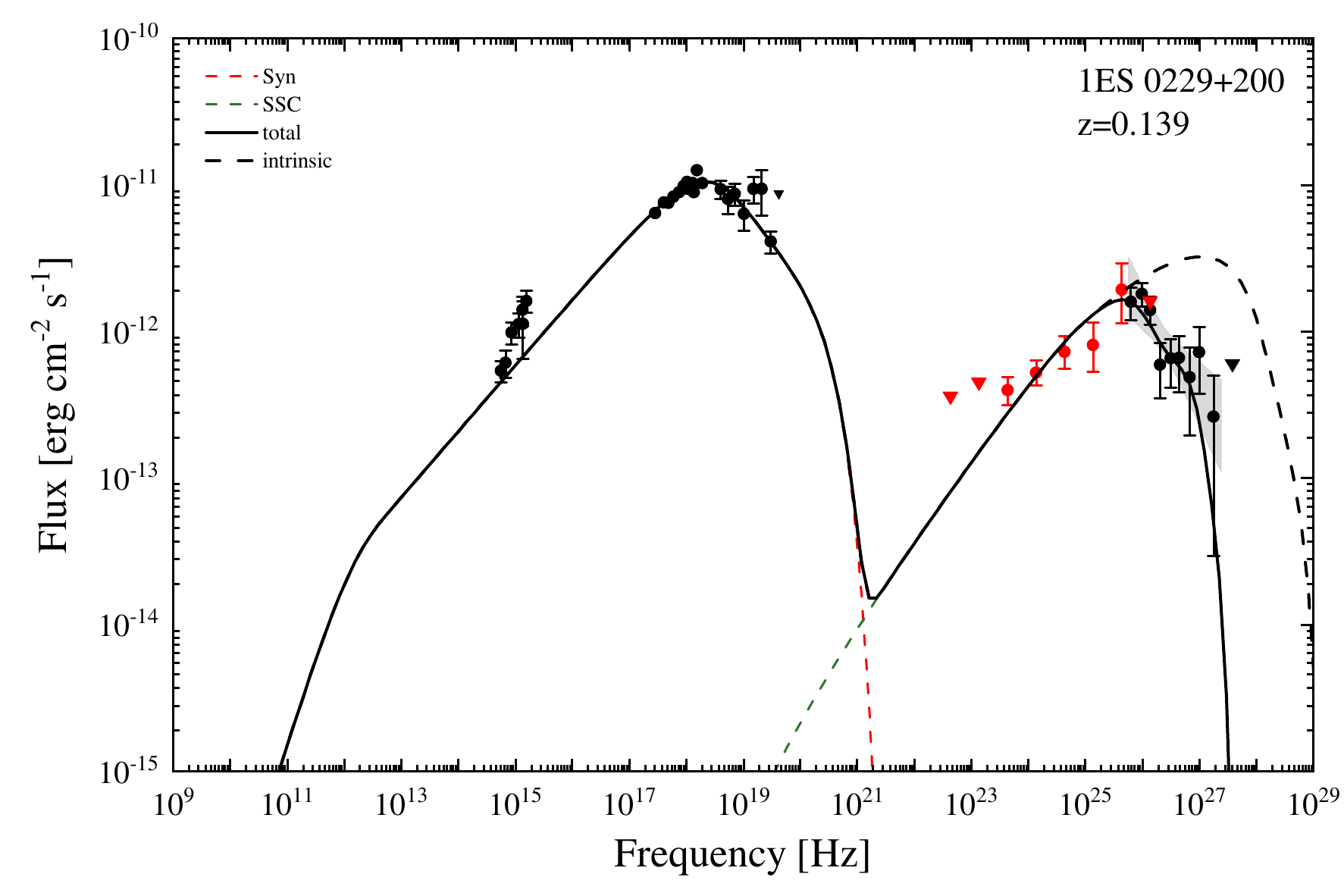}
        \includegraphics[angle=0,width=0.33\linewidth]{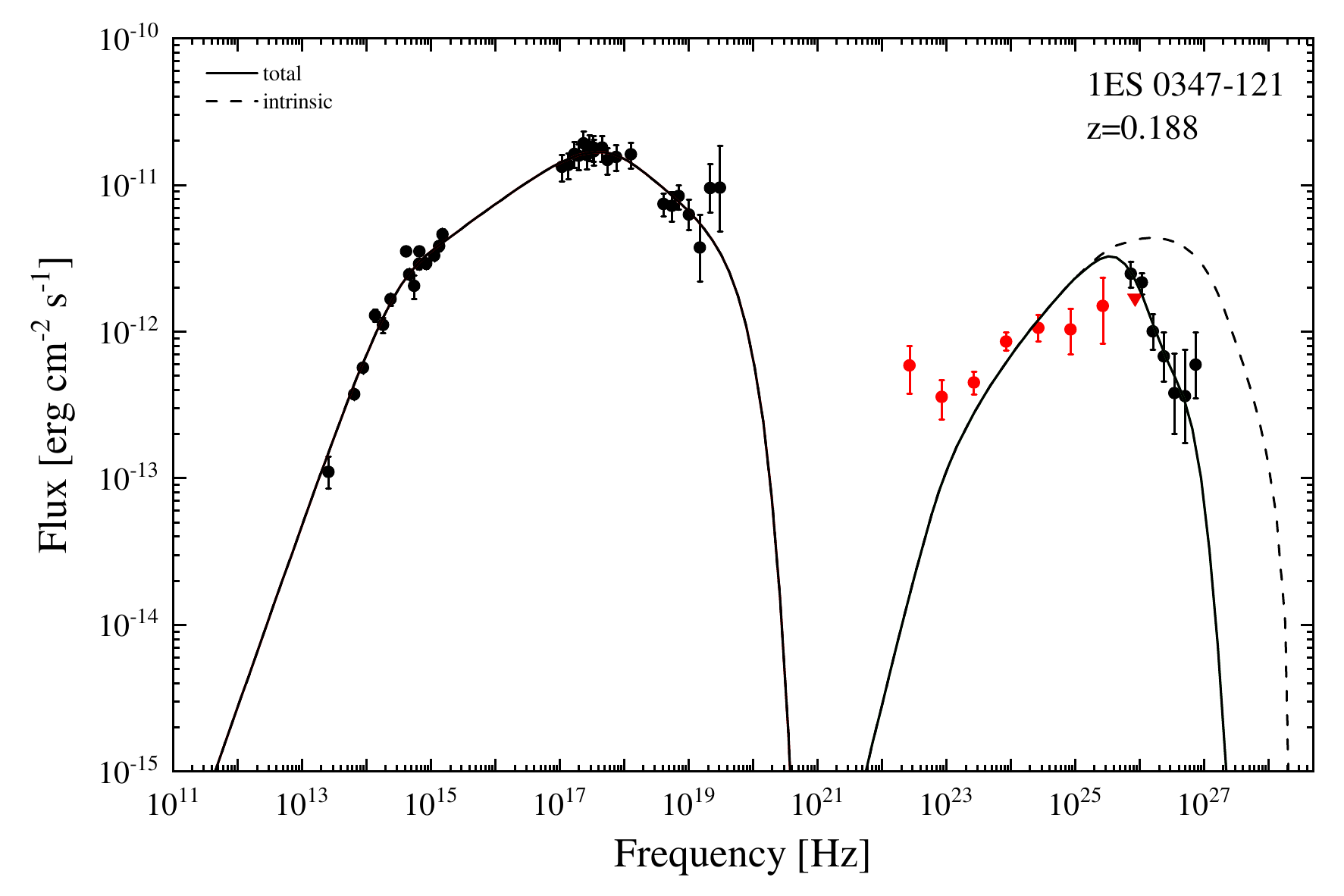}
       \includegraphics[angle=0,width=0.33\linewidth]{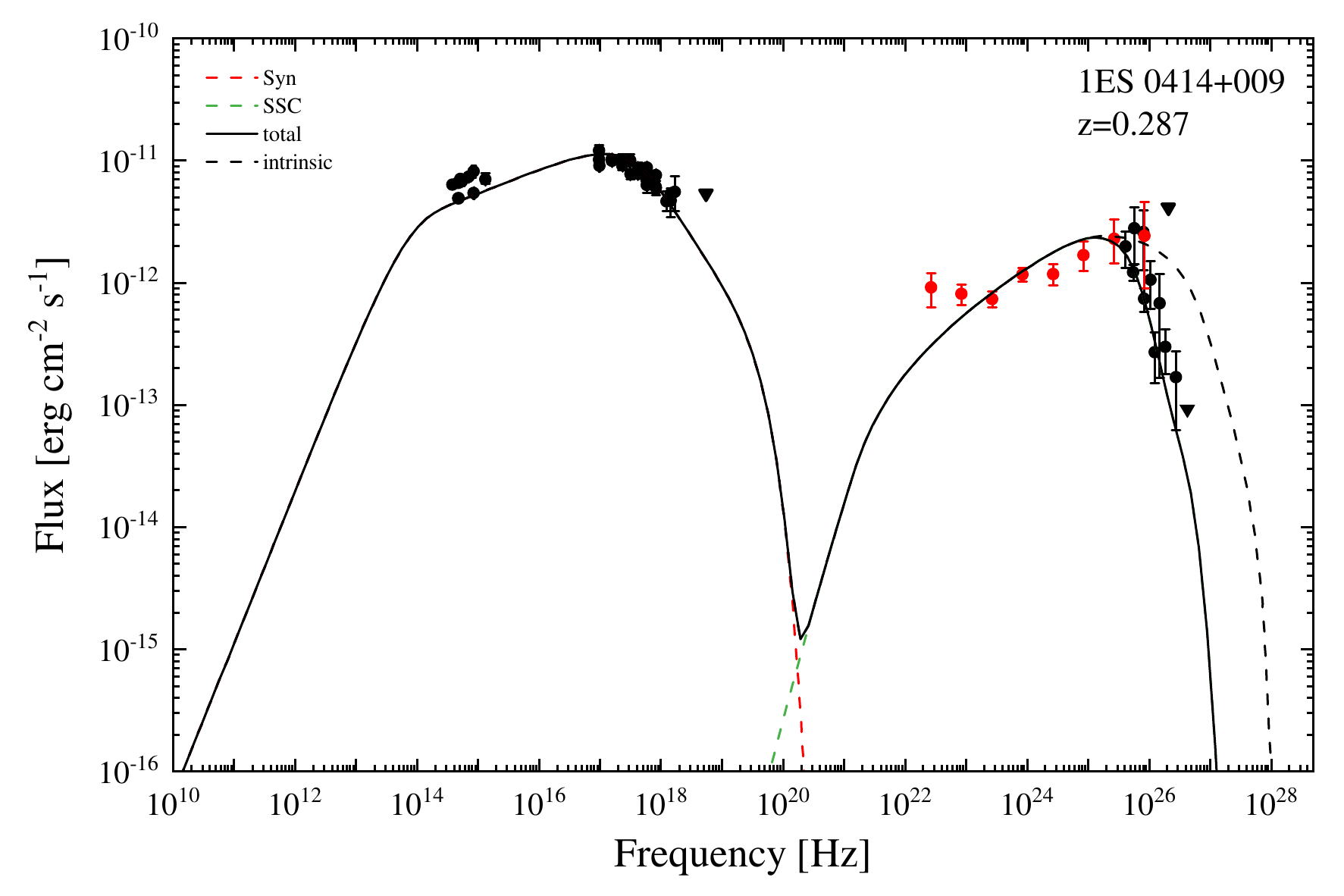}\\
       \includegraphics[angle=0,width=0.33\linewidth]{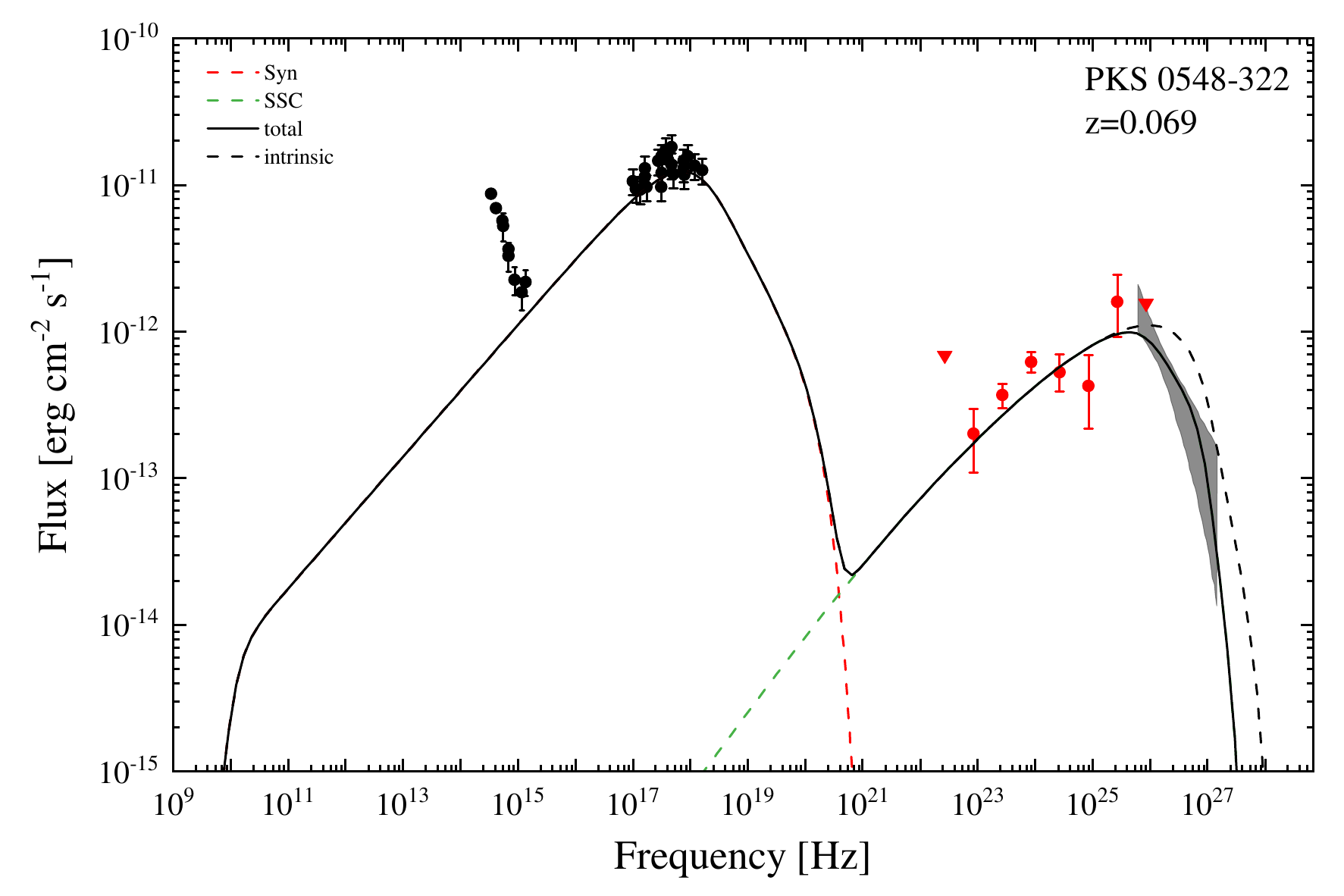}
        \includegraphics[angle=0,width=0.33\linewidth]{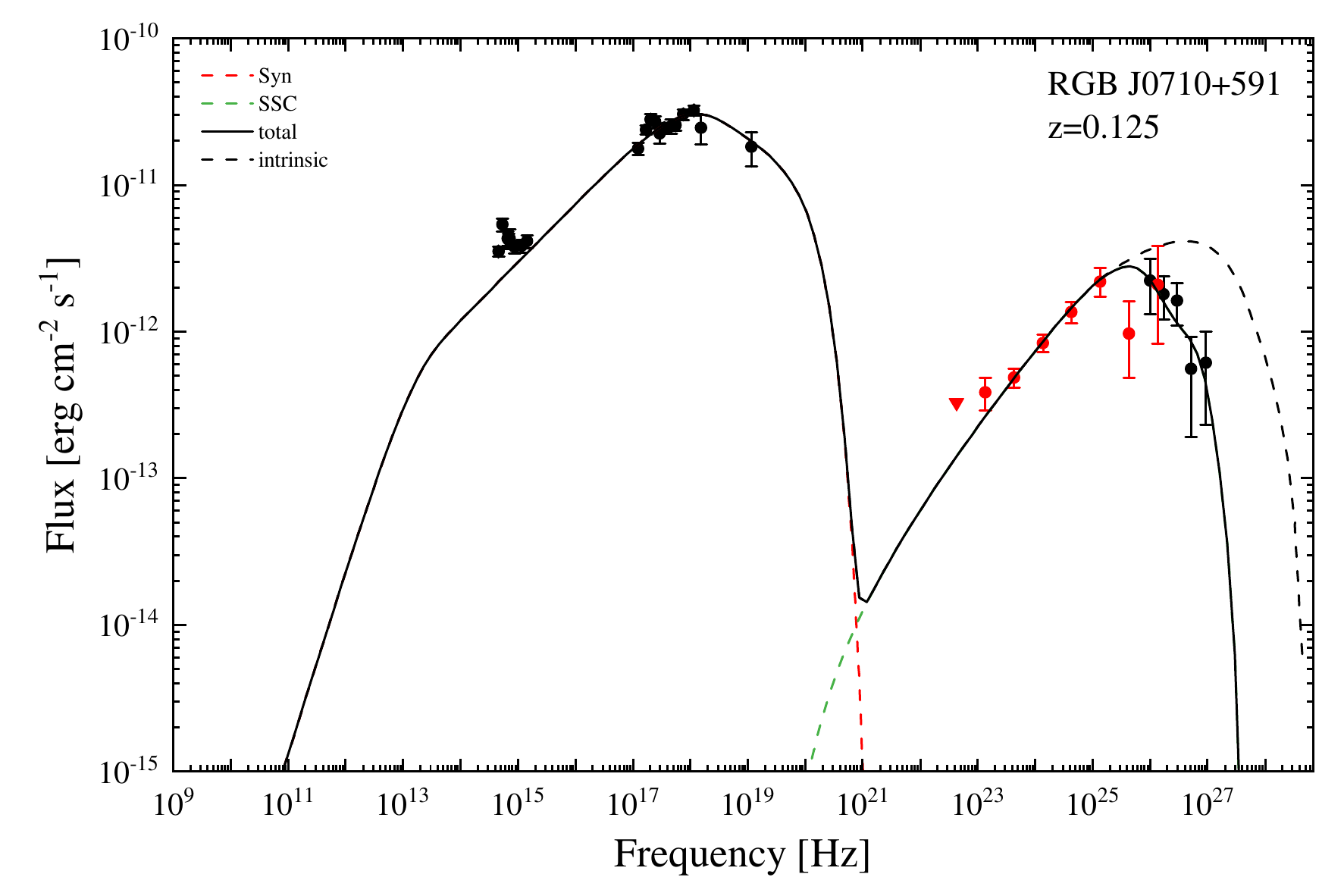}    
        \includegraphics[angle=0,width=0.33\linewidth]{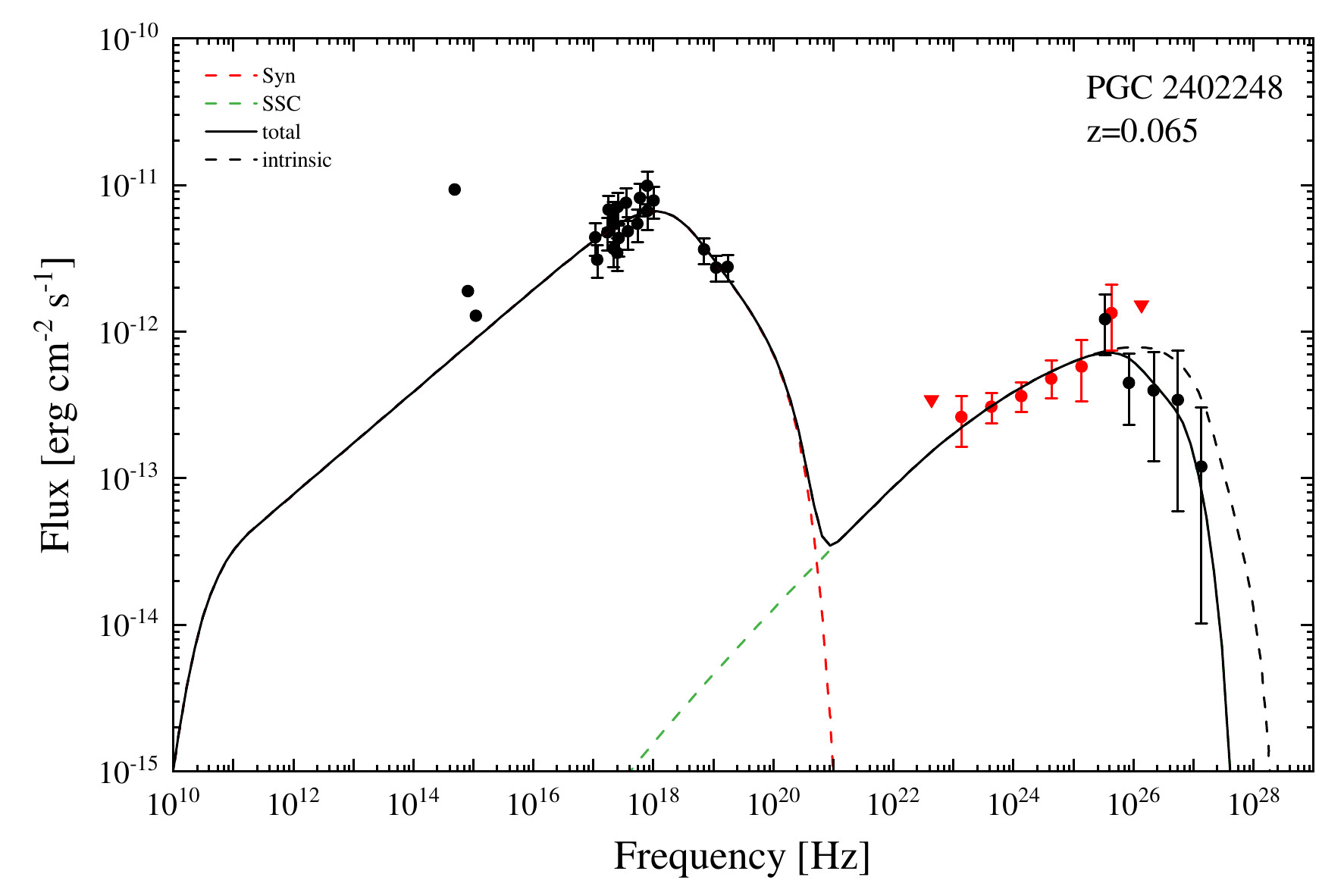}\\  
       \includegraphics[angle=0,width=0.33\linewidth]{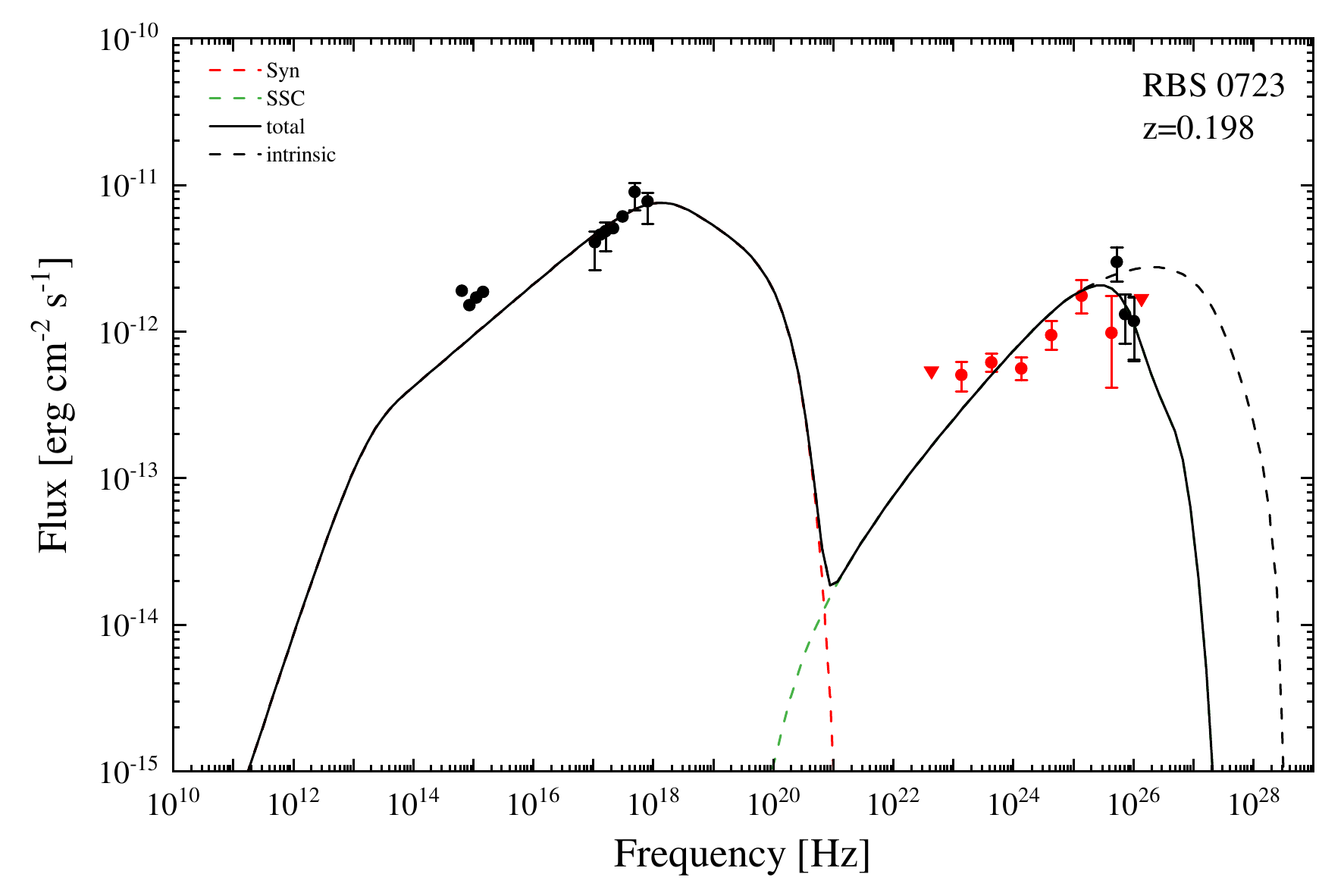}    
        \includegraphics[angle=0,width=0.33\linewidth]{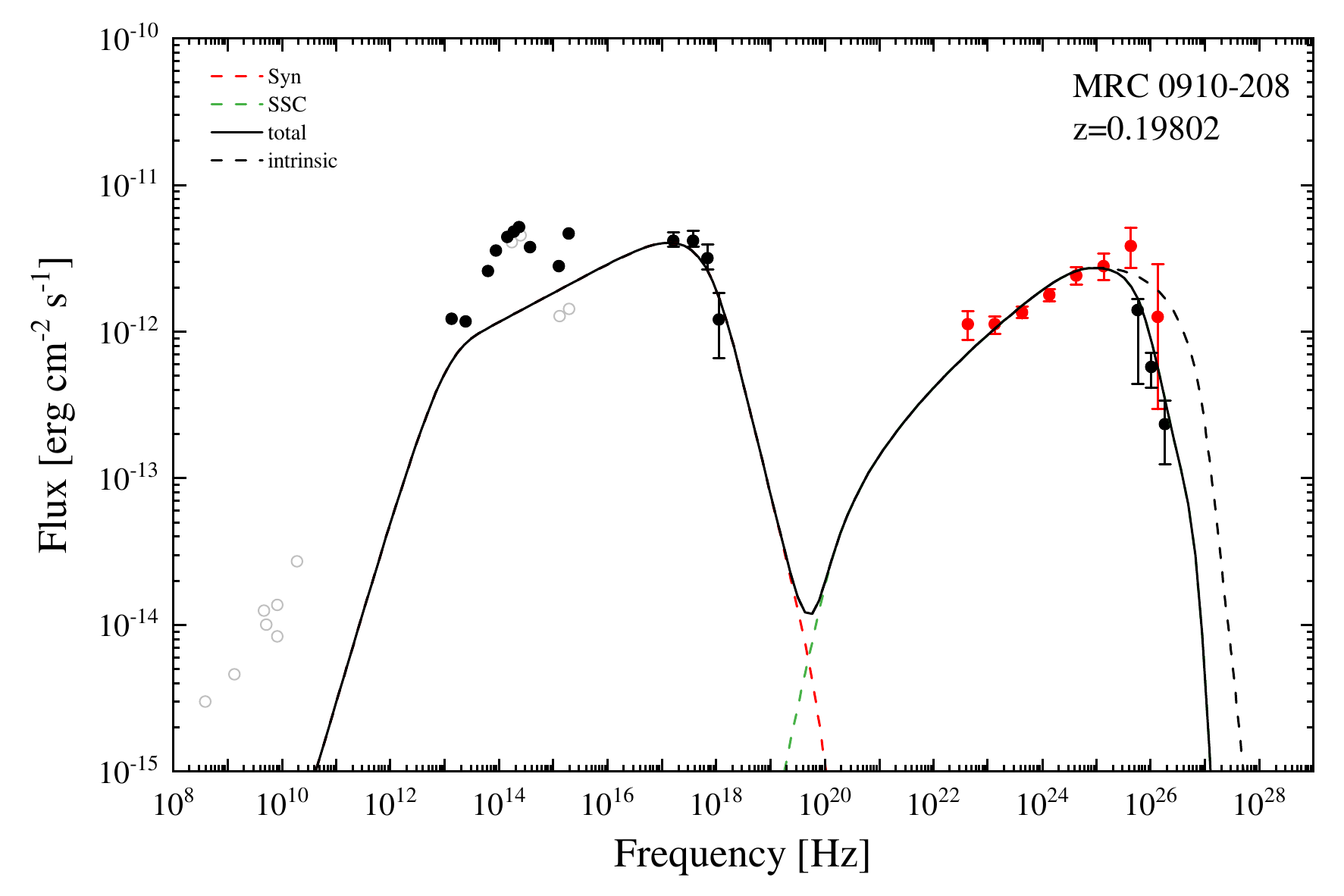} 
        \includegraphics[angle=0,width=0.33\linewidth]{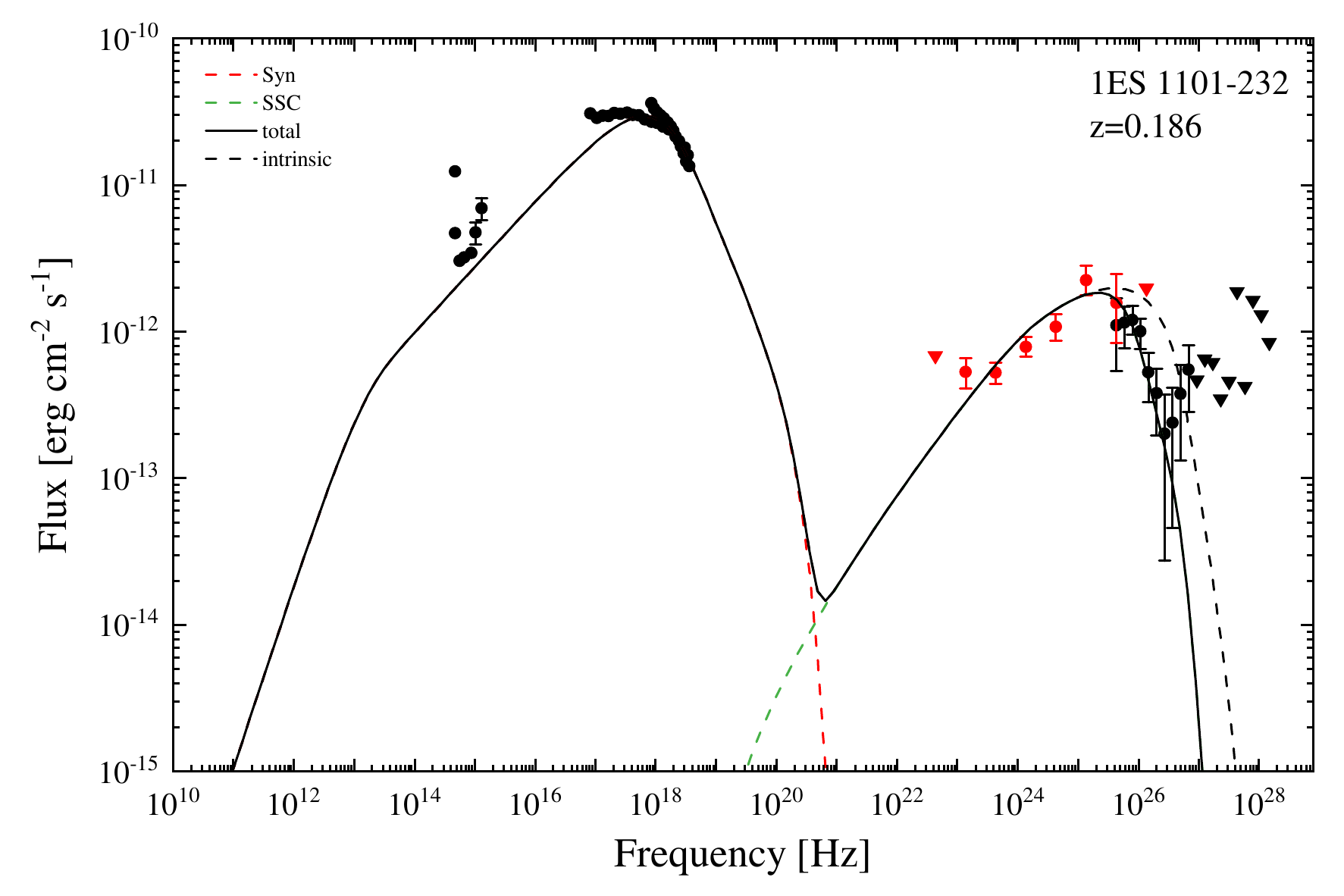} \\
       \includegraphics[angle=0,width=0.33\linewidth]{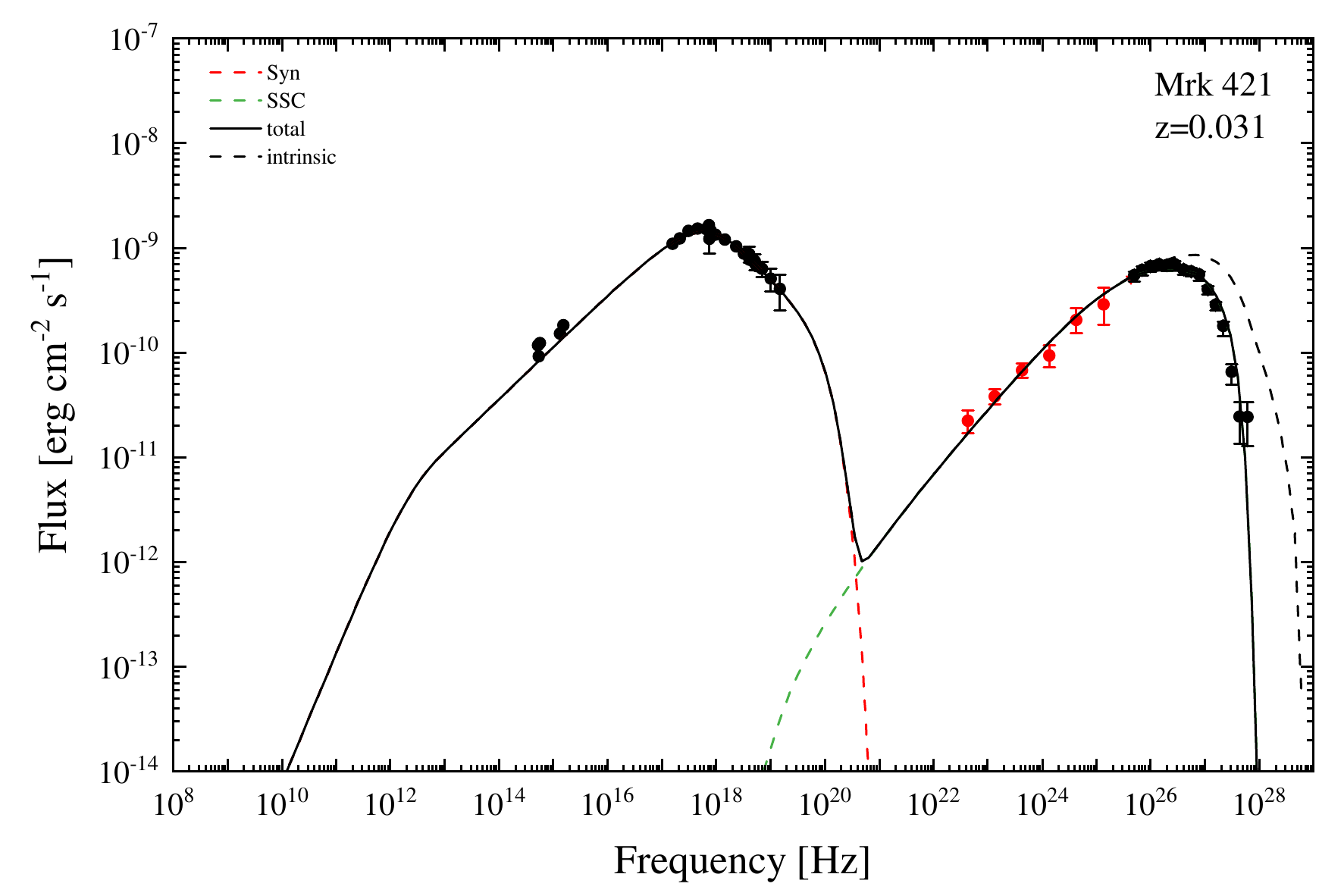}       
       \includegraphics[angle=0,width=0.33\linewidth]{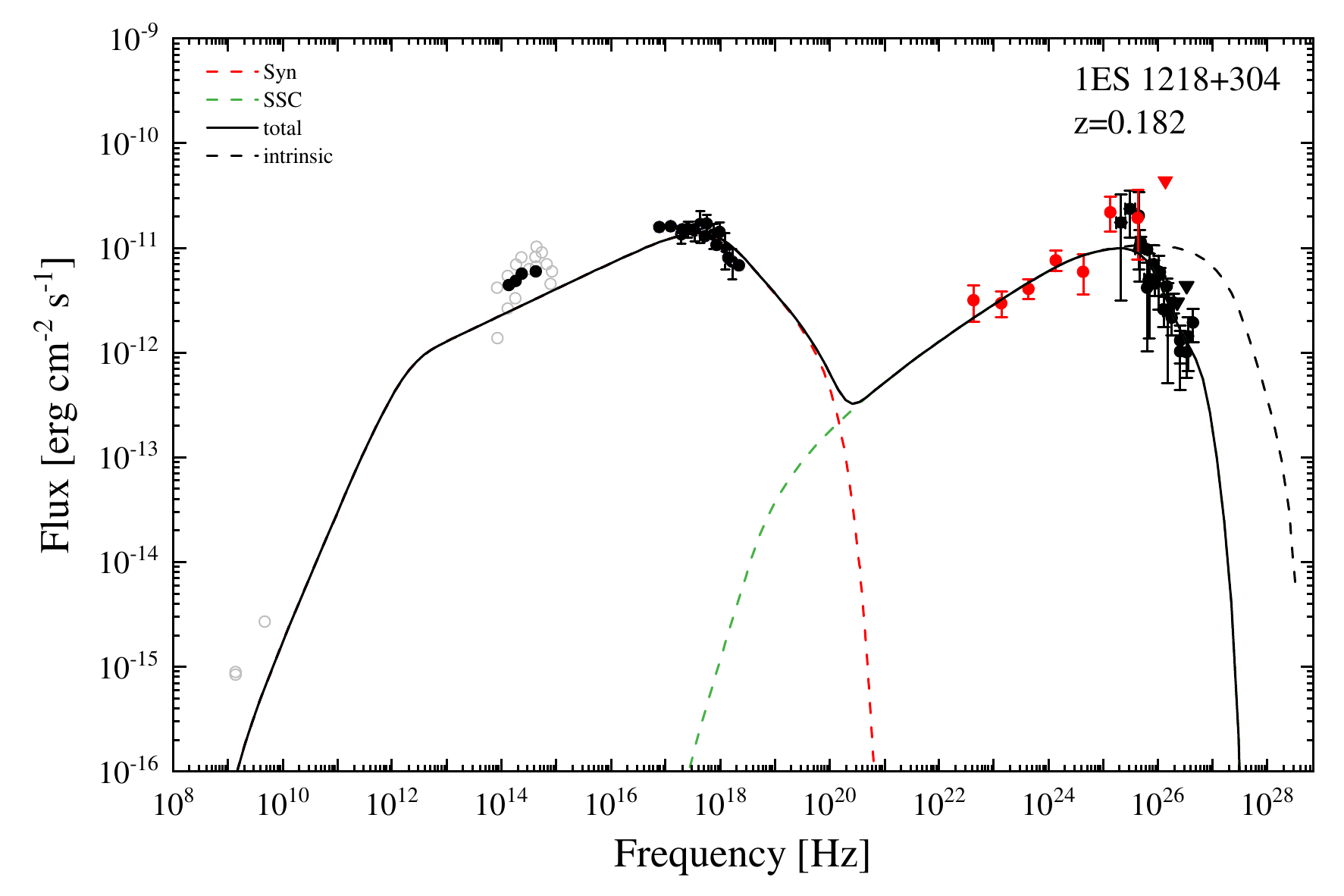}
        \includegraphics[angle=0,width=0.33\linewidth]{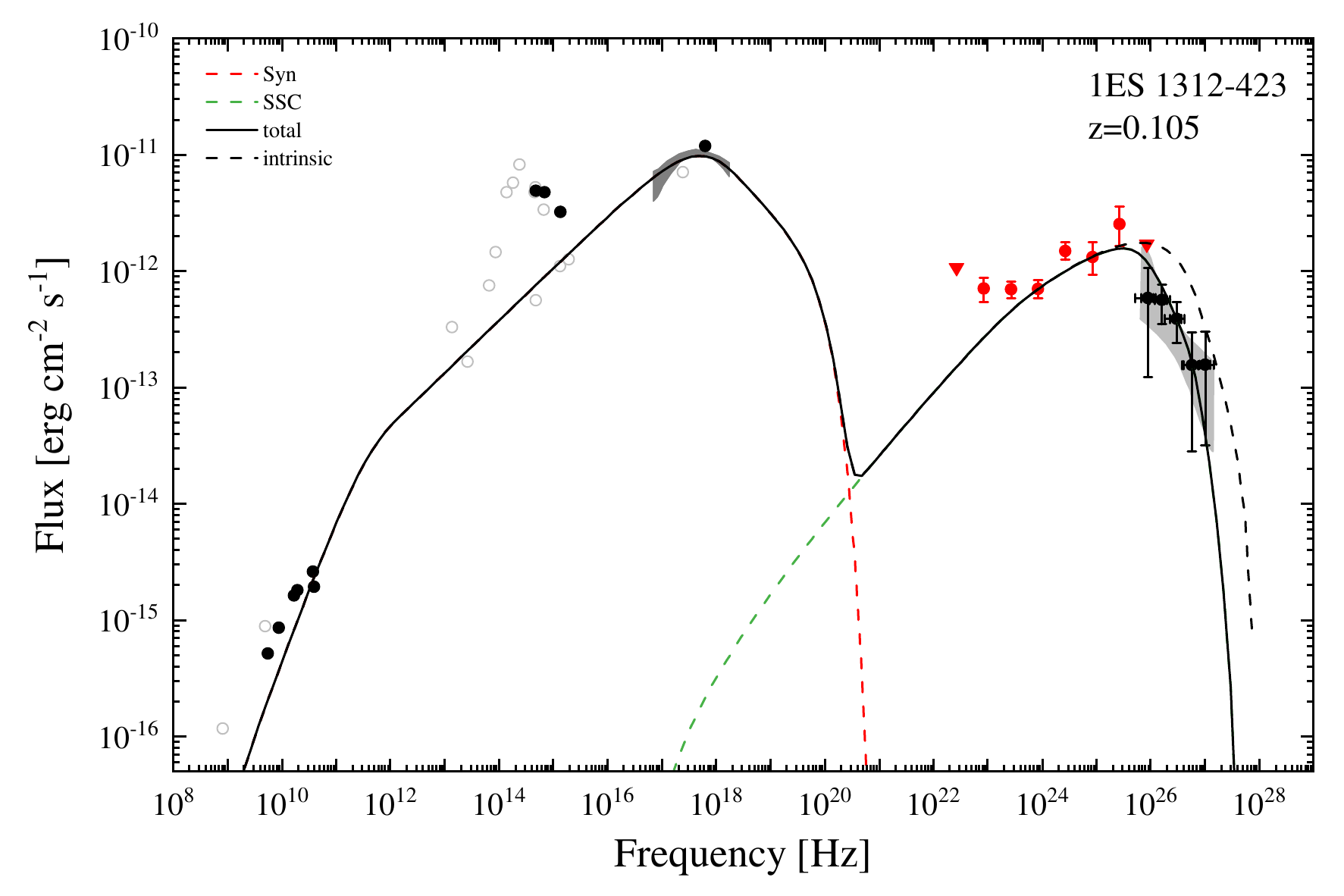}                
    \caption{Observed SEDs (scattered data points) with the one-zone SSC model fits (lines) for the 25 EHBLs. The red solid circles and inverted triangles represent Fermi-LAT data obtained in this study. Black solid circles and inverted triangles denote the simultaneous or quasi-simultaneous observation data from the previous works (see Table \ref{tab:Fit Parameters} and Appendix \ref{appendix: Details} for details on each source), along with the archived data (gray hollow dots). Inverted triangles denote upper limits. The TeV $\gamma$-ray data have been corrected to the observer's frame using the EBL model in \citet{2022ApJ...941...33F}. The black solid lines represent the total modeled flux, comprising the sum of the synchrotron radiation component (red dashed line) and the SSC contribution (green dashed line). The black dashed lines in the TeV energy range show the intrinsic model-predicted fluxes before EBL absorption.}
    \label{Fig: SED fitting}
\end{figure*}

\figsetgrpstart
\figsetgrpnum{2.1}
\figsetgrptitle{Image of SHBL J001355.9--185406
}
\figsetplot{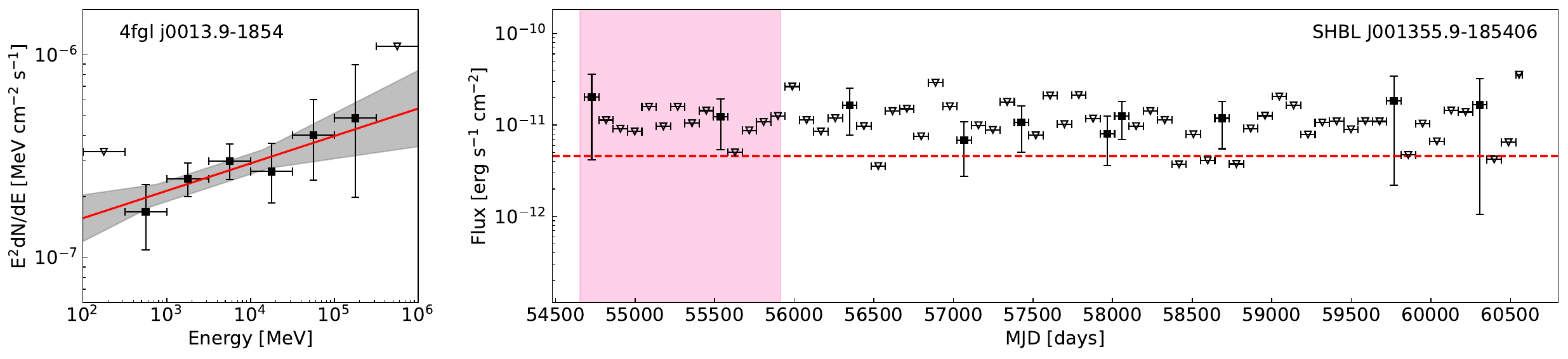}
\figsetgrpnote{Left column: the $\sim$16-yr Fermi-LAT average spectra along with the fitting results, where the open inverted triangles indicate TS<4 for that energy bin. Right column: the long-term Fermi-LAT light curves with time bins of 90 days, where the open inverted triangles indicate the TS value of that time bin less than 9. The horizontal red dash lines represent the $\sim$16 yr average $\gamma$-ray flux of the sources. The pink-shaded regions denote the period covered by the TeV observations, as well as other multiwavelength observations performed during the TeV campaigns. For sources such as 1ES 0347--121, PKS 0548--322, 1ES 1101--232, and 1ES 2356--309, whose TeV observations were conducted prior to the launch of the Fermi satellite, no such shaded regions appear in their Fermi-LAT light curves. For the four BL Lacs (Mrk 421, 1ES 1218+304, 1ES 1727+502 and 1ES 1959+650; marked with a red star in the light-curve figure panels) with significant variability, the contemporaneous Fermi-LAT spectra with the TeV observations are derived to construct their broadband SEDs in Figure \ref{Fig: SED fitting}, that is, their spectra in both the GeV and TeV bands from the observation periods marked as the pink-shaded regions in their light curves. For the other two sources (Mrk 501 and 1ES 2344+514, marked with a red rhombus in the light-curve figure panels) with significant variability, slightly extended observation periods (the green-shaded regions in their light curves) relative to their TeV campaigns (the pink-shaded regions in their light curves) are employed to generate the GeV spectrum for their broadband SED construction. For the remaining 19 objects, the 16-year Fermi-LAT averaged spectra are used to construct their broadband SEDs in Figure \ref{Fig: SED fitting}.}
\figsetgrpend

\figsetgrpstart
\figsetgrpnum{2.2}
\figsetgrptitle{Image of RGB J0152+017
}
\figsetplot{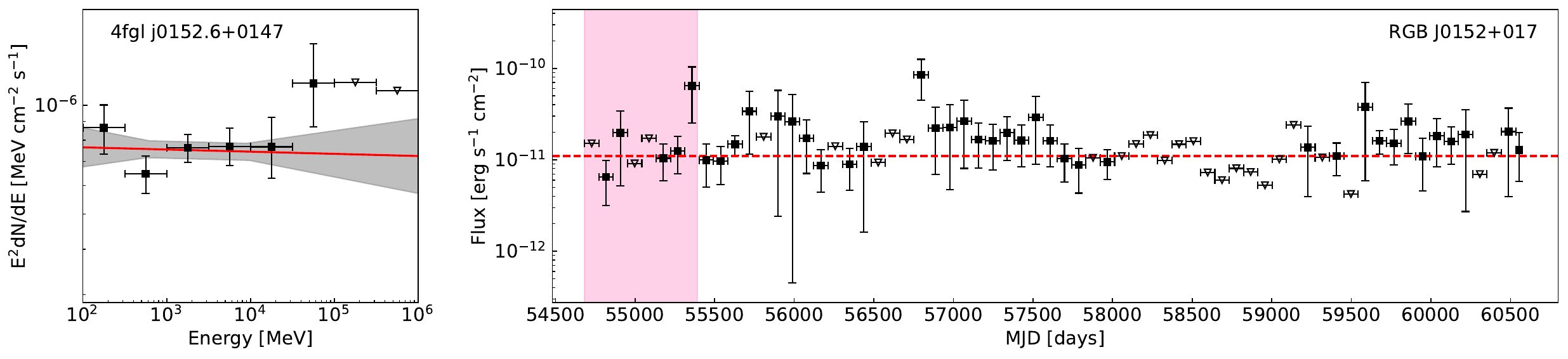}
\figsetgrpnote{Left column: the $\sim$16-yr Fermi-LAT average spectra along with the fitting results, where the open inverted triangles indicate TS<4 for that energy bin. Right column: the long-term Fermi-LAT light curves with time bins of 90 days, where the open inverted triangles indicate the TS value of that time bin less than 9. The horizontal red dash lines represent the $\sim$16 yr average $\gamma$-ray flux of the sources. The pink-shaded regions denote the period covered by the TeV observations, as well as other multiwavelength observations performed during the TeV campaigns. For sources such as 1ES 0347--121, PKS 0548--322, 1ES 1101--232, and 1ES 2356--309, whose TeV observations were conducted prior to the launch of the Fermi satellite, no such shaded regions appear in their Fermi-LAT light curves. For the four BL Lacs (Mrk 421, 1ES 1218+304, 1ES 1727+502 and 1ES 1959+650; marked with a red star in the light-curve figure panels) with significant variability, the contemporaneous Fermi-LAT spectra with the TeV observations are derived to construct their broadband SEDs in Figure \ref{Fig: SED fitting}, that is, their spectra in both the GeV and TeV bands from the observation periods marked as the pink-shaded regions in their light curves. For the other two sources (Mrk 501 and 1ES 2344+514, marked with a red rhombus in the light-curve figure panels) with significant variability, slightly extended observation periods (the green-shaded regions in their light curves) relative to their TeV campaigns (the pink-shaded regions in their light curves) are employed to generate the GeV spectrum for their broadband SED construction. For the remaining 19 objects, the 16-year Fermi-LAT averaged spectra are used to construct their broadband SEDs in Figure \ref{Fig: SED fitting}.}
\figsetgrpend

\figsetgrpstart
\figsetgrpnum{2.3}
\figsetgrptitle{Image of TXS 0210+515
}
\figsetplot{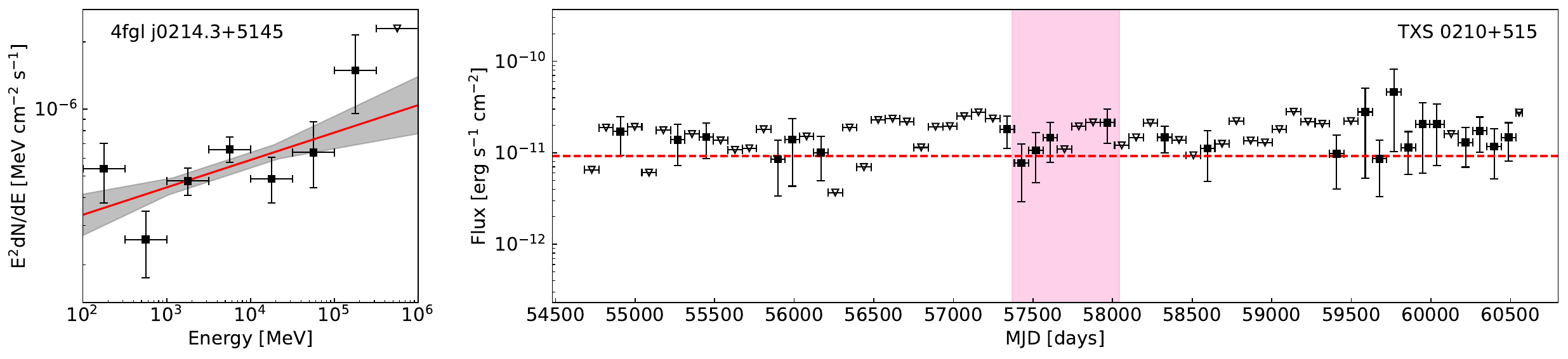}
\figsetgrpnote{Left column: the $\sim$16-yr Fermi-LAT average spectra along with the fitting results, where the open inverted triangles indicate TS<4 for that energy bin. Right column: the long-term Fermi-LAT light curves with time bins of 90 days, where the open inverted triangles indicate the TS value of that time bin less than 9. The horizontal red dash lines represent the $\sim$16 yr average $\gamma$-ray flux of the sources. The pink-shaded regions denote the period covered by the TeV observations, as well as other multiwavelength observations performed during the TeV campaigns. For sources such as 1ES 0347--121, PKS 0548--322, 1ES 1101--232, and 1ES 2356--309, whose TeV observations were conducted prior to the launch of the Fermi satellite, no such shaded regions appear in their Fermi-LAT light curves. For the four BL Lacs (Mrk 421, 1ES 1218+304, 1ES 1727+502 and 1ES 1959+650; marked with a red star in the light-curve figure panels) with significant variability, the contemporaneous Fermi-LAT spectra with the TeV observations are derived to construct their broadband SEDs in Figure \ref{Fig: SED fitting}, that is, their spectra in both the GeV and TeV bands from the observation periods marked as the pink-shaded regions in their light curves. For the other two sources (Mrk 501 and 1ES 2344+514, marked with a red rhombus in the light-curve figure panels) with significant variability, slightly extended observation periods (the green-shaded regions in their light curves) relative to their TeV campaigns (the pink-shaded regions in their light curves) are employed to generate the GeV spectrum for their broadband SED construction. For the remaining 19 objects, the 16-year Fermi-LAT averaged spectra are used to construct their broadband SEDs in Figure \ref{Fig: SED fitting}.}
\figsetgrpend

\figsetgrpstart
\figsetgrpnum{2.4}
\figsetgrptitle{Image of 1ES 0229+200
}
\figsetplot{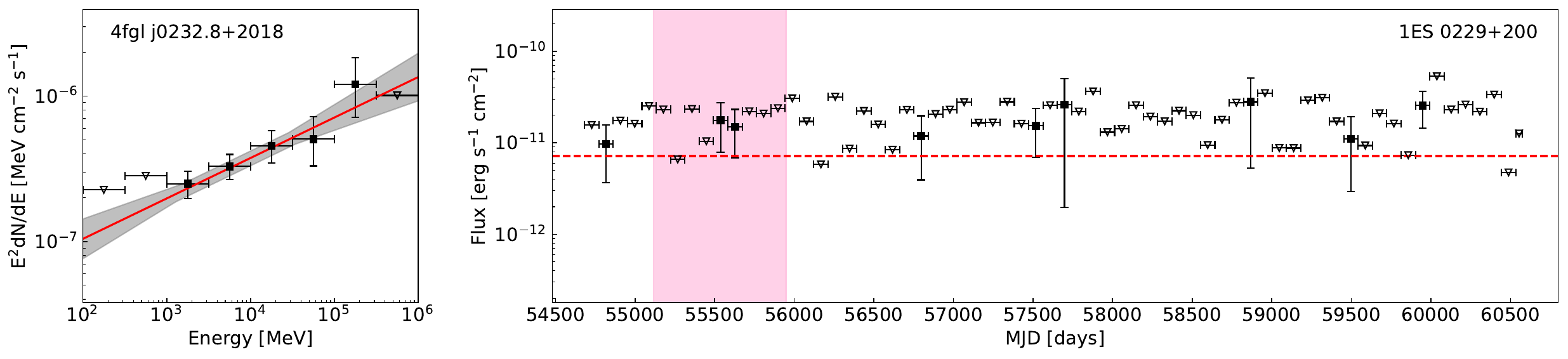}
\figsetgrpnote{Left column: the $\sim$16-yr Fermi-LAT average spectra along with the fitting results, where the open inverted triangles indicate TS<4 for that energy bin. Right column: the long-term Fermi-LAT light curves with time bins of 90 days, where the open inverted triangles indicate the TS value of that time bin less than 9. The horizontal red dash lines represent the $\sim$16 yr average $\gamma$-ray flux of the sources. The pink-shaded regions denote the period covered by the TeV observations, as well as other multiwavelength observations performed during the TeV campaigns. For sources such as 1ES 0347--121, PKS 0548--322, 1ES 1101--232, and 1ES 2356--309, whose TeV observations were conducted prior to the launch of the Fermi satellite, no such shaded regions appear in their Fermi-LAT light curves. For the four BL Lacs (Mrk 421, 1ES 1218+304, 1ES 1727+502 and 1ES 1959+650; marked with a red star in the light-curve figure panels) with significant variability, the contemporaneous Fermi-LAT spectra with the TeV observations are derived to construct their broadband SEDs in Figure \ref{Fig: SED fitting}, that is, their spectra in both the GeV and TeV bands from the observation periods marked as the pink-shaded regions in their light curves. For the other two sources (Mrk 501 and 1ES 2344+514, marked with a red rhombus in the light-curve figure panels) with significant variability, slightly extended observation periods (the green-shaded regions in their light curves) relative to their TeV campaigns (the pink-shaded regions in their light curves) are employed to generate the GeV spectrum for their broadband SED construction. For the remaining 19 objects, the 16-year Fermi-LAT averaged spectra are used to construct their broadband SEDs in Figure \ref{Fig: SED fitting}.}
\figsetgrpend

\figsetgrpstart
\figsetgrpnum{2.5}
\figsetgrptitle{Image of 1ES 0229+200
}
\figsetplot{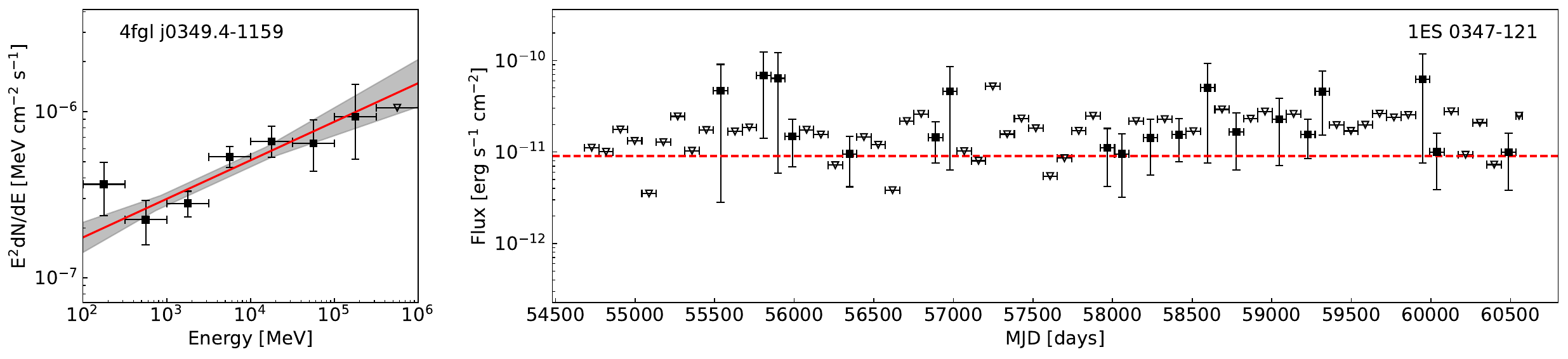}
\figsetgrpnote{Left column: the $\sim$16-yr Fermi-LAT average spectra along with the fitting results, where the open inverted triangles indicate TS<4 for that energy bin. Right column: the long-term Fermi-LAT light curves with time bins of 90 days, where the open inverted triangles indicate the TS value of that time bin less than 9. The horizontal red dash lines represent the $\sim$16 yr average $\gamma$-ray flux of the sources. The pink-shaded regions denote the period covered by the TeV observations, as well as other multiwavelength observations performed during the TeV campaigns. For sources such as 1ES 0347--121, PKS 0548--322, 1ES 1101--232, and 1ES 2356--309, whose TeV observations were conducted prior to the launch of the Fermi satellite, no such shaded regions appear in their Fermi-LAT light curves. For the four BL Lacs (Mrk 421, 1ES 1218+304, 1ES 1727+502 and 1ES 1959+650; marked with a red star in the light-curve figure panels) with significant variability, the contemporaneous Fermi-LAT spectra with the TeV observations are derived to construct their broadband SEDs in Figure \ref{Fig: SED fitting}, that is, their spectra in both the GeV and TeV bands from the observation periods marked as the pink-shaded regions in their light curves. For the other two sources (Mrk 501 and 1ES 2344+514, marked with a red rhombus in the light-curve figure panels) with significant variability, slightly extended observation periods (the green-shaded regions in their light curves) relative to their TeV campaigns (the pink-shaded regions in their light curves) are employed to generate the GeV spectrum for their broadband SED construction. For the remaining 19 objects, the 16-year Fermi-LAT averaged spectra are used to construct their broadband SEDs in Figure \ref{Fig: SED fitting}.}
\figsetgrpend

\figsetgrpstart
\figsetgrpnum{2.6}
\figsetgrptitle{Image of 1ES 0414+009
}
\figsetplot{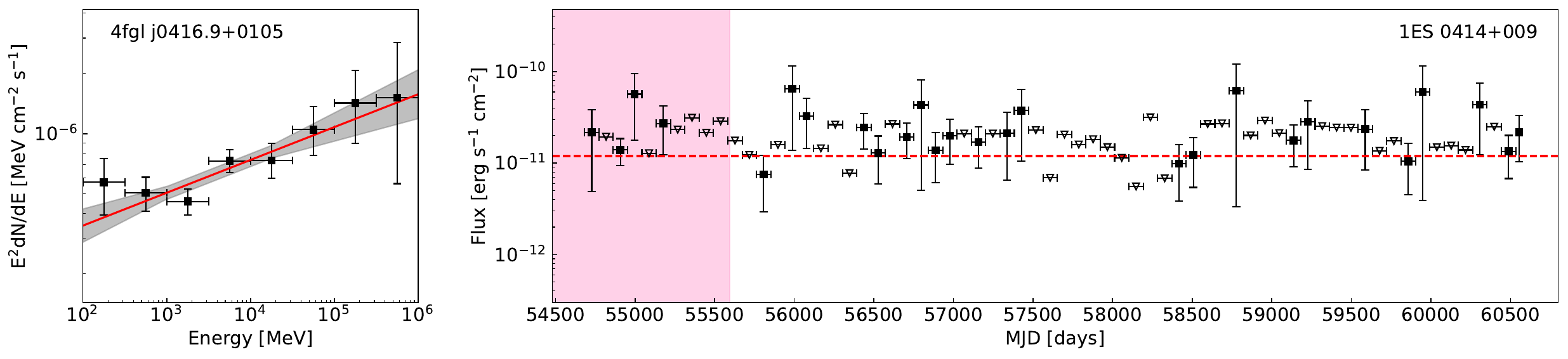}
\figsetgrpnote{Left column: the $\sim$16-yr Fermi-LAT average spectra along with the fitting results, where the open inverted triangles indicate TS<4 for that energy bin. Right column: the long-term Fermi-LAT light curves with time bins of 90 days, where the open inverted triangles indicate the TS value of that time bin less than 9. The horizontal red dash lines represent the $\sim$16 yr average $\gamma$-ray flux of the sources. The pink-shaded regions denote the period covered by the TeV observations, as well as other multiwavelength observations performed during the TeV campaigns. For sources such as 1ES 0347--121, PKS 0548--322, 1ES 1101--232, and 1ES 2356--309, whose TeV observations were conducted prior to the launch of the Fermi satellite, no such shaded regions appear in their Fermi-LAT light curves. For the four BL Lacs (Mrk 421, 1ES 1218+304, 1ES 1727+502 and 1ES 1959+650; marked with a red star in the light-curve figure panels) with significant variability, the contemporaneous Fermi-LAT spectra with the TeV observations are derived to construct their broadband SEDs in Figure \ref{Fig: SED fitting}, that is, their spectra in both the GeV and TeV bands from the observation periods marked as the pink-shaded regions in their light curves. For the other two sources (Mrk 501 and 1ES 2344+514, marked with a red rhombus in the light-curve figure panels) with significant variability, slightly extended observation periods (the green-shaded regions in their light curves) relative to their TeV campaigns (the pink-shaded regions in their light curves) are employed to generate the GeV spectrum for their broadband SED construction. For the remaining 19 objects, the 16-year Fermi-LAT averaged spectra are used to construct their broadband SEDs in Figure \ref{Fig: SED fitting}.}
\figsetgrpend

\figsetgrpstart
\figsetgrpnum{2.7}
\figsetgrptitle{Image of 1ES 0548--322
}
\figsetplot{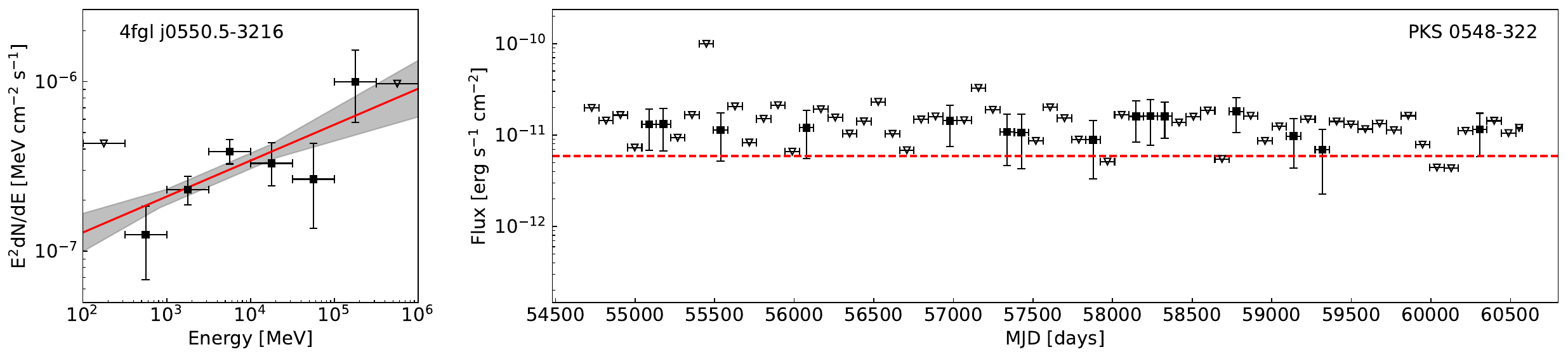}
\figsetgrpnote{Left column: the $\sim$16-yr Fermi-LAT average spectra along with the fitting results, where the open inverted triangles indicate TS<4 for that energy bin. Right column: the long-term Fermi-LAT light curves with time bins of 90 days, where the open inverted triangles indicate the TS value of that time bin less than 9. The horizontal red dash lines represent the $\sim$16 yr average $\gamma$-ray flux of the sources. The pink-shaded regions denote the period covered by the TeV observations, as well as other multiwavelength observations performed during the TeV campaigns. For sources such as 1ES 0347--121, PKS 0548--322, 1ES 1101--232, and 1ES 2356--309, whose TeV observations were conducted prior to the launch of the Fermi satellite, no such shaded regions appear in their Fermi-LAT light curves. For the four BL Lacs (Mrk 421, 1ES 1218+304, 1ES 1727+502 and 1ES 1959+650; marked with a red star in the light-curve figure panels) with significant variability, the contemporaneous Fermi-LAT spectra with the TeV observations are derived to construct their broadband SEDs in Figure \ref{Fig: SED fitting}, that is, their spectra in both the GeV and TeV bands from the observation periods marked as the pink-shaded regions in their light curves. For the other two sources (Mrk 501 and 1ES 2344+514, marked with a red rhombus in the light-curve figure panels) with significant variability, slightly extended observation periods (the green-shaded regions in their light curves) relative to their TeV campaigns (the pink-shaded regions in their light curves) are employed to generate the GeV spectrum for their broadband SED construction. For the remaining 19 objects, the 16-year Fermi-LAT averaged spectra are used to construct their broadband SEDs in Figure \ref{Fig: SED fitting}.}
\figsetgrpend

\figsetgrpstart
\figsetgrpnum{2.8}
\figsetgrptitle{Image of RGB J0710+591
}
\figsetplot{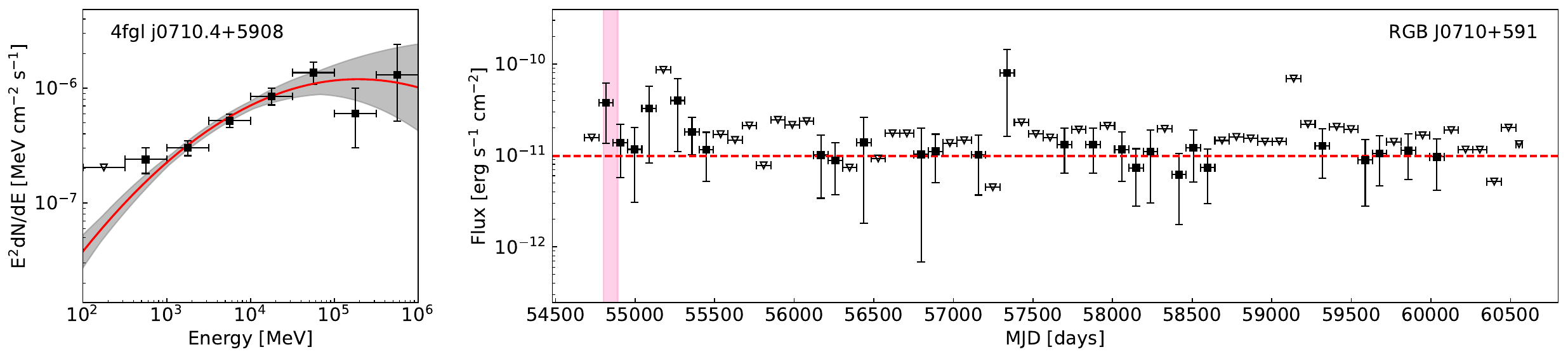}
\figsetgrpnote{Left column: the $\sim$16-yr Fermi-LAT average spectra along with the fitting results, where the open inverted triangles indicate TS<4 for that energy bin. Right column: the long-term Fermi-LAT light curves with time bins of 90 days, where the open inverted triangles indicate the TS value of that time bin less than 9. The horizontal red dash lines represent the $\sim$16 yr average $\gamma$-ray flux of the sources. The pink-shaded regions denote the period covered by the TeV observations, as well as other multiwavelength observations performed during the TeV campaigns. For sources such as 1ES 0347--121, PKS 0548--322, 1ES 1101--232, and 1ES 2356--309, whose TeV observations were conducted prior to the launch of the Fermi satellite, no such shaded regions appear in their Fermi-LAT light curves. For the four BL Lacs (Mrk 421, 1ES 1218+304, 1ES 1727+502 and 1ES 1959+650; marked with a red star in the light-curve figure panels) with significant variability, the contemporaneous Fermi-LAT spectra with the TeV observations are derived to construct their broadband SEDs in Figure \ref{Fig: SED fitting}, that is, their spectra in both the GeV and TeV bands from the observation periods marked as the pink-shaded regions in their light curves. For the other two sources (Mrk 501 and 1ES 2344+514, marked with a red rhombus in the light-curve figure panels) with significant variability, slightly extended observation periods (the green-shaded regions in their light curves) relative to their TeV campaigns (the pink-shaded regions in their light curves) are employed to generate the GeV spectrum for their broadband SED construction. For the remaining 19 objects, the 16-year Fermi-LAT averaged spectra are used to construct their broadband SEDs in Figure \ref{Fig: SED fitting}.}
\figsetgrpend

\figsetgrpstart
\figsetgrpnum{2.9}
\figsetgrptitle{Image of PGC 2402248
}
\figsetplot{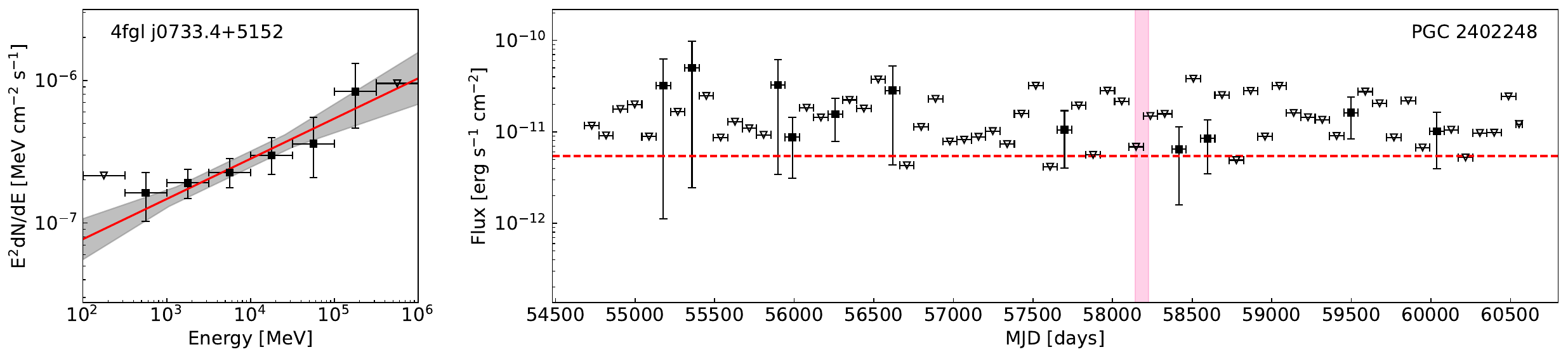}
\figsetgrpnote{Left column: the $\sim$16-yr Fermi-LAT average spectra along with the fitting results, where the open inverted triangles indicate TS<4 for that energy bin. Right column: the long-term Fermi-LAT light curves with time bins of 90 days, where the open inverted triangles indicate the TS value of that time bin less than 9. The horizontal red dash lines represent the $\sim$16 yr average $\gamma$-ray flux of the sources. The pink-shaded regions denote the period covered by the TeV observations, as well as other multiwavelength observations performed during the TeV campaigns. For sources such as 1ES 0347--121, PKS 0548--322, 1ES 1101--232, and 1ES 2356--309, whose TeV observations were conducted prior to the launch of the Fermi satellite, no such shaded regions appear in their Fermi-LAT light curves. For the four BL Lacs (Mrk 421, 1ES 1218+304, 1ES 1727+502 and 1ES 1959+650; marked with a red star in the light-curve figure panels) with significant variability, the contemporaneous Fermi-LAT spectra with the TeV observations are derived to construct their broadband SEDs in Figure \ref{Fig: SED fitting}, that is, their spectra in both the GeV and TeV bands from the observation periods marked as the pink-shaded regions in their light curves. For the other two sources (Mrk 501 and 1ES 2344+514, marked with a red rhombus in the light-curve figure panels) with significant variability, slightly extended observation periods (the green-shaded regions in their light curves) relative to their TeV campaigns (the pink-shaded regions in their light curves) are employed to generate the GeV spectrum for their broadband SED construction. For the remaining 19 objects, the 16-year Fermi-LAT averaged spectra are used to construct their broadband SEDs in Figure \ref{Fig: SED fitting}.}
\figsetgrpend

\figsetgrpstart
\figsetgrpnum{2.10}
\figsetgrptitle{Image of RBS 0723
}
\figsetplot{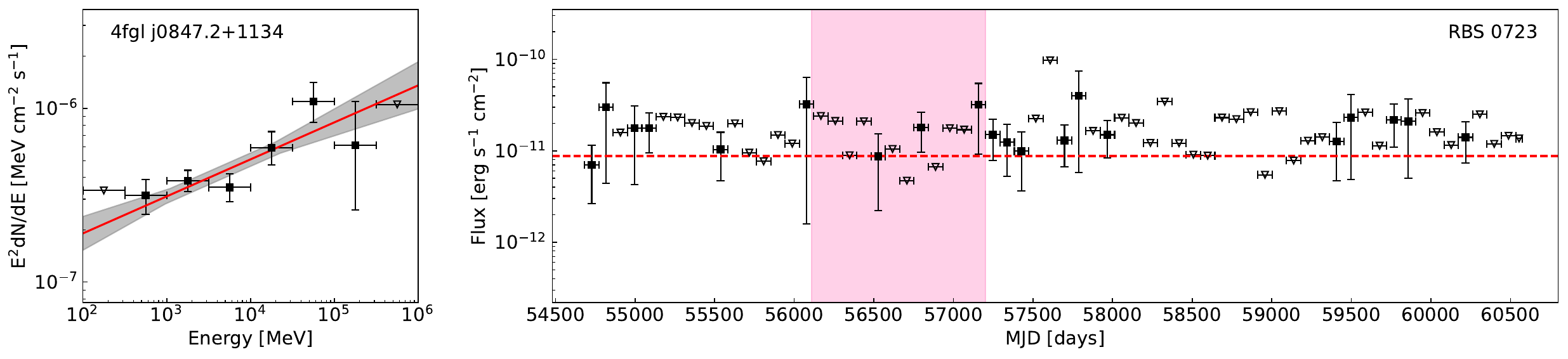}
\figsetgrpnote{Left column: the $\sim$16-yr Fermi-LAT average spectra along with the fitting results, where the open inverted triangles indicate TS<4 for that energy bin. Right column: the long-term Fermi-LAT light curves with time bins of 90 days, where the open inverted triangles indicate the TS value of that time bin less than 9. The horizontal red dash lines represent the $\sim$16 yr average $\gamma$-ray flux of the sources. The pink-shaded regions denote the period covered by the TeV observations, as well as other multiwavelength observations performed during the TeV campaigns. For sources such as 1ES 0347--121, PKS 0548--322, 1ES 1101--232, and 1ES 2356--309, whose TeV observations were conducted prior to the launch of the Fermi satellite, no such shaded regions appear in their Fermi-LAT light curves. For the four BL Lacs (Mrk 421, 1ES 1218+304, 1ES 1727+502 and 1ES 1959+650; marked with a red star in the light-curve figure panels) with significant variability, the contemporaneous Fermi-LAT spectra with the TeV observations are derived to construct their broadband SEDs in Figure \ref{Fig: SED fitting}, that is, their spectra in both the GeV and TeV bands from the observation periods marked as the pink-shaded regions in their light curves. For the other two sources (Mrk 501 and 1ES 2344+514, marked with a red rhombus in the light-curve figure panels) with significant variability, slightly extended observation periods (the green-shaded regions in their light curves) relative to their TeV campaigns (the pink-shaded regions in their light curves) are employed to generate the GeV spectrum for their broadband SED construction. For the remaining 19 objects, the 16-year Fermi-LAT averaged spectra are used to construct their broadband SEDs in Figure \ref{Fig: SED fitting}.}
\figsetgrpend

\figsetgrpstart
\figsetgrpnum{2.11}
\figsetgrptitle{Image of MRC 0910-208
}
\figsetplot{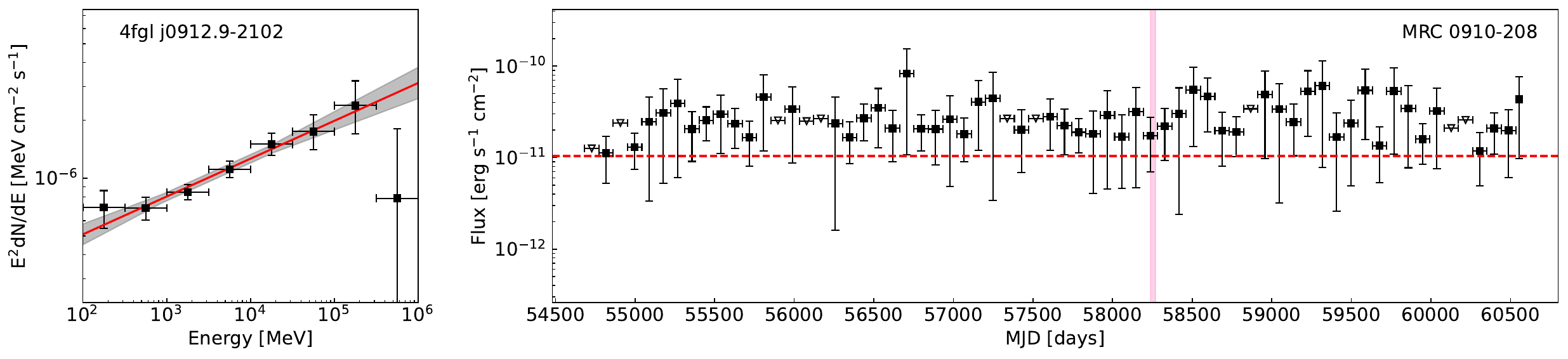}
\figsetgrpnote{Left column: the $\sim$16-yr Fermi-LAT average spectra along with the fitting results, where the open inverted triangles indicate TS<4 for that energy bin. Right column: the long-term Fermi-LAT light curves with time bins of 90 days, where the open inverted triangles indicate the TS value of that time bin less than 9. The horizontal red dash lines represent the $\sim$16 yr average $\gamma$-ray flux of the sources. The pink-shaded regions denote the period covered by the TeV observations, as well as other multiwavelength observations performed during the TeV campaigns. For sources such as 1ES 0347--121, PKS 0548--322, 1ES 1101--232, and 1ES 2356--309, whose TeV observations were conducted prior to the launch of the Fermi satellite, no such shaded regions appear in their Fermi-LAT light curves. For the four BL Lacs (Mrk 421, 1ES 1218+304, 1ES 1727+502 and 1ES 1959+650; marked with a red star in the light-curve figure panels) with significant variability, the contemporaneous Fermi-LAT spectra with the TeV observations are derived to construct their broadband SEDs in Figure \ref{Fig: SED fitting}, that is, their spectra in both the GeV and TeV bands from the observation periods marked as the pink-shaded regions in their light curves. For the other two sources (Mrk 501 and 1ES 2344+514, marked with a red rhombus in the light-curve figure panels) with significant variability, slightly extended observation periods (the green-shaded regions in their light curves) relative to their TeV campaigns (the pink-shaded regions in their light curves) are employed to generate the GeV spectrum for their broadband SED construction. For the remaining 19 objects, the 16-year Fermi-LAT averaged spectra are used to construct their broadband SEDs in Figure \ref{Fig: SED fitting}.}
\figsetgrpend

\figsetgrpstart
\figsetgrpnum{2.12}
\figsetgrptitle{Image of 1ES 1101--232
}
\figsetplot{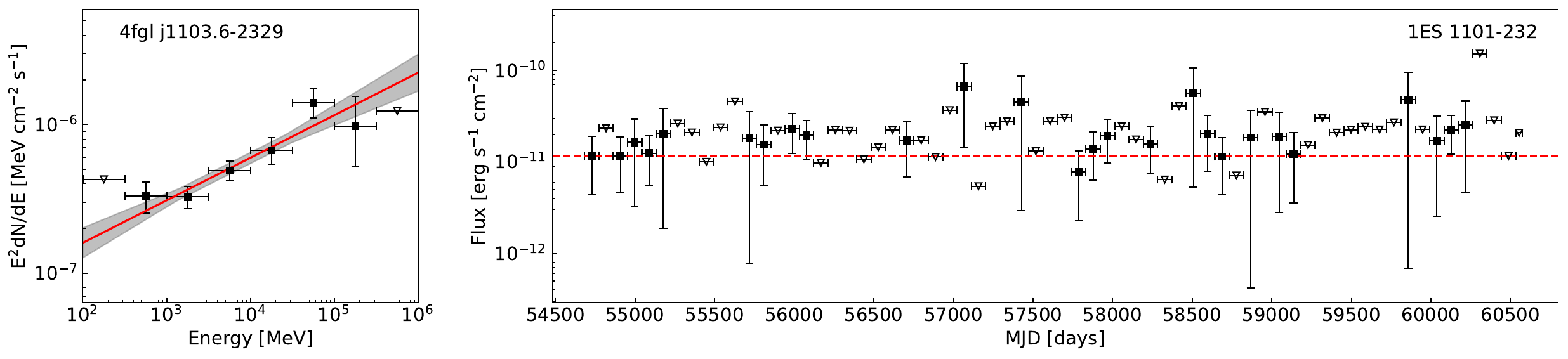}
\figsetgrpnote{Left column: the $\sim$16-yr Fermi-LAT average spectra along with the fitting results, where the open inverted triangles indicate TS<4 for that energy bin. Right column: the long-term Fermi-LAT light curves with time bins of 90 days, where the open inverted triangles indicate the TS value of that time bin less than 9. The horizontal red dash lines represent the $\sim$16 yr average $\gamma$-ray flux of the sources. The pink-shaded regions denote the period covered by the TeV observations, as well as other multiwavelength observations performed during the TeV campaigns. For sources such as 1ES 0347--121, PKS 0548--322, 1ES 1101--232, and 1ES 2356--309, whose TeV observations were conducted prior to the launch of the Fermi satellite, no such shaded regions appear in their Fermi-LAT light curves. For the four BL Lacs (Mrk 421, 1ES 1218+304, 1ES 1727+502 and 1ES 1959+650; marked with a red star in the light-curve figure panels) with significant variability, the contemporaneous Fermi-LAT spectra with the TeV observations are derived to construct their broadband SEDs in Figure \ref{Fig: SED fitting}, that is, their spectra in both the GeV and TeV bands from the observation periods marked as the pink-shaded regions in their light curves. For the other two sources (Mrk 501 and 1ES 2344+514, marked with a red rhombus in the light-curve figure panels) with significant variability, slightly extended observation periods (the green-shaded regions in their light curves) relative to their TeV campaigns (the pink-shaded regions in their light curves) are employed to generate the GeV spectrum for their broadband SED construction. For the remaining 19 objects, the 16-year Fermi-LAT averaged spectra are used to construct their broadband SEDs in Figure \ref{Fig: SED fitting}.}
\figsetgrpend

\figsetgrpstart
\figsetgrpnum{2.13}
\figsetgrptitle{Image of Mrk 421
}
\figsetplot{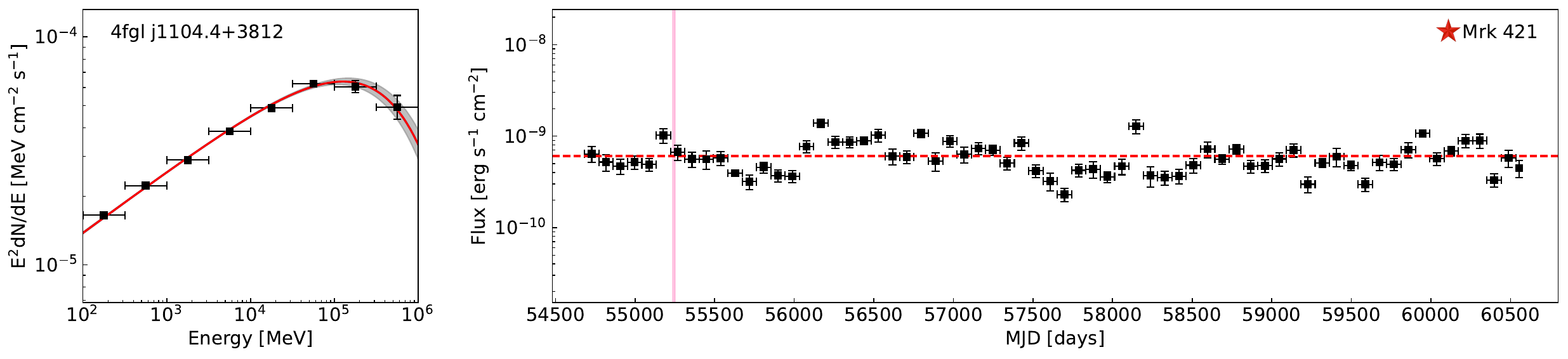}
\figsetgrpnote{Left column: the $\sim$16-yr Fermi-LAT average spectra along with the fitting results, where the open inverted triangles indicate TS<4 for that energy bin. Right column: the long-term Fermi-LAT light curves with time bins of 90 days, where the open inverted triangles indicate the TS value of that time bin less than 9. The horizontal red dash lines represent the $\sim$16 yr average $\gamma$-ray flux of the sources. The pink-shaded regions denote the period covered by the TeV observations, as well as other multiwavelength observations performed during the TeV campaigns. For sources such as 1ES 0347--121, PKS 0548--322, 1ES 1101--232, and 1ES 2356--309, whose TeV observations were conducted prior to the launch of the Fermi satellite, no such shaded regions appear in their Fermi-LAT light curves. For the four BL Lacs (Mrk 421, 1ES 1218+304, 1ES 1727+502 and 1ES 1959+650; marked with a red star in the light-curve figure panels) with significant variability, the contemporaneous Fermi-LAT spectra with the TeV observations are derived to construct their broadband SEDs in Figure \ref{Fig: SED fitting}, that is, their spectra in both the GeV and TeV bands from the observation periods marked as the pink-shaded regions in their light curves. For the other two sources (Mrk 501 and 1ES 2344+514, marked with a red rhombus in the light-curve figure panels) with significant variability, slightly extended observation periods (the green-shaded regions in their light curves) relative to their TeV campaigns (the pink-shaded regions in their light curves) are employed to generate the GeV spectrum for their broadband SED construction. For the remaining 19 objects, the 16-year Fermi-LAT averaged spectra are used to construct their broadband SEDs in Figure \ref{Fig: SED fitting}.}
\figsetgrpend

\figsetgrpstart
\figsetgrpnum{2.14}
\figsetgrptitle{Image of 1ES 1218+304
}
\figsetplot{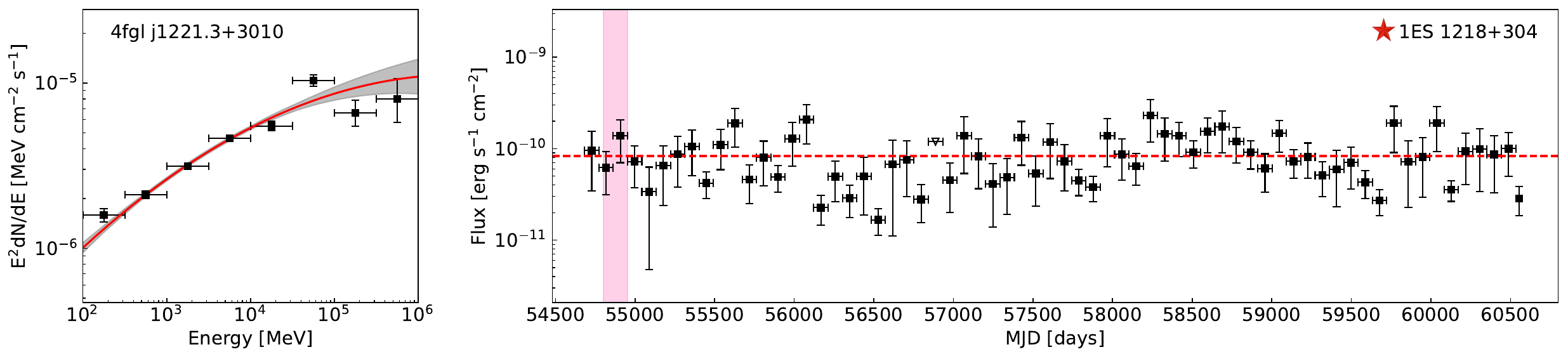}
\figsetgrpnote{Left column: the $\sim$16-yr Fermi-LAT average spectra along with the fitting results, where the open inverted triangles indicate TS<4 for that energy bin. Right column: the long-term Fermi-LAT light curves with time bins of 90 days, where the open inverted triangles indicate the TS value of that time bin less than 9. The horizontal red dash lines represent the $\sim$16 yr average $\gamma$-ray flux of the sources. The pink-shaded regions denote the period covered by the TeV observations, as well as other multiwavelength observations performed during the TeV campaigns. For sources such as 1ES 0347--121, PKS 0548--322, 1ES 1101--232, and 1ES 2356--309, whose TeV observations were conducted prior to the launch of the Fermi satellite, no such shaded regions appear in their Fermi-LAT light curves. For the four BL Lacs (Mrk 421, 1ES 1218+304, 1ES 1727+502 and 1ES 1959+650; marked with a red star in the light-curve figure panels) with significant variability, the contemporaneous Fermi-LAT spectra with the TeV observations are derived to construct their broadband SEDs in Figure \ref{Fig: SED fitting}, that is, their spectra in both the GeV and TeV bands from the observation periods marked as the pink-shaded regions in their light curves. For the other two sources (Mrk 501 and 1ES 2344+514, marked with a red rhombus in the light-curve figure panels) with significant variability, slightly extended observation periods (the green-shaded regions in their light curves) relative to their TeV campaigns (the pink-shaded regions in their light curves) are employed to generate the GeV spectrum for their broadband SED construction. For the remaining 19 objects, the 16-year Fermi-LAT averaged spectra are used to construct their broadband SEDs in Figure \ref{Fig: SED fitting}.}
\figsetgrpend

\figsetgrpstart
\figsetgrpnum{2.15}
\figsetgrptitle{Image of 1ES 1312--423
}
\figsetplot{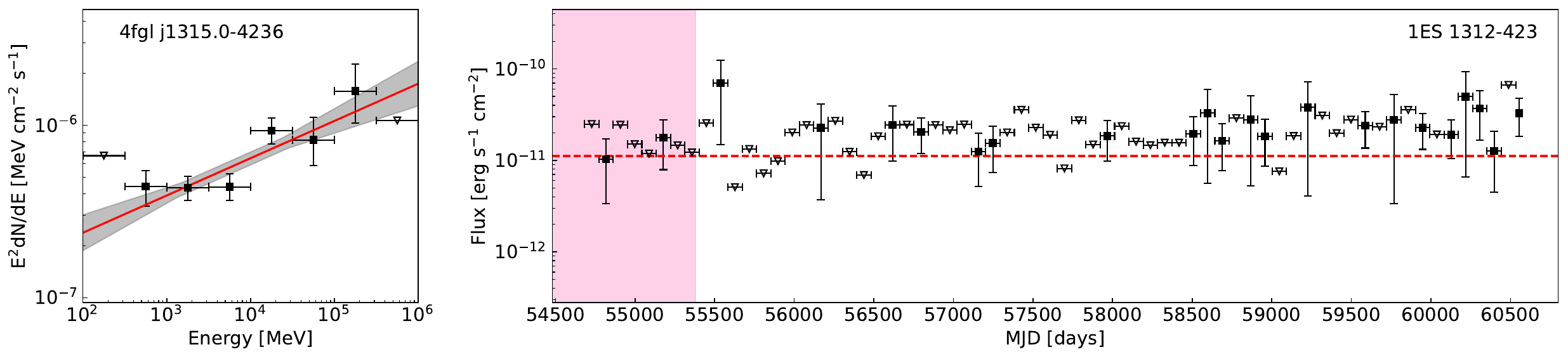}
\figsetgrpnote{Left column: the $\sim$16-yr Fermi-LAT average spectra along with the fitting results, where the open inverted triangles indicate TS<4 for that energy bin. Right column: the long-term Fermi-LAT light curves with time bins of 90 days, where the open inverted triangles indicate the TS value of that time bin less than 9. The horizontal red dash lines represent the $\sim$16 yr average $\gamma$-ray flux of the sources. The pink-shaded regions denote the period covered by the TeV observations, as well as other multiwavelength observations performed during the TeV campaigns. For sources such as 1ES 0347--121, PKS 0548--322, 1ES 1101--232, and 1ES 2356--309, whose TeV observations were conducted prior to the launch of the Fermi satellite, no such shaded regions appear in their Fermi-LAT light curves. For the four BL Lacs (Mrk 421, 1ES 1218+304, 1ES 1727+502 and 1ES 1959+650; marked with a red star in the light-curve figure panels) with significant variability, the contemporaneous Fermi-LAT spectra with the TeV observations are derived to construct their broadband SEDs in Figure \ref{Fig: SED fitting}, that is, their spectra in both the GeV and TeV bands from the observation periods marked as the pink-shaded regions in their light curves. For the other two sources (Mrk 501 and 1ES 2344+514, marked with a red rhombus in the light-curve figure panels) with significant variability, slightly extended observation periods (the green-shaded regions in their light curves) relative to their TeV campaigns (the pink-shaded regions in their light curves) are employed to generate the GeV spectrum for their broadband SED construction. For the remaining 19 objects, the 16-year Fermi-LAT averaged spectra are used to construct their broadband SEDs in Figure \ref{Fig: SED fitting}.}
\figsetgrpend

\figsetgrpstart
\figsetgrpnum{2.16}
\figsetgrptitle{Image of 1ES 1426+428
}
\figsetplot{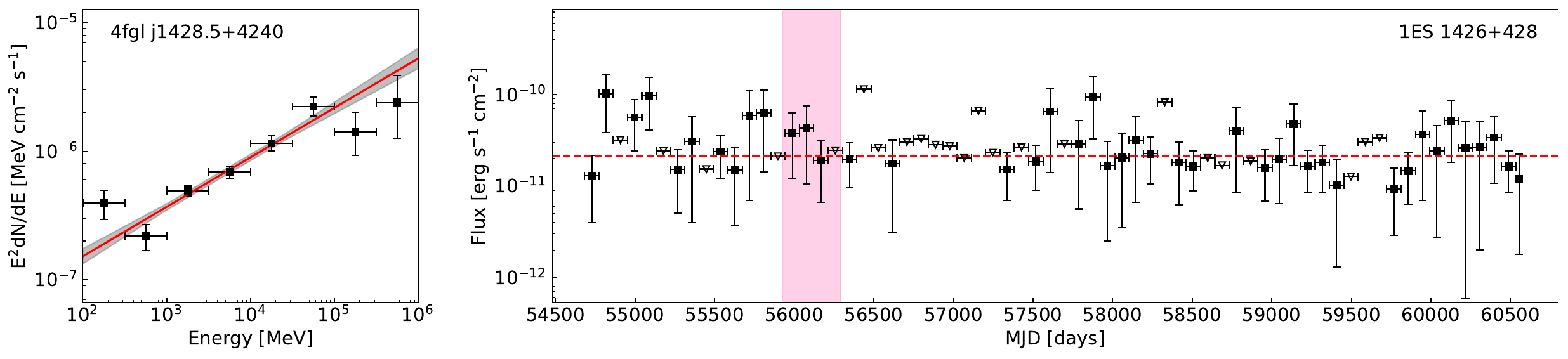}
\figsetgrpnote{Left column: the $\sim$16-yr Fermi-LAT average spectra along with the fitting results, where the open inverted triangles indicate TS<4 for that energy bin. Right column: the long-term Fermi-LAT light curves with time bins of 90 days, where the open inverted triangles indicate the TS value of that time bin less than 9. The horizontal red dash lines represent the $\sim$16 yr average $\gamma$-ray flux of the sources. The pink-shaded regions denote the period covered by the TeV observations, as well as other multiwavelength observations performed during the TeV campaigns. For sources such as 1ES 0347--121, PKS 0548--322, 1ES 1101--232, and 1ES 2356--309, whose TeV observations were conducted prior to the launch of the Fermi satellite, no such shaded regions appear in their Fermi-LAT light curves. For the four BL Lacs (Mrk 421, 1ES 1218+304, 1ES 1727+502 and 1ES 1959+650; marked with a red star in the light-curve figure panels) with significant variability, the contemporaneous Fermi-LAT spectra with the TeV observations are derived to construct their broadband SEDs in Figure \ref{Fig: SED fitting}, that is, their spectra in both the GeV and TeV bands from the observation periods marked as the pink-shaded regions in their light curves. For the other two sources (Mrk 501 and 1ES 2344+514, marked with a red rhombus in the light-curve figure panels) with significant variability, slightly extended observation periods (the green-shaded regions in their light curves) relative to their TeV campaigns (the pink-shaded regions in their light curves) are employed to generate the GeV spectrum for their broadband SED construction. For the remaining 19 objects, the 16-year Fermi-LAT averaged spectra are used to construct their broadband SEDs in Figure \ref{Fig: SED fitting}.}
\figsetgrpend

\figsetgrpstart
\figsetgrpnum{2.17}
\figsetgrptitle{Image of 1ES 1440+122
}
\figsetplot{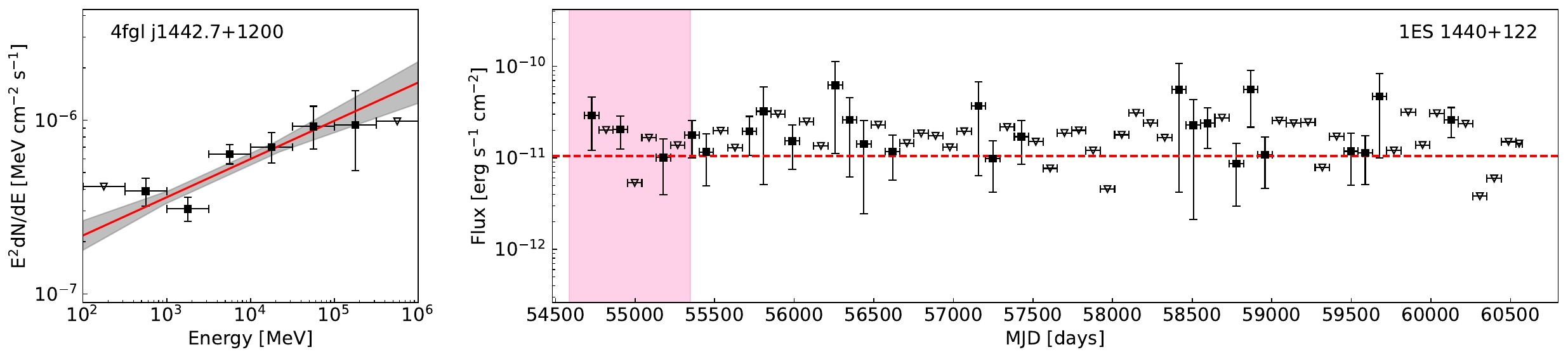}
\figsetgrpnote{Left column: the $\sim$16-yr Fermi-LAT average spectra along with the fitting results, where the open inverted triangles indicate TS<4 for that energy bin. Right column: the long-term Fermi-LAT light curves with time bins of 90 days, where the open inverted triangles indicate the TS value of that time bin less than 9. The horizontal red dash lines represent the $\sim$16 yr average $\gamma$-ray flux of the sources. The pink-shaded regions denote the period covered by the TeV observations, as well as other multiwavelength observations performed during the TeV campaigns. For sources such as 1ES 0347--121, PKS 0548--322, 1ES 1101--232, and 1ES 2356--309, whose TeV observations were conducted prior to the launch of the Fermi satellite, no such shaded regions appear in their Fermi-LAT light curves. For the four BL Lacs (Mrk 421, 1ES 1218+304, 1ES 1727+502 and 1ES 1959+650; marked with a red star in the light-curve figure panels) with significant variability, the contemporaneous Fermi-LAT spectra with the TeV observations are derived to construct their broadband SEDs in Figure \ref{Fig: SED fitting}, that is, their spectra in both the GeV and TeV bands from the observation periods marked as the pink-shaded regions in their light curves. For the other two sources (Mrk 501 and 1ES 2344+514, marked with a red rhombus in the light-curve figure panels) with significant variability, slightly extended observation periods (the green-shaded regions in their light curves) relative to their TeV campaigns (the pink-shaded regions in their light curves) are employed to generate the GeV spectrum for their broadband SED construction. For the remaining 19 objects, the 16-year Fermi-LAT averaged spectra are used to construct their broadband SEDs in Figure \ref{Fig: SED fitting}.}
\figsetgrpend

\figsetgrpstart
\figsetgrpnum{2.18}
\figsetgrptitle{Image of Mrk 501
}
\figsetplot{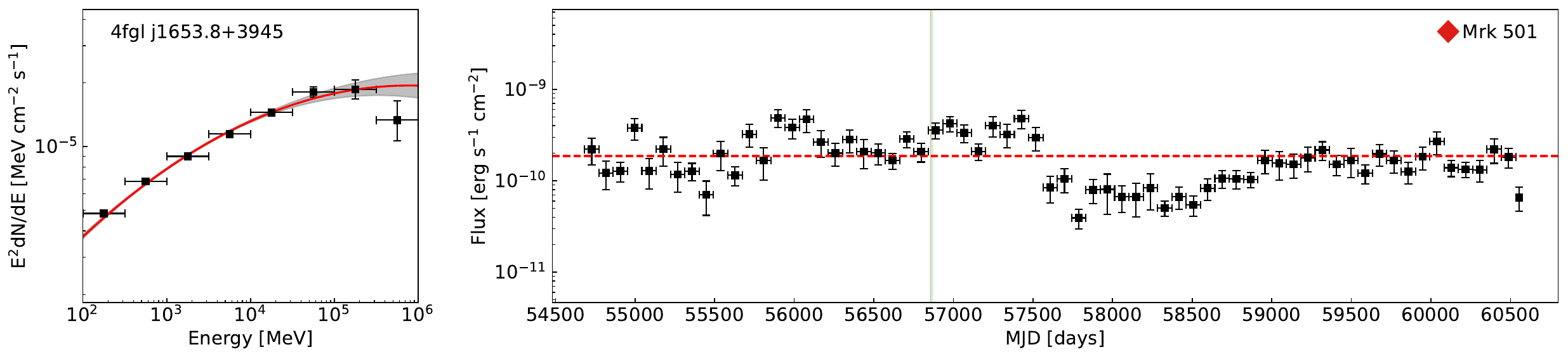}
\figsetgrpnote{Left column: the $\sim$16-yr Fermi-LAT average spectra along with the fitting results, where the open inverted triangles indicate TS<4 for that energy bin. Right column: the long-term Fermi-LAT light curves with time bins of 90 days, where the open inverted triangles indicate the TS value of that time bin less than 9. The horizontal red dash lines represent the $\sim$16 yr average $\gamma$-ray flux of the sources. The pink-shaded regions denote the period covered by the TeV observations, as well as other multiwavelength observations performed during the TeV campaigns. For sources such as 1ES 0347--121, PKS 0548--322, 1ES 1101--232, and 1ES 2356--309, whose TeV observations were conducted prior to the launch of the Fermi satellite, no such shaded regions appear in their Fermi-LAT light curves. For the four BL Lacs (Mrk 421, 1ES 1218+304, 1ES 1727+502 and 1ES 1959+650; marked with a red star in the light-curve figure panels) with significant variability, the contemporaneous Fermi-LAT spectra with the TeV observations are derived to construct their broadband SEDs in Figure \ref{Fig: SED fitting}, that is, their spectra in both the GeV and TeV bands from the observation periods marked as the pink-shaded regions in their light curves. For the other two sources (Mrk 501 and 1ES 2344+514, marked with a red rhombus in the light-curve figure panels) with significant variability, slightly extended observation periods (the green-shaded regions in their light curves) relative to their TeV campaigns (the pink-shaded regions in their light curves) are employed to generate the GeV spectrum for their broadband SED construction. For the remaining 19 objects, the 16-year Fermi-LAT averaged spectra are used to construct their broadband SEDs in Figure \ref{Fig: SED fitting}.}
\figsetgrpend

\figsetgrpstart
\figsetgrpnum{2.19}
\figsetgrptitle{Image of 1ES 1727+502
}
\figsetplot{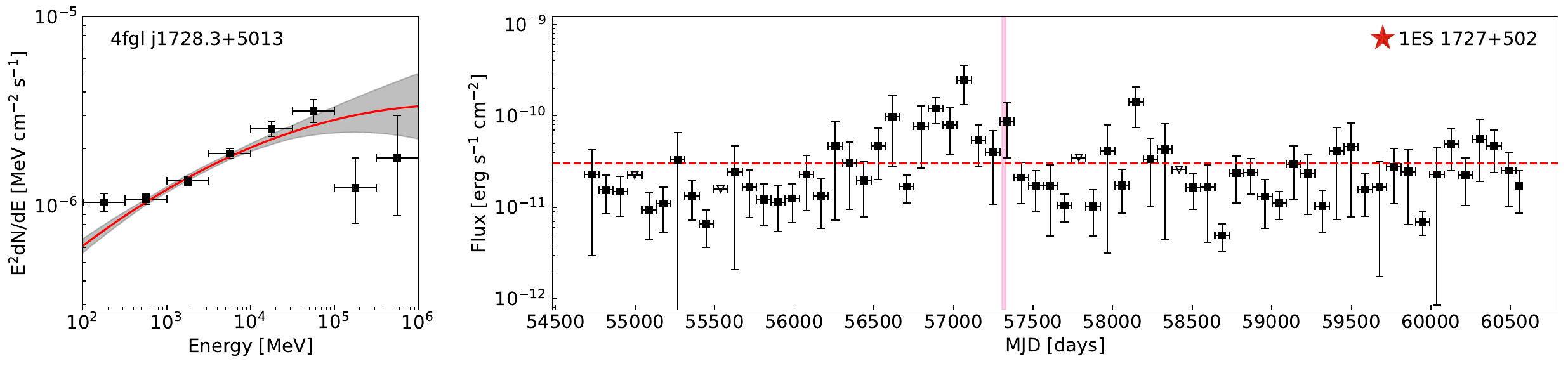}
\figsetgrpnote{Left column: the $\sim$16-yr Fermi-LAT average spectra along with the fitting results, where the open inverted triangles indicate TS<4 for that energy bin. Right column: the long-term Fermi-LAT light curves with time bins of 90 days, where the open inverted triangles indicate the TS value of that time bin less than 9. The horizontal red dash lines represent the $\sim$16 yr average $\gamma$-ray flux of the sources. The pink-shaded regions denote the period covered by the TeV observations, as well as other multiwavelength observations performed during the TeV campaigns. For sources such as 1ES 0347--121, PKS 0548--322, 1ES 1101--232, and 1ES 2356--309, whose TeV observations were conducted prior to the launch of the Fermi satellite, no such shaded regions appear in their Fermi-LAT light curves. For the four BL Lacs (Mrk 421, 1ES 1218+304, 1ES 1727+502 and 1ES 1959+650; marked with a red star in the light-curve figure panels) with significant variability, the contemporaneous Fermi-LAT spectra with the TeV observations are derived to construct their broadband SEDs in Figure \ref{Fig: SED fitting}, that is, their spectra in both the GeV and TeV bands from the observation periods marked as the pink-shaded regions in their light curves. For the other two sources (Mrk 501 and 1ES 2344+514, marked with a red rhombus in the light-curve figure panels) with significant variability, slightly extended observation periods (the green-shaded regions in their light curves) relative to their TeV campaigns (the pink-shaded regions in their light curves) are employed to generate the GeV spectrum for their broadband SED construction. For the remaining 19 objects, the 16-year Fermi-LAT averaged spectra are used to construct their broadband SEDs in Figure \ref{Fig: SED fitting}.}
\figsetgrpend

\figsetgrpstart
\figsetgrpnum{2.20}
\figsetgrptitle{Image of 1ES 1741+196
}
\figsetplot{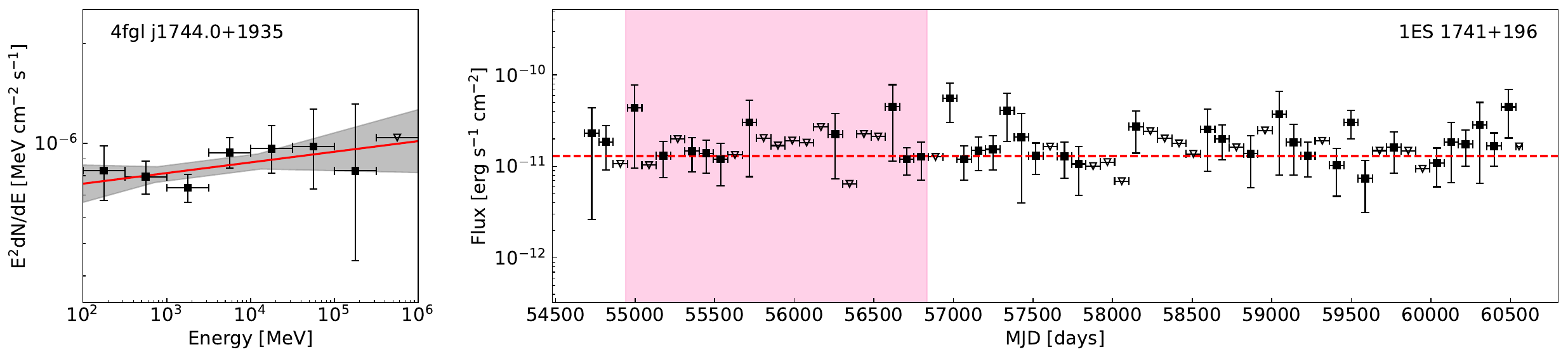}
\figsetgrpnote{Left column: the $\sim$16-yr Fermi-LAT average spectra along with the fitting results, where the open inverted triangles indicate TS<4 for that energy bin. Right column: the long-term Fermi-LAT light curves with time bins of 90 days, where the open inverted triangles indicate the TS value of that time bin less than 9. The horizontal red dash lines represent the $\sim$16 yr average $\gamma$-ray flux of the sources. The pink-shaded regions denote the period covered by the TeV observations, as well as other multiwavelength observations performed during the TeV campaigns. For sources such as 1ES 0347--121, PKS 0548--322, 1ES 1101--232, and 1ES 2356--309, whose TeV observations were conducted prior to the launch of the Fermi satellite, no such shaded regions appear in their Fermi-LAT light curves. For the four BL Lacs (Mrk 421, 1ES 1218+304, 1ES 1727+502 and 1ES 1959+650; marked with a red star in the light-curve figure panels) with significant variability, the contemporaneous Fermi-LAT spectra with the TeV observations are derived to construct their broadband SEDs in Figure \ref{Fig: SED fitting}, that is, their spectra in both the GeV and TeV bands from the observation periods marked as the pink-shaded regions in their light curves. For the other two sources (Mrk 501 and 1ES 2344+514, marked with a red rhombus in the light-curve figure panels) with significant variability, slightly extended observation periods (the green-shaded regions in their light curves) relative to their TeV campaigns (the pink-shaded regions in their light curves) are employed to generate the GeV spectrum for their broadband SED construction. For the remaining 19 objects, the 16-year Fermi-LAT averaged spectra are used to construct their broadband SEDs in Figure \ref{Fig: SED fitting}.}
\figsetgrpend

\figsetgrpstart
\figsetgrpnum{2.21}
\figsetgrptitle{Image of 1RXS J195815.6--301119
}
\figsetplot{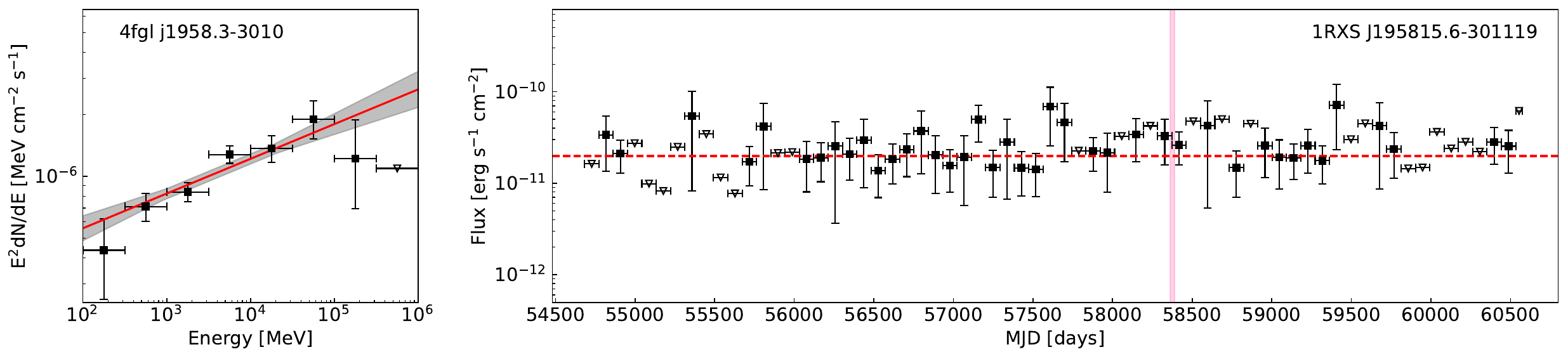}
\figsetgrpnote{Left column: the $\sim$16-yr Fermi-LAT average spectra along with the fitting results, where the open inverted triangles indicate TS<4 for that energy bin. Right column: the long-term Fermi-LAT light curves with time bins of 90 days, where the open inverted triangles indicate the TS value of that time bin less than 9. The horizontal red dash lines represent the $\sim$16 yr average $\gamma$-ray flux of the sources. The pink-shaded regions denote the period covered by the TeV observations, as well as other multiwavelength observations performed during the TeV campaigns. For sources such as 1ES 0347--121, PKS 0548--322, 1ES 1101--232, and 1ES 2356--309, whose TeV observations were conducted prior to the launch of the Fermi satellite, no such shaded regions appear in their Fermi-LAT light curves. For the four BL Lacs (Mrk 421, 1ES 1218+304, 1ES 1727+502 and 1ES 1959+650; marked with a red star in the light-curve figure panels) with significant variability, the contemporaneous Fermi-LAT spectra with the TeV observations are derived to construct their broadband SEDs in Figure \ref{Fig: SED fitting}, that is, their spectra in both the GeV and TeV bands from the observation periods marked as the pink-shaded regions in their light curves. For the other two sources (Mrk 501 and 1ES 2344+514, marked with a red rhombus in the light-curve figure panels) with significant variability, slightly extended observation periods (the green-shaded regions in their light curves) relative to their TeV campaigns (the pink-shaded regions in their light curves) are employed to generate the GeV spectrum for their broadband SED construction. For the remaining 19 objects, the 16-year Fermi-LAT averaged spectra are used to construct their broadband SEDs in Figure \ref{Fig: SED fitting}.}
\figsetgrpend

\figsetgrpstart
\figsetgrpnum{2.22}
\figsetgrptitle{Image of 1ES 1959+650
}
\figsetplot{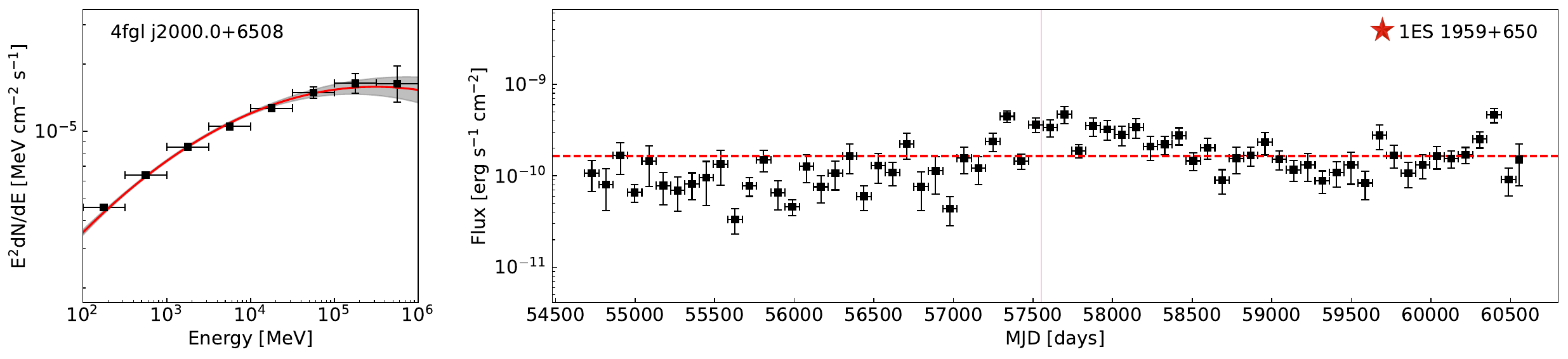}
\figsetgrpnote{Left column: the $\sim$16-yr Fermi-LAT average spectra along with the fitting results, where the open inverted triangles indicate TS<4 for that energy bin. Right column: the long-term Fermi-LAT light curves with time bins of 90 days, where the open inverted triangles indicate the TS value of that time bin less than 9. The horizontal red dash lines represent the $\sim$16 yr average $\gamma$-ray flux of the sources. The pink-shaded regions denote the period covered by the TeV observations, as well as other multiwavelength observations performed during the TeV campaigns. For sources such as 1ES 0347--121, PKS 0548--322, 1ES 1101--232, and 1ES 2356--309, whose TeV observations were conducted prior to the launch of the Fermi satellite, no such shaded regions appear in their Fermi-LAT light curves. For the four BL Lacs (Mrk 421, 1ES 1218+304, 1ES 1727+502 and 1ES 1959+650; marked with a red star in the light-curve figure panels) with significant variability, the contemporaneous Fermi-LAT spectra with the TeV observations are derived to construct their broadband SEDs in Figure \ref{Fig: SED fitting}, that is, their spectra in both the GeV and TeV bands from the observation periods marked as the pink-shaded regions in their light curves. For the other two sources (Mrk 501 and 1ES 2344+514, marked with a red rhombus in the light-curve figure panels) with significant variability, slightly extended observation periods (the green-shaded regions in their light curves) relative to their TeV campaigns (the pink-shaded regions in their light curves) are employed to generate the GeV spectrum for their broadband SED construction. For the remaining 19 objects, the 16-year Fermi-LAT averaged spectra are used to construct their broadband SEDs in Figure \ref{Fig: SED fitting}.}
\figsetgrpend

\figsetgrpstart
\figsetgrpnum{2.23}
\figsetgrptitle{Image of 1ES 2037+521
}
\figsetplot{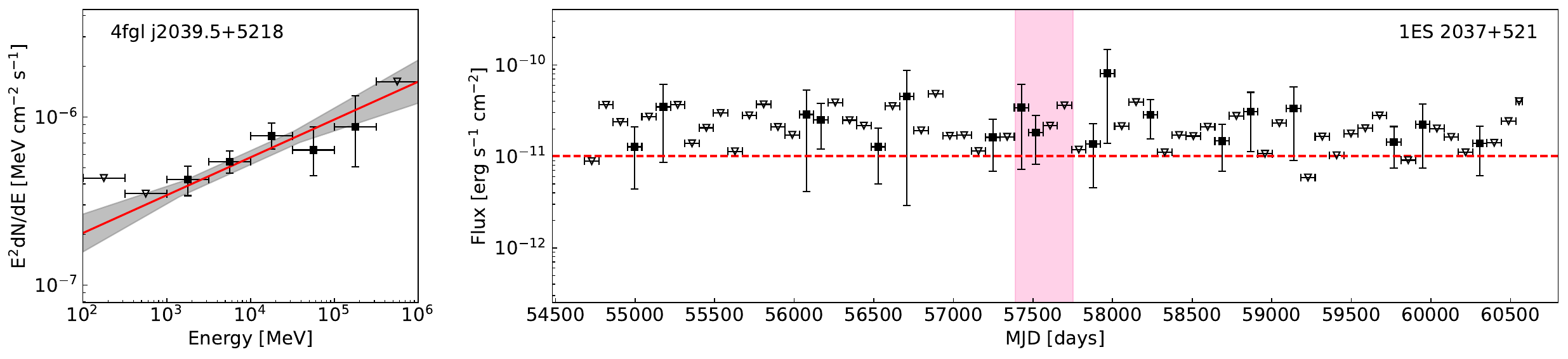}
\figsetgrpnote{Left column: the $\sim$16-yr Fermi-LAT average spectra along with the fitting results, where the open inverted triangles indicate TS<4 for that energy bin. Right column: the long-term Fermi-LAT light curves with time bins of 90 days, where the open inverted triangles indicate the TS value of that time bin less than 9. The horizontal red dash lines represent the $\sim$16 yr average $\gamma$-ray flux of the sources. The pink-shaded regions denote the period covered by the TeV observations, as well as other multiwavelength observations performed during the TeV campaigns. For sources such as 1ES 0347--121, PKS 0548--322, 1ES 1101--232, and 1ES 2356--309, whose TeV observations were conducted prior to the launch of the Fermi satellite, no such shaded regions appear in their Fermi-LAT light curves. For the four BL Lacs (Mrk 421, 1ES 1218+304, 1ES 1727+502 and 1ES 1959+650; marked with a red star in the light-curve figure panels) with significant variability, the contemporaneous Fermi-LAT spectra with the TeV observations are derived to construct their broadband SEDs in Figure \ref{Fig: SED fitting}, that is, their spectra in both the GeV and TeV bands from the observation periods marked as the pink-shaded regions in their light curves. For the other two sources (Mrk 501 and 1ES 2344+514, marked with a red rhombus in the light-curve figure panels) with significant variability, slightly extended observation periods (the green-shaded regions in their light curves) relative to their TeV campaigns (the pink-shaded regions in their light curves) are employed to generate the GeV spectrum for their broadband SED construction. For the remaining 19 objects, the 16-year Fermi-LAT averaged spectra are used to construct their broadband SEDs in Figure \ref{Fig: SED fitting}.}
\figsetgrpend

\figsetgrpstart
\figsetgrpnum{2.24}
\figsetgrptitle{Image of 1ES 2344+514
}
\figsetplot{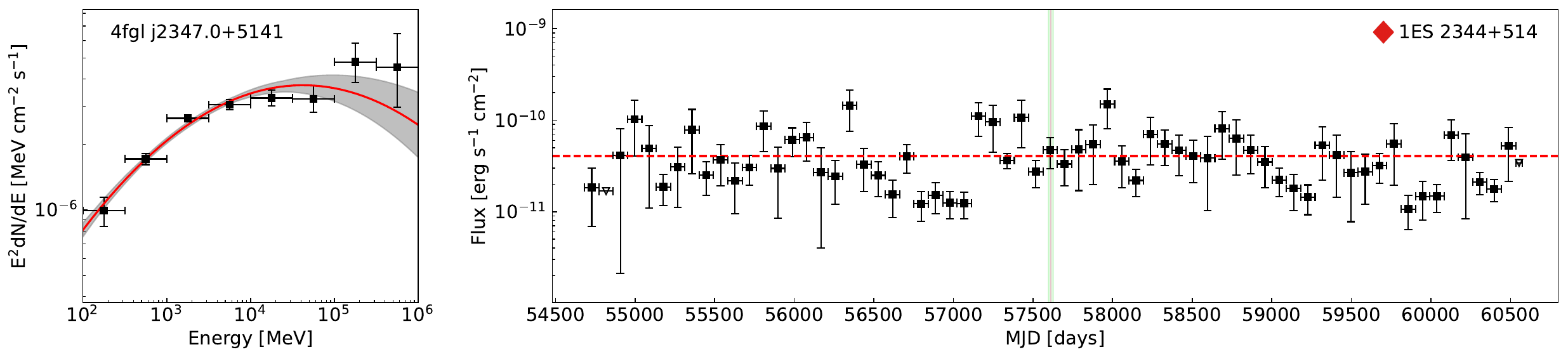}
\figsetgrpnote{Left column: the $\sim$16-yr Fermi-LAT average spectra along with the fitting results, where the open inverted triangles indicate TS<4 for that energy bin. Right column: the long-term Fermi-LAT light curves with time bins of 90 days, where the open inverted triangles indicate the TS value of that time bin less than 9. The horizontal red dash lines represent the $\sim$16 yr average $\gamma$-ray flux of the sources. The pink-shaded regions denote the period covered by the TeV observations, as well as other multiwavelength observations performed during the TeV campaigns. For sources such as 1ES 0347--121, PKS 0548--322, 1ES 1101--232, and 1ES 2356--309, whose TeV observations were conducted prior to the launch of the Fermi satellite, no such shaded regions appear in their Fermi-LAT light curves. For the four BL Lacs (Mrk 421, 1ES 1218+304, 1ES 1727+502 and 1ES 1959+650; marked with a red star in the light-curve figure panels) with significant variability, the contemporaneous Fermi-LAT spectra with the TeV observations are derived to construct their broadband SEDs in Figure \ref{Fig: SED fitting}, that is, their spectra in both the GeV and TeV bands from the observation periods marked as the pink-shaded regions in their light curves. For the other two sources (Mrk 501 and 1ES 2344+514, marked with a red rhombus in the light-curve figure panels) with significant variability, slightly extended observation periods (the green-shaded regions in their light curves) relative to their TeV campaigns (the pink-shaded regions in their light curves) are employed to generate the GeV spectrum for their broadband SED construction. For the remaining 19 objects, the 16-year Fermi-LAT averaged spectra are used to construct their broadband SEDs in Figure \ref{Fig: SED fitting}.}
\figsetgrpend

\figsetgrpstart
\figsetgrpnum{2.25}
\figsetgrptitle{Image of H 2356--309}
\figsetplot{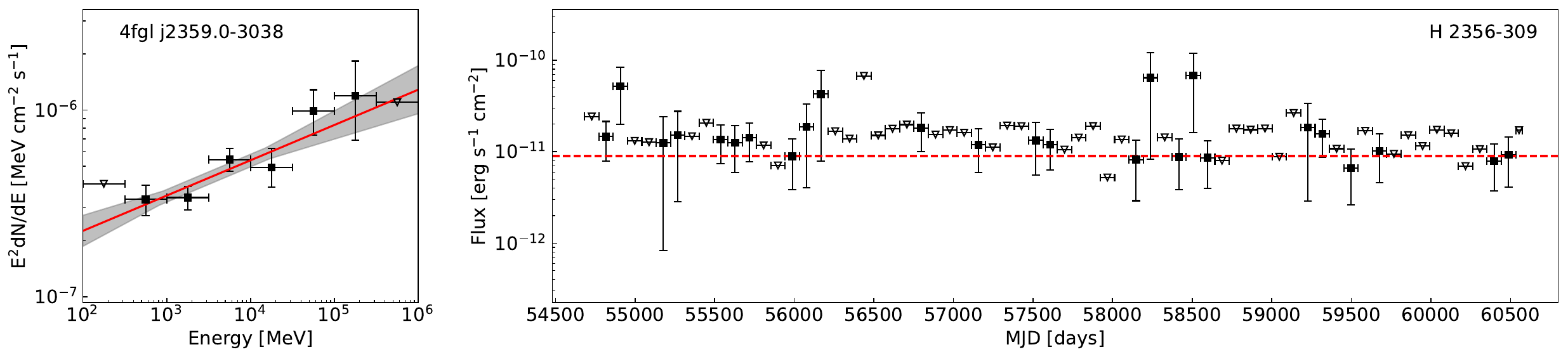}
\figsetgrpnote{Left column: the $\sim$16-yr Fermi-LAT average spectra along with the fitting results, where the open inverted triangles indicate TS<4 for that energy bin. Right column: the long-term Fermi-LAT light curves with time bins of 90 days, where the open inverted triangles indicate the TS value of that time bin less than 9. The horizontal red dash lines represent the $\sim$16 yr average $\gamma$-ray flux of the sources. The pink-shaded regions denote the period covered by the TeV observations, as well as other multiwavelength observations performed during the TeV campaigns. For sources such as 1ES 0347--121, PKS 0548--322, 1ES 1101--232, and 1ES 2356--309, whose TeV observations were conducted prior to the launch of the Fermi satellite, no such shaded regions appear in their Fermi-LAT light curves. For the four BL Lacs (Mrk 421, 1ES 1218+304, 1ES 1727+502 and 1ES 1959+650; marked with a red star in the light-curve figure panels) with significant variability, the contemporaneous Fermi-LAT spectra with the TeV observations are derived to construct their broadband SEDs in Figure \ref{Fig: SED fitting}, that is, their spectra in both the GeV and TeV bands from the observation periods marked as the pink-shaded regions in their light curves. For the other two sources (Mrk 501 and 1ES 2344+514, marked with a red rhombus in the light-curve figure panels) with significant variability, slightly extended observation periods (the green-shaded regions in their light curves) relative to their TeV campaigns (the pink-shaded regions in their light curves) are employed to generate the GeV spectrum for their broadband SED construction. For the remaining 19 objects, the 16-year Fermi-LAT averaged spectra are used to construct their broadband SEDs in Figure \ref{Fig: SED fitting}.}
\figsetgrpend

\figsetend

\begin{figure*} 
        \includegraphics[angle=0,width=0.33\linewidth]{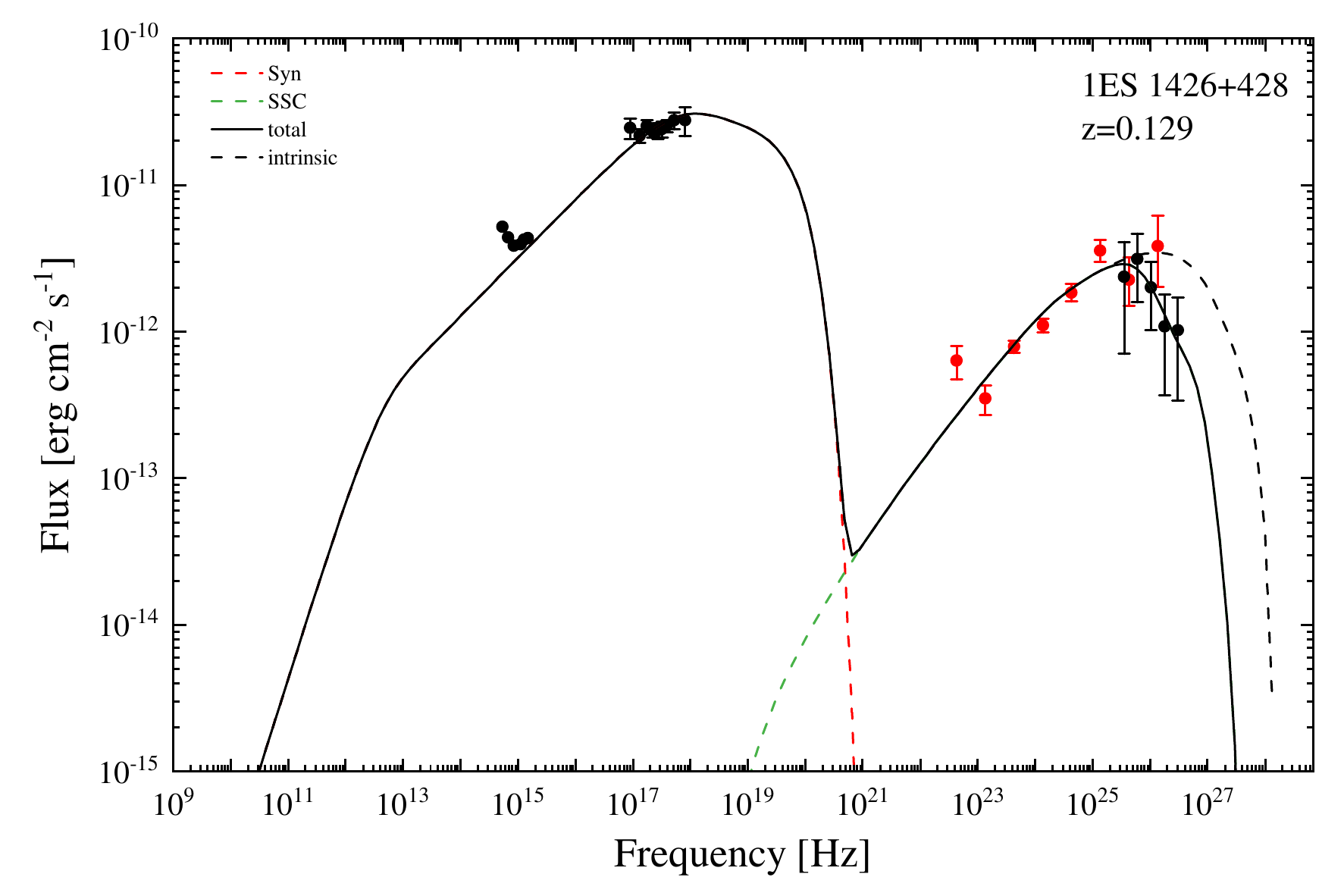} 
        \includegraphics[angle=0,width=0.33\linewidth]{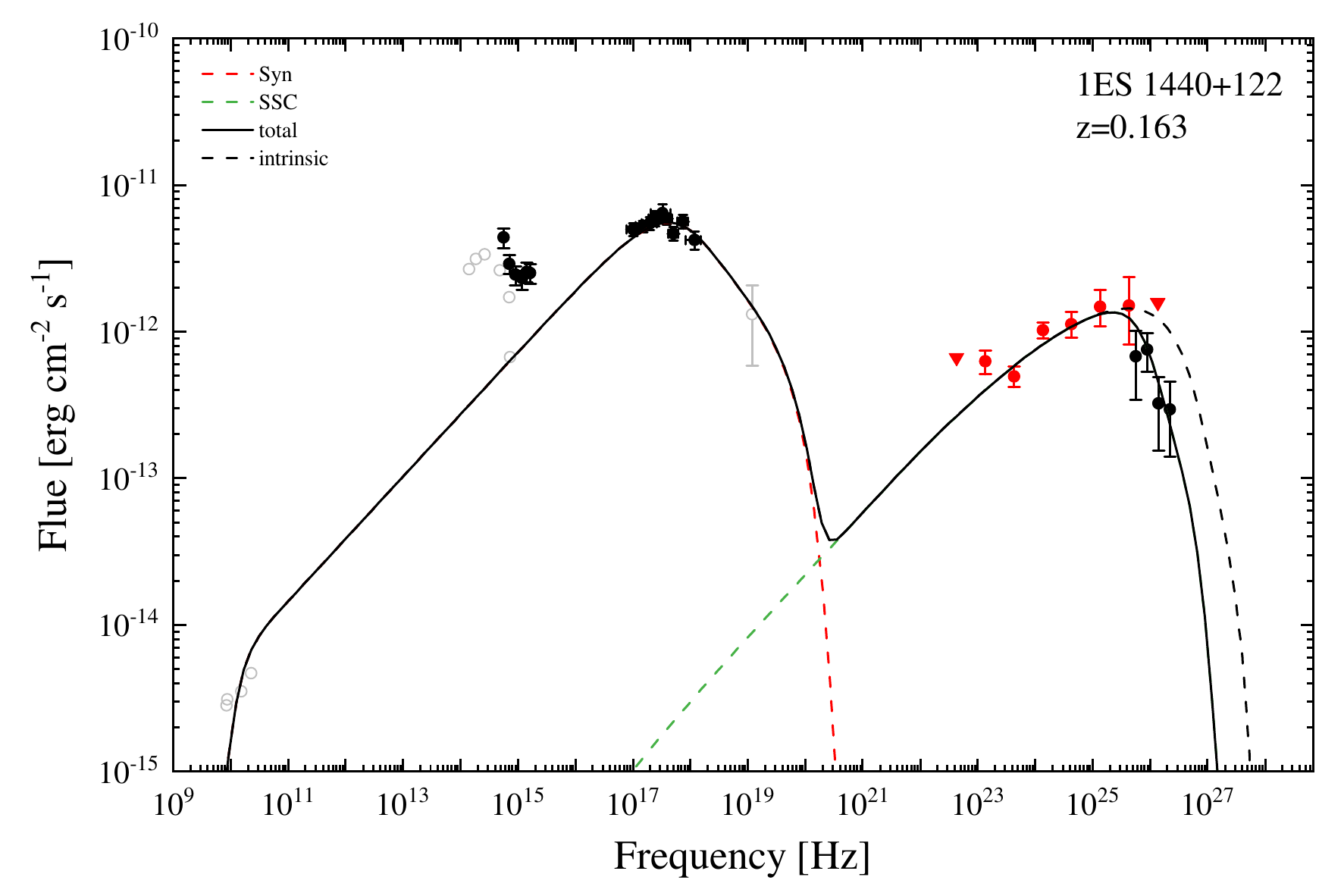}     
        \includegraphics[angle=0,width=0.33\linewidth]{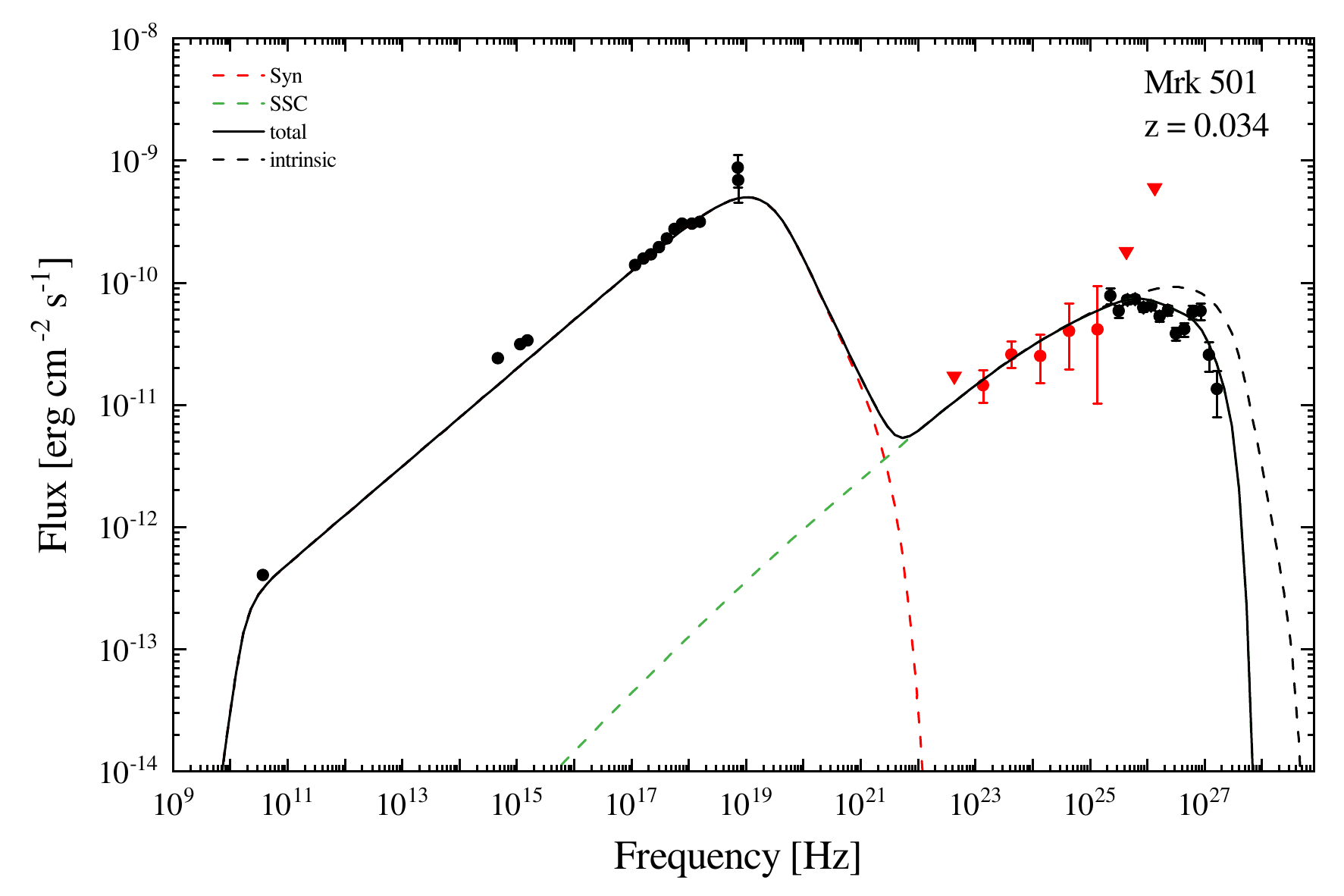}\\  
        \includegraphics[angle=0,width=0.33\linewidth]{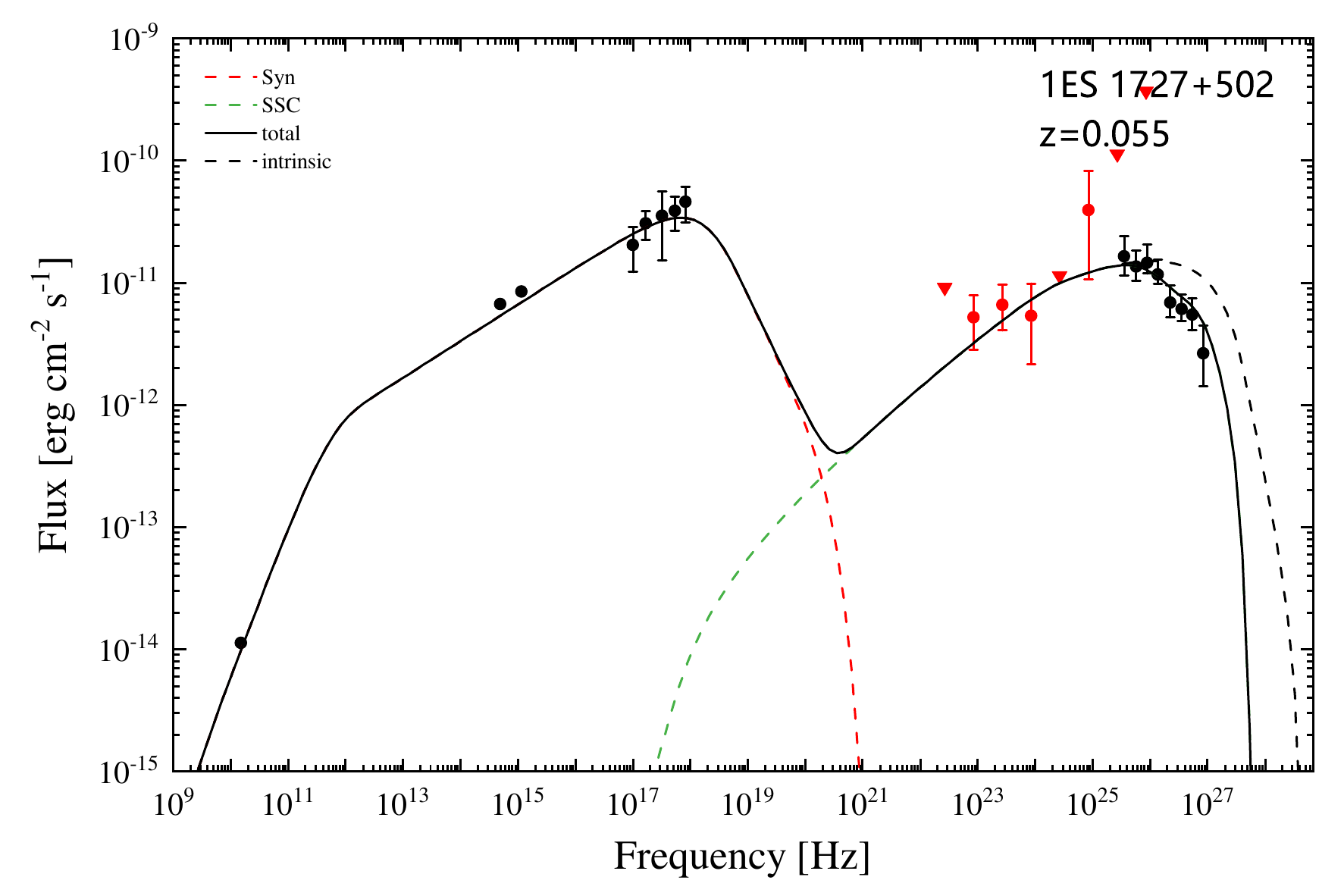}
        \includegraphics[angle=0,width=0.33\linewidth]{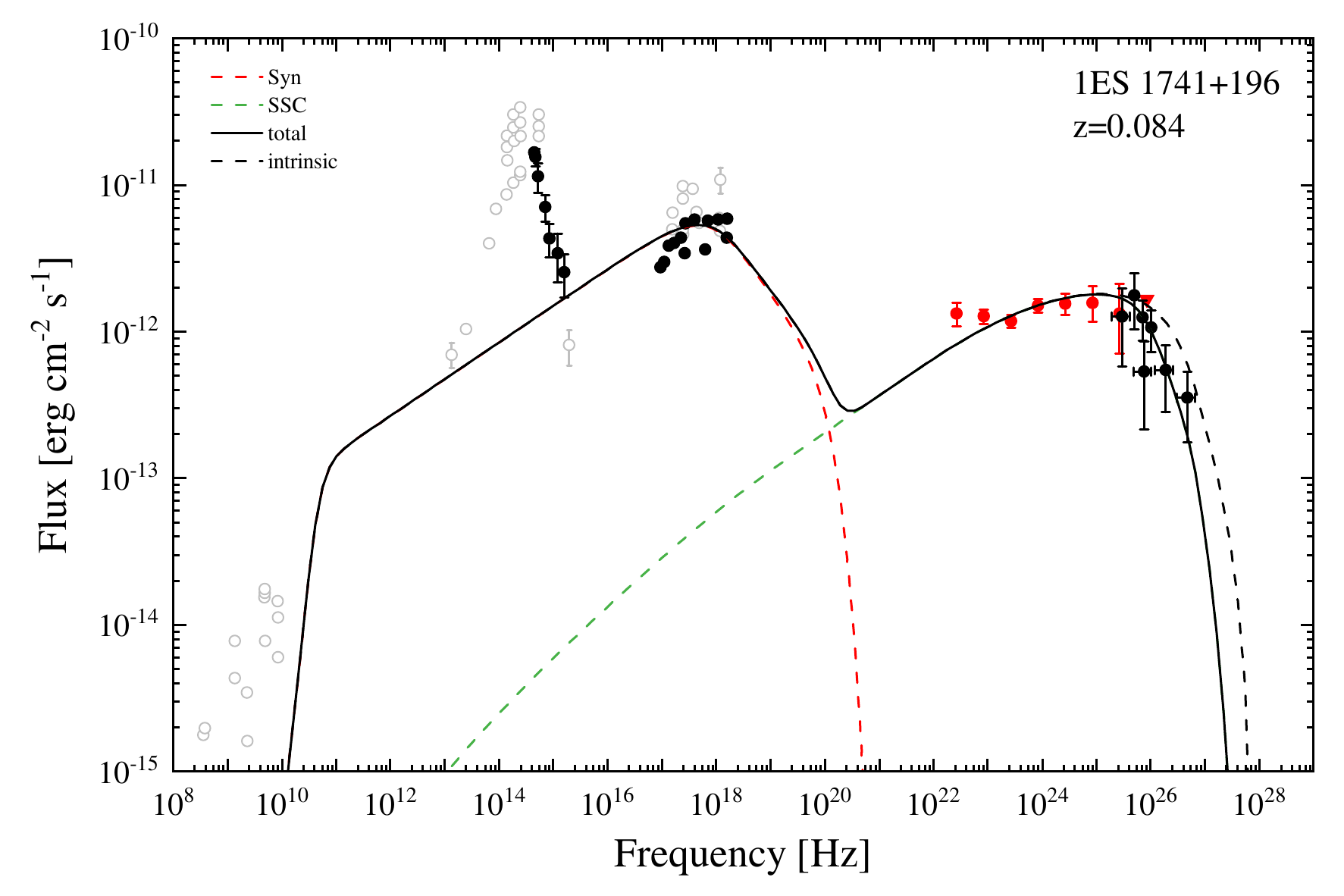}
        \includegraphics[angle=0,width=0.33\linewidth]{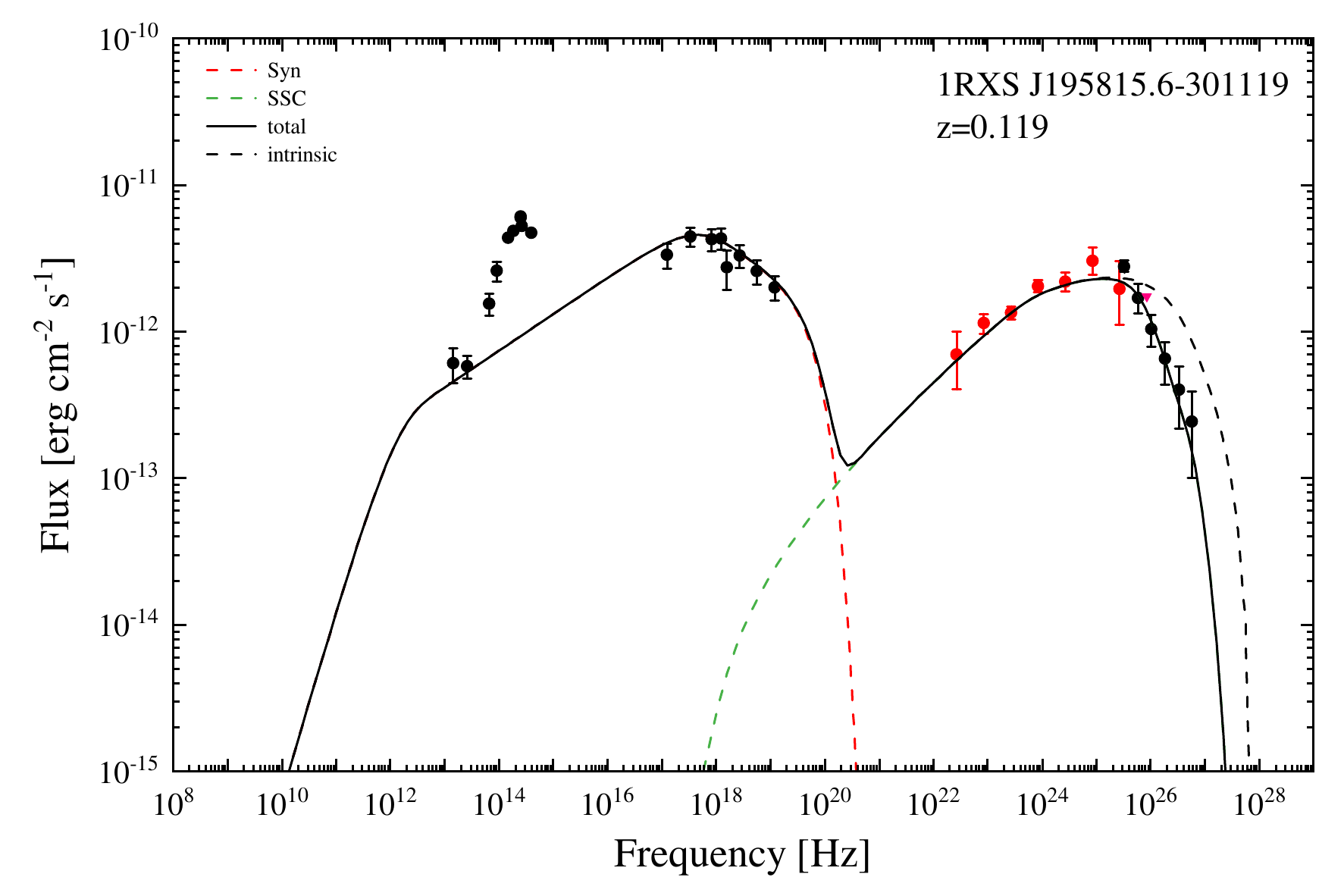}\\
       \includegraphics[angle=0,width=0.33\linewidth]{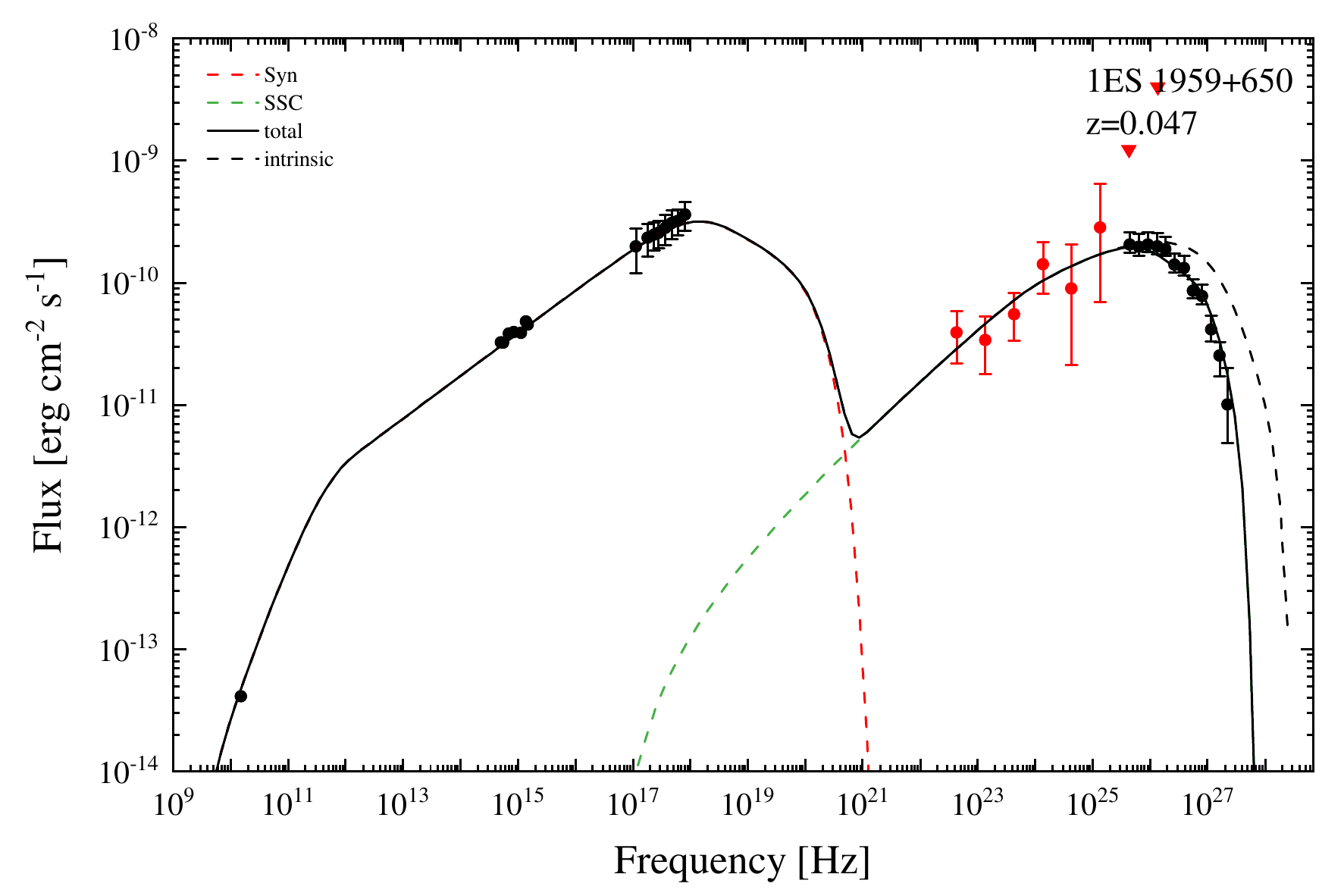}
       \includegraphics[angle=0,width=0.33\linewidth]{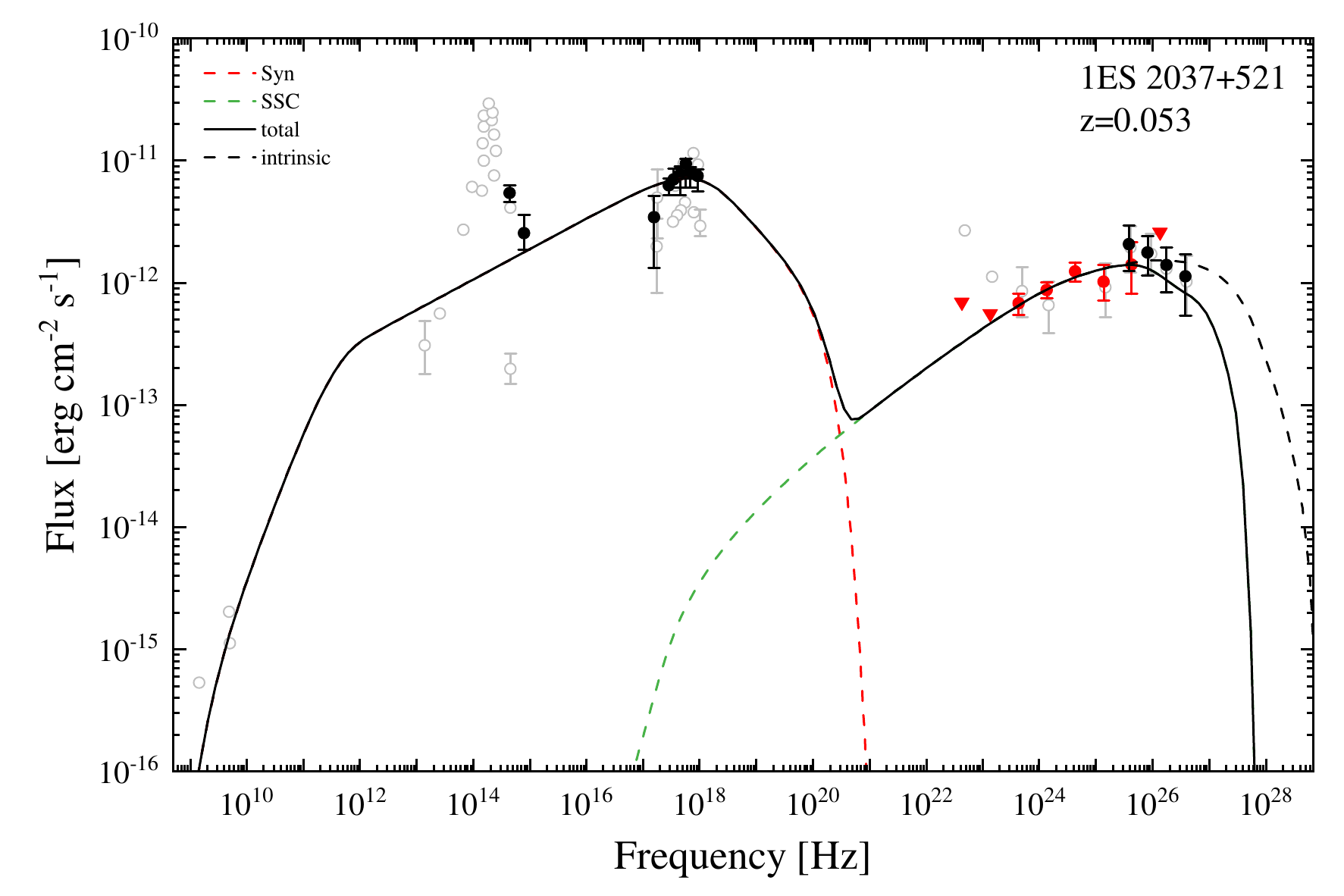}
        \includegraphics[angle=0,width=0.33\linewidth]{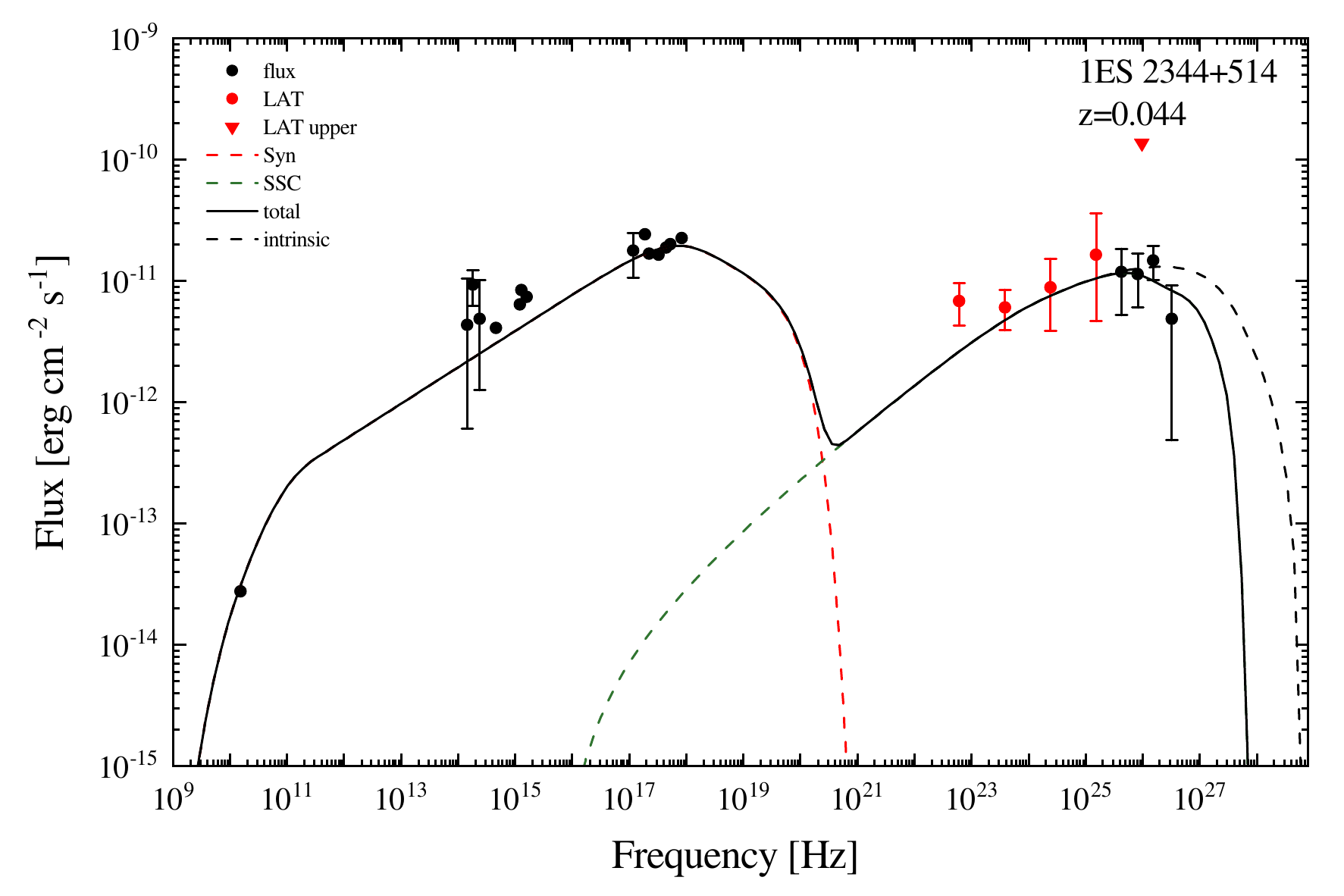}\\  
        \includegraphics[angle=0,width=0.33\linewidth]{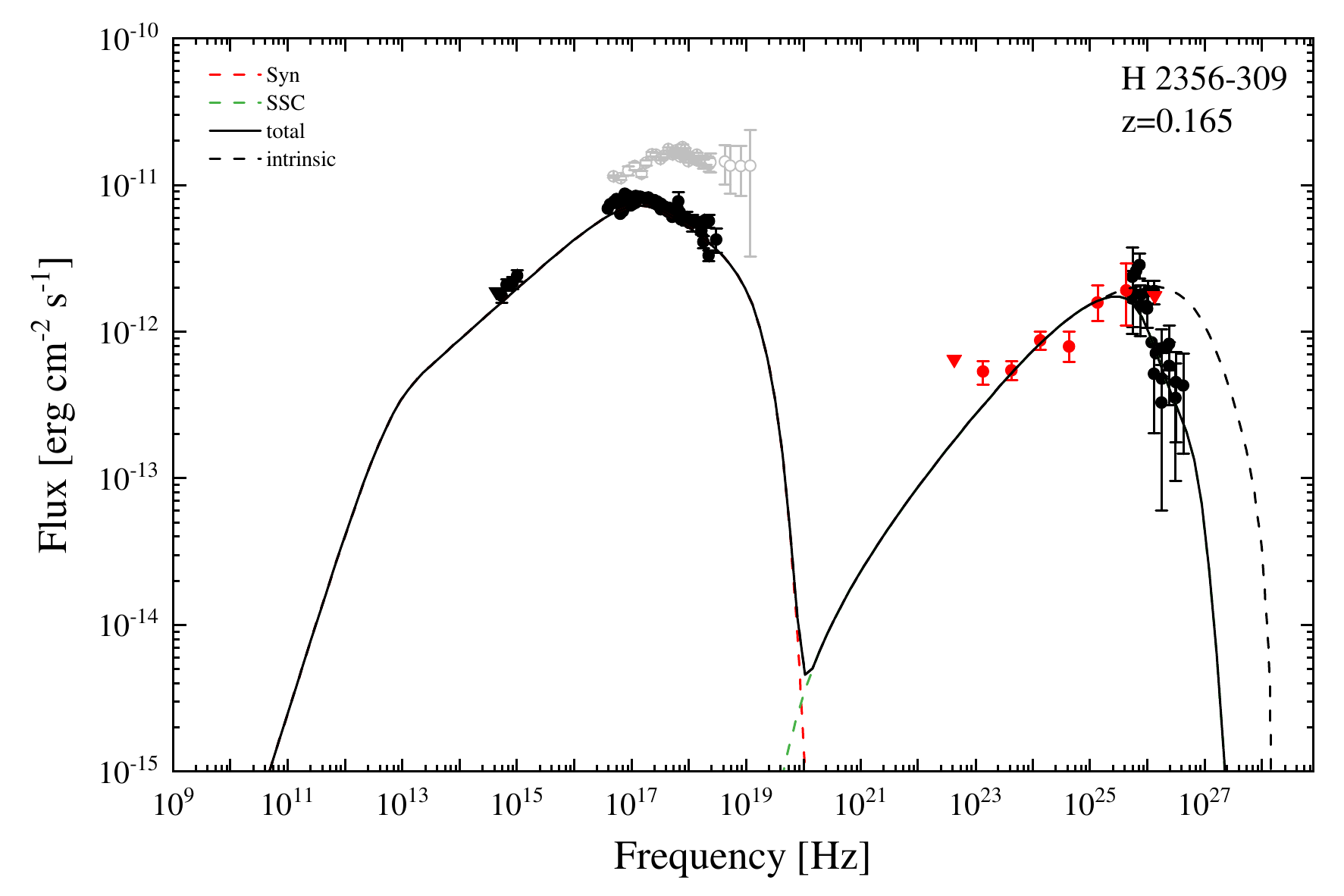}
    \center{Figure \ref{Fig: SED fitting}. (Continued.)}   
\end{figure*}

\begin{figure*}
    \centering   
    \includegraphics[angle=0,width=0.9\linewidth]{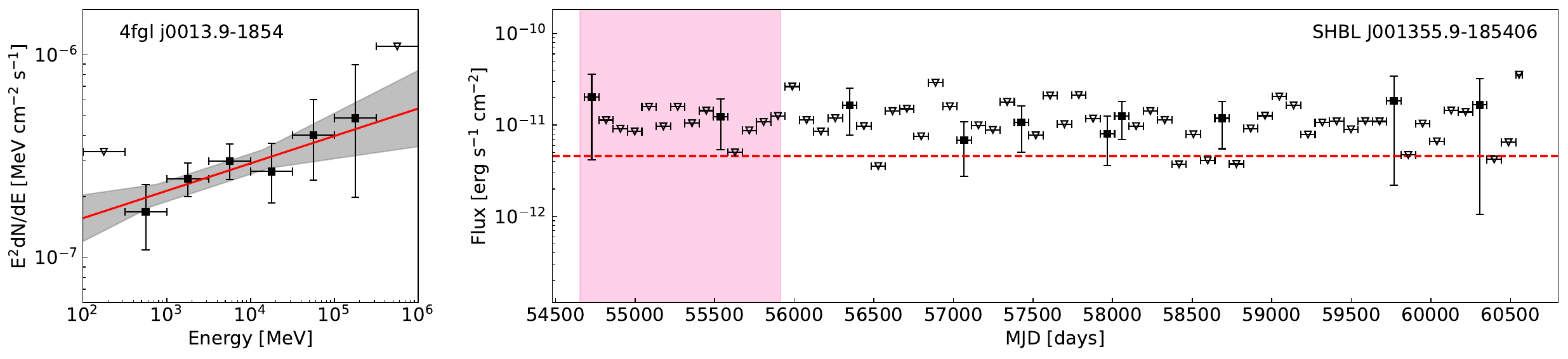} \\ 
    \includegraphics[angle=0,width=0.9\linewidth]{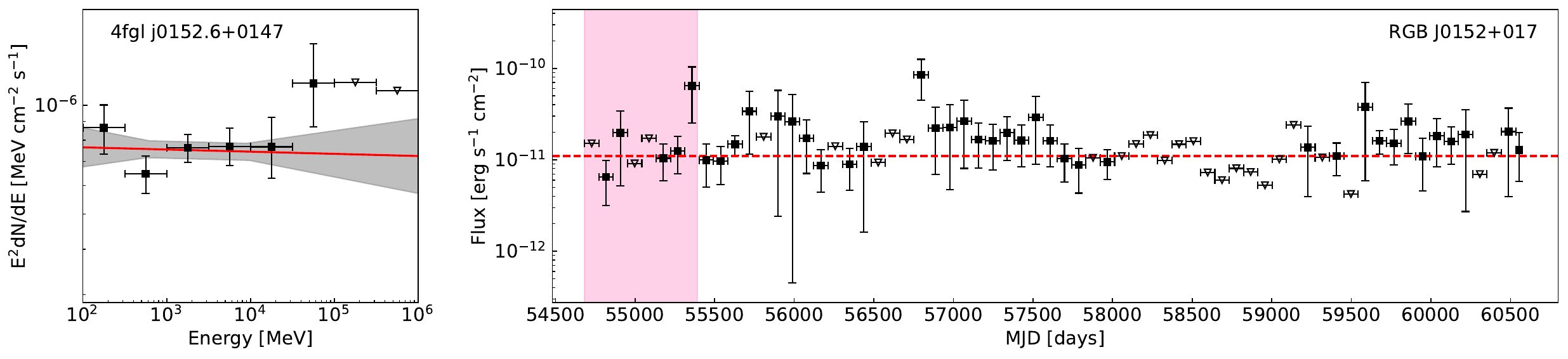}         \\  
    \includegraphics[angle=0,width=0.9\linewidth]{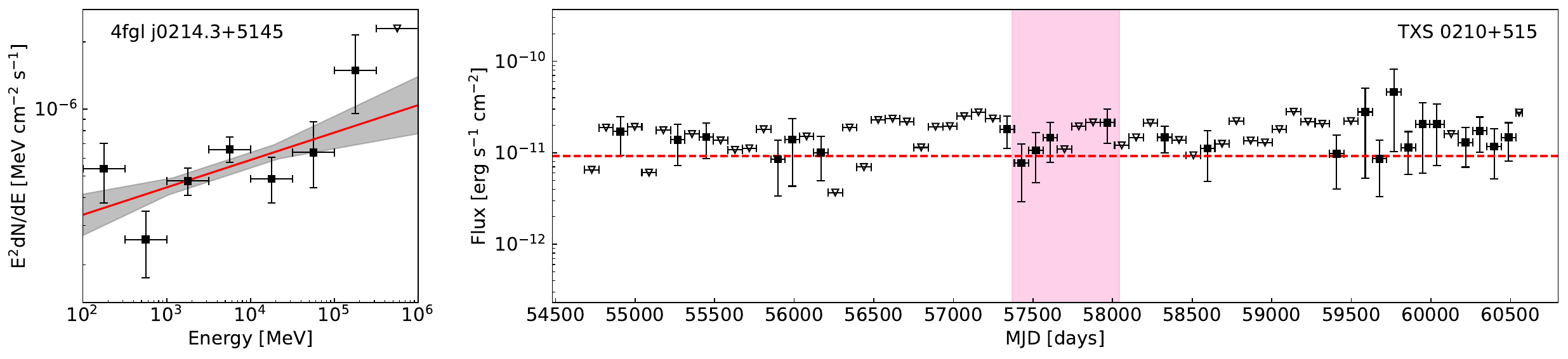}         \\      
    \includegraphics[angle=0,width=0.9\linewidth]{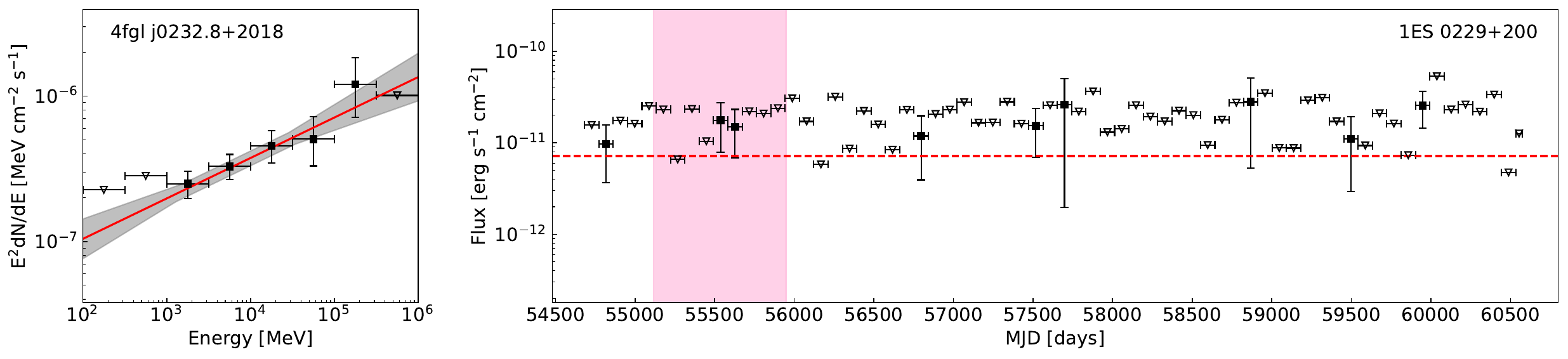}        \\
    \includegraphics[angle=0,width=0.9\linewidth]{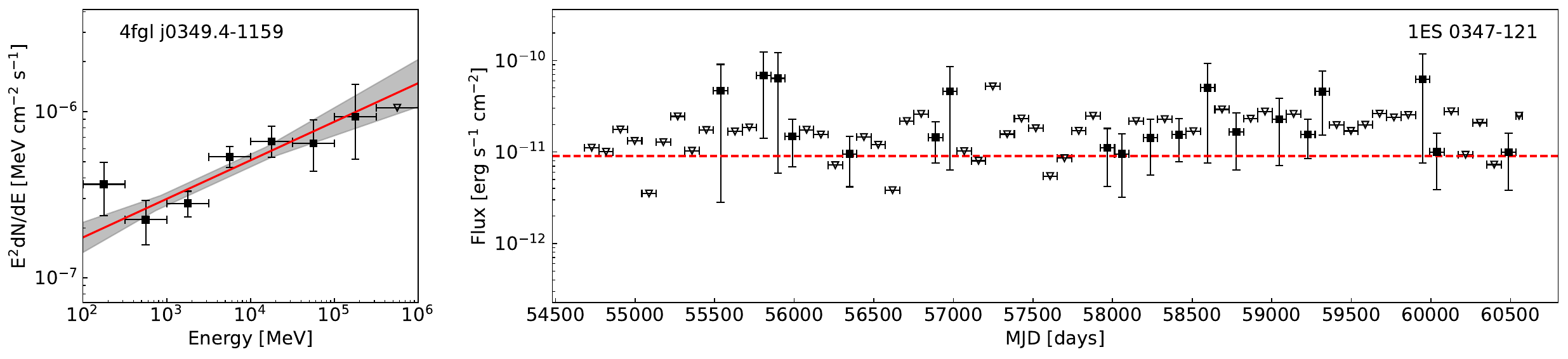}        \\
\end{figure*}     
\begin{figure*}
    \centering    
    \includegraphics[angle=0,width=\textwidth]{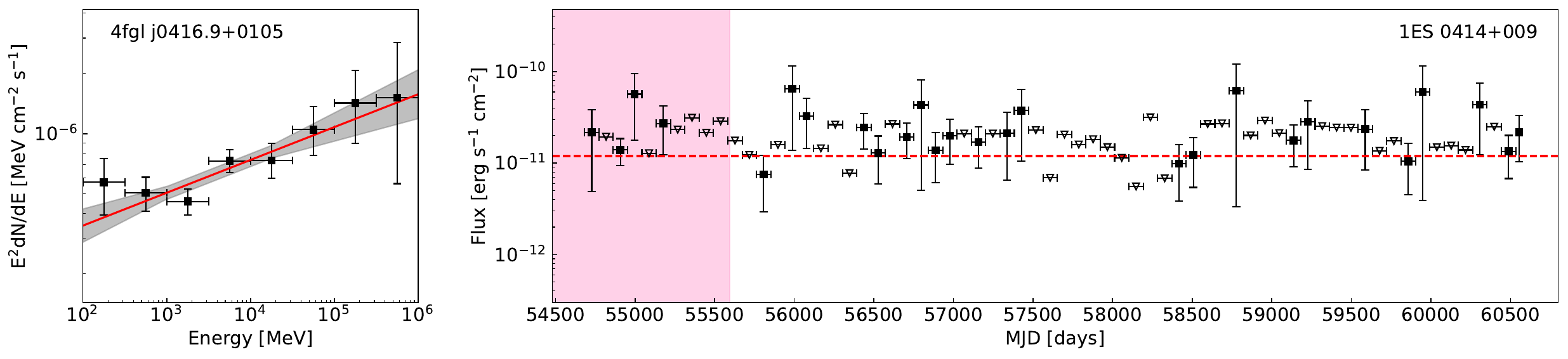}        \\
       \includegraphics[angle=0,width=\textwidth]{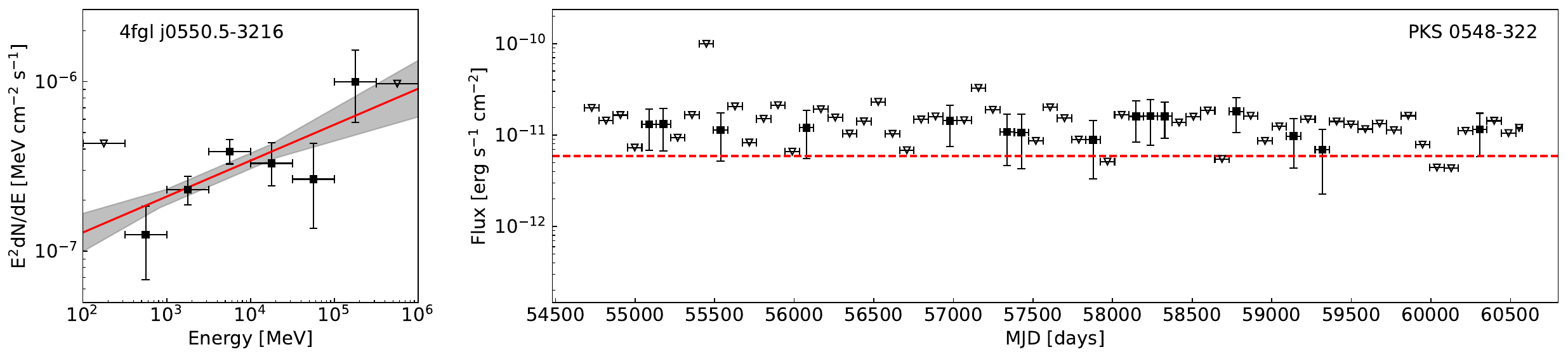}    \\   
        \includegraphics[angle=0,width=\textwidth]{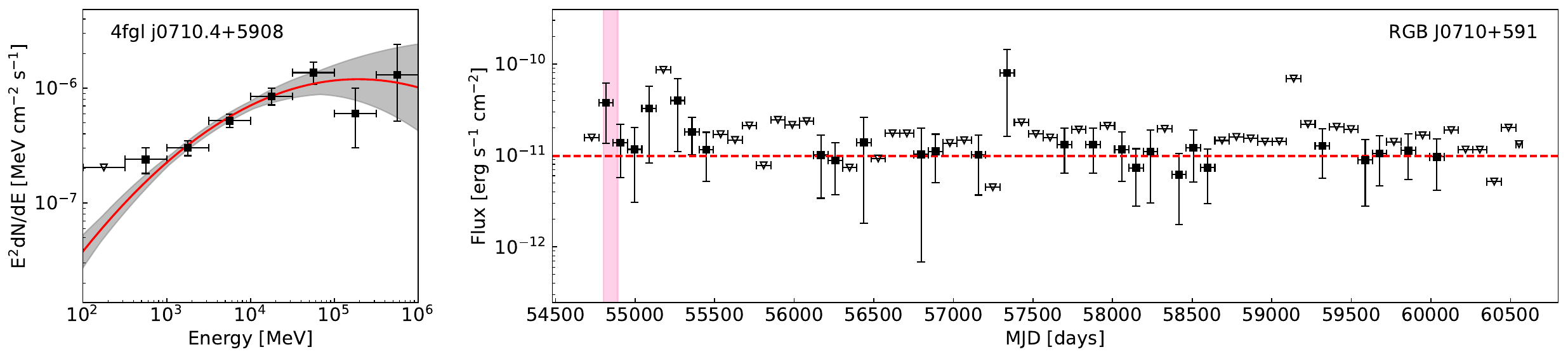} \\       
        \includegraphics[angle=0,width=\textwidth]{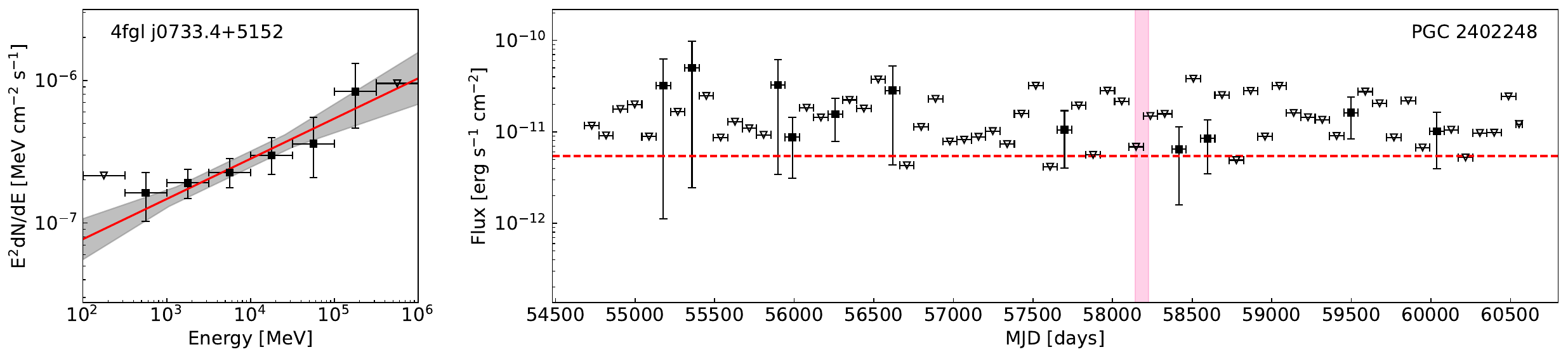}   \\       
       \includegraphics[angle=0,width=\textwidth]{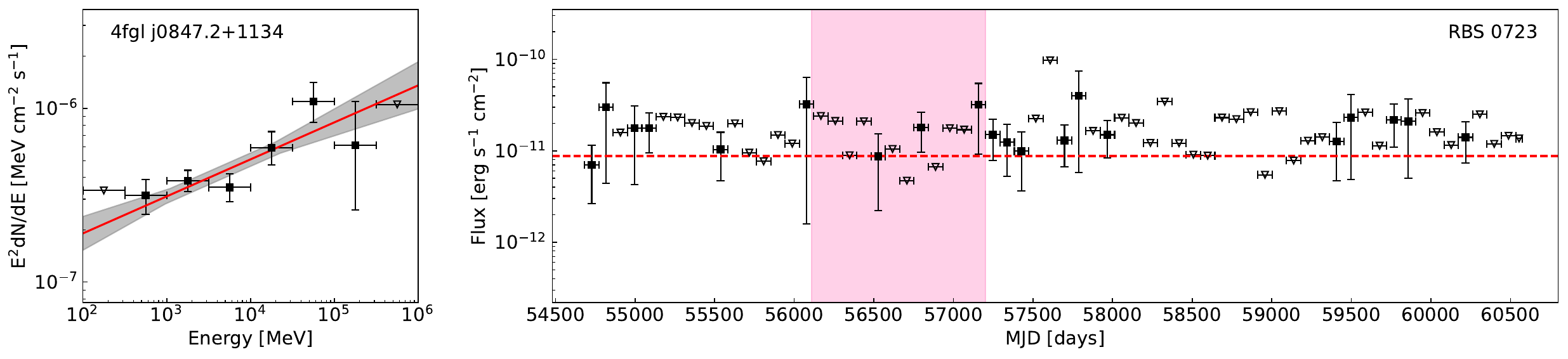}   \\  
\end{figure*}   
\begin{figure*}
    \centering
        \includegraphics[angle=0,width=\textwidth]{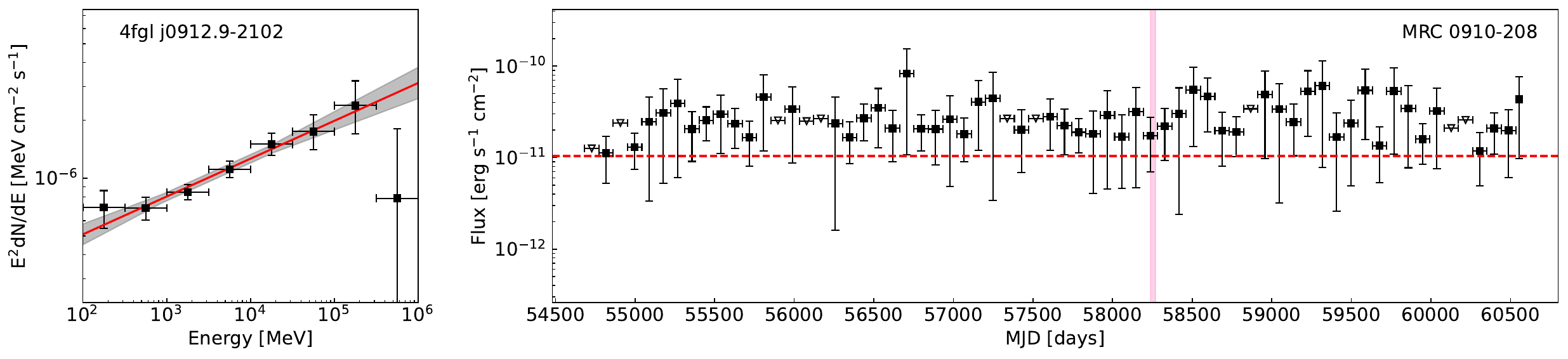}   \\  
        \includegraphics[angle=0,width=\textwidth]{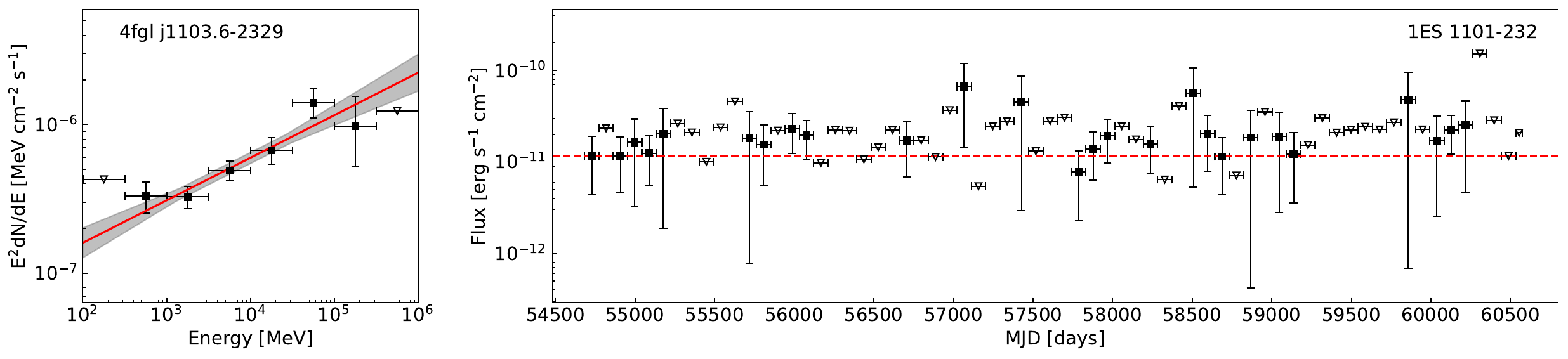} \\     
        \includegraphics[angle=0,width=\textwidth]{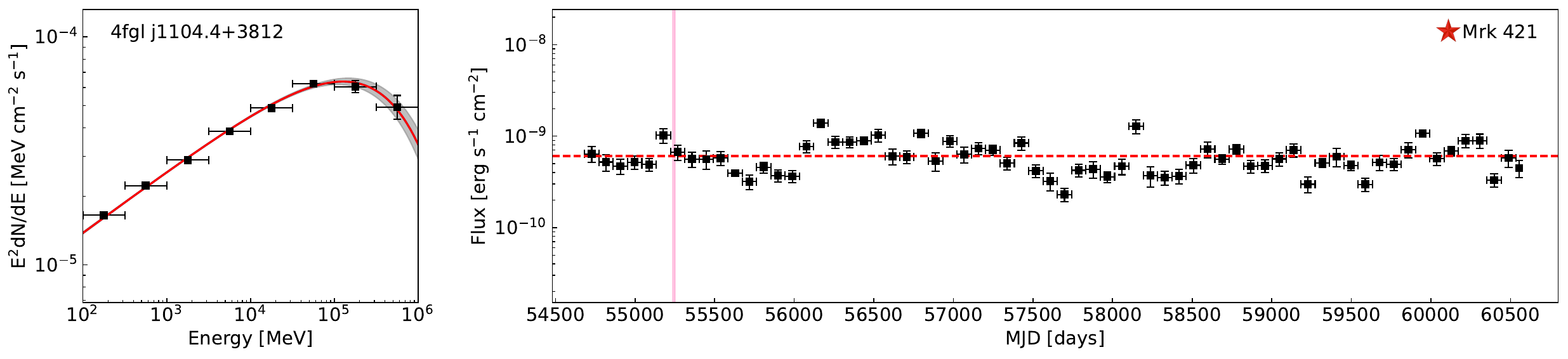}  \\   
        \includegraphics[angle=0,width=\textwidth]{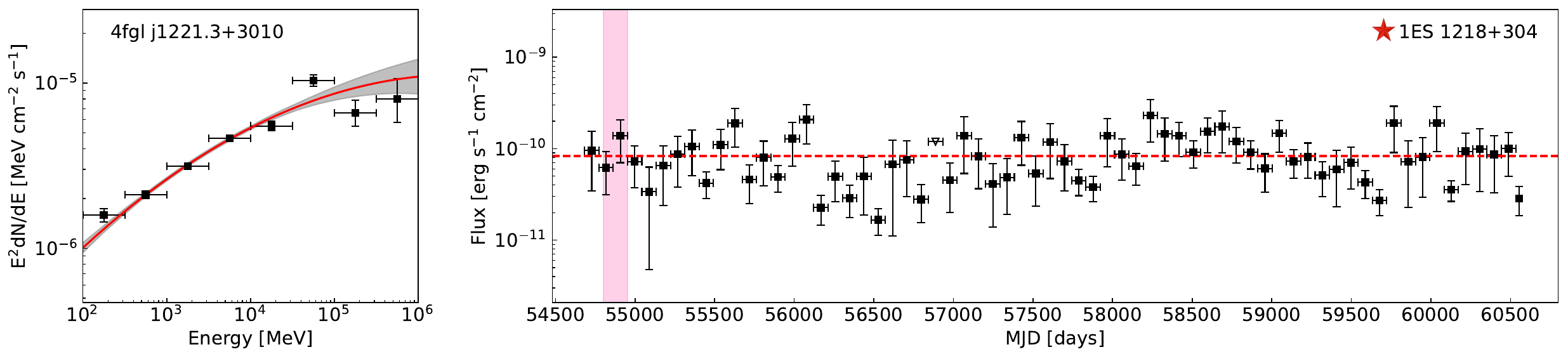}    \\ 
        \includegraphics[angle=0,width=\textwidth]{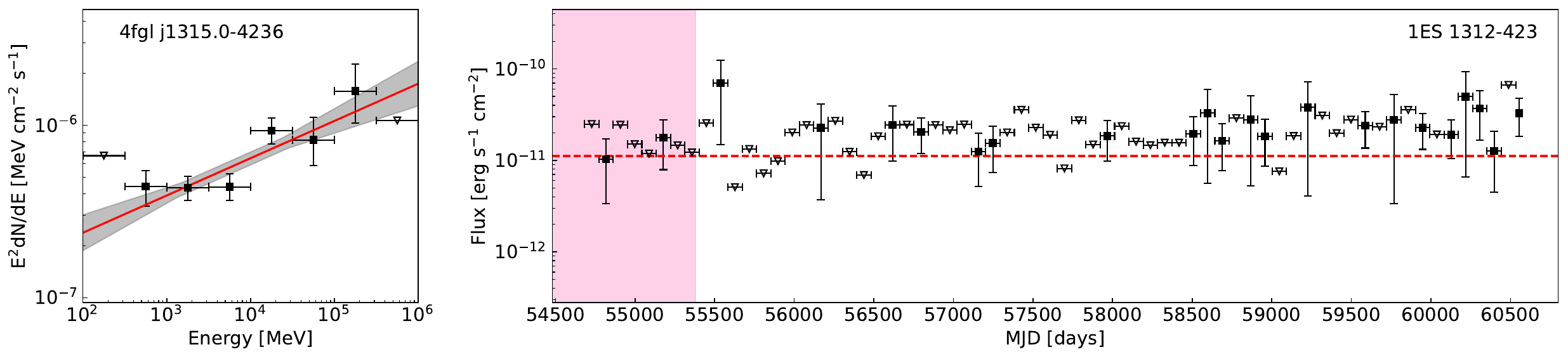}   \\ 
\end{figure*}     
\begin{figure*}
        \includegraphics[angle=0,width=\textwidth]{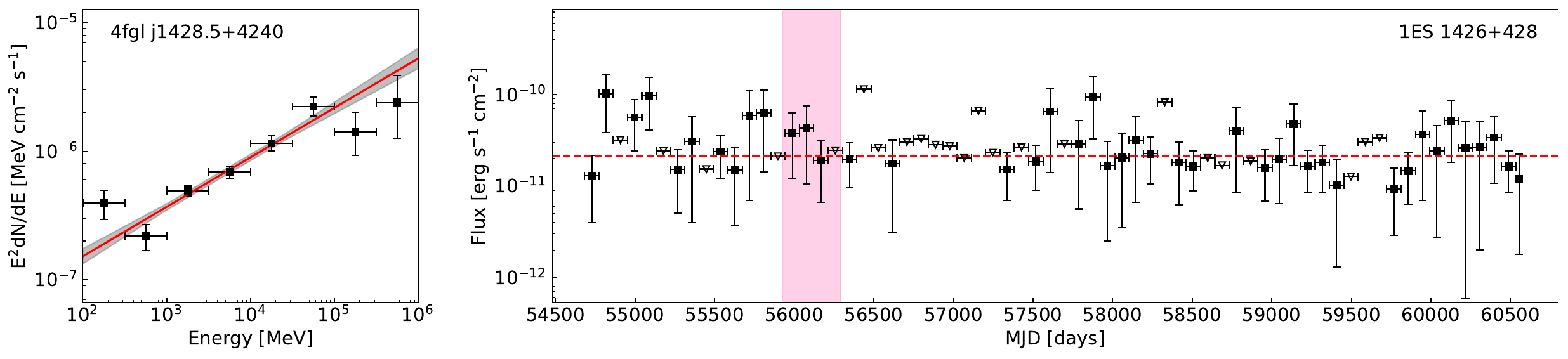} \\      
        \includegraphics[angle=0,width=\textwidth]{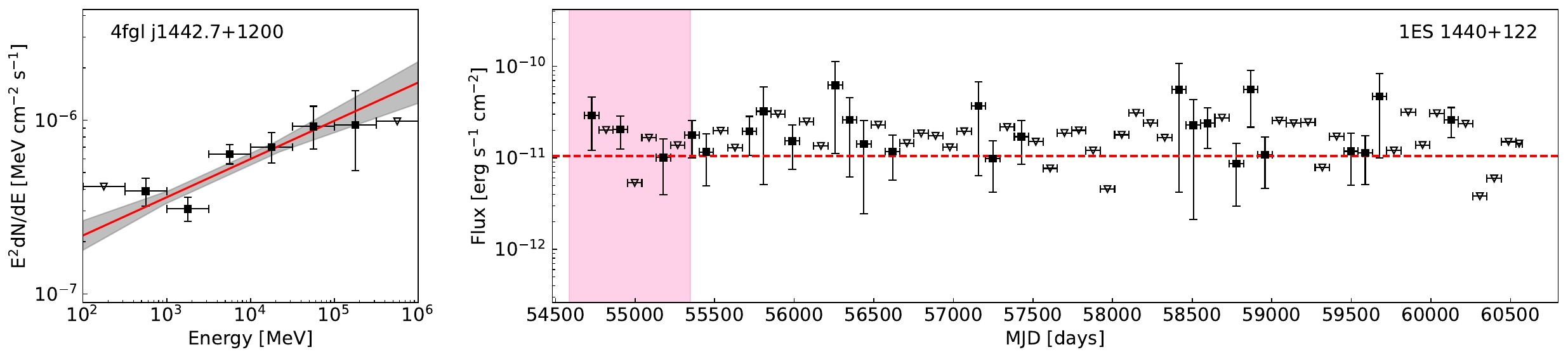} \\        
        \includegraphics[angle=0,width=\textwidth]{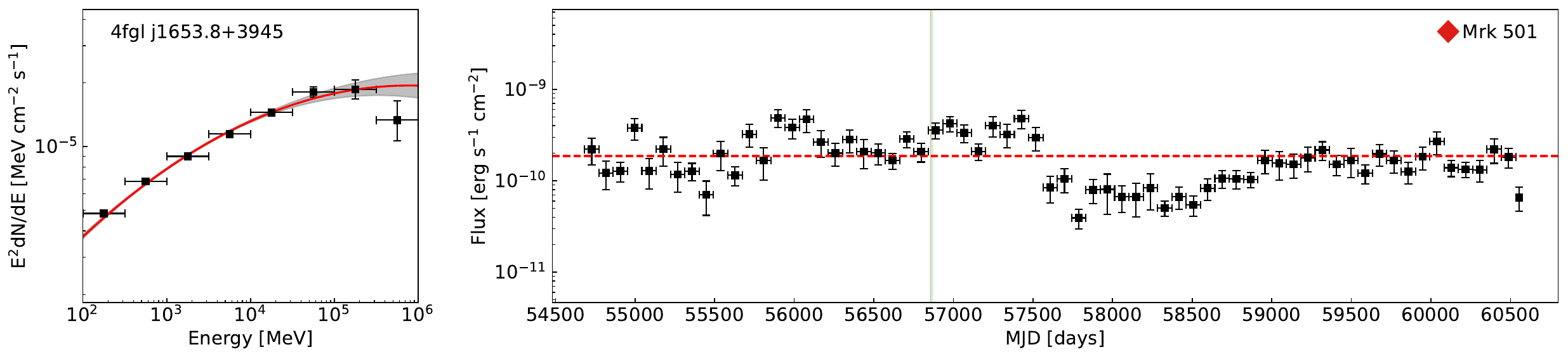}    \\        
        \includegraphics[angle=0,width=\textwidth]{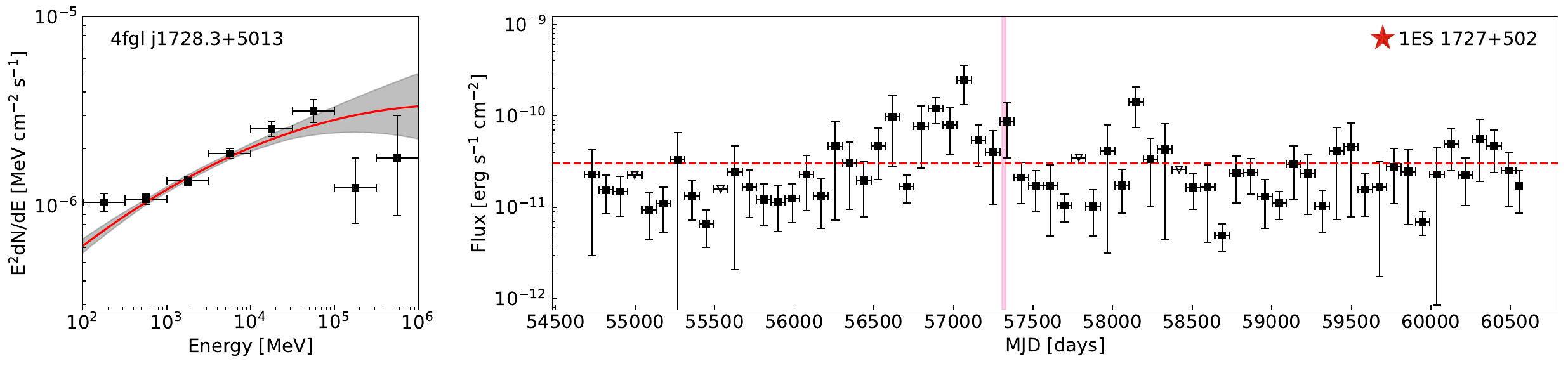}  \\         
        \includegraphics[angle=0,width=\textwidth]{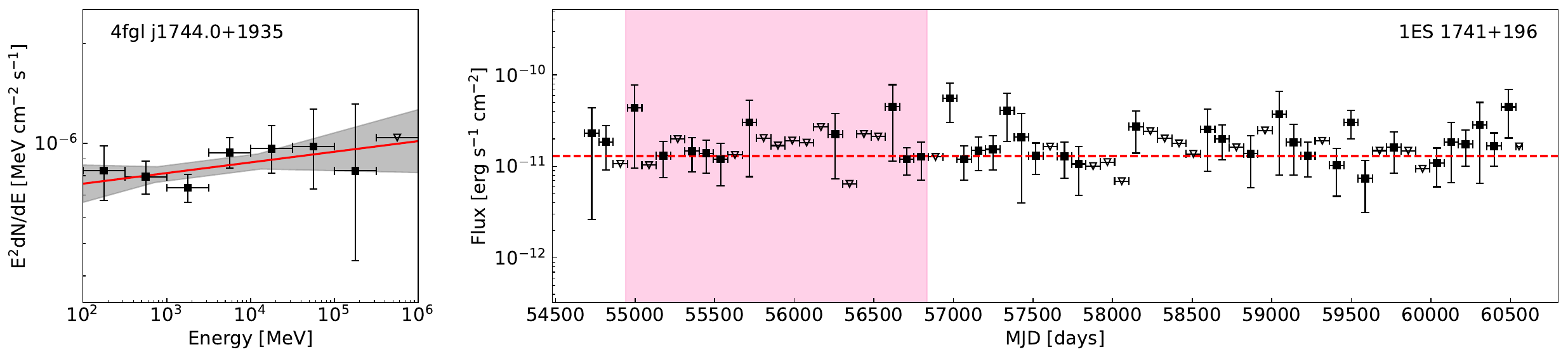} \\    
\end{figure*}
\begin{figure*}  
        \includegraphics[angle=0,width=\textwidth]{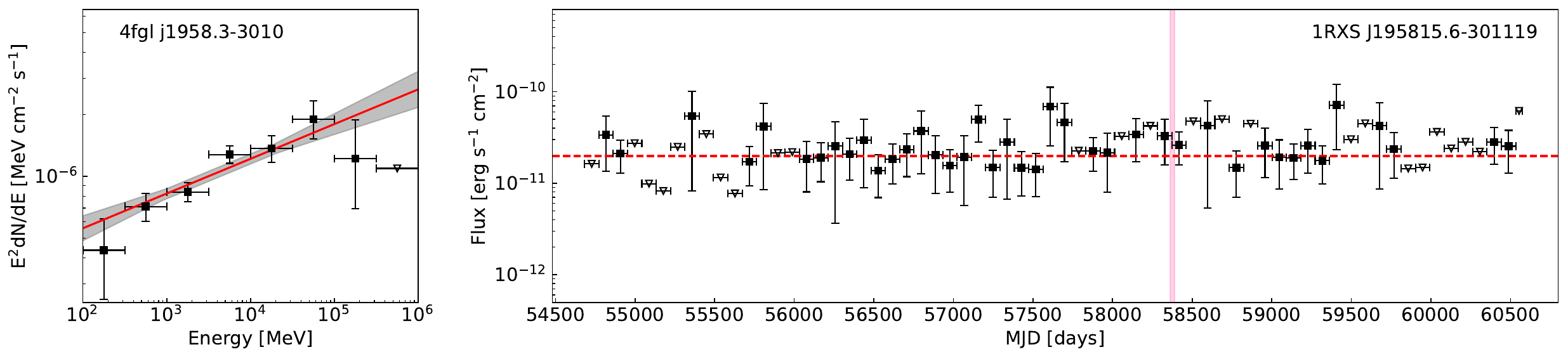} \\ 
       \includegraphics[angle=0,width=\textwidth]{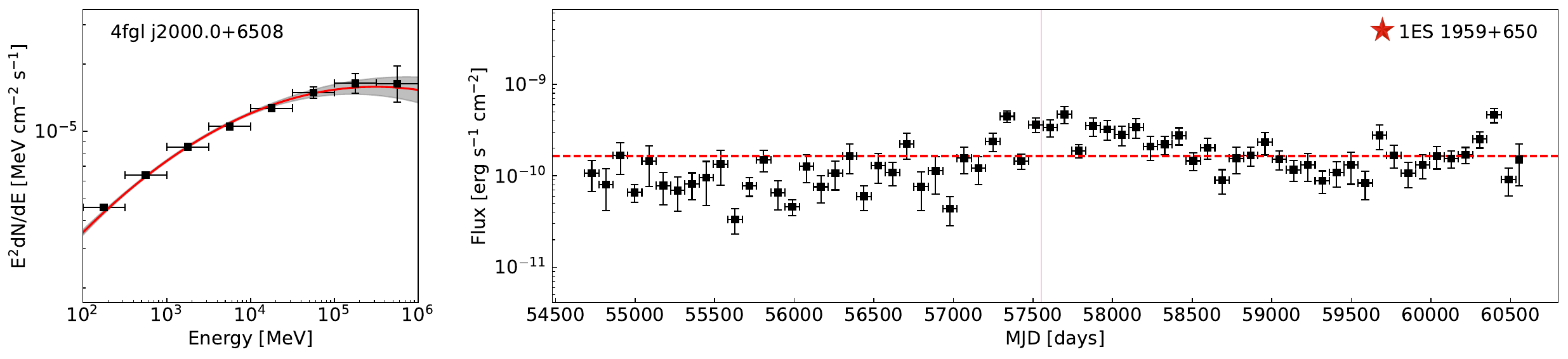}  \\     
       \includegraphics[angle=0,width=\textwidth]{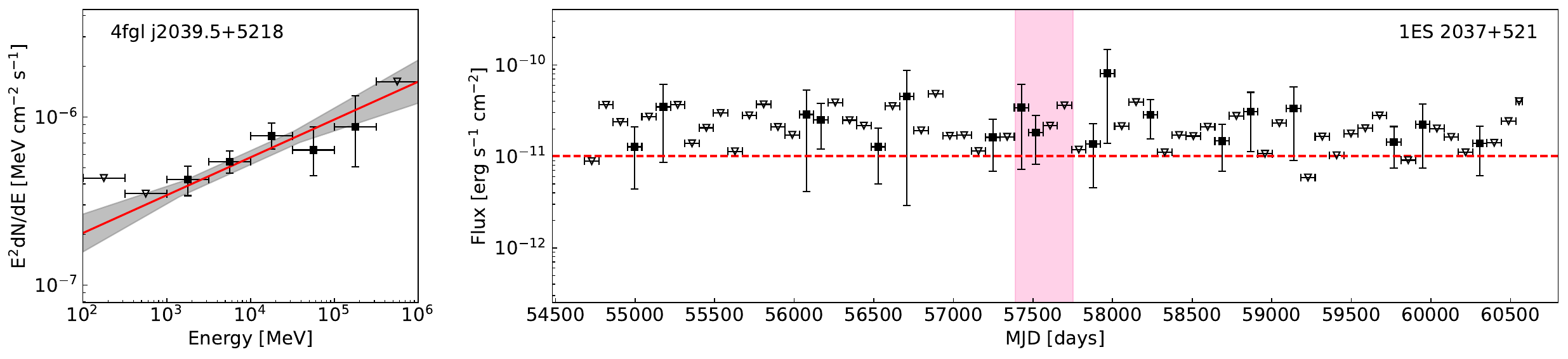} \\   
        \includegraphics[angle=0,width=\textwidth]{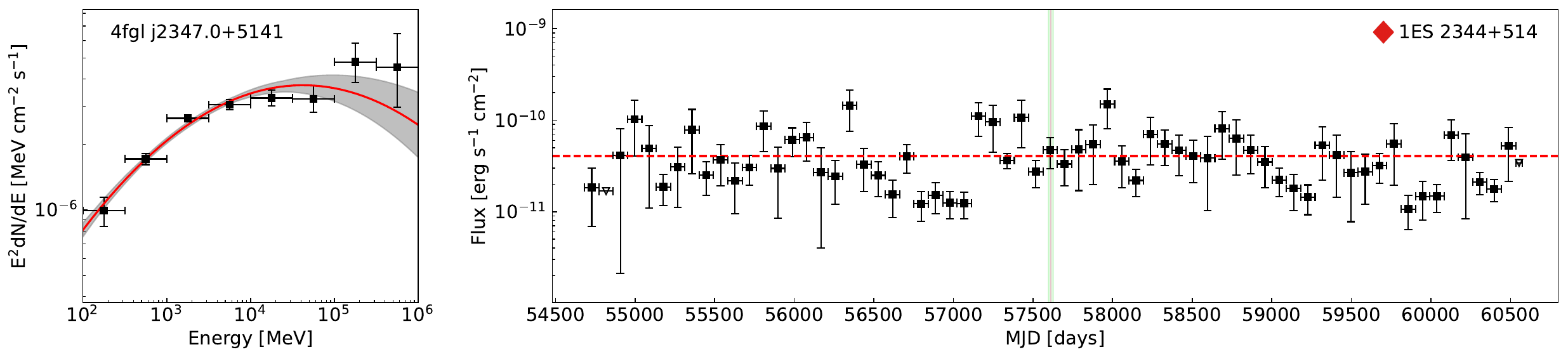}  \\      
        \includegraphics[angle=0,width=\textwidth]{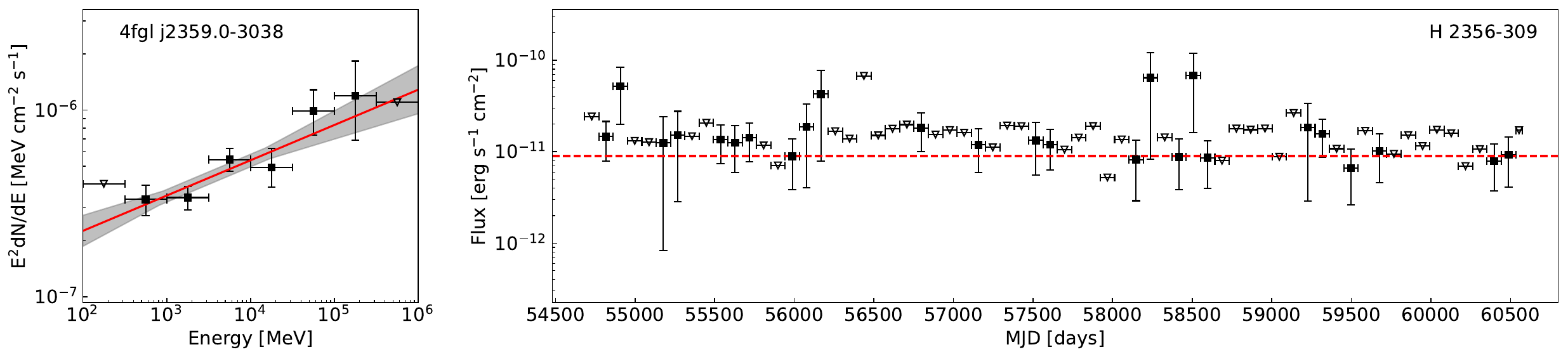}
\end{figure*}

\clearpage
\bibliographystyle{aasjournal}
\bibliography{biblist.bib}

@ARTICLE{2019ApJ...884...91P,
       author = {{Peng}, Fang-Kun and {Zhang}, Hai-Ming and {Wang}, Xiang-Yu and {Wang}, Jun-Feng and {Zhi}, Qi-Jun},
        title = "{Evidence of AGN Activity in the Gamma-Ray Emission from Two Starburst Galaxies}",
      journal = {\apj},
         year = 2019,
        month = oct,
       volume = {884},
       number = {1},
          eid = {91},
        pages = {91},
          doi = {10.3847/1538-4357/ab3e6f},
archivePrefix = {arXiv},
       eprint = {1906.06720},
 primaryClass = {astro-ph.HE},
       adsurl = {https://ui.adsabs.harvard.edu/abs/2019ApJ...884...91P}
}

@ARTICLE{2020ApJS..247...33A,
       author = {{Abdollahi}, S. and {Acero}, F. and {Ackermann}, M. and {Ajello}, M. and {Atwood}, W.~B. and {Axelsson}, M. and {Baldini}, L. and {Ballet}, J. and {Barbiellini}, G. and {Bastieri}, D. and {Becerra Gonzalez}, J. and {Bellazzini}, R. and {Berretta}, A. and {Bissaldi}, E. and {Blandford}, R.~D. and {Bloom}, E.~D. and {Bonino}, R. and {Bottacini}, E. and {Brandt}, T.~J. and {Bregeon}, J. and {Bruel}, P. and {Buehler}, R. and {Burnett}, T.~H. and {Buson}, S. and {Cameron}, R.~A. and {Caputo}, R. and {Caraveo}, P.~A. and {Casandjian}, J.~M. and {Castro}, D. and {Cavazzuti}, E. and {Charles}, E. and {Chaty}, S. and {Chen}, S. and {Cheung}, C.~C. and {Chiaro}, G. and {Ciprini}, S. and {Cohen-Tanugi}, J. and {Cominsky}, L.~R. and {Coronado-Bl{\'a}zquez}, J. and {Costantin}, D. and {Cuoco}, A. and {Cutini}, S. and {D'Ammando}, F. and {DeKlotz}, M. and {de la Torre Luque}, P. and {de Palma}, F. and {Desai}, A. and {Digel}, S.~W. and {Di Lalla}, N. and {Di Mauro}, M. and {Di Venere}, L. and {Dom{\'\i}nguez}, A. and {Dumora}, D. and {Fana Dirirsa}, F. and {Fegan}, S.~J. and {Ferrara}, E.~C. and {Franckowiak}, A. and {Fukazawa}, Y. and {Funk}, S. and {Fusco}, P. and {Gargano}, F. and {Gasparrini}, D. and {Giglietto}, N. and {Giommi}, P. and {Giordano}, F. and {Giroletti}, M. and {Glanzman}, T. and {Green}, D. and {Grenier}, I.~A. and {Griffin}, S. and {Grondin}, M. -H. and {Grove}, J.~E. and {Guiriec}, S. and {Harding}, A.~K. and {Hayashi}, K. and {Hays}, E. and {Hewitt}, J.~W. and {Horan}, D. and {J{\'o}hannesson}, G. and {Johnson}, T.~J. and {Kamae}, T. and {Kerr}, M. and {Kocevski}, D. and {Kovac'evic'}, M. and {Kuss}, M. and {Landriu}, D. and {Larsson}, S. and {Latronico}, L. and {Lemoine-Goumard}, M. and {Li}, J. and {Liodakis}, I. and {Longo}, F. and {Loparco}, F. and {Lott}, B. and {Lovellette}, M.~N. and {Lubrano}, P. and {Madejski}, G.~M. and {Maldera}, S. and {Malyshev}, D. and {Manfreda}, A. and {Marchesini}, E.~J. and {Marcotulli}, L. and {Mart{\'\i}-Devesa}, G. and {Martin}, P. and {Massaro}, F. and {Mazziotta}, M.~N. and {McEnery}, J.~E. and {Mereu}, I. and {Meyer}, M. and {Michelson}, P.~F. and {Mirabal}, N. and {Mizuno}, T. and {Monzani}, M.~E. and {Morselli}, A. and {Moskalenko}, I.~V. and {Negro}, M. and {Nuss}, E. and {Ojha}, R. and {Omodei}, N. and {Orienti}, M. and {Orlando}, E. and {Ormes}, J.~F. and {Palatiello}, M. and {Paliya}, V.~S. and {Paneque}, D. and {Pei}, Z. and {Pe{\~n}a-Herazo}, H. and {Perkins}, J.~S. and {Persic}, M. and {Pesce-Rollins}, M. and {Petrosian}, V. and {Petrov}, L. and {Piron}, F. and {Poon}, H. and {Porter}, T.~A. and {Principe}, G. and {Rain{\`o}}, S. and {Rando}, R. and {Razzano}, M. and {Razzaque}, S. and {Reimer}, A. and {Reimer}, O. and {Remy}, Q. and {Reposeur}, T. and {Romani}, R.~W. and {Saz Parkinson}, P.~M. and {Schinzel}, F.~K. and {Serini}, D. and {Sgr{\`o}}, C. and {Siskind}, E.~J. and {Smith}, D.~A. and {Spandre}, G. and {Spinelli}, P. and {Strong}, A.~W. and {Suson}, D.~J. and {Tajima}, H. and {Takahashi}, M.~N. and {Tak}, D. and {Thayer}, J.~B. and {Thompson}, D.~J. and {Tibaldo}, L. and {Torres}, D.~F. and {Torresi}, E. and {Valverde}, J. and {Van Klaveren}, B. and {van Zyl}, P. and {Wood}, K. and {Yassine}, M. and {Zaharijas}, G.},
        title = "{Fermi Large Area Telescope Fourth Source Catalog}",
      journal = {\apjs},
         year = 2020,
        month = mar,
       volume = {247},
       number = {1},
          eid = {33},
        pages = {33},
          doi = {10.3847/1538-4365/ab6bcb},
archivePrefix = {arXiv},
       eprint = {1902.10045},
 primaryClass = {astro-ph.HE},
       adsurl = {https://ui.adsabs.harvard.edu/abs/2020ApJS..247...33A}
}

@INPROCEEDINGS{Wood2017,
       author = {{Wood}, M. and {Caputo}, R. and {Charles}, E. and {Di Mauro}, M. and {Magill}, J. and {Perkins}, J.~S. and {Fermi-LAT Collaboration}},
        title = "{Fermipy: An open-source Python package for analysis of Fermi-LAT Data}",
    booktitle = {35th International Cosmic Ray Conference (ICRC2017)},
         year = 2017,
       series = {International Cosmic Ray Conference},
       volume = {301},
        month = jul,
          eid = {824},
        pages = {824},
          doi = {10.22323/1.301.0824},
archivePrefix = {arXiv},
       eprint = {1707.09551},
 primaryClass = {astro-ph.IM},
       adsurl = {https://ui.adsabs.harvard.edu/abs/2017ICRC...35..824W}
}

@ARTICLE{1987PhR...154....1B,
       author = {{Blandford}, Roger and {Eichler}, David},
        title = "{Particle acceleration at astrophysical shocks: A theory of cosmic ray origin}",
      journal = {\physrep},
         year = 1987,
        month = oct,
       volume = {154},
       number = {1},
        pages = {1-75},
          doi = {10.1016/0370-1573(87)90134-7},
       adsurl = {https://ui.adsabs.harvard.edu/abs/1987PhR...154....1B}
}

@ARTICLE{1995PASP..107..803U,
       author = {{Urry}, C. Megan and {Padovani}, Paolo},
        title = "{Unified Schemes for Radio-Loud Active Galactic Nuclei}",
      journal = {\pasp},
         year = 1995,
        month = sep,
       volume = {107},
        pages = {803},
          doi = {10.1086/133630},
archivePrefix = {arXiv},
       eprint = {astro-ph/9506063},
 primaryClass = {astro-ph},
       adsurl = {https://ui.adsabs.harvard.edu/abs/1995PASP..107..803U}
}

@BOOK{2013LNP...873.....G,
       author = {{Ghisellini}, Gabriele},
        title = "{Radiative Processes in High Energy Astrophysics}",
         year = 2013,
       volume = {873},
          doi = {10.1007/978-3-319-00612-3},
       adsurl = {https://ui.adsabs.harvard.edu/abs/2013LNP...873.....G}
}

@BOOK{1979rpa..book.....R,
       author = {{Rybicki}, George B. and {Lightman}, Alan P.},
        title = "{Radiative processes in astrophysics}",
         year = 1979,
       adsurl = {https://ui.adsabs.harvard.edu/abs/1979rpa..book.....R}
}

@ARTICLE{1998MNRAS.301..451G,
       author = {{Ghisellini}, G. and {Celotti}, A. and {Fossati}, G. and {Maraschi}, L. and {Comastri}, A.},
        title = "{A theoretical unifying scheme for gamma-ray bright blazars}",
      journal = {\mnras},
         year = 1998,
        month = dec,
       volume = {301},
       number = {2},
        pages = {451-468},
          doi = {10.1046/j.1365-8711.1998.02032.x},
archivePrefix = {arXiv},
       eprint = {astro-ph/9807317},
 primaryClass = {astro-ph},
       adsurl = {https://ui.adsabs.harvard.edu/abs/1998MNRAS.301..451G}
}

@ARTICLE{2022PhRvD.106j3021X,
       author = {{Xue}, Rui and {Wang}, Ze-Rui and {Li}, Wei-Jian},
        title = "{Hadronuclear interactions in the jet of low TeV luminosity AGN: Implications for the low-state very-high-energy gamma-ray emission}",
      journal = {\prd},
         year = 2022,
        month = nov,
       volume = {106},
       number = {10},
          eid = {103021},
        pages = {103021},
          doi = {10.1103/PhysRevD.106.103021},
archivePrefix = {arXiv},
       eprint = {2210.09797},
 primaryClass = {astro-ph.HE},
       adsurl = {https://ui.adsabs.harvard.edu/abs/2022PhRvD.106j3021X}
}

@ARTICLE{2022ApJ...941...33F,
       author = {{Finke}, Justin D. and {Ajello}, Marco and {Dom{\'\i}nguez}, Alberto and {Desai}, Abhishek and {Hartmann}, Dieter H. and {Paliya}, Vaidehi S. and {Saldana-Lopez}, Alberto},
        title = "{Modeling the Extragalactic Background Light and the Cosmic Star Formation History}",
      journal = {\apj},
         year = 2022,
        month = dec,
       volume = {941},
       number = {1},
          eid = {33},
        pages = {33},
          doi = {10.3847/1538-4357/ac9843},
archivePrefix = {arXiv},
       eprint = {2210.01157},
 primaryClass = {astro-ph.GA},
       adsurl = {https://ui.adsabs.harvard.edu/abs/2022ApJ...941...33F}
}

@ARTICLE{2020NatAs...4..124B,
       author = {{Biteau}, J. and {Prandini}, E. and {Costamante}, L. and {Lemoine}, M. and {Padovani}, P. and {Pueschel}, E. and {Resconi}, E. and {Tavecchio}, F. and {Taylor}, A. and {Zech}, A.},
        title = "{Progress in unveiling extreme particle acceleration in persistent astrophysical jets}",
      journal = {Nature Astronomy},
         year = 2020,
        month = feb,
       volume = {4},
        pages = {124-131},
          doi = {10.1038/s41550-019-0988-4},
archivePrefix = {arXiv},
       eprint = {2001.09222},
 primaryClass = {astro-ph.HE},
       adsurl = {https://ui.adsabs.harvard.edu/abs/2020NatAs...4..124B}
}

@ARTICLE{2022MNRAS.517L..16T,
       author = {{Tavecchio}, Fabrizio and {Costa}, Agnese and {Sciaccaluga}, Alberto},
        title = "{Extreme blazars: the result of unstable recollimated jets?}",
      journal = {\mnras},
         year = 2022,
        month = nov,
       volume = {517},
       number = {1},
        pages = {L16-L20},
          doi = {10.1093/mnrasl/slac084},
archivePrefix = {arXiv},
       eprint = {2207.12766},
 primaryClass = {astro-ph.HE},
       adsurl = {https://ui.adsabs.harvard.edu/abs/2022MNRAS.517L..16T}
}

@ARTICLE{2019MNRAS.490.2284M,
       author = {{MAGIC Collaboration} and {Acciari}, V.~A. and {Ansoldi}, S. and {Antonelli}, L.~A. and {Arbet Engels}, A. and {Baack}, D. and {Babi{\'c}} and {}, A. and {Banerjee}, B. and {Barres de Almeida}, U. and {Barrio}, J.~A. and {Becerra Gonz{\'a}lez}, J. and {Bednarek}, W. and {Bellizzi}, L. and {Bernardini}, E. and {Berti}, A. and {Besenrieder}, J. and {Bhattacharyya}, W. and {Bigongiari}, C. and {Biland}, A. and {Blanch}, O. and {Bonnoli}, G. and {Bo{\v{s}}njak}, {\v{Z}}. and {Busetto}, G. and {Carosi}, R. and {Ceribella}, G. and {Cerruti}, M. and {Chai}, Y. and {Chilingaryan}, A. and {Cikota}, S. and {Colak}, S.~M. and {Colin}, U. and {Colombo}, E. and {Contreras}, J.~L. and {Cortina}, J. and {Covino}, S. and {D'Elia}, V. and {da Vela}, P. and {Dazzi}, F. and {de Angelis}, A. and {de Lotto}, B. and {Delfino}, M. and {Delgado}, J. and {Depaoli}, D. and {di Pierro}, F. and {di Venere}, L. and {Do Souto Espi{\~n}eira}, E. and {Dominis Prester}, D. and {Donini}, A. and {Dorner}, D. and {Doro}, M. and {Elsaesser}, D. and {Fallah Ramazani}, V. and {Fattorini}, A. and {Ferrara}, G. and {Fidalgo}, D. and {Foffano}, L. and {Fonseca}, M.~V. and {Font}, L. and {Fruck}, C. and {Fukami}, S. and {Garc{\'\i}a L{\'o}pez}, R.~J. and {Garczarczyk}, M. and {Gasparyan}, S. and {Gaug}, M. and {Giglietto}, N. and {Giordano}, F. and {Godinovi{\'c}} and {}, N. and {Green}, D. and {Guberman}, D. and {Hadasch}, D. and {Hahn}, A. and {Herrera}, J. and {Hoang}, J. and {Hrupec}, D. and {H{\"u}tten}, M. and {Inada}, T. and {Inoue}, S. and {Ishio}, K. and {Iwamura}, Y. and {Jouvin}, L. and {Kerszberg}, D. and {Kubo}, H. and {Kushida}, J. and {Lamastra}, A. and {Lelas}, D. and {Leone}, F. and {Lindfors}, E. and {Lombardi}, S. and {Longo}, F. and {L{\'o}pez}, M. and {L{\'o}pez-Coto}, R. and {L{\'o}pez-Oramas}, A. and {Loporchio}, S. and {Machado de Oliveira Fraga}, B. and {Maggio}, C. and {Majumdar}, P. and {Makariev}, M. and {Mallamaci}, M. and {Maneva}, G. and {Manganaro}, M. and {Mannheim}, K. and {Maraschi}, L. and {Mariotti}, M. and {Mart{\'\i}nez}, M. and {Mazin}, D. and {Mi{\'c}} and {Anovi{\'c}} and {}, S. and {Miceli}, D. and {Minev}, M. and {Miranda}, J.~M. and {Mirzoyan}, R. and {Molina}, E. and {Moralejo}, A. and {Morcuende}, D. and {Moreno}, V. and {Moretti}, E. and {Munar-Adrover}, P. and {Neustroev}, V. and {Nigro}, C. and {Nilsson}, K. and {Ninci}, D. and {Nishijima}, K. and {Noda}, K. and {Nogu{\'e}s}, L. and {Nozaki}, S. and {Paiano}, S. and {Palacio}, J. and {Palatiello}, M. and {Paneque}, D. and {Paoletti}, R. and {Paredes}, J.~M. and {Pe{\~n}il}, P. and {Peresano}, M. and {Persic}, M. and {Prada Moroni}, P.~G. and {Prandini}, E. and {Puljak}, I. and {Rhode}, W. and {Rib{\'o}}, M. and {Rico}, J. and {Righi}, C. and {Rugliancich}, A. and {Saha}, L. and {Sahakyan}, N. and {Saito}, T. and {Sakurai}, S. and {Satalecka}, K. and {Schmidt}, K. and {Schweizer}, T. and {Sitarek}, J. and {{\v{S}}nidari{\'c}} and {}, I. and {Sobczynska}, D. and {Somero}, A. and {Stamerra}, A. and {Strom}, D. and {Strzys}, M. and {Suda}, Y. and {Suri{\'c}} and {}, T. and {Takahashi}, M. and {Tavecchio}, F. and {Temnikov}, P. and {Terzi{\'c}} and {}, T. and {Teshima}, M. and {Torres-Alb{\`a}}, N. and {Tosti}, L. and {Vagelli}, V. and {van Scherpenberg}, J. and {Vanzo}, G. and {Vazquez Acosta}, M. and {Vigorito}, C.~F. and {Vitale}, V. and {Vovk}, I. and {Will}, M. and {Zari{\'c}} and {}, D. and {Asano}, K. and {D'Ammando}, F. and {Clavero}, R.},
        title = "{Testing emission models on the extreme blazar 2WHSP J073326.7+515354 detected at very high energies with the MAGIC telescopes}",
      journal = {\mnras},
         year = 2019,
        month = dec,
       volume = {490},
       number = {2},
        pages = {2284-2299},
          doi = {10.1093/mnras/stz2725},
archivePrefix = {arXiv},
       eprint = {1909.11621},
 primaryClass = {astro-ph.HE},
       adsurl = {https://ui.adsabs.harvard.edu/abs/2019MNRAS.490.2284M}
}

@ARTICLE{2001A&A...371..512C,
       author = {{Costamante}, L. and {Ghisellini}, G. and {Giommi}, P. and {Tagliaferri}, G. and {Celotti}, A. and {Chiaberge}, M. and {Fossati}, G. and {Maraschi}, L. and {Tavecchio}, F. and {Treves}, A. and {Wolter}, A.},
        title = "{Extreme synchrotron BL Lac objects. Stretching the blazar sequence}",
      journal = {\aap},
         year = 2001,
        month = may,
       volume = {371},
        pages = {512-526},
          doi = {10.1051/0004-6361:20010412},
archivePrefix = {arXiv},
       eprint = {astro-ph/0103343},
 primaryClass = {astro-ph},
       adsurl = {https://ui.adsabs.harvard.edu/abs/2001A&A...371..512C}
}

@ARTICLE{2010ApJ...716...30A,
       author = {{Abdo}, A.~A. and {Ackermann}, M. and {Agudo}, I. and {Ajello}, M. and {Aller}, H.~D. and {Aller}, M.~F. and {Angelakis}, E. and {Arkharov}, A.~A. and {Axelsson}, M. and {Bach}, U. and {Baldini}, L. and {Ballet}, J. and {Barbiellini}, G. and {Bastieri}, D. and {Baughman}, B.~M. and {Bechtol}, K. and {Bellazzini}, R. and {Benitez}, E. and {Berdyugin}, A. and {Berenji}, B. and {Blandford}, R.~D. and {Bloom}, E.~D. and {Boettcher}, M. and {Bonamente}, E. and {Borgland}, A.~W. and {Bregeon}, J. and {Brez}, A. and {Brigida}, M. and {Bruel}, P. and {Burnett}, T.~H. and {Burrows}, D. and {Buson}, S. and {Caliandro}, G.~A. and {Calzoletti}, L. and {Cameron}, R.~A. and {Capalbi}, M. and {Caraveo}, P.~A. and {Carosati}, D. and {Casandjian}, J.~M. and {Cavazzuti}, E. and {Cecchi}, C. and {{\c{C}}elik}, {\"O}. and {Charles}, E. and {Chaty}, S. and {Chekhtman}, A. and {Chen}, W.~P. and {Chiang}, J. and {Chincarini}, G. and {Ciprini}, S. and {Claus}, R. and {Cohen-Tanugi}, J. and {Colafrancesco}, S. and {Cominsky}, L.~R. and {Conrad}, J. and {Costamante}, L. and {Cutini}, S. and {D'ammando}, F. and {Deitrick}, R. and {D'Elia}, V. and {Dermer}, C.~D. and {de Angelis}, A. and {de Palma}, F. and {Digel}, S.~W. and {Donnarumma}, I. and {Silva}, E. do Couto e. and {Drell}, P.~S. and {Dubois}, R. and {Dultzin}, D. and {Dumora}, D. and {Falcone}, A. and {Farnier}, C. and {Favuzzi}, C. and {Fegan}, S.~J. and {Focke}, W.~B. and {Forn{\'e}}, E. and {Fortin}, P. and {Frailis}, M. and {Fuhrmann}, L. and {Fukazawa}, Y. and {Funk}, S. and {Fusco}, P. and {G{\'o}mez}, J.~L. and {Gargano}, F. and {Gasparrini}, D. and {Gehrels}, N. and {Germani}, S. and {Giebels}, B. and {Giglietto}, N. and {Giommi}, P. and {Giordano}, F. and {Giuliani}, A. and {Glanzman}, T. and {Godfrey}, G. and {Grenier}, I.~A. and {Gronwall}, C. and {Grove}, J.~E. and {Guillemot}, L. and {Guiriec}, S. and {Gurwell}, M.~A. and {Hadasch}, D. and {Hanabata}, Y. and {Harding}, A.~K. and {Hayashida}, M. and {Hays}, E. and {Healey}, S.~E. and {Heidt}, J. and {Hiriart}, D. and {Horan}, D. and {Hoversten}, E.~A. and {Hughes}, R.~E. and {Itoh}, R. and {Jackson}, M.~S. and {J{\'o}hannesson}, G. and {Johnson}, A.~S. and {Johnson}, W.~N. and {Jorstad}, S.~G. and {Kadler}, M. and {Kamae}, T. and {Katagiri}, H. and {Kataoka}, J. and {Kawai}, N. and {Kennea}, J. and {Kerr}, M. and {Kimeridze}, G. and {Kn{\"o}dlseder}, J. and {Kocian}, M.~L. and {Kopatskaya}, E.~N. and {Koptelova}, E. and {Konstantinova}, T.~S. and {Kovalev}, Y.~Y. and {Kovalev}, Yu. A. and {Kurtanidze}, O.~M. and {Kuss}, M. and {Lande}, J. and {Larionov}, V.~M. and {Latronico}, L. and {Leto}, P. and {Lindfors}, E. and {Longo}, F. and {Loparco}, F. and {Lott}, B. and {Lovellette}, M.~N. and {Lubrano}, P. and {Madejski}, G.~M. and {Makeev}, A. and {Marchegiani}, P. and {Marscher}, A.~P. and {Marshall}, F. and {Max-Moerbeck}, W. and {Mazziotta}, M.~N. and {McConville}, W. and {McEnery}, J.~E. and {Meurer}, C. and {Michelson}, P.~F. and {Mitthumsiri}, W. and {Mizuno}, T. and {Moiseev}, A.~A. and {Monte}, C. and {Monzani}, M.~E. and {Morselli}, A. and {Moskalenko}, I.~V. and {Murgia}, S. and {Nestoras}, I. and {Nilsson}, K. and {Nizhelsky}, N.~A. and {Nolan}, P.~L. and {Norris}, J.~P. and {Nuss}, E. and {Ohsugi}, T. and {Ojha}, R. and {Omodei}, N. and {Orlando}, E. and {Ormes}, J.~F. and {Osborne}, J. and {Ozaki}, M. and {Pacciani}, L. and {Padovani}, P. and {Pagani}, C. and {Page}, K. and {Paneque}, D. and {Panetta}, J.~H. and {Parent}, D. and {Pasanen}, M. and {Pavlidou}, V. and {Pelassa}, V. and {Pepe}, M. and {Perri}, M. and {Pesce-Rollins}, M. and {Piranomonte}, S. and {Piron}, F. and {Pittori}, C. and {Porter}, T.~A. and {Puccetti}, S. and {Rahoui}, F. and {Rain{\`o}}, S. and {Raiteri}, C. and {Rando}, R. and {Razzano}, M. and {Reimer}, A. and {Reimer}, O. and {Reposeur}, T. and {Richards}, J.~L. and {Ritz}, S. and {Rochester}, L.~S. and {Rodriguez}, A.~Y. and {Romani}, R.~W. and {Ros}, J.~A. and {Roth}, M. and {Roustazadeh}, P. and {Ryde}, F. and {Sadrozinski}, H.~F. -W. and {Sadun}, A. and {Sanchez}, D. and {Sander}, A. and {Saz Parkinson}, P.~M. and {Scargle}, J.~D. and {Sellerholm}, A. and {Sgr{\`o}}, C. and {Shaw}, M.~S. and {Sigua}, L.~A. and {Siskind}, E.~J. and {Smith}, D.~A. and {Smith}, P.~D. and {Spandre}, G. and {Spinelli}, P. and {Starck}, J. -L. and {Stevenson}, M. and {Stratta}, G. and {Strickman}, M.~S. and {Suson}, D.~J. and {Tajima}, H. and {Takahashi}, H. and {Takahashi}, T. and {Takalo}, L.~O. and {Tanaka}, T. and {Thayer}, J.~B. and {Thayer}, J.~G. and {Thompson}, D.~J. and {Tibaldo}, L. and {Torres}, D.~F. and {Tosti}, G. and {Tramacere}, A. and {Uchiyama}, Y. and {Usher}, T.~L. and {Vasileiou}, V. and {Verrecchia}, F. and {Vilchez}, N. and {Villata}, M. and {Vitale}, V. and {Waite}, A.~P. and {Wang}, P. and {Winer}, B.~L. and {Wood}, K.~S. and {Ylinen}, T. and {Zensus}, J.~A. and {Zhekanis}, G.~V. and {Ziegler}, M.},
        title = "{The Spectral Energy Distribution of Fermi Bright Blazars}",
      journal = {\apj},
         year = 2010,
        month = jun,
       volume = {716},
       number = {1},
        pages = {30-70},
          doi = {10.1088/0004-637X/716/1/30},
archivePrefix = {arXiv},
       eprint = {0912.2040},
 primaryClass = {astro-ph.CO},
       adsurl = {https://ui.adsabs.harvard.edu/abs/2010ApJ...716...30A}
}

@ARTICLE{2019NewA...7301278S,
       author = {{Singh}, KK and {Meintjes}, PJ and {Ramamonjisoa}, FA and {Tolamatti}, A.},
        title = "{Extremely High energy peaked BL Lac nature of the TeV blazar Mrk 501}",
      journal = {\na},
         year = 2019,
        month = nov,
       volume = {73},
          eid = {101278},
        pages = {101278},
          doi = {10.1016/j.newast.2019.101278},
archivePrefix = {arXiv},
       eprint = {1906.04486},
 primaryClass = {astro-ph.GA},
       adsurl = {https://ui.adsabs.harvard.edu/abs/2019NewA...7301278S}
}

@ARTICLE{2022Galax..10..105S,
       author = {{Sol}, H{\'e}l{\`e}ne and {Zech}, Andreas},
        title = "{Blazars at Very High Energies: Emission Modelling}",
      journal = {Galaxies},
         year = 2022,
        month = nov,
       volume = {10},
       number = {6},
          eid = {105},
        pages = {105},
          doi = {10.3390/galaxies10060105},
archivePrefix = {arXiv},
       eprint = {2211.03580},
 primaryClass = {astro-ph.HE},
       adsurl = {https://ui.adsabs.harvard.edu/abs/2022Galax..10..105S}
}

@ARTICLE{2018MNRAS.477.4257C,
       author = {{Costamante}, L. and {Bonnoli}, G. and {Tavecchio}, F. and {Ghisellini}, G. and {Tagliaferri}, G. and {Khangulyan}, D.},
        title = "{The NuSTAR view on hard-TeV BL Lacs}",
      journal = {\mnras},
         year = 2018,
        month = jul,
       volume = {477},
       number = {3},
        pages = {4257-4268},
          doi = {10.1093/mnras/sty857},
archivePrefix = {arXiv},
       eprint = {1711.06282},
 primaryClass = {astro-ph.HE},
       adsurl = {https://ui.adsabs.harvard.edu/abs/2018MNRAS.477.4257C}
}

@ARTICLE{1992ApJ...397L...5M,
       author = {{Maraschi}, L. and {Ghisellini}, G. and {Celotti}, A.},
        title = "{A Jet Model for the Gamma-Ray--emitting Blazar 3C 279}",
      journal = {\apjl},
         year = 1992,
        month = sep,
       volume = {397},
        pages = {L5},
          doi = {10.1086/186531},
       adsurl = {https://ui.adsabs.harvard.edu/abs/1992ApJ...397L...5M}
}

@ARTICLE{2006MNRAS.368L..52K,
       author = {{Katarzy{\'n}ski}, K. and {Ghisellini}, G. and {Tavecchio}, F. and {Gracia}, J. and {Maraschi}, L.},
        title = "{Hard TeV spectra of blazars and the constraints to the infrared intergalactic background}",
      journal = {\mnras},
         year = 2006,
        month = may,
       volume = {368},
       number = {1},
        pages = {L52-L56},
          doi = {10.1111/j.1745-3933.2006.00156.x},
archivePrefix = {arXiv},
       eprint = {astro-ph/0603030},
 primaryClass = {astro-ph},
       adsurl = {https://ui.adsabs.harvard.edu/abs/2006MNRAS.368L..52K}
}

@ARTICLE{2010MNRAS.401.1570T,
       author = {{Tavecchio}, F. and {Ghisellini}, G. and {Ghirlanda}, G. and {Foschini}, L. and {Maraschi}, L.},
        title = "{TeV BL Lac objects at the dawn of the Fermi era}",
      journal = {\mnras},
         year = 2010,
        month = jan,
       volume = {401},
       number = {3},
        pages = {1570-1586},
          doi = {10.1111/j.1365-2966.2009.15784.x},
archivePrefix = {arXiv},
       eprint = {0909.0651},
 primaryClass = {astro-ph.HE},
       adsurl = {https://ui.adsabs.harvard.edu/abs/2010MNRAS.401.1570T}
}

@ARTICLE{2015MNRAS.448..910C,
       author = {{Cerruti}, M. and {Zech}, A. and {Boisson}, C. and {Inoue}, S.},
        title = "{A hadronic origin for ultra-high-frequency-peaked BL Lac objects}",
      journal = {\mnras},
         year = 2015,
        month = mar,
       volume = {448},
       number = {1},
        pages = {910-927},
          doi = {10.1093/mnras/stu2691},
archivePrefix = {arXiv},
       eprint = {1411.5968},
 primaryClass = {astro-ph.HE},
       adsurl = {https://ui.adsabs.harvard.edu/abs/2015MNRAS.448..910C}
}

@ARTICLE{2015MNRAS.451..611B,
       author = {{Bonnoli}, G. and {Tavecchio}, F. and {Ghisellini}, G. and {Sbarrato}, T.},
        title = "{An emerging population of BL Lacs with extreme properties: towards a class of EBL and cosmic magnetic field probes?}",
      journal = {\mnras},
         year = 2015,
        month = jul,
       volume = {451},
       number = {1},
        pages = {611-621},
          doi = {10.1093/mnras/stv953},
archivePrefix = {arXiv},
       eprint = {1501.01974},
 primaryClass = {astro-ph.HE},
       adsurl = {https://ui.adsabs.harvard.edu/abs/2015MNRAS.451..611B}
}

@ARTICLE{2022MNRAS.512.1557A,
       author = {{Aguilar-Ruiz}, E. and {Fraija}, N. and {Galv{\'a}n-G{\'a}mez}, A. and {Ben{\'\i}tez}, E.},
        title = "{A two-zone model as origin of hard TeV spectrum in extreme BL lacs}",
      journal = {\mnras},
         year = 2022,
        month = may,
       volume = {512},
       number = {2},
        pages = {1557-1566},
          doi = {10.1093/mnras/stac591},
archivePrefix = {arXiv},
       eprint = {2203.00880},
 primaryClass = {astro-ph.HE},
       adsurl = {https://ui.adsabs.harvard.edu/abs/2022MNRAS.512.1557A}
}

@ARTICLE{2024A&A...682A.134G,
       author = {{Goswami}, P. and {Zacharias}, M. and {Zech}, A. and {Chandra}, S. and {Boettcher}, M. and {Sushch}, I.},
        title = "{The variety of extreme blazars in the AstroSat view}",
      journal = {\aap},
         year = 2024,
        month = feb,
       volume = {682},
          eid = {A134},
        pages = {A134},
          doi = {10.1051/0004-6361/202348121},
archivePrefix = {arXiv},
       eprint = {2311.12695},
 primaryClass = {astro-ph.HE},
       adsurl = {https://ui.adsabs.harvard.edu/abs/2024A&A...682A.134G}
}

@ARTICLE{2020ApJS..247...16A,
       author = {{Acciari}, V.~A. and {Ansoldi}, S. and {Antonelli}, L.~A. and {Engels}, A. Arbet and {Asano}, K. and {Baack}, D. and {Babi{\'c}}, A. and {Banerjee}, B. and {Barres de Almeida}, U. and {Barrio}, J.~A. and {Becerra Gonz{\'a}lez}, J. and {Bednarek}, W. and {Bellizzi}, L. and {Bernardini}, E. and {Berti}, A. and {Besenrieder}, J. and {Bhattacharyya}, W. and {Bigongiari}, C. and {Biland}, A. and {Blanch}, O. and {Bonnoli}, G. and {Bo{\v{s}}njak}, {\v{Z}}. and {Busetto}, G. and {Carosi}, R. and {Ceribella}, G. and {Cerruti}, M. and {Chai}, Y. and {Chilingaryan}, A. and {Cikota}, S. and {Colak}, S.~M. and {Colin}, U. and {Colombo}, E. and {Contreras}, J.~L. and {Cortina}, J. and {Covino}, S. and {D'Elia}, V. and {Da Vela}, P. and {Dazzi}, F. and {De Angelis}, A. and {De Lotto}, B. and {Delfino}, M. and {Delgado}, J. and {Depaoli}, D. and {Di Pierro}, F. and {Di Venere}, L. and {Do Souto Espi{\~n}eira}, E. and {Dominis Prester}, D. and {Donini}, A. and {Dorner}, D. and {Doro}, M. and {Elsaesser}, D. and {Ramazani}, V. Fallah and {Fattorini}, A. and {Ferrara}, G. and {Fidalgo}, D. and {Foffano}, L. and {Fonseca}, M.~V. and {Font}, L. and {Fruck}, C. and {Fukami}, S. and {Garc{\'\i}a L{\'o}pez}, R.~J. and {Garczarczyk}, M. and {Gasparyan}, S. and {Gaug}, M. and {Giglietto}, N. and {Giordano}, F. and {Godinovi{\'c}}, N. and {Green}, D. and {Guberman}, D. and {Hadasch}, D. and {Hahn}, A. and {Herrera}, J. and {Hoang}, J. and {Hrupec}, D. and {H{\"u}tten}, M. and {Inada}, T. and {Inoue}, S. and {Ishio}, K. and {Iwamura}, Y. and {Jouvin}, L. and {Kerszberg}, D. and {Kubo}, H. and {Kushida}, J. and {Lamastra}, A. and {Lelas}, D. and {Leone}, F. and {Lindfors}, E. and {Lombardi}, S. and {Longo}, F. and {L{\'o}pez}, M. and {L{\'o}pez-Coto}, R. and {L{\'o}pez-Oramas}, A. and {Loporchio}, S. and {Machado de Oliveira Fraga}, B. and {Maggio}, C. and {Majumdar}, P. and {Makariev}, M. and {Mallamaci}, M. and {Maneva}, G. and {Manganaro}, M. and {Mannheim}, K. and {Maraschi}, L. and {Mariotti}, M. and {Mart{\'\i}nez}, M. and {Mazin}, D. and {Mi{\'c}anovi{\'c}}, S. and {Miceli}, D. and {Minev}, M. and {Miranda}, J.~M. and {Mirzoyan}, R. and {Molina}, E. and {Moralejo}, A. and {Morcuende}, D. and {Moreno}, V. and {Moretti}, E. and {Munar-Adrover}, P. and {Neustroev}, V. and {Nigro}, C. and {Nilsson}, K. and {Ninci}, D. and {Nishijima}, K. and {Noda}, K. and {Nogu{\'e}s}, L. and {Nozaki}, S. and {Paiano}, S. and {Palatiello}, M. and {Paneque}, D. and {Paoletti}, R. and {Paredes}, J.~M. and {Pe{\~n}il}, P. and {Peresano}, M. and {Persic}, M. and {Prada Moroni}, P.~G. and {Prandini}, E. and {Puljak}, I. and {Rhode}, W. and {Rib{\'o}}, M. and {Rico}, J. and {Righi}, C. and {Rugliancich}, A. and {Saha}, L. and {Sahakyan}, N. and {Saito}, T. and {Sakurai}, S. and {Satalecka}, K. and {Schmidt}, K. and {Schweizer}, T. and {Sitarek}, J. and {{\v{S}}nidari{\'c}}, I. and {Sobczynska}, D. and {Somero}, A. and {Stamerra}, A. and {Strom}, D. and {Strzys}, M. and {Suda}, Y. and {Suri{\'c}}, T. and {Takahashi}, M. and {Tavecchio}, F. and {Temnikov}, P. and {Terzi{\'c}}, T. and {Teshima}, M. and {Torres-Alb{\`a}}, N. and {Tosti}, L. and {Vagelli}, V. and {van Scherpenberg}, J. and {Vanzo}, G. and {Vazquez Acosta}, M. and {Vigorito}, C.~F. and {Vitale}, V. and {Vovk}, I. and {Will}, M. and {Zari{\'c}}, D. and {Arcaro}, C. and {Carosi}, A. and {D'Ammando}, F. and {Tombesi}, F. and {Lohfink}, A.},
        title = "{New Hard-TeV Extreme Blazars Detected with the MAGIC Telescopes}",
      journal = {\apjs},
         year = 2020,
        month = mar,
       volume = {247},
       number = {1},
          eid = {16},
        pages = {16},
          doi = {10.3847/1538-4365/ab5b98},
archivePrefix = {arXiv},
       eprint = {1911.06680},
 primaryClass = {astro-ph.HE},
       adsurl = {https://ui.adsabs.harvard.edu/abs/2020ApJS..247...16A}
}

@ARTICLE{2003ApJ...585L..23F,
       author = {{Fan}, J.~H.},
        title = "{Relation between BL Lacertae Objects and Flat-Spectrum Radio Quasars}",
      journal = {\apjl},
         year = 2003,
        month = mar,
       volume = {585},
       number = {1},
        pages = {L23-L24},
          doi = {10.1086/374033},
       adsurl = {https://ui.adsabs.harvard.edu/abs/2003ApJ...585L..23F}
}

@ARTICLE{2012ApJ...752..157Z,
       author = {{Zhang}, Jin and {Liang}, En-Wei and {Zhang}, Shuang-Nan and {Bai}, J.~M.},
        title = "{Radiation Mechanisms and Physical Properties of GeV-TeV BL Lac Objects}",
      journal = {\apj},
         year = 2012,
        month = jun,
       volume = {752},
       number = {2},
          eid = {157},
        pages = {157},
          doi = {10.1088/0004-637X/752/2/157},
archivePrefix = {arXiv},
       eprint = {1108.0607},
 primaryClass = {astro-ph.HE},
       adsurl = {https://ui.adsabs.harvard.edu/abs/2012ApJ...752..157Z}
}

@ARTICLE{2014ApJ...782...13A,
       author = {{Aliu}, E. and {Archambault}, S. and {Arlen}, T. and {Aune}, T. and {Behera}, B. and {Beilicke}, M. and {Benbow}, W. and {Berger}, K. and {Bird}, R. and {Bouvier}, A. and {Buckley}, J.~H. and {Bugaev}, V. and {Byrum}, K. and {Cerruti}, M. and {Chen}, X. and {Ciupik}, L. and {Connolly}, M.~P. and {Cui}, W. and {Duke}, C. and {Dumm}, J. and {Errando}, M. and {Falcone}, A. and {Federici}, S. and {Feng}, Q. and {Finley}, J.~P. and {Fleischhack}, H. and {Fortin}, P. and {Fortson}, L. and {Furniss}, A. and {Galante}, N. and {Gillanders}, G.~H. and {Griffin}, S. and {Griffiths}, S.~T. and {Grube}, J. and {Gyuk}, G. and {Hanna}, D. and {Holder}, J. and {Hughes}, G. and {Humensky}, T.~B. and {Johnson}, C.~A. and {Kaaret}, P. and {Kertzman}, M. and {Khassen}, Y. and {Kieda}, D. and {Krawczynski}, H. and {Krennrich}, F. and {Lang}, M.~J. and {Madhavan}, A.~S. and {Maier}, G. and {Majumdar}, P. and {McArthur}, S. and {McCann}, A. and {Meagher}, K. and {Millis}, J. and {Moriarty}, P. and {Mukherjee}, R. and {Nieto}, D. and {O'Faol{\'a}in de Bhr{\'o}ithe}, A. and {Ong}, R.~A. and {Otte}, A.~N. and {Park}, N. and {Perkins}, J.~S. and {Pohl}, M. and {Popkow}, A. and {Prokoph}, H. and {Quinn}, J. and {Ragan}, K. and {Reyes}, L.~C. and {Reynolds}, P.~T. and {Richards}, G.~T. and {Roache}, E. and {Sembroski}, G.~H. and {Smith}, A.~W. and {Staszak}, D. and {Telezhinsky}, I. and {Theiling}, M. and {Varlotta}, A. and {Vassiliev}, V.~V. and {Vincent}, S. and {Wakely}, S.~P. and {Weekes}, T.~C. and {Weinstein}, A. and {Welsing}, R. and {Williams}, D.~A. and {Zajczyk}, A. and {Zitzer}, B.},
        title = "{A Three-year Multi-wavelength Study of the Very-high-energy {\ensuremath{\gamma}}-Ray Blazar 1ES 0229+200}",
      journal = {\apj},
         year = 2014,
        month = feb,
       volume = {782},
       number = {1},
          eid = {13},
        pages = {13},
          doi = {10.1088/0004-637X/782/1/13},
archivePrefix = {arXiv},
       eprint = {1312.6592},
 primaryClass = {astro-ph.HE},
       adsurl = {https://ui.adsabs.harvard.edu/abs/2014ApJ...782...13A}
}

@ARTICLE{2014ApJ...787..155T,
       author = {{Tanaka}, Y.~T. and {Stawarz}, {\L}. and {Finke}, J. and {Cheung}, C.~C. and {Dermer}, C.~D. and {Kataoka}, J. and {Bamba}, A. and {Dubus}, G. and {De Naurois}, M. and {Wagner}, S.~J. and {Fukazawa}, Y. and {Thompson}, D.~J.},
        title = "{Extreme Blazars Studied with Fermi-LAT and Suzaku: 1ES 0347-121 and Blazar Candidate HESS J1943+213}",
      journal = {\apj},
         year = 2014,
        month = jun,
       volume = {787},
       number = {2},
          eid = {155},
        pages = {155},
          doi = {10.1088/0004-637X/787/2/155},
archivePrefix = {arXiv},
       eprint = {1404.3727},
 primaryClass = {astro-ph.HE},
       adsurl = {https://ui.adsabs.harvard.edu/abs/2014ApJ...787..155T}
}

@ARTICLE{2012A&A...538A.103H,
       author = {{H.~E.~S.~S. Collaboration} and {Abramowski}, A. and {Acero}, F. and {Aharonian}, F. and {Akhperjanian}, A.~G. and {Anton}, G. and {Balzer}, A. and {Barnacka}, A. and {Barres de Almeida}, U. and {Becherini}, Y. and {Becker}, J. and {Behera}, B. and {Bernloehr}, K. and {Birsin}, E. and {Biteau}, J. and {Bochow}, A. and {Boisson}, C. and {Bolmont}, J. and {Bordas}, P. and {Brucker}, J. and {Brun}, F. and {Brun}, P. and {Bulik}, T. and {Buesching}, I. and {Carrigan}, S. and {Casanova}, S. and {Cerruti}, M. and {Chadwick}, P.~M. and {Charbonnier}, A. and {Chaves}, R.~C.~G. and {Cheesebrough}, A. and {Chounet}, L. -M. and {Clapson}, A.~C. and {Coignet}, G. and {Cologna}, G. and {Conrad}, J. and {Dalton}, M. and {Daniel}, M.~K. and {Davids}, I.~D. and {Degrange}, B. and {Deil}, C. and {Dickinson}, H.~J. and {Djannati-Ataie}, A. and {Domainko}, W. and {Drury}, L. O'c. and {Dubois}, F. and {Dubus}, G. and {Dutson}, K. and {Dyks}, J. and {Dyrda}, M. and {Egberts}, K. and {Eger}, P. and {Espigat}, P. and {Fallon}, L. and {Farnier}, C. and {Feinstein}, F. and {Fernandes}, M.~V. and {Fiasson}, A. and {Fontaine}, G. and {Foerster}, A. and {Fuesling}, M. and {Gallant}, Y.~A. and {Gast}, H. and {Gerard}, L. and {Gerbig}, D. and {Giebels}, B. and {Glicenstein}, J.~F. and {Glueck}, B. and {Goret}, P. and {Goering}, D. and {Haeffner}, S. and {Hague}, J.~D. and {Hampf}, D. and {Hauser}, M. and {Heinz}, S. and {Heinzelmann}, G. and {Henri}, G. and {Hermann}, G. and {Hinton}, J.~A. and {Hoffmann}, A. and {Hofmann}, W. and {Hofverberg}, P. and {Holler}, M. and {Horns}, D. and {Jacholkowska}, A. and {de Jager}, O.~C. and {Jahn}, C. and {Jamrozy}, M. and {Jung}, I. and {Kastendieck}, M.~A. and {Katarzynski}, K. and {Katz}, U. and {Kaufmann}, S. and {Keogh}, D. and {Khangulyan}, D. and {Khelifi}, B. and {Klochkov}, D. and {Kluzniak}, W. and {Kneiske}, T. and {Komin}, Nu. and {Kosack}, K. and {Kossakowski}, R. and {Laffon}, H. and {Lamanna}, G. and {Lennarz}, D. and {Lohse}, T. and {Lopatin}, A. and {Lu}, C. -C. and {Marandon}, V. and {Marcowith}, A. and {Masbou}, J. and {Maurin}, D. and {Maxted}, N. and {Mayer}, M. and {McComb}, T.~J.~L. and {Medina}, M.~C. and {Mehault}, J. and {Moderski}, R. and {Moulin}, E. and {Naumann}, C.~L. and {Naumann-Godo}, M. and {de Naurois}, M. and {Nedbal}, D. and {Nekrassov}, D. and {Nguyen}, N. and {Nicholas}, B. and {Niemiec}, J. and {Nolan}, S.~J. and {Ohm}, S. and {de Ona Wilhelmi}, E. and {Opitz}, B. and {Ostrowski}, M. and {Oya}, I. and {Panter}, M. and {Paz Arribas}, M. and {Pedaletti}, G. and {Pelletier}, G. and {Petrucci}, P. -O. and {Pita}, S. and {Puehlhofer}, G. and {Punch}, M. and {Quirrenbach}, A. and {Raue}, M. and {Rayner}, S.~M. and {Reimer}, A. and {Reimer}, O. and {Renaud}, M. and {de Los Reyes}, R. and {Rieger}, F. and {Ripken}, J. and {Rob}, L. and {Rosier-Lees}, S. and {Rowell}, G. and {Rudak}, B. and {Rulten}, C.~B. and {Ruppel}, J. and {Sahakian}, V. and {Sanchez}, D.~A. and {Santangelo}, A. and {Schlickeiser}, R. and {Schoeck}, F.~M. and {Schulz}, A. and {Schwanke}, U. and {Schwarzburg}, S. and {Schwemmer}, S. and {Sheidaei}, F. and {Sikora}, M. and {Skilton}, J.~L. and {Sol}, H. and {Spengler}, G. and {Stawarz}, L. and {Steenkamp}, R. and {Stegmann}, C. and {Stinzing}, F. and {Stycz}, K. and {Sushch}, I. and {Szostek}, A. and {Tavernet}, J. -P. and {Terrier}, R. and {Tluczykont}, M. and {Valerius}, K. and {van Eldik}, C. and {Vasileiadis}, G. and {Venter}, C. and {Vialle}, J.~P. and {Viana}, A. and {Vincent}, P. and {Voelk}, H.~J. and {Volpe}, F. and {Vorobiov}, S. and {Vorster}, M. and {Wagner}, S.~J. and {Ward}, M. and {White}, R. and {Wierzcholska}, A. and {Zacharias}, M. and {Zajczyk}, A. and {Zdziarski}, A.~A. and {Zech}, A. and {Zechlin}, H. -S.~L. and {Costamante}, L. and {Fegan}, S. and {Ajello}, M.},
        title = "{Discovery of hard-spectrum {\ensuremath{\gamma}}-ray emission from the BL Lacertae object 1ES 0414+009}",
      journal = {\aap},
         year = 2012,
        month = feb,
       volume = {538},
          eid = {A103},
        pages = {A103},
          doi = {10.1051/0004-6361/201118406},
archivePrefix = {arXiv},
       eprint = {1201.2044},
 primaryClass = {astro-ph.HE},
       adsurl = {https://ui.adsabs.harvard.edu/abs/2012A&A...538A.103H}
}

@ARTICLE{2012ApJ...755..118A,
       author = {{Aliu}, E. and {Archambault}, S. and {Arlen}, T. and {Aune}, T. and {Beilicke}, M. and {Benbow}, W. and {B{\"o}ttcher}, M. and {Bouvier}, A. and {Bugaev}, V. and {Cannon}, A. and {Cesarini}, A. and {Ciupik}, L. and {Collins-Hughes}, E. and {Connolly}, M.~P. and {Cui}, W. and {Dickherber}, R. and {Dumm}, J. and {Errando}, M. and {Falcone}, A. and {Federici}, S. and {Feng}, Q. and {Finley}, J.~P. and {Finnegan}, G. and {Fortson}, L. and {Furniss}, A. and {Galante}, N. and {Gall}, D. and {Godambe}, S. and {Griffin}, S. and {Grube}, J. and {Gyuk}, G. and {Hanna}, D. and {Holder}, J. and {Huan}, H. and {Hughes}, G. and {Hui}, C.~M. and {Imran}, A. and {Jameil}, O. and {Kaaret}, P. and {Karlsson}, N. and {Kertzman}, M. and {Kerr}, J. and {Khassen}, Y. and {Kieda}, D. and {Krawczynski}, H. and {Krennrich}, F. and {Lang}, M.~J. and {Lee}, K. and {Madhavan}, A.~S. and {Majumdar}, P. and {McArthur}, S. and {McCann}, A. and {Moriarty}, P. and {Mukherjee}, R. and {Nelson}, T. and {O'Faol{\'a}in de Bhr{\'o}ithe}, A. and {Ong}, R.~A. and {Orr}, M. and {Otte}, A.~N. and {Park}, N. and {Perkins}, J.~S. and {Pichel}, A. and {Pohl}, M. and {Quinn}, J. and {Ragan}, K. and {Reynolds}, P.~T. and {Roache}, E. and {Ruppel}, J. and {Saxon}, D.~B. and {Schroedter}, M. and {Sembroski}, G.~H. and {{\c{S}}ent{\"u}rk}, G.~D. and {Smith}, A.~W. and {Staszak}, D. and {Stroh}, M. and {Telezhinsky}, I. and {Te{\v{s}}i{\'c}}, G. and {Theiling}, M. and {Thibadeau}, S. and {Tsurusaki}, K. and {Varlotta}, A. and {Vassiliev}, V.~V. and {Vivier}, M. and {Wakely}, S.~P. and {Ward}, J.~E. and {Weinstein}, A. and {Welsing}, R. and {Williams}, D.~A. and {Zitzer}, B.},
        title = "{Multiwavelength Observations of the AGN 1ES 0414+009 with VERITAS, Fermi-LAT, Swift-XRT, and MDM}",
      journal = {\apj},
         year = 2012,
        month = aug,
       volume = {755},
       number = {2},
          eid = {118},
        pages = {118},
          doi = {10.1088/0004-637X/755/2/118},
archivePrefix = {arXiv},
       eprint = {1206.4080},
 primaryClass = {astro-ph.HE},
       adsurl = {https://ui.adsabs.harvard.edu/abs/2012ApJ...755..118A}
}

@ARTICLE{1991AJ....101..818H,
       author = {{Halpern}, Jules P. and {Chen}, Vera S. and {Madejski}, Greg M. and {Chanan}, Gary A.},
        title = "{The Redshift of the X-ray Selected BL Lacertae Object H0414+009}",
      journal = {\aj},
         year = 1991,
        month = mar,
       volume = {101},
        pages = {818},
          doi = {10.1086/115725},
       adsurl = {https://ui.adsabs.harvard.edu/abs/1991AJ....101..818H}
}

@ARTICLE{2010A&A...521A..69A,
       author = {{Aharonian}, F. and {Akhperjanian}, A.~G. and {Anton}, G. and {Barres de Almeida}, U. and {Bazer-Bachi}, A.~R. and {Becherini}, Y. and {Behera}, B. and {Benbow}, W. and {Bernl{\"o}hr}, K. and {Bochow}, A. and {Boisson}, C. and {Bolmont}, J. and {Borrel}, V. and {Brucker}, J. and {Brun}, F. and {Brun}, P. and {B{\"u}hler}, R. and {Bulik}, T. and {B{\"u}sching}, I. and {Boutelier}, T. and {Chadwick}, P.~M. and {Charbonnier}, A. and {Chaves}, R.~C.~G. and {Cheesebrough}, A. and {Chounet}, L. -M. and {Clapson}, A.~C. and {Coignet}, G. and {Dalton}, M. and {Daniel}, M.~K. and {Davids}, I.~D. and {Degrange}, B. and {Deil}, C. and {Dickinson}, H.~J. and {Djannati-Ata{\"\i}}, A. and {Domainko}, W. and {O'C. Drury}, L. and {Dubois}, F. and {Dubus}, G. and {Dyks}, J. and {Dyrda}, M. and {Egberts}, K. and {Emmanoulopoulos}, D. and {Espigat}, P. and {Farnier}, C. and {Feinstein}, F. and {Fiasson}, A. and {F{\"o}rster}, A. and {Fontaine}, G. and {F{\"u}{\ss}ling}, M. and {Gabici}, S. and {Gallant}, Y.~A. and {G{\'e}rard}, L. and {Gerbig}, D. and {Giebels}, B. and {Glicenstein}, J.~F. and {Gl{\"u}ck}, B. and {Goret}, P. and {G{\"o}ring}, D. and {Hauser}, D. and {Hauser}, M. and {Heinz}, S. and {Heinzelmann}, G. and {Henri}, G. and {Hermann}, G. and {Hinton}, J.~A. and {Hoffmann}, A. and {Hofmann}, W. and {Holleran}, M. and {Hoppe}, S. and {Horns}, D. and {Jacholkowska}, A. and {de Jager}, O.~C. and {Jahn}, C. and {Jung}, I. and {Katarzy{\'n}ski}, K. and {Katz}, U. and {Kaufmann}, S. and {Kendziorra}, E. and {Kerschhaggl}, M. and {Khangulyan}, D. and {Kh{\'e}lifi}, B. and {Keogh}, D. and {Klu{\'z}niak}, W. and {Kneiske}, T. and {Komin}, Nu. and {Kosack}, K. and {Lamanna}, G. and {Lenain}, J. -P. and {Lohse}, T. and {Marandon}, V. and {Martin}, J.~M. and {Martineau-Huynh}, O. and {Marcowith}, A. and {Masbou}, J. and {Maurin}, D. and {McComb}, T.~J.~L. and {Medina}, M.~C. and {Moderski}, R. and {Moulin}, E. and {Naumann-Godo}, M. and {de Naurois}, M. and {Nedbal}, D. and {Nekrassov}, D. and {Nicholas}, B. and {Niemiec}, J. and {Nolan}, S.~J. and {Ohm}, S. and {Olive}, J. -F. and {de O{\~n}a Wilhelmi}, E. and {Orford}, K.~J. and {Ostrowski}, M. and {Panter}, M. and {Paz Arribas}, M. and {Pedaletti}, G. and {Pelletier}, G. and {Petrucci}, P. -O. and {Pita}, S. and {P{\"u}hlhofer}, G. and {Punch}, M. and {Quirrenbach}, A. and {Raubenheimer}, B.~C. and {Raue}, M. and {Rayner}, S.~M. and {Renaud}, M. and {Rieger}, F. and {Ripken}, J. and {Rob}, L. and {Rosier-Lees}, S. and {Rowell}, G. and {Rudak}, B. and {Rulten}, C.~B. and {Ruppel}, J. and {Sahakian}, V. and {Santangelo}, A. and {Schlickeiser}, R. and {Sch{\"o}ck}, F.~M. and {Schr{\"o}der}, R. and {Schwanke}, U. and {Schwarzburg}, S. and {Schwemmer}, S. and {Shalchi}, A. and {Sikora}, M. and {Skilton}, J.~L. and {Sol}, H. and {Spangler}, D. and {Stawarz}, {\L}. and {Steenkamp}, R. and {Stegmann}, C. and {Stinzing}, F. and {Superina}, G. and {Szostek}, A. and {Tam}, P.~H. and {Tavernet}, J. -P. and {Terrier}, R. and {Tibolla}, O. and {Tluczykont}, M. and {van Eldik}, C. and {Vasileiadis}, G. and {Venter}, C. and {Venter}, L. and {Vialle}, J.~P. and {Vincent}, P. and {Vivier}, H.~J. and {Volpe}, F. and {Wagner}, S.~J. and {Ward}, M. and {Zdziarski}, A.~A. and {Zech}, A.},
        title = "{Discovery of VHE {\ensuremath{\gamma}}-rays from the BL Lacertae object PKS 0548-322}",
      journal = {\aap},
         year = 2010,
        month = oct,
       volume = {521},
          eid = {A69},
        pages = {A69},
          doi = {10.1051/0004-6361/200912363},
archivePrefix = {arXiv},
       eprint = {1006.5289},
 primaryClass = {astro-ph.HE},
       adsurl = {https://ui.adsabs.harvard.edu/abs/2010A&A...521A..69A}
}

@ARTICLE{2007A&A...462..889P,
       author = {{Perri}, M. and {Maselli}, A. and {Giommi}, P. and {Massaro}, E. and {Nesci}, R. and {Tramacere}, A. and {Capalbi}, M. and {Cusumano}, G. and {Chincarini}, G. and {Tagliaferri}, G. and {Burrows}, D.~N. and {Berk}, D.~V. and {Gehrels}, N. and {Sambruna}, R.~M.},
        title = "{Swift XRT and UVOT deep observations of the high-energy peaked BL Lacertae object PKS 0548-322 close to its brightest state}",
      journal = {\aap},
         year = 2007,
        month = feb,
       volume = {462},
       number = {3},
        pages = {889-893},
          doi = {10.1051/0004-6361:20066063},
archivePrefix = {arXiv},
       eprint = {astro-ph/0610416},
 primaryClass = {astro-ph},
       adsurl = {https://ui.adsabs.harvard.edu/abs/2007A&A...462..889P}
}

@ARTICLE{2010ApJ...715L..49A,
       author = {{Acciari}, V.~A. and {Aliu}, E. and {Arlen}, T. and {Aune}, T. and {Bautista}, M. and {Beilicke}, M. and {Benbow}, W. and {B{\"o}ttcher}, M. and {Boltuch}, D. and {Bradbury}, S.~M. and {Buckley}, J.~H. and {Bugaev}, V. and {Byrum}, K. and {Cannon}, A. and {Cesarini}, A. and {Ciupik}, L. and {Cui}, W. and {Dickherber}, R. and {Duke}, C. and {Falcone}, A. and {Finley}, J.~P. and {Finnegan}, G. and {Fortson}, L. and {Furniss}, A. and {Galante}, N. and {Gall}, D. and {Gibbs}, K. and {Gillanders}, G.~H. and {Godambe}, S. and {Grube}, J. and {Guenette}, R. and {Gyuk}, G. and {Hanna}, D. and {Holder}, J. and {Hui}, C.~M. and {Humensky}, T.~B. and {Imran}, A. and {Kaaret}, P. and {Karlsson}, N. and {Kertzman}, M. and {Kieda}, D. and {Konopelko}, A. and {Krawczynski}, H. and {Krennrich}, F. and {Lang}, M.~J. and {Lamerato}, A. and {LeBohec}, S. and {Maier}, G. and {McArthur}, S. and {McCann}, A. and {McCutcheon}, M. and {Moriarty}, P. and {Mukherjee}, R. and {Ong}, R.~A. and {Otte}, A.~N. and {Pandel}, D. and {Perkins}, J.~S. and {Petry}, D. and {Pichel}, A. and {Pohl}, M. and {Quinn}, J. and {Ragan}, K. and {Reyes}, L.~C. and {Reynolds}, P.~T. and {Roache}, E. and {Rose}, H.~J. and {Roustazadeh}, P. and {Schroedter}, M. and {Sembroski}, G.~H. and {Senturk}, G. Demet and {Smith}, A.~W. and {Steele}, D. and {Swordy}, S.~P. and {Te{\v{s}}i{\'c}}, G. and {Theiling}, M. and {Thibadeau}, S. and {Varlotta}, A. and {Vassiliev}, V.~V. and {Vincent}, S. and {Wagner}, R.~G. and {Wakely}, S.~P. and {Ward}, J.~E. and {Weekes}, T.~C. and {Weinstein}, A. and {Weisgarber}, T. and {Williams}, D.~A. and {Wissel}, S. and {Wood}, M. and {Zitzer}, B. and {Ackermann}, M. and {Ajello}, M. and {Antolini}, E. and {Baldini}, L. and {Ballet}, J. and {Barbiellini}, G. and {Bastieri}, D. and {Bechtol}, K. and {Bellazzini}, R. and {Berenji}, B. and {Blandford}, R.~D. and {Bloom}, E.~D. and {Bonamente}, E. and {Borgland}, A.~W. and {Bouvier}, A. and {Bregeon}, J. and {Brigida}, M. and {Bruel}, P. and {Buehler}, R. and {Buson}, S. and {Caliandro}, G.~A. and {Cameron}, R.~A. and {Caraveo}, P.~A. and {Carrigan}, S. and {Casandjian}, J.~M. and {Cavazzuti}, E. and {Cecchi}, C. and {{\c{C}}elik}, {\"O}. and {Charles}, E. and {Chekhtman}, A. and {Cheung}, C.~C. and {Chiang}, J. and {Ciprini}, S. and {Claus}, R. and {Cohen-Tanugi}, J. and {Conrad}, J. and {Dermer}, C.~D. and {de Palma}, F. and {Silva}, E. do Couto e. and {Drell}, P.~S. and {Dubois}, R. and {Dumora}, D. and {Farnier}, C. and {Favuzzi}, C. and {Fegan}, S.~J. and {Fortin}, P. and {Frailis}, M. and {Fukazawa}, Y. and {Funk}, S. and {Fusco}, P. and {Gargano}, F. and {Gasparrini}, D. and {Gehrels}, N. and {Germani}, S. and {Giebels}, B. and {Giglietto}, N. and {Giordano}, F. and {Giroletti}, M. and {Glanzman}, T. and {Godfrey}, G. and {Grenier}, I.~A. and {Grove}, J.~E. and {Guiriec}, S. and {Hays}, E. and {Horan}, D. and {Hughes}, R.~E. and {J{\'o}hannesson}, G. and {Johnson}, A.~S. and {Johnson}, W.~N. and {Kamae}, T. and {Katagiri}, H. and {Kataoka}, J. and {Kn{\"o}dlseder}, J. and {Kuss}, M. and {Lande}, J. and {Latronico}, L. and {Lee}, S. -H. and {Llena Garde}, M. and {Longo}, F. and {Loparco}, F. and {Lott}, B. and {Lovellette}, M.~N. and {Lubrano}, P. and {Makeev}, A. and {Mazziotta}, M.~N. and {Michelson}, P.~F. and {Mitthumsiri}, W. and {Mizuno}, T. and {Moiseev}, A.~A. and {Monte}, C. and {Monzani}, M.~E. and {Morselli}, A. and {Moskalenko}, I.~V. and {Murgia}, S. and {Nolan}, P.~L. and {Norris}, J.~P. and {Nuss}, E. and {Ohno}, M. and {Ohsugi}, T. and {Omodei}, N. and {Orlando}, E. and {Ormes}, J.~F. and {Paneque}, D. and {Panetta}, J.~H. and {Pelassa}, V. and {Pepe}, M. and {Pesce-Rollins}, M. and {Piron}, F. and {Porter}, T.~A. and {Rain{\`o}}, S. and {Rando}, R. and {Razzano}, M. and {Reimer}, A. and {Reimer}, O. and {Ripken}, J. and {Rodriguez}, A.~Y. and {Roth}, M. and {Sadrozinski}, H.~F. -W. and {Sanchez}, D. and {Sander}, A. and {Scargle}, J.~D. and {Sgr{\`o}}, C. and {Siskind}, E.~J. and {Smith}, P.~D. and {Spandre}, G. and {Spinelli}, P. and {Strickman}, M.~S. and {Suson}, D.~J. and {Takahashi}, H. and {Tanaka}, T. and {Thayer}, J.~B. and {Thayer}, J.~G. and {Thompson}, D.~J. and {Tibaldo}, L. and {Torres}, D.~F. and {Tosti}, G. and {Tramacere}, A. and {Usher}, T.~L. and {Vasileiou}, V. and {Vilchez}, N. and {Vitale}, V. and {Waite}, A.~P. and {Wang}, P. and {Winer}, B.~L. and {Wood}, K.~S. and {Yang}, Z. and {Ylinen}, T. and {Ziegler}, M.},
        title = "{The Discovery of {\ensuremath{\gamma}}-Ray Emission from the Blazar RGB J0710+591}",
      journal = {\apjl},
         year = 2010,
        month = may,
       volume = {715},
       number = {1},
        pages = {L49-L55},
          doi = {10.1088/2041-8205/715/1/L49},
archivePrefix = {arXiv},
       eprint = {1005.0041},
 primaryClass = {astro-ph.HE},
       adsurl = {https://ui.adsabs.harvard.edu/abs/2010ApJ...715L..49A}
}

@ARTICLE{2023MNRAS.519..854M,
       author = {{Medina-Carrillo}, B. and {Sahu}, Sarira and {S{\'a}nchez-Col{\'o}n}, G. and {Rajpoot}, Subhash},
        title = "{Very high energy emission mechanism in the extreme blazar PGC 2402248}",
      journal = {\mnras},
         year = 2023,
        month = feb,
       volume = {519},
       number = {1},
        pages = {854-860},
          doi = {10.1093/mnras/stac3591},
archivePrefix = {arXiv},
       eprint = {2212.03064},
 primaryClass = {astro-ph.HE},
       adsurl = {https://ui.adsabs.harvard.edu/abs/2023MNRAS.519..854M}
}

@INPROCEEDINGS{2022icrc.confE.823B,
       author = {{Bony (de)}, M. and {Bylund}, T. and {Meyer}, M. and {Noel}, A.~P. and {Sanchez}, D. and {Abdalla}, H. and {Aharonian}, F. and {Ait-Benkhali}, F. and {Anguener}, O. and {Arcaro}, C. and {Armand}, C. and {Armstrong}, T. and {Ashkar}, H. and {Backes}, M. and {Baghmanyan}, V. and {Barbosa Martins}, V. and {Barnacka}, A. and {Barnard}, M. and {Batzofin}, R. and {Becherini}, Y. and {Berge}, D. and {Bernloehr}, K. and {Bi}, B. and {B{\"o}ttcher}, M. and {Boisson}, C. and {Bolmont}, J. and {Breuhaus}, M. and {Brose}, R. and {Brun}, F. and {Bulik}, T. and {Cangemi}, F. and {Caroff}, S. and {Casanova}, S. and {Catalano}, J. and {Chambery}, P. and {Chand}, T.~B. and {Chen}, A. and {Cotter}, G. and {Curlo}, M. and {Dalgleish}, H. and {Damascene Mbarubucyeye}, J. and {Davids}, I.~D. and {Davies}, J. and {Devin}, J. and {Djannati-Ata{\"\i}}, A. and {Dmytriiev}, A. and {Donath}, A. and {Doroshenko}, V. and {Dreyer}, L. and {Du Plessis}, L. and {Duffy}, C. and {Egberts}, K. and {Einecke}, S. and {Ernenwein}, J.~P. and {Fegan}, S. and {Feijen}, K. and {Fiasson}, A. and {Fichet de Clairfontaine}, G. and {Fontaine}, G. and {Frans}, L. and {Fuessling}, M. and {Funk}, S. and {Gabici}, S. and {Gallant}, Y. and {Giavitto}, G. and {Giunti}, L. and {Glawion}, D. and {Glicenstein}, J.~F. and {Grondin}, M.~H. and {Hattingh}, S. and {Haupt}, M. and {Hermann}, G. and {Hinton}, J. and {Hofmann}, W. and {Hoischen}, C. and {Holch}, T. and {Holler}, M. and {Horns}, D. and {Huang}, Z. and {Huber}, D. and {H{\"o}rbe}, M. and {Jamrozy}, M. and {Jankowsky}, F. and {Joshi}, V. and {Jung}, I. and {Kasai}, E. and {Katarzynski}, K. and {Katz}, U. and {Khangulyan}, D. and {Khelifi}, B. and {Klepser}, S. and {Kluzniak}, W. and {Komin}, N. and {Konno}, R. and {Kosack}, K. and {Kostunin}, D. and {Kreter}, M. and {Kukec Mezek}, G. and {Kundu}, A. and {Lamanna}, G. and {Le Stum}, S. and {Lemiere}, A. and {Lemoine-Goumard}, M. and {Lenain}, J.~P. and {Leuschner}, F. and {Levy}, C. and {Lohse}, T. and {Luashvili}, A. and {Lypova}, I. and {Mackey}, J. and {Majumdar}, J. and {Malyshev}, D. and {Malyshev}, D. and {Marandon}, V. and {Marchegiani}, P. and {Marcowith}, A. and {Mares}, A. and {Marti'i-Devesa}, G. and {Marx}, R. and {Maurin}, G. and {Meintjes}, P. and {Mitchell}, A. and {Moderski}, R. and {Mohrmann}, L. and {Montanari}, A. and {Moore}, C. and {Morris}, P. and {Moulin}, E. and {Muller}, J. and {Murach}, T. and {Nakashima}, K. and {Naurois (de)}, M. and {Nayerhoda}, A. and {Davids}, H. and {Niemiec}, J. and {O'Brien}, P. and {Oberholzer}, L.~L. and {Ohm}, S. and {Olivera-Nieto}, L. and {Ona-Wilhelmi (de)}, E. and {Ostrowski}, M. and {Panny}, S. and {Panter}, M. and {Parsons}, D. and {Peron}, G. and {Pita}, S. and {Poireau}, V. and {Prokhorov}, D. and {Prokoph}, H. and {Puehlhofer}, G. and {Punch}, M. and {Quirrenbach}, A. and {Reichherzer}, P. and {Reimer}, A. and {Reimer}, O. and {Remy}, Q. and {Renaud}, M. and {Reville}, B. and {Rieger}, F. and {Romoli}, C. and {Rowell}, G. and {Rudak}, B. and {Rueda Ricarte}, H. and {Ruiz Velasco}, E. and {Sahakian}, V. and {Sailer}, S. and {Salzmann}, H. and {Santangelo}, A. and {Sasaki}, M. and {Schaefer}, J. and {Schutte}, H. and {Schwanke}, U. and {Sch{\"u}ssler}, F. and {Senniappan}, M. and {Seyffert}, A. and {Shapopi}, J.~N.~S. and {Shiningayamwe}, K. and {Simoni}, R. and {Sinha}, A. and {Sol}, H. and {Spackman}, H. and {Specovius}, A. and {Spencer}, S.~T. and {Spir-Jacob}, M. and {Stawarz}, L. and {Steenkamp}, R. and {Stegmann}, C. and {Steinmassl}, S. and {Steppa}, C. and {Sun}, L. and {Takahashi}, T. and {Tanaka}, T. and {Tavernier}, T. and {Taylor}, A. and {Terrier}, R. and {Thiersen}, H. and {Thorpe-Morgan}, C. and {Tluczykont}, M. and {Tomankova}, L. and {Tsirou}, M. and {Tsuji}, N. and {Tuffs}, R. and {Uchiyama}, Y. and {van der Walt}, J. and {van Eldik}, C. and {van Rensburg}, C. and {van Soelen}, B. and {Vasileiadis}, G. and {Veh}, J. and {Venter}, C. and {Vincent}, P. and {Vink}, J. H.~J. and {Wagner}, S. and {Watson}, J.~J. and {Werner}, F. and {White}, R. and {Wierzcholska}, A. and {Wong}, Y.~W. and {Yassin}, H. and {Yusafzai}, A. and {Zacharias}, M. and {Zanin}, R. and {Zargaryan}, D. and {Zdziarski}, A. and {Zech}, A. and {Zhu}, S. and {Zmija}, A. and {Zouari}, S. and {{\.Z}ywucka}, N.},
        title = "{Detection of new Extreme BL Lac objects with H.E.S.S. and Swift XRT}",
    booktitle = {37th International Cosmic Ray Conference},
         year = 2022,
        month = mar,
          eid = {823},
        pages = {823},
          doi = {10.22323/1.395.0823},
archivePrefix = {arXiv},
       eprint = {2108.02232},
 primaryClass = {astro-ph.HE},
       adsurl = {https://ui.adsabs.harvard.edu/abs/2022icrc.confE.823B}
}

@ARTICLE{2007A&A...470..475A,
       author = {{Aharonian}, F. and {Akhperjanian}, A.~G. and {Bazer-Bachi}, A.~R. and {Beilicke}, M. and {Benbow}, W. and {Berge}, D. and {Bernl{\"o}hr}, K. and {Boisson}, C. and {Bolz}, O. and {Borrel}, V. and {Braun}, I. and {Brion}, E. and {Brown}, A.~M. and {B{\"u}hler}, R. and {B{\"u}sching}, I. and {Boutelier}, T. and {Carrigan}, S. and {Chadwick}, P.~M. and {Chounet}, L. -M. and {Coignet}, G. and {Cornils}, R. and {Costamante}, L. and {Degrange}, B. and {Dickinson}, H.~J. and {Djannati-Ata{\"\i}}, A. and {O'C. Drury}, L. and {Dubus}, G. and {Egberts}, K. and {Emmanoulopoulos}, D. and {Espigat}, P. and {Farnier}, C. and {Feinstein}, F. and {Ferrero}, E. and {Fiasson}, A. and {Fontaine}, G. and {Funk}, Seb. and {Funk}, S. and {F{\"u}{\ss}ling}, M. and {Gallant}, Y.~A. and {Giebels}, B. and {Glicenstein}, J.~F. and {Gl{\"u}ck}, B. and {Goret}, P. and {Hadjichristidis}, C. and {Hauser}, D. and {Hauser}, M. and {Heinzelmann}, G. and {Henri}, G. and {Hermann}, G. and {Hinton}, J.~A. and {Hoffmann}, A. and {Hofmann}, W. and {Holleran}, M. and {Hoppe}, S. and {Horns}, D. and {Jacholkowska}, A. and {de Jager}, O.~C. and {Kendziorra}, E. and {Kerschhaggl}, M. and {Kh{\'e}lifi}, B. and {Komin}, Nu. and {Kosack}, K. and {Lamanna}, G. and {Latham}, I.~J. and {Le Gallou}, R. and {Lemi{\`e}re}, A. and {Lemoine-Goumard}, M. and {Lohse}, T. and {Martin}, J.~M. and {Martineau-Huynh}, O. and {Marcowith}, A. and {Masterson}, C. and {Maurin}, G. and {McComb}, T.~J.~L. and {Moulin}, E. and {de Naurois}, M. and {Nedbal}, D. and {Nolan}, S.~J. and {Noutsos}, A. and {Olive}, J. -P. and {Orford}, K.~J. and {Osborne}, J.~L. and {Panter}, M. and {Pelletier}, G. and {Petrucci}, P. -O. and {Pita}, S. and {P{\"u}hlhofer}, G. and {Punch}, M. and {Ranchon}, S. and {Raubenheimer}, B.~C. and {Raue}, M. and {Rayner}, S.~M. and {Ripken}, J. and {Rob}, L. and {Rolland}, L. and {Rosier-Lees}, S. and {Rowell}, G. and {Sahakian}, V. and {Santangelo}, A. and {Saug{\'e}}, L. and {Schlenker}, S. and {Schlickeiser}, R. and {Schr{\"o}der}, R. and {Schwanke}, U. and {Schwarzburg}, S. and {Schwemmer}, S. and {Shalchi}, A. and {Sol}, H. and {Spangler}, D. and {Spanier}, F. and {Steenkamp}, R. and {Stegmann}, C. and {Superina}, G. and {Tam}, P.~H. and {Tavernet}, J. -P. and {Terrier}, R. and {Tluczykont}, M. and {van Eldik}, C. and {Vasileiadis}, G. and {Venter}, C. and {Vialle}, J.~P. and {Vincent}, P. and and {Wagner}, S.~J. and {Ward}, M.},
        title = "{Detection of VHE gamma-ray emission from the distant blazar 1ES{\,}1101-232 with HESS and broadband characterisation}",
      journal = {\aap},
         year = 2007,
        month = aug,
       volume = {470},
       number = {2},
        pages = {475-489},
          doi = {10.1051/0004-6361:20077057},
archivePrefix = {arXiv},
       eprint = {0705.2946},
 primaryClass = {astro-ph},
       adsurl = {https://ui.adsabs.harvard.edu/abs/2007A&A...470..475A}
}

@ARTICLE{2020ApJ...889..149D,
       author = {{Das}, Saikat and {Gupta}, Nayantara and {Razzaque}, Soebur},
        title = "{Ultrahigh-energy Cosmic-Ray Interactions as the Origin of Very High-energy {\ensuremath{\gamma}}-Rays from BL Lacertae Objects}",
      journal = {\apj},
         year = 2020,
        month = feb,
       volume = {889},
       number = {2},
          eid = {149},
        pages = {149},
          doi = {10.3847/1538-4357/ab6131},
archivePrefix = {arXiv},
       eprint = {1911.06011},
 primaryClass = {astro-ph.HE},
       adsurl = {https://ui.adsabs.harvard.edu/abs/2020ApJ...889..149D}
}

@ARTICLE{2008A&A...481L.103A,
       author = {{Aharonian}, F. and {Akhperjanian}, A.~G. and {Barres de Almeida}, U. and {Bazer-Bachi}, A.~R. and {Behera}, B. and {Beilicke}, M. and {Benbow}, W. and {Bernl{\"o}hr}, K. and {Boisson}, C. and {Borrel}, V. and {Braun}, I. and {Brion}, E. and {Brucker}, J. and {B{\"u}hler}, R. and {Bulik}, T. and {B{\"u}sching}, I. and {Boutelier}, T. and {Carrigan}, S. and {Chadwick}, P.~M. and {Chaves}, R.~C.~G. and {Chounet}, L. -M. and {Clapson}, A.~C. and {Coignet}, G. and {Cornils}, R. and {Costamante}, L. and {Dalton}, M. and {Degrange}, B. and {Dickinson}, H.~J. and {Djannati-Ata{\"\i}}, A. and {Domainko}, W. and {O'C. Drury}, L. and {Dubois}, F. and {Dubus}, G. and {Dyks}, J. and {Egberts}, K. and {Emmanoulopoulos}, D. and {Espigat}, P. and {Farnier}, C. and {Feinstein}, F. and {Fiasson}, A. and {F{\"o}rster}, A. and {Fontaine}, G. and {F{\"u}{\ss}ling}, M. and {Gabici}, S. and {Gallant}, Y.~A. and {Giebels}, B. and {Glicenstein}, J. -F. and {Gl{\"u}ck}, B. and {Goret}, P. and {Hadjichristidis}, C. and {Hauser}, D. and {Hauser}, M. and {Heinzelmann}, G. and {Henri}, G. and {Hermann}, G. and {Hinton}, J.~A. and {Hoffmann}, A. and {Hofmann}, W. and {Holleran}, M. and {Hoppe}, S. and {Horns}, D. and {Jacholkowska}, A. and {de Jager}, O.~C. and {Jung}, I. and {Katarzy{\'n}ski}, K. and {Kaufmann}, S. and {Kendziorra}, E. and {Kerschhaggl}, M. and {Khangulyan}, D. and {Kh{\'e}lifi}, B. and {Keogh}, D. and {Komin}, Nu. and {Kosack}, K. and {Lamanna}, G. and {Latham}, I.~J. and {Lenain}, J. -P. and {Lohse}, T. and {Martin}, J. -M. and {Martineau-Huynh}, O. and {Marcowith}, A. and {Masterson}, C. and {Maurin}, D. and {McComb}, T.~J.~L. and {Moderski}, R. and {Moulin}, E. and {Naumann-Godo}, M. and {de Naurois}, M. and {Nedbal}, D. and {Nekrassov}, D. and {Nolan}, S.~J. and {Ohm}, S. and {Olive}, J. -P. and {de O{\~n}a Wilhelmi}, E. and {Orford}, K.~J. and {Osborne}, J.~L. and {Ostrowski}, M. and {Panter}, M. and {Pedaletti}, G. and {Pelletier}, G. and {Petrucci}, P. -O. and {Pita}, S. and {P{\"u}hlhofer}, G. and {Punch}, M. and {Quirrenbach}, A. and {Raubenheimer}, B.~C. and {Raue}, M. and {Rayner}, S.~M. and {Renaud}, M. and {Rieger}, F. and {Ripken}, J. and {Rob}, L. and {Rosier-Lees}, S. and {Rowell}, G. and {Rudak}, B. and {Ruppel}, J. and {Sahakian}, V. and {Santangelo}, A. and {Schlickeiser}, R. and {Sch{\"o}ck}, F.~M. and {Schr{\"o}der}, R. and {Schwanke}, U. and {Schwarzburg}, S. and {Schwemmer}, S. and {Shalchi}, A. and {Sol}, H. and {Spangler}, D. and {Stawarz}, {\L}. and {Steenkamp}, R. and {Stegmann}, C. and {Superina}, G. and {Tam}, P.~H. and {Tavernet}, J. -P. and {Terrier}, R. and {van Eldik}, C. and {Vasileiadis}, G. and {Venter}, C. and {Vialle}, J. -P. and {Vincent}, P. and {Vivier}, . and {Volpe}, F. and {Wagner}, S.~J. and {Ward}, M. and {Zdziarski}, A.~A. and {Zech}, A.},
        title = "{Discovery of VHE {\ensuremath{\gamma}}-rays from the high-frequency-peaked BL Lacertae object RGB J0152+017}",
      journal = {\aap},
         year = 2008,
        month = apr,
       volume = {481},
       number = {3},
        pages = {L103-L107},
          doi = {10.1051/0004-6361:200809603},
archivePrefix = {arXiv},
       eprint = {0802.4021},
 primaryClass = {astro-ph},
       adsurl = {https://ui.adsabs.harvard.edu/abs/2008A&A...481L.103A}
}

@ARTICLE{2019MmSAI..90..164P,
       author = {{Prandini}, E. and {MAGIC Collaboration}},
        title = "{MAGIC extragalactic highlights from a MeV perspective}",
      journal = {\memsai},
         year = 2019,
        month = jan,
       volume = {90},
        pages = {164},
          doi = {10.48550/arXiv.1908.02154},
archivePrefix = {arXiv},
       eprint = {1908.02154},
 primaryClass = {astro-ph.HE},
       adsurl = {https://ui.adsabs.harvard.edu/abs/2019MmSAI..90..164P}
}

@ARTICLE{2010A&A...516A..56H,
       author = {{H.~E.~S.~S. Collaboration} and {Abramowski}, A. and {Acero}, F. and {Aharonian}, F. and {Akhperjanian}, A.~G. and {Anton}, G. and {Barres de Almeida}, U. and {Bazer-Bachi}, A.~R. and {Becherini}, Y. and {Behera}, B. and {Benbow}, W. and {Bernl{\"o}hr}, K. and {Bochow}, A. and {Boisson}, C. and {Bolmont}, J. and {Borrel}, V. and {Brucker}, J. and {Brun}, F. and {Brun}, P. and {B{\"u}hler}, R. and {Bulik}, T. and {B{\"u}sching}, I. and {Boutelier}, T. and {Chadwick}, P.~M. and {Charbonnier}, A. and {Chaves}, R.~C.~G. and {Cheesebrough}, A. and {Conrad}, J. and {Chounet}, L. -M. and {Clapson}, A.~C. and {Coignet}, G. and {Costamante}, L. and {Dalton}, M. and {Daniel}, M.~K. and {Davids}, I.~D. and {Degrange}, B. and {Deil}, C. and {Dickinson}, H.~J. and {Djannati-Ata{\"\i}}, A. and {Domainko}, W. and {O'C. Drury}, L. and {Dubois}, F. and {Dubus}, G. and {Dyks}, J. and {Dyrda}, M. and {Egberts}, K. and {Eger}, P. and {Espigat}, P. and {Fallon}, L. and {Farnier}, C. and {Fegan}, S. and {Feinstein}, F. and {Fernandes}, M.~V. and {Fiasson}, A. and {F{\"o}rster}, A. and {Fontaine}, G. and {F{\"u}{\ss}ling}, M. and {Gabici}, S. and {Gallant}, Y.~A. and {G{\'e}rard}, L. and {Gerbig}, D. and {Giebels}, B. and {Glicenstein}, J.~F. and {Gl{\"u}ck}, B. and {Goret}, P. and {G{\"o}ring}, D. and {Hampf}, D. and {Hauser}, M. and {Heinz}, S. and {Heinzelmann}, G. and {Henri}, G. and {Hermann}, G. and {Hinton}, J.~A. and {Hoffmann}, A. and {Hofmann}, W. and {Hofverberg}, P. and {Holleran}, M. and {Hoppe}, S. and {Horns}, D. and {Jacholkowska}, A. and {de Jager}, O.~C. and {Jahn}, C. and {Jung}, I. and {Katarzy{\'n}ski}, K. and {Katz}, U. and {Kaufmann}, S. and {Kerschhaggl}, M. and {Khangulyan}, D. and {Kh{\'e}lifi}, B. and {Keogh}, D. and {Klochkov}, D. and {Klu{\v{z}}niak}, W. and {Kneiske}, T. and {Komin}, Nu. and {Kosack}, K. and {Kossakowski}, R. and {Lamanna}, G. and {Lenain}, J. -P. and {Lohse}, T. and {Lu}, C. -C. and {Marandon}, V. and {Marcowith}, A. and {Masbou}, J. and {Maurin}, D. and {McComb}, T.~J.~L. and {Medina}, M.~C. and {M{\'e}hault}, J. and {Moderski}, R. and {Moulin}, E. and {Naumann-Godo}, M. and {de Naurois}, M. and {Nedbal}, D. and {Nekrassov}, D. and {Nguyen}, N. and {Nicholas}, B. and {Niemiec}, J. and {Nolan}, S.~J. and {Ohm}, S. and {Olive}, J. -F. and {de O{\~n}a Wilhelmi}, E. and {Opitz}, B. and {Orford}, K.~J. and {Ostrowski}, M. and {Panter}, M. and {Paz Arribas}, M. and {Pedaletti}, G. and {Pelletier}, G. and {Petrucci}, P. -O. and {Pita}, S. and {P{\"u}hlhofer}, G. and {Punch}, M. and {Quirrenbach}, A. and {Raubenheimer}, B.~C. and {Raue}, M. and {Rayner}, S.~M. and {Reimer}, O. and {Renaud}, M. and {de Los Reyes}, R. and {Rieger}, F. and {Ripken}, J. and {Rob}, L. and {Rosier-Lees}, S. and {Rowell}, G. and {Rudak}, B. and {Rulten}, C.~B. and {Ruppel}, J. and {Ryde}, F. and {Sahakian}, V. and {Santangelo}, A. and {Schlickeiser}, R. and {Sch{\"o}ck}, F.~M. and {Sch{\"o}nwald}, A. and {Schwanke}, U. and {Schwarzburg}, S. and {Schwemmer}, S. and {Shalchi}, A. and {Sushch}, I. and {Sikora}, M. and {Skilton}, J.~L. and {Sol}, H. and {Stawarz}, {\L}. and {Steenkamp}, R. and {Stegmann}, C. and {Stinzing}, F. and {Szostek}, A. and {Tam}, P.~H. and {Tavernet}, J. -P. and {Terrier}, R. and {Tibolla}, O. and {Tluczykont}, M. and {Valerius}, K. and {van Eldik}, C. and {Vasileiadis}, G. and {Venter}, C. and {Venter}, L. and {Vialle}, J.~P. and {Viana}, A. and {Vincent}, P. and {Vivier}, M. and {Volpe}, F. and {Vorobiov}, S. and {Wagner}, S.~J. and {Ward}, M. and {Zdziarski}, A.~A. and {Zech}, A. and {Zechlin}, H. -S.},
        title = "{Multi-wavelength observations of H 2356-309}",
      journal = {\aap},
         year = 2010,
        month = jun,
       volume = {516},
          eid = {A56},
        pages = {A56},
          doi = {10.1051/0004-6361/201014321},
archivePrefix = {arXiv},
       eprint = {1004.2089},
 primaryClass = {astro-ph.HE},
       adsurl = {https://ui.adsabs.harvard.edu/abs/2010A&A...516A..56H}
}

@ARTICLE{2021MNRAS.505.2712D,
       author = {{Dmytriiev}, A. and {Sol}, H. and {Zech}, A.},
        title = "{Connecting steady emission and very high energy flaring states in blazars: the case of Mrk 421}",
      journal = {\mnras},
         year = 2021,
        month = aug,
       volume = {505},
       number = {2},
        pages = {2712-2730},
          doi = {10.1093/mnras/stab1445},
archivePrefix = {arXiv},
       eprint = {2105.12480},
 primaryClass = {astro-ph.HE},
       adsurl = {https://ui.adsabs.harvard.edu/abs/2021MNRAS.505.2712D}
}

@ARTICLE{1992Natur.358..477P,
       author = {{Punch}, M. and {Akerlof}, C.~W. and {Cawley}, M.~F. and {Chantell}, M. and {Fegan}, D.~J. and {Fennell}, S. and {Gaidos}, J.~A. and {Hagan}, J. and {Hillas}, A.~M. and {Jiang}, Y. and {Kerrick}, A.~D. and {Lamb}, R.~C. and {Lawrence}, M.~A. and {Lewis}, D.~A. and {Meyer}, D.~I. and {Mohanty}, G. and {O'Flaherty}, K.~S. and {Reynolds}, P.~T. and {Rovero}, A.~C. and {Schubnell}, M.~S. and {Sembroski}, G. and {Weekes}, T.~C. and {Whitaker}, T. and {Wilson}, C.},
        title = "{Detection of TeV photons from the active galaxy Markarian 421}",
      journal = {\nat},
         year = 1992,
        month = aug,
       volume = {358},
       number = {6386},
        pages = {477-478},
          doi = {10.1038/358477a0},
       adsurl = {https://ui.adsabs.harvard.edu/abs/1992Natur.358..477P}
}

@ARTICLE{1996ApJ...456L..83Q,
       author = {{Quinn}, J. and {Akerlof}, C.~W. and {Biller}, S. and {Buckley}, J. and {Carter-Lewis}, D.~A. and {Cawley}, M.~F. and {Catanese}, M. and {Connaughton}, V. and {Fegan}, D.~J. and {Finley}, J.~P. and {Gaidos}, J. and {Hillas}, A.~M. and {Lamb}, R.~C. and {Krennrich}, F. and {Lessard}, R. and {McEnery}, J.~E. and {Meyer}, D.~I. and {Mohanty}, G. and {Rodgers}, A.~J. and {Rose}, H.~J. and {Sembroski}, G. and {Schubnell}, M.~S. and {Weekes}, T.~C. and {Wilson}, C. and {Zweerink}, J.},
        title = "{Detection of Gamma Rays with E > 300 GeV from Markarian 501}",
      journal = {\apjl},
         year = 1996,
        month = jan,
       volume = {456},
        pages = {L83},
          doi = {10.1086/309878},
       adsurl = {https://ui.adsabs.harvard.edu/abs/1996ApJ...456L..83Q}
}

@ARTICLE{2022ApJ...929..125A,
       author = {{Albert}, A. and {Alfaro}, R. and {Alvarez}, C. and {Angeles Camacho}, J.~R. and {Arteaga-Vel{\'a}zquez}, J.~C. and {Arunbabu}, K.~P. and {Avila Rojas}, D. and {Ayala Solares}, H.~A. and {Baghmanyan}, V. and {Belmont-Moreno}, E. and {Caballero-Mora}, K.~S. and {Capistr{\'a}n}, T. and {Carrami{\~n}ana}, A. and {Casanova}, S. and {Cotti}, U. and {Cotzomi}, J. and {Couti{\~n}o de Le{\'o}n}, S. and {de La Fuente}, E. and {Diaz Hernandez}, R. and {Duvernois}, M.~A. and {Durocher}, M. and {D{\'\i}az-V{\'e}lez}, J.~C. and {Engel}, K. and {Espinoza}, C. and {Fan}, K.~L. and {Fern{\'a}ndez Alonso}, M. and {Fraija}, N. and {Garcia}, D. and {Garc{\'\i}a-Gonz{\'a}lez}, J.~A. and {Garfias}, F. and {Gonz{\'a}lez}, M.~M. and {Goodman}, J.~A. and {Harding}, J.~P. and {Hona}, B. and {Huang}, D. and {Hueyotl-Zahuantitla}, F. and {H{\"u}ntemeyer}, P. and {Iriarte}, A. and {Joshi}, V. and {Lara}, A. and {Lee}, W.~H. and {Lee}, J. and {Vargas}, H. Le{\'o}n and {Linneman}, J.~T. and {Longinotti}, A.~L. and {Luis-Raya}, G. and {Malone}, K. and {Martinez}, O. and {Mart{\'\i}nez-Castro}, J. and {Matthews}, J.~A. and {Miranda-Romagnoli}, P. and {Moreno}, E. and {Mostaf{\'a}}, M. and {Nayerhoda}, A. and {Nellen}, L. and {Newbold}, M. and {Noriega-Papaqui}, R. and {Peisker}, A. and {P{\'e}rez Araujo}, Y. and {P{\'e}rez-P{\'e}rez}, E.~G. and {Rho}, C.~D. and {Rosa-Gonz{\'a}lez}, D. and {Salazar}, H. and {Salesa Greus}, F. and {Sandoval}, A. and {Schneider}, M. and {Serna-Franco}, J. and {Smith}, A.~J. and {Springer}, R.~W. and {Tollefson}, K. and {Torres}, I. and {Torres-Escobedo}, R. and {Ure{\~n}a-Mena}, F. and {Villase{\~n}or}, L. and {Wang}, X. and {Weisgarber}, T. and {Willox}, E. and {Zhou}, H. and {de Le{\'o}n}, C. and {HAWC Collaboration}},
        title = "{Long-term Spectra of the Blazars Mrk 421 and Mrk 501 at TeV Energies Seen by HAWC}",
      journal = {\apj},
         year = 2022,
        month = apr,
       volume = {929},
       number = {2},
          eid = {125},
        pages = {125},
          doi = {10.3847/1538-4357/ac58f6},
archivePrefix = {arXiv},
       eprint = {2106.03946},
 primaryClass = {astro-ph.HE},
       adsurl = {https://ui.adsabs.harvard.edu/abs/2022ApJ...929..125A}
}

@ARTICLE{2020A&A...637A..86M,
       author = {{MAGIC Collaboration} and {Acciari}, V.~A. and {Ansoldi}, S. and {Antonelli}, L.~A. and {Babi{\'c}}, A. and {Banerjee}, B. and {Barres de Almeida}, U. and {Barrio}, J.~A. and {Becerra Gonz{\'a}lez}, J. and {Bednarek}, W. and {Bernardini}, E. and {Berti}, A. and {Besenrieder}, J. and {Bhattacharyya}, W. and {Bigongiari}, C. and {Blanch}, O. and {Bonnoli}, G. and {Busetto}, G. and {Carosi}, R. and {Ceribella}, G. and {Cikota}, S. and {Colak}, S.~M. and {Colin}, P. and {Colombo}, E. and {Contreras}, J.~L. and {Cortina}, J. and {Covino}, S. and {D'Elia}, V. and {da Vela}, P. and {Dazzi}, F. and {de Angelis}, A. and {de Lotto}, B. and {Delfino}, M. and {Delgado}, J. and {di Pierro}, F. and {Do Souto Espi{\~n}era}, E. and {Dom{\'\i}nguez}, A. and {Dominis Prester}, D. and {Doro}, M. and {Fallah Ramazani}, V. and {Fattorini}, A. and {Fern{\'a}ndez-Barral}, A. and {Ferrara}, G. and {Fidalgo}, D. and {Foffano}, L. and {Fonseca}, M.~V. and {Font}, L. and {Fruck}, C. and {Galindo}, D. and {Gallozzi}, S. and {Garc{\'\i}a L{\'o}pez}, R.~J. and {Garczarczyk}, M. and {Gasparyan}, S. and {Gaug}, M. and {Giammaria}, P. and {Godinovi{\'c}}, N. and {Guberman}, D. and {Hadasch}, D. and {Hahn}, A. and {Hassan}, T. and {Herrera}, J. and {Hoang}, J. and {Hrupec}, D. and {Inoue}, S. and {Ishio}, K. and {Iwamura}, Y. and {Kubo}, H. and {Kushida}, J. and {Kuve{\v{z}}di{\'c}}, D. and {Lamastra}, A. and {Lelas}, D. and {Leone}, F. and {Lindfors}, E. and {Lombardi}, S. and {Longo}, F. and {L{\'o}pez}, M. and {L{\'o}pez-Oramas}, A. and {Machado de Oliveira Fraga}, B. and {Maggio}, C. and {Majumdar}, P. and {Makariev}, M. and {Mallamaci}, M. and {Maneva}, G. and {Manganaro}, M. and {Maraschi}, L. and {Mariotti}, M. and {Mart{\'\i}nez}, M. and {Masuda}, S. and {Mazin}, D. and {Minev}, M. and {Miranda}, J.~M. and {Mirzoyan}, R. and {Molina}, E. and {Moralejo}, A. and {Moreno}, V. and {Moretti}, E. and {Munar-Adrover}, P. and {Neustroev}, V. and {Niedzwiecki}, A. and {Nievas Rosillo}, M. and {Nigro}, C. and {Nilsson}, K. and {Ninci}, D. and {Nishijima}, K. and {Noda}, K. and {Nogu{\'e}s}, L. and {Paiano}, S. and {Palacio}, J. and {Paneque}, D. and {Paoletti}, R. and {Paredes}, J.~M. and {Pedaletti}, G. and {Pe{\~n}il}, P. and {Peresano}, M. and {Persic}, M. and {Prada Moroni}, P.~G. and {Prandini}, E. and {Puljak}, I. and {Garcia}, J.~R. and {Rib{\'o}}, M. and {Rico}, J. and {Righi}, C. and {Rugliancich}, A. and {Saha}, L. and {Sahakyan}, N. and {Saito}, T. and {Satalecka}, K. and {Schweizer}, T. and {Sitarek}, J. and {{\v{S}}nidari{\'c}}, I. and {Sobczynska}, D. and {Somero}, A. and {Stamerra}, A. and {Strzys}, M. and {Suri{\'c}}, T. and {Tavecchio}, F. and {Temnikov}, P. and {Terzi{\'c}}, T. and {Teshima}, M. and {Torres-Alb{\`a}}, N. and {Tsujimoto}, S. and {van Scherpenberg}, J. and {Vanzo}, G. and {Vazquez Acosta}, M. and {Vovk}, I. and {Will}, M. and {Zari{\'c}}, D. and {Fact Collaboration} and {Arbet-Engels}, A. and {Baack}, D. and {Balbo}, M. and {Biland}, A. and {Blank}, M. and {Bretz}, T. and {Bruegge}, K. and {Bulinski}, M. and {Buss}, J. and {Doerr}, M. and {Dorner}, D. and {Einecke}, S. and {Elsaesser}, D. and {Hildebrand}, D. and {Linhoff}, L. and {Mannheim}, K. and {Mueller}, S. and {Neise}, D. and {Neronov}, A. and {Noethe}, M. and {Paravac}, A. and {Rhode}, W. and {Schleicher}, B. and {Schulz}, F. and {Sedlaczek}, K. and {Shukla}, A. and {Sliusar}, V. and {von Willert}, E. and {Walter}, R. and {Wendel}, C. and {Tramacere}, A. and {Lien}, A. and {Perri}, M. and {Verrecchia}, F. and {Armas Padilla}, M. and {Leto}, C. and {L{\"a}hteenm{\"a}ki}, A. and {Tornikoski}, M. and {Tammi}, J.},
        title = "{Study of the variable broadband emission of Markarian 501 during the most extreme Swift X-ray activity}",
      journal = {\aap},
         year = 2020,
        month = may,
       volume = {637},
          eid = {A86},
        pages = {A86},
          doi = {10.1051/0004-6361/201834603},
archivePrefix = {arXiv},
       eprint = {2001.07729},
 primaryClass = {astro-ph.HE},
       adsurl = {https://ui.adsabs.harvard.edu/abs/2020A&A...637A..86M}
}

@INPROCEEDINGS{1999ICRC....3..370N,
       author = {{Nishiyama}, T.},
        title = "{Detection of a new TeV gamma-ray source of BL Lac object 1ES 1959+650}",
    booktitle = {26th International Cosmic Ray Conference (ICRC26), Volume 3},
         year = 1999,
       editor = {{Kieda}, D. and {Salamon}, M. and {Dingus}, B.},
       series = {International Cosmic Ray Conference},
       volume = {3},
        month = aug,
        pages = {370},
       adsurl = {https://ui.adsabs.harvard.edu/abs/1999ICRC....3..370N}
}

@ARTICLE{2018A&A...611A..44P,
       author = {{Patel}, S.~R. and {Shukla}, A. and {Chitnis}, V.~R. and {Dorner}, D. and {Mannheim}, K. and {Acharya}, B.~S. and {Nagare}, B.~J.},
        title = "{Broadband study of blazar 1ES 1959+650 during flaring state in 2016}",
      journal = {\aap},
         year = 2018,
        month = mar,
       volume = {611},
          eid = {A44},
        pages = {A44},
          doi = {10.1051/0004-6361/201731987},
archivePrefix = {arXiv},
       eprint = {1711.08583},
 primaryClass = {astro-ph.HE},
       adsurl = {https://ui.adsabs.harvard.edu/abs/2018A&A...611A..44P}
}

@ARTICLE{2014ApJ...797...89A,
       author = {{Aliu}, E. and {Archambault}, S. and {Arlen}, T. and {Aune}, T. and {Barnacka}, A. and {Beilicke}, M. and {Benbow}, W. and {Berger}, K. and {Bird}, R. and {Bouvier}, A. and {Buckley}, J.~H. and {Bugaev}, V. and {Cerruti}, M. and {Chen}, X. and {Ciupik}, L. and {Collins-Hughes}, E. and {Connolly}, M.~P. and {Cui}, W. and {Dumm}, J. and {Eisch}, J.~D. and {Falcone}, A. and {Federici}, S. and {Feng}, Q. and {Finley}, J.~P. and {Fleischhack}, H. and {Fortin}, P. and {Fortson}, L. and {Furniss}, A. and {Galante}, N. and {Gillanders}, G.~H. and {Griffin}, S. and {Griffiths}, S.~T. and {Grube}, J. and {Gyuk}, G. and {H{\r{a}}kansson}, N. and {Hanna}, D. and {Holder}, J. and {Hughes}, G. and {Hughes}, Z. and {Humensky}, T.~B. and {Johnson}, C.~A. and {Kaaret}, P. and {Kar}, P. and {Kertzman}, M. and {Khassen}, Y. and {Kieda}, D. and {Krawczynski}, H. and {Krennrich}, F. and {Lang}, M.~J. and {Madhavan}, A.~S. and {Majumdar}, P. and {McArthur}, S. and {McCann}, A. and {Meagher}, K. and {Millis}, J. and {Moriarty}, P. and {Mukherjee}, R. and {Nelson}, T. and {Nieto}, D. and {O'Faol{\'a}in de Bhr{\'o}ithe}, A. and {Ong}, R.~A. and {Otte}, A.~N. and {Park}, N. and {Perkins}, J.~S. and {Pohl}, M. and {Popkow}, A. and {Prokoph}, H. and {Quinn}, J. and {Ragan}, K. and {Rajotte}, J. and {Reyes}, L.~C. and {Reynolds}, P.~T. and {Richards}, G.~T. and {Roache}, E. and {Sadun}, A. and {Santander}, M. and {Sembroski}, G.~H. and {Shahinyan}, K. and {Sheidaei}, F. and {Smith}, A.~W. and {Staszak}, D. and {Telezhinsky}, I. and {Theiling}, M. and {Tyler}, J. and {Varlotta}, A. and {Vassiliev}, V.~V. and {Vincent}, S. and {Wakely}, S.~P. and {Weekes}, T.~C. and {Weinstein}, A. and {Welsing}, R. and {Wilhelm}, A. and {Williams}, D.~A. and {Zitzer}, B. and {VERITAS Collaboration} and {B{\"o}ttcher}, M. and {Fumagalli}, M.},
        title = "{Investigating Broadband Variability of the TeV Blazar 1ES 1959+650}",
      journal = {\apj},
         year = 2014,
        month = dec,
       volume = {797},
       number = {2},
          eid = {89},
        pages = {89},
          doi = {10.1088/0004-637X/797/2/89},
archivePrefix = {arXiv},
       eprint = {1412.1031},
 primaryClass = {astro-ph.HE},
       adsurl = {https://ui.adsabs.harvard.edu/abs/2014ApJ...797...89A}
}

@ARTICLE{2014ChA&A..38..233Y,
       author = {{Yuan}, Yu-hai and {Liu}, Fu-qing},
        title = "{The Long-term Light Variation of BL Lac Object 1ES 1959+650}",
      journal = {\caa},
         year = 2014,
        month = jul,
       volume = {38},
       number = {3},
        pages = {233-238},
          doi = {10.1016/j.chinastron.2014.07.001},
       adsurl = {https://ui.adsabs.harvard.edu/abs/2014ChA&A..38..233Y}
}

@ARTICLE{2015AJ....150...67Y,
       author = {{Yuan}, Y.~H. and {Fan}, J.~H. and {Pan}, H.~J.},
        title = "{Optical Photometry of the BL Lac Object 1ES 1959+650}",
      journal = {\aj},
         year = 2015,
        month = sep,
       volume = {150},
       number = {3},
          eid = {67},
        pages = {67},
          doi = {10.1088/0004-6256/150/3/67},
       adsurl = {https://ui.adsabs.harvard.edu/abs/2015AJ....150...67Y}
}

@ARTICLE{2016MNRAS.457..704K,
       author = {{Kapanadze}, B. and {Romano}, P. and {Vercellone}, S. and {Kapanadze}, S. and {Mdzinarishvili}, T. and {Kharshiladze}, G.},
        title = "{The long-term Swift observations of the high-energy peaked BL Lacertae source 1ES 1959+650}",
      journal = {\mnras},
         year = 2016,
        month = mar,
       volume = {457},
       number = {1},
        pages = {704-722},
          doi = {10.1093/mnras/stv3004},
       adsurl = {https://ui.adsabs.harvard.edu/abs/2016MNRAS.457..704K}
}

@ARTICLE{2018MNRAS.473.2542K,
       author = {{Kapanadze}, B. and {Dorner}, D. and {Vercellone}, S. and {Romano}, P. and {Hughes}, P. and {Aller}, M. and {Aller}, H. and {Reynolds}, M. and {Kapanadze}, S. and {Tabagari}, L.},
        title = "{The second strong X-ray flare and multifrequency variability of 1ES 1959+650 in 2016 January-August}",
      journal = {\mnras},
         year = 2018,
        month = jan,
       volume = {473},
       number = {2},
        pages = {2542-2564},
          doi = {10.1093/mnras/stx2492},
       adsurl = {https://ui.adsabs.harvard.edu/abs/2018MNRAS.473.2542K}
}

@ARTICLE{2017ApJ...846..158K,
       author = {{Kaur}, Navpreet and {Chandra}, S. and {Baliyan}, Kiran S. and {Sameer} and {Ganesh}, S.},
        title = "{A Multiwavelength Study of Flaring Activity in the High-energy Peaked BL Lac Object 1ES 1959+650 During 2015-2016}",
      journal = {\apj},
         year = 2017,
        month = sep,
       volume = {846},
       number = {2},
          eid = {158},
        pages = {158},
          doi = {10.3847/1538-4357/aa86b0},
archivePrefix = {arXiv},
       eprint = {1706.04411},
 primaryClass = {astro-ph.GA},
       adsurl = {https://ui.adsabs.harvard.edu/abs/2017ApJ...846..158K}
}

@ARTICLE{2017ApJ...847....8L,
       author = {{Li}, Xiao-Pan and {Luo}, Yu-Hui and {Yang}, Hai-Yan and {Yang}, Cheng and {Cai}, Yan and {Yang}, Hai-Tao},
        title = "{A Search for Quasi-periodic Oscillations in the Blazar 1ES 1959+650}",
      journal = {\apj},
         year = 2017,
        month = sep,
       volume = {847},
       number = {1},
          eid = {8},
        pages = {8},
          doi = {10.3847/1538-4357/aa86ee},
       adsurl = {https://ui.adsabs.harvard.edu/abs/2017ApJ...847....8L}
}

@ARTICLE{2017MNRAS.469.1682Z,
       author = {{Zhang}, You-Hong and {Li}, Jia-Chen},
        title = "{Optical variability of the high synchrotron energy peaked blazar 1ES 1959+650 on various time-scales}",
      journal = {\mnras},
         year = 2017,
        month = aug,
       volume = {469},
       number = {2},
        pages = {1682-1690},
          doi = {10.1093/mnras/stx942},
       adsurl = {https://ui.adsabs.harvard.edu/abs/2017MNRAS.469.1682Z}
}

@ARTICLE{2020A&A...640A.132M,
       author = {{MAGIC Collaboration} and {Acciari}, V.~A. and {Ansoldi}, S. and {Antonelli}, L.~A. and {Arbet Engels}, A. and {Baack}, D. and {Babi{\'c}}, A. and {Banerjee}, B. and {Barres de Almeida}, U. and {Barrio}, J.~A. and {Becerra Gonz{\'a}lez}, J. and {Bednarek}, W. and {Bellizzi}, L. and {Bernardini}, E. and {Berti}, A. and {Besenrieder}, J. and {Bhattacharyya}, W. and {Bigongiari}, C. and {Biland}, A. and {Blanch}, O. and {Bonnoli}, G. and {Bo{\v{s}}njak}, {\v{Z}}. and {Busetto}, G. and {Carosi}, R. and {Ceribella}, G. and {Cerruti}, M. and {Chai}, Y. and {Chilingarian}, A. and {Cikota}, S. and {Colak}, S.~M. and {Colin}, U. and {Colombo}, E. and {Contreras}, J.~L. and {Cortina}, J. and {Covino}, S. and {D'Amico}, G. and {D'Elia}, V. and {da Vela}, P. and {Dazzi}, F. and {de Angelis}, A. and {de Lotto}, B. and {Delfino}, M. and {Delgado}, J. and {Depaoli}, D. and {di Pierro}, F. and {di Venere}, L. and {Do Souto Espi{\~n}eira}, E. and {Dominis Prester}, D. and {Donini}, A. and {Dorner}, D. and {Doro}, M. and {Elsaesser}, D. and {Fallah Ramazani}, V. and {Fattorini}, A. and {Ferrara}, G. and {Foffano}, L. and {Fonseca}, M.~V. and {Font}, L. and {Fruck}, C. and {Fukami}, S. and {Garc{\'\i}a L{\'o}pez}, R.~J. and {Garczarczyk}, M. and {Gasparyan}, S. and {Gaug}, M. and {Giglietto}, N. and {Giordano}, F. and {Gliwny}, P. and {Godinovi{\'c}}, N. and {Green}, D. and {Hadasch}, D. and {Hahn}, A. and {Herrera}, J. and {Hoang}, J. and {Hrupec}, D. and {H{\"u}tten}, M. and {Inada}, T. and {Inoue}, S. and {Ishio}, K. and {Iwamura}, Y. and {Jouvin}, L. and {Kajiwara}, Y. and {Karjalainen}, M. and {Kerszberg}, D. and {Kobayashi}, Y. and {Kubo}, H. and {Kushida}, J. and {Lamastra}, A. and {Lelas}, D. and {Leone}, F. and {Lindfors}, E. and {Lombardi}, S. and {Longo}, F. and {L{\'o}pez}, M. and {L{\'o}pez-Coto}, R. and {L{\'o}pez-Oramas}, A. and {Loporchio}, S. and {Machado de Oliveira Fraga}, B. and {Maggio}, C. and {Majumdar}, P. and {Makariev}, M. and {Mallamaci}, M. and {Maneva}, G. and {Manganaro}, M. and {Mannheim}, K. and {Maraschi}, L. and {Mariotti}, M. and {Mart{\'\i}nez}, M. and {Mazin}, D. and {Mender}, S. and {Mi{\'c}anovi{\'c}}, S. and {Miceli}, D. and {Miener}, T. and {Minev}, M. and {Miranda}, J.~M. and {Mirzoyan}, R. and {Molina}, E. and {Moralejo}, A. and {Morcuende}, D. and {Moreno}, V. and {Moretti}, E. and {Munar-Adrover}, P. and {Neustroev}, V. and {Nigro}, C. and {Nilsson}, K. and {Ninci}, D. and {Nishijima}, K. and {Noda}, K. and {Nogu{\'e}s}, L. and {Nozaki}, S. and {Ohtani}, Y. and {Oka}, T. and {Otero-Santos}, J. and {Palatiello}, M. and {Paneque}, D. and {Paoletti}, R. and {Paredes}, J.~M. and {Pavleti{\'c}}, L. and {Pe{\~n}il}, P. and {Peresano}, M. and {Persic}, M. and {Prada Moroni}, P.~G. and {Prandini}, E. and {Puljak}, I. and {Rhode}, W. and {Rib{\'o}}, M. and {Rico}, J. and {Righi}, C. and {Rugliancich}, A. and {Saha}, L. and {Sahakyan}, N. and {Saito}, T. and {Sakurai}, S. and {Satalecka}, K. and {Schleicher}, B. and {Schmidt}, K. and {Schweizer}, T. and {Sitarek}, J. and {{\v{S}}nidari{\'c}}, I. and {Sobczynska}, D. and {Spolon}, A. and {Stamerra}, A. and {Strom}, D. and {Strzys}, M. and {Suda}, Y. and {Suri{\'c}}, T. and {Takahashi}, M. and {Tavecchio}, F. and {Temnikov}, P. and {Terzi{\'c}}, T. and {Teshima}, M. and {Torres-Alb{\`a}}, N. and {Tosti}, L. and {van Scherpenberg}, J. and {Vanzo}, G. and {Vazquez Acosta}, M. and {Ventura}, S. and {Verguilov}, V. and {Vigorito}, C.~F. and {Vitale}, V. and {Vovk}, I. and {Will}, M. and {Zari{\'c}}, D. and {Nievas-Rosillo}, M. and {Arcaro}, C. and {D'Ammando}, F. and {de Palma}, F. and {Hodges}, M. and {Hovatta}, T. and {Kiehlmann}, S. and {Max-Moerbeck}, W. and {Readhead}, A.~C.~S. and {Reeves}, R. and {Takalo}, L. and {Reinthal}, R. and {Jormanainen}, J. and {Wierda}, F. and {Wagner}, S.~M. and {Berdyugin}, A. and {Nabizadeh}, A. and {Talebpour Sheshvan}, N. and {Oksanen}, A. and {Bachev}, R. and {Strigachev}, A. and {Kehusmaa}, P.},
        title = "{Testing two-component models on very high-energy gamma-ray-emitting BL Lac objects}",
      journal = {\aap},
         year = 2020,
        month = aug,
       volume = {640},
          eid = {A132},
        pages = {A132},
          doi = {10.1051/0004-6361/202037811},
archivePrefix = {arXiv},
       eprint = {2006.04493},
 primaryClass = {astro-ph.HE},
       adsurl = {https://ui.adsabs.harvard.edu/abs/2020A&A...640A.132M}
}

@BOOK{1991rc3..book.....D,
       author = {{de Vaucouleurs}, Gerard and {de Vaucouleurs}, Antoinette and {Corwin}, Herold G., Jr. and {Buta}, Ronald J. and {Paturel}, Georges and {Fouque}, Pascal},
        title = "{Third Reference Catalogue of Bright Galaxies}",
         year = 1991,
       adsurl = {https://ui.adsabs.harvard.edu/abs/1991rc3..book.....D}
}

@ARTICLE{2015ApJ...808..110A,
       author = {{Archambault}, S. and {Archer}, A. and {Beilicke}, M. and {Benbow}, W. and {Bird}, R. and {Biteau}, J. and {Bouvier}, A. and {Bugaev}, V. and {Cardenzana}, J.~V. and {Cerruti}, M. and {Chen}, X. and {Ciupik}, L. and {Connolly}, M.~P. and {Cui}, W. and {Dickinson}, H.~J. and {Dumm}, J. and {Eisch}, J.~D. and {Errando}, M. and {Falcone}, A. and {Feng}, Q. and {Finley}, J.~P. and {Fleischhack}, H. and {Fortin}, P. and {Fortson}, L. and {Furniss}, A. and {Gillanders}, G.~H. and {Griffin}, S. and {Griffiths}, S.~T. and {Grube}, J. and {Gyuk}, G. and {H{\r{a}}kansson}, N. and {Hanna}, D. and {Holder}, J. and {Humensky}, T.~B. and {Johnson}, C.~A. and {Kaaret}, P. and {Kar}, P. and {Kertzman}, M. and {Khassen}, Y. and {Kieda}, D. and {Krause}, M. and {Krennrich}, F. and {Kumar}, S. and {Lang}, M.~J. and {Maier}, G. and {McArthur}, S. and {McCann}, A. and {Meagher}, K. and {Millis}, J. and {Moriarty}, P. and {Mukherjee}, R. and {Nieto}, D. and {O'Faol{\'a}in de Bhr{\'o}ithe}, A. and {Ong}, R.~A. and {Otte}, A.~N. and {Park}, N. and {Pohl}, M. and {Popkow}, A. and {Prokoph}, H. and {Pueschel}, E. and {Quinn}, J. and {Ragan}, K. and {Reyes}, L.~C. and {Reynolds}, P.~T. and {Richards}, G.~T. and {Roache}, E. and {Santander}, M. and {Sembroski}, G.~H. and {Shahinyan}, K. and {Smith}, A.~W. and {Staszak}, D. and {Telezhinsky}, I. and {Tucci}, J.~V. and {Tyler}, J. and {Varlotta}, A. and {Vincent}, S. and {Wakely}, S.~P. and {Weinstein}, A. and {Welsing}, R. and {Wilhelm}, A. and {Williams}, D.~A. and {Zitzer}, B. and {Veritas Collaboration} and {Hughes}, Z.~D.},
        title = "{VERITAS Detection of {\ensuremath{\gamma}}-Ray Flaring Activity From the BL Lac Object 1ES 1727+502 During Bright Moonlight Observations}",
      journal = {\apj},
         year = 2015,
        month = aug,
       volume = {808},
       number = {2},
          eid = {110},
        pages = {110},
          doi = {10.1088/0004-637X/808/2/110},
archivePrefix = {arXiv},
       eprint = {1506.06246},
 primaryClass = {astro-ph.HE},
       adsurl = {https://ui.adsabs.harvard.edu/abs/2015ApJ...808..110A}
}

@ARTICLE{2014A&A...563A..90A,
       author = {{Aleksi{\'c}}, J. and {Antonelli}, L.~A. and {Antoranz}, P. and {Asensio}, M. and {Backes}, M. and {Barres de Almeida}, U. and {Barrio}, J.~A. and {Becerra Gonz{\'a}lez}, J. and {Bednarek}, W. and {Berger}, K. and {Bernardini}, E. and {Biland}, A. and {Blanch}, O. and {Bock}, R.~K. and {Boller}, A. and {Bonnefoy}, S. and {Bonnoli}, G. and {Borla Tridon}, D. and {Borracci}, F. and {Bretz}, T. and {Carmona}, E. and {Carosi}, A. and {Carreto Fidalgo}, D. and {Colin}, P. and {Colombo}, E. and {Contreras}, J.~L. and {Cortina}, J. and {Cossio}, L. and {Covino}, S. and {da Vela}, P. and {Dazzi}, F. and {de Angelis}, A. and {de Caneva}, G. and {de Lotto}, B. and {Delgado Mendez}, C. and {Doert}, M. and {Dom{\'\i}nguez}, A. and {Dominis Prester}, D. and {Dorner}, D. and {Doro}, M. and {Eisenacher}, D. and {Elsaesser}, D. and {Farina}, E. and {Ferenc}, D. and {Fonseca}, M.~V. and {Font}, L. and {Fruck}, C. and {Garc{\'\i}a L{\'o}pez}, R.~J. and {Garczarczyk}, M. and {Garrido Terrats}, D. and {Gaug}, M. and {Giavitto}, G. and {Godinovi{\'c}}, N. and {Gonz{\'a}lez Mu{\~n}oz}, A. and {Gozzini}, S.~R. and {Hadamek}, A. and {Hadasch}, D. and {H{\"a}fner}, D. and {Herrero}, A. and {Hose}, J. and {Hrupec}, D. and {Idec}, W. and {Jankowski}, F. and {Kadenius}, V. and {Klepser}, S. and {Knoetig}, M.~L. and {Kr{\"a}henb{\"u}hl}, T. and {Krause}, J. and {Kushida}, J. and {La Barbera}, A. and {Lelas}, D. and {Lewandowska}, N. and {Lindfors}, E. and {Lombardi}, S. and {L{\'o}pez}, M. and {L{\'o}pez-Coto}, R. and {L{\'o}pez-Oramas}, A. and {Lorenz}, E. and {Lozano}, I. and {Makariev}, M. and {Mallot}, K. and {Maneva}, G. and {Mankuzhiyil}, N. and {Mannheim}, K. and {Maraschi}, L. and {Marcote}, B. and {Mariotti}, M. and {Mart{\'\i}nez}, M. and {Masbou}, J. and {Mazin}, D. and {Meucci}, M. and {Miranda}, J.~M. and {Mirzoyan}, R. and {Mold{\'o}n}, J. and {Moralejo}, A. and {Munar-Adrover}, P. and {Nakajima}, D. and {Niedzwiecki}, A. and {Nilsson}, K. and {Nowak}, N. and {Orito}, R. and {Paiano}, S. and {Palatiello}, M. and {Paneque}, D. and {Paoletti}, R. and {Paredes}, J.~M. and {Partini}, S. and {Persic}, M. and {Prada}, F. and {Prada Moroni}, P.~G. and {Prandini}, E. and {Puljak}, I. and {Reichardt}, I. and {Reinthal}, R. and {Rhode}, W. and {Rib{\'o}}, M. and {Rico}, J. and {R{\"u}gamer}, S. and {Saggion}, A. and {Saito}, K. and {Saito}, T.~Y. and {Salvati}, M. and {Satalecka}, K. and {Scalzotto}, V. and {Scapin}, V. and {Schultz}, C. and {Schweizer}, T. and {Shore}, S.~N. and {Sillanp{\"a}{\"a}}, A. and {Sitarek}, J. and {Snidaric}, I. and {Sobczynska}, D. and {Spanier}, F. and {Spiro}, S. and {Stamatescu}, V. and {Stamerra}, A. and {Steinke}, B. and {Storz}, J. and {Sun}, S. and {Suri{\'c}}, T. and {Takalo}, L. and {Takami}, H. and {Tavecchio}, F. and {Temnikov}, P. and {Terzi{\'c}}, T. and {Tescaro}, D. and {Teshima}, M. and {Tibolla}, O. and {Torres}, D.~F. and {Toyama}, T. and {Treves}, A. and {Uellenbeck}, M. and {Vogler}, P. and {Wagner}, R.~M. and {Weitzel}, Q. and {Zandanel}, F. and {Zanin}, R. and {MAGIC Collaboration}},
        title = "{Discovery of very high energy gamma-ray emission from the blazar 1ES 1727+502 with the MAGIC Telescopes}",
      journal = {\aap},
         year = 2014,
        month = mar,
       volume = {563},
          eid = {A90},
        pages = {A90},
          doi = {10.1051/0004-6361/201321360},
archivePrefix = {arXiv},
       eprint = {1302.6140},
 primaryClass = {astro-ph.HE},
       adsurl = {https://ui.adsabs.harvard.edu/abs/2014A&A...563A..90A}
}

@ARTICLE{2016MNRAS.459.2550A,
       author = {{Abeysekara}, A.~U. and {Archambault}, S. and {Archer}, A. and {Benbow}, W. and {Bird}, R. and {Biteau}, J. and {Buchovecky}, M. and {Buckley}, J.~H. and {Bugaev}, V. and {Byrum}, K. and {Cardenzana}, J.~V. and {Cerruti}, M. and {Chen}, X. and {Christiansen}, J.~L. and {Ciupik}, L. and {Connolly}, M.~P. and {Cui}, W. and {Dickinson}, H.~J. and {Dumm}, J. and {Eisch}, J.~D. and {Errando}, M. and {Falcone}, A. and {Feng}, Q. and {Finley}, J.~P. and {Fleischhack}, H. and {Flinders}, A. and {Fortin}, P. and {Fortson}, L. and {Furniss}, A. and {Gillanders}, G.~H. and {Griffin}, S. and {Grube}, J. and {Gyuk}, G. and {Huetten}, M. and {Hanna}, D. and {Holder}, J. and {Humensky}, T.~B. and {Johnson}, C.~A. and {Kaaret}, P. and {Kar}, P. and {Kelley-Hoskins}, N. and {Kertzman}, M. and {Kieda}, D. and {Krause}, M. and {Krennrich}, F. and {Lang}, M.~J. and {Maier}, G. and {McArthur}, S. and {McCann}, A. and {Meagher}, K. and {Moriarty}, P. and {Mukherjee}, R. and {Nieto}, D. and {O'Brien}, S. and {O'Faol{\'a}in de Bhr{\'o}ithe}, A. and {Ong}, R.~A. and {Otte}, A.~N. and {Park}, N. and {Pelassa}, V. and {Petrashyk}, A. and {Petry}, D. and {Pohl}, M. and {Popkow}, A. and {Pueschel}, E. and {Quinn}, J. and {Ragan}, K. and {Ratliff}, G. and {Reyes}, L.~C. and {Reynolds}, P.~T. and {Reynolds}, K. and {Richards}, G.~T. and {Roache}, E. and {Rulten}, C. and {Santander}, M. and {Sembroski}, G.~H. and {Shahinyan}, K. and {Smith}, A.~W. and {Staszak}, D. and {Telezhinsky}, I. and {Tucci}, J.~V. and {Tyler}, J. and {Vincent}, S. and {Wakely}, S.~P. and {Weiner}, O.~M. and {Weinstein}, A. and {Wilhelm}, A. and {Williams}, D.~A. and {Zitzer}, B.},
        title = "{VERITAS and multiwavelength observations of the BL Lacertae object 1ES 1741+196}",
      journal = {\mnras},
         year = 2016,
        month = jul,
       volume = {459},
       number = {3},
        pages = {2550-2557},
          doi = {10.1093/mnras/stw664},
archivePrefix = {arXiv},
       eprint = {1603.07286},
 primaryClass = {astro-ph.HE},
       adsurl = {https://ui.adsabs.harvard.edu/abs/2016MNRAS.459.2550A}
}

@ARTICLE{2017MNRAS.468.1534A,
       author = {{Ahnen}, M.~L. and {Ansoldi}, S. and {Antonelli}, L.~A. and {Antoranz}, P. and {Arcaro}, C. and {Babic}, A. and {Banerjee}, B. and {Bangale}, P. and {Barres de Almeida}, U. and {Barrio}, J.~A. and {Bednarek}, W. and {Bernardini}, E. and {Berti}, A. and {Biasuzzi}, B. and {Biland}, A. and {Blanch}, O. and {Bonnefoy}, S. and {Bonnoli}, G. and {Borracci}, F. and {Bretz}, T. and {Buson}, S. and {Carosi}, A. and {Chatterjee}, A. and {Clavero}, R. and {Colin}, P. and {Colombo}, E. and {Contreras}, J.~L. and {Cortina}, J. and {Covino}, S. and {Da Vela}, P. and {Dazzi}, F. and {De Angelis}, A. and {De Lotto}, B. and {de O{\~n}a Wilhelmi}, E. and {Di Pierro}, F. and {Doert}, M. and {Dom{\'\i}nguez}, A. and {Dominis Prester}, D. and {Dorner}, D. and {Doro}, M. and {Einecke}, S. and {Eisenacher Glawion}, D. and {Elsaesser}, D. and {Engelkemeier}, M. and {Fallah Ramazani}, V. and {Fern{\'a}ndez-Barral}, A. and {Fidalgo}, D. and {Fonseca}, M.~V. and {Font}, L. and {Frantzen}, K. and {Fruck}, C. and {Galindo}, D. and {Garc{\'\i}a L{\'o}pez}, R.~J. and {Garczarczyk}, M. and {Garrido Terrats}, D. and {Gaug}, M. and {Giammaria}, P. and {Godinovi{\'c}}, N. and {Gora}, D. and {Guberman}, D. and {Hadasch}, D. and {Hahn}, A. and {Hayashida}, M. and {Herrera}, J. and {Hose}, J. and {Hrupec}, D. and {Hughes}, G. and {Idec}, W. and {Kodani}, K. and {Konno}, Y. and {Kubo}, H. and {Kushida}, J. and {La Barbera}, A. and {Lelas}, D. and {Lindfors}, E. and {Lombardi}, S. and {Longo}, F. and {L{\'o}pez}, M. and {L{\'o}pez-Coto}, R. and {Majumdar}, P. and {Makariev}, M. and {Mallot}, K. and {Maneva}, G. and {Manganaro}, M. and {Mankuzhiyil}, N. and {Mannheim}, K. and {Maraschi}, L. and {Marcote}, B. and {Mariotti}, M. and {Mart{\'\i}nez}, M. and {Mazin}, D. and {Menzel}, U. and {Miranda}, J.~M. and {Mirzoyan}, R. and {Moralejo}, A. and {Moretti}, E. and {Nakajima}, D. and {Neustroev}, V. and {Niedzwiecki}, A. and {Nievas Rosillo}, M. and {Nilsson}, K. and {Nishijima}, K. and {Noda}, K. and {Nogu{\'e}s}, L. and {Paiano}, S. and {Palacio}, J. and {Palatiello}, M. and {Paneque}, D. and {Paoletti}, R. and {Paredes}, J.~M. and {Paredes-Fortuny}, X. and {Pedaletti}, G. and {Peresano}, M. and {Perri}, L. and {Persic}, M. and {Poutanen}, J. and {Prada Moroni}, P.~G. and {Prandini}, E. and {Puljak}, I. and {Garcia}, J.~R. and {Reichardt}, I. and {Rhode}, W. and {Rib{\'o}}, M. and {Rico}, J. and {Saito}, T. and {Satalecka}, K. and {Schroeder}, S. and {Schweizer}, T. and {Shore}, S.~N. and {Sillanp{\"a}{\"a}}, A. and {Sitarek}, J. and {Snidaric}, I. and {Sobczynska}, D. and {Stamerra}, A. and {Strzys}, M. and {Suri{\'c}}, T. and {Takalo}, L. and {Takami}, H. and {Tavecchio}, F. and {Temnikov}, P. and {Terzi{\'c}}, T. and {Tescaro}, D. and {Teshima}, M. and {Torres}, D.~F. and {Toyama}, T. and {Treves}, A. and {Vanzo}, G. and {Verguilov}, V. and {Vovk}, I. and {Ward}, J.~E. and {Will}, M. and {Wu}, M.~H. and {Zanin}, R. and {Becerra Gonz{\'a}lez}, J. and {Rani}, B. and {Krauss}, F. and {Perri}, M. and {Verrecchia}, F. and {Reinthal}, R.},
        title = "{MAGIC detection of very high energy {\ensuremath{\gamma}}-ray emission from the low-luminosity blazar 1ES 1741+196}",
      journal = {\mnras},
         year = 2017,
        month = jun,
       volume = {468},
       number = {2},
        pages = {1534-1541},
          doi = {10.1093/mnras/stx472},
archivePrefix = {arXiv},
       eprint = {1702.06795},
 primaryClass = {astro-ph.HE},
       adsurl = {https://ui.adsabs.harvard.edu/abs/2017MNRAS.468.1534A}
}

@ARTICLE{2014A&A...563A.135Z,
       author = {{{\.Z}ywucka}, Natalia and {Goyal}, Arti and {Jamrozy}, Marek and {Ostrowski}, Micha{\l} and {Stawarz}, {\L}ukasz},
        title = "{Low-frequency high-resolution radio observations of the TeV-emitting blazar SHBL J001355.9-185406}",
      journal = {\aap},
         year = 2014,
        month = mar,
       volume = {563},
          eid = {A135},
        pages = {A135},
          doi = {10.1051/0004-6361/201423500},
archivePrefix = {arXiv},
       eprint = {1402.1320},
 primaryClass = {astro-ph.GA},
       adsurl = {https://ui.adsabs.harvard.edu/abs/2014A&A...563A.135Z}
}

@ARTICLE{2012ApJS..199...31N,
       author = {{Nolan}, P.~L. and {Abdo}, A.~A. and {Ackermann}, M. and {Ajello}, M. and {Allafort}, A. and {Antolini}, E. and {Atwood}, W.~B. and {Axelsson}, M. and {Baldini}, L. and {Ballet}, J. and {Barbiellini}, G. and {Bastieri}, D. and {Bechtol}, K. and {Belfiore}, A. and {Bellazzini}, R. and {Berenji}, B. and {Bignami}, G.~F. and {Blandford}, R.~D. and {Bloom}, E.~D. and {Bonamente}, E. and {Bonnell}, J. and {Borgland}, A.~W. and {Bottacini}, E. and {Bouvier}, A. and {Brandt}, T.~J. and {Bregeon}, J. and {Brigida}, M. and {Bruel}, P. and {Buehler}, R. and {Burnett}, T.~H. and {Buson}, S. and {Caliandro}, G.~A. and {Cameron}, R.~A. and {Campana}, R. and {Ca{\~n}adas}, B. and {Cannon}, A. and {Caraveo}, P.~A. and {Casandjian}, J.~M. and {Cavazzuti}, E. and {Ceccanti}, M. and {Cecchi}, C. and {{\c{C}}elik}, {\"O}. and {Charles}, E. and {Chekhtman}, A. and {Cheung}, C.~C. and {Chiang}, J. and {Chipaux}, R. and {Ciprini}, S. and {Claus}, R. and {Cohen-Tanugi}, J. and {Cominsky}, L.~R. and {Conrad}, J. and {Corbet}, R. and {Cutini}, S. and {D'Ammando}, F. and {Davis}, D.~S. and {de Angelis}, A. and {DeCesar}, M.~E. and {DeKlotz}, M. and {De Luca}, A. and {den Hartog}, P.~R. and {de Palma}, F. and {Dermer}, C.~D. and {Digel}, S.~W. and {Silva}, E. do Couto e. and {Drell}, P.~S. and {Drlica-Wagner}, A. and {Dubois}, R. and {Dumora}, D. and {Enoto}, T. and {Escande}, L. and {Fabiani}, D. and {Falletti}, L. and {Favuzzi}, C. and {Fegan}, S.~J. and {Ferrara}, E.~C. and {Focke}, W.~B. and {Fortin}, P. and {Frailis}, M. and {Fukazawa}, Y. and {Funk}, S. and {Fusco}, P. and {Gargano}, F. and {Gasparrini}, D. and {Gehrels}, N. and {Germani}, S. and {Giebels}, B. and {Giglietto}, N. and {Giommi}, P. and {Giordano}, F. and {Giroletti}, M. and {Glanzman}, T. and {Godfrey}, G. and {Grenier}, I.~A. and {Grondin}, M. -H. and {Grove}, J.~E. and {Guillemot}, L. and {Guiriec}, S. and {Gustafsson}, M. and {Hadasch}, D. and {Hanabata}, Y. and {Harding}, A.~K. and {Hayashida}, M. and {Hays}, E. and {Hill}, A.~B. and {Horan}, D. and {Hou}, X. and {Hughes}, R.~E. and {Iafrate}, G. and {Itoh}, R. and {J{\'o}hannesson}, G. and {Johnson}, R.~P. and {Johnson}, T.~E. and {Johnson}, A.~S. and {Johnson}, T.~J. and {Kamae}, T. and {Katagiri}, H. and {Kataoka}, J. and {Katsuta}, J. and {Kawai}, N. and {Kerr}, M. and {Kn{\"o}dlseder}, J. and {Kocevski}, D. and {Kuss}, M. and {Lande}, J. and {Landriu}, D. and {Latronico}, L. and {Lemoine-Goumard}, M. and {Lionetto}, A.~M. and {Llena Garde}, M. and {Longo}, F. and {Loparco}, F. and {Lott}, B. and {Lovellette}, M.~N. and {Lubrano}, P. and {Madejski}, G.~M. and {Marelli}, M. and {Massaro}, E. and {Mazziotta}, M.~N. and {McConville}, W. and {McEnery}, J.~E. and {Mehault}, J. and {Michelson}, P.~F. and {Minuti}, M. and {Mitthumsiri}, W. and {Mizuno}, T. and {Moiseev}, A.~A. and {Mongelli}, M. and {Monte}, C. and {Monzani}, M.~E. and {Morselli}, A. and {Moskalenko}, I.~V. and {Murgia}, S. and {Nakamori}, T. and {Naumann-Godo}, M. and {Norris}, J.~P. and {Nuss}, E. and {Nymark}, T. and {Ohno}, M. and {Ohsugi}, T. and {Okumura}, A. and {Omodei}, N. and {Orlando}, E. and {Ormes}, J.~F. and {Ozaki}, M. and {Paneque}, D. and {Panetta}, J.~H. and {Parent}, D. and {Perkins}, J.~S. and {Pesce-Rollins}, M. and {Pierbattista}, M. and {Pinchera}, M. and {Piron}, F. and {Pivato}, G. and {Porter}, T.~A. and {Racusin}, J.~L. and {Rain{\`o}}, S. and {Rando}, R. and {Razzano}, M. and {Razzaque}, S. and {Reimer}, A. and {Reimer}, O. and {Reposeur}, T. and {Ritz}, S. and {Rochester}, L.~S. and {Romani}, R.~W. and {Roth}, M. and {Rousseau}, R. and {Ryde}, F. and {Sadrozinski}, H.~F. -W. and {Salvetti}, D. and {Sanchez}, D.~A. and {Saz Parkinson}, P.~M. and {Sbarra}, C. and {Scargle}, J.~D. and {Schalk}, T.~L. and {Sgr{\`o}}, C. and {Shaw}, M.~S. and {Shrader}, C. and {Siskind}, E.~J. and {Smith}, D.~A. and {Spandre}, G. and {Spinelli}, P. and {Stephens}, T.~E. and {Strickman}, M.~S. and {Suson}, D.~J. and {Tajima}, H. and {Takahashi}, H. and {Takahashi}, T. and {Tanaka}, T. and {Thayer}, J.~G. and {Thayer}, J.~B. and {Thompson}, D.~J. and {Tibaldo}, L. and {Tibolla}, O. and {Tinebra}, F. and {Tinivella}, M. and {Torres}, D.~F. and {Tosti}, G. and {Troja}, E. and {Uchiyama}, Y. and {Vandenbroucke}, J. and {Van Etten}, A. and {Van Klaveren}, B. and {Vasileiou}, V. and {Vianello}, G. and {Vitale}, V. and {Waite}, A.~P. and {Wallace}, E. and {Wang}, P. and {Werner}, M. and {Winer}, B.~L. and {Wood}, D.~L. and {Wood}, K.~S. and {Wood}, M. and {Yang}, Z. and {Zimmer}, S.},
        title = "{Fermi Large Area Telescope Second Source Catalog}",
      journal = {\apjs},
         year = 2012,
        month = apr,
       volume = {199},
       number = {2},
          eid = {31},
        pages = {31},
          doi = {10.1088/0067-0049/199/2/31},
archivePrefix = {arXiv},
       eprint = {1108.1435},
 primaryClass = {astro-ph.HE},
       adsurl = {https://ui.adsabs.harvard.edu/abs/2012ApJS..199...31N}
}

@ARTICLE{2016MNRAS.461..202A,
       author = {{Archambault}, S. and {Archer}, A. and {Barnacka}, A. and {Behera}, B. and {Beilicke}, M. and {Benbow}, W. and {Berger}, K. and {Bird}, R. and {B{\"o}ttcher}, M. and {Buckley}, J.~H. and {Bugaev}, V. and {Cardenzana}, J.~V. and {Cerruti}, M. and {Chen}, X. and {Christiansen}, J.~L. and {Ciupik}, L. and {Collins-Hughes}, E. and {Connolly}, M.~P. and {Cui}, W. and {Dickinson}, H.~J. and {Dumm}, J. and {Eisch}, J.~D. and {Errando}, M. and {Falcone}, A. and {Federici}, S. and {Feng}, Q. and {Finley}, J.~P. and {Fleischhack}, H. and {Fortson}, L. and {Furniss}, A. and {Gillanders}, G.~H. and {Godambe}, S. and {Griffin}, S. and {Griffiths}, S.~T. and {Grube}, J. and {Gyuk}, G. and {H{\r{a}}kansson}, N. and {Hanna}, D. and {Holder}, J. and {Hughes}, G. and {Johnson}, C.~A. and {Kaaret}, P. and {Kar}, P. and {Kertzman}, M. and {Khassen}, Y. and {Kieda}, D. and {Krawczynski}, H. and {Kumar}, S. and {Lang}, M.~J. and {Madhavan}, A.~S. and {Maier}, G. and {McArthur}, S. and {McCann}, A. and {Meagher}, K. and {Millis}, J. and {Moriarty}, P. and {Nelson}, T. and {Nieto}, D. and {de Bhr{\'o}ithe}, A. O'Faol{\'a}in and {Ong}, R.~A. and {Otte}, A.~N. and {Park}, N. and {Perkins}, J.~S. and {Pohl}, M. and {Popkow}, A. and {Prokoph}, H. and {Pueschel}, E. and {Quinn}, J. and {Ragan}, K. and {Rajotte}, J. and {Reyes}, L.~C. and {Reynolds}, P.~T. and {Richards}, G.~T. and {Roache}, E. and {Sembroski}, G.~H. and {Shahinyan}, K. and {Smith}, A.~W. and {Staszak}, D. and {Sweeney}, K. and {Telezhinsky}, I. and {Tucci}, J.~V. and {Tyler}, J. and {Varlotta}, A. and {Vassiliev}, V.~V. and {Wakely}, S.~P. and {Welsing}, R. and {Wilhelm}, A. and {Williams}, D.~A. and {Zitzer}, B.},
        title = "{Discovery of very high energy gamma rays from 1ES 1440+122}",
      journal = {\mnras},
         year = 2016,
        month = sep,
       volume = {461},
       number = {1},
        pages = {202-208},
          doi = {10.1093/mnras/stw1319},
archivePrefix = {arXiv},
       eprint = {1608.02769},
 primaryClass = {astro-ph.HE},
       adsurl = {https://ui.adsabs.harvard.edu/abs/2016MNRAS.461..202A}
}

@ARTICLE{2024A&A...682A.114M,
       author = {{MAGIC Collaboration} and {Abe}, H. and {Abe}, S. and {Acciari}, V.~A. and {Agudo}, I. and {Aniello}, T. and {Ansoldi}, S. and {Antonelli}, L.~A. and {Arbet Engels}, A. and {Arcaro}, C. and {Artero}, M. and {Asano}, K. and {Baack}, D. and {Babi{\'c}}, A. and {Baquero}, A. and {Barres de Almeida}, U. and {Batkovi{\'c}}, I. and {Baxter}, J. and {Becerra Gonz{\'a}lez}, J. and {Bernardini}, E. and {Bernete}, J. and {Berti}, A. and {Besenrieder}, J. and {Bigongiari}, C. and {Biland}, A. and {Blanch}, O. and {Bonnoli}, G. and {Bo{\v{s}}njak}, {\v{Z}}. and {Burelli}, I. and {Busetto}, G. and {Campoy-Ordaz}, A. and {Carosi}, A. and {Carosi}, R. and {Carretero-Castrillo}, M. and {Castro-Tirado}, A.~J. and {Chai}, Y. and {Cifuentes}, A. and {Cikota}, S. and {Colombo}, E. and {Contreras}, J.~L. and {Cortina}, J. and {Covino}, S. and {D'Amico}, G. and {D'Ammando}, F. and {D'Elia}, V. and {Da Vela}, P. and {Dazzi}, F. and {De Angelis}, A. and {De Lotto}, B. and {Del Popolo}, A. and {Delfino}, M. and {Delgado}, J. and {Delgado Mendez}, C. and {Depaoli}, D. and {Di Pierro}, F. and {Di Venere}, L. and {Dominis Prester}, D. and {Dorner}, D. and {Doro}, M. and {Elsaesser}, D. and {Emery}, G. and {Escudero}, J. and {Fari{\~n}a}, L. and {Fattorini}, A. and {Foffano}, L. and {Font}, L. and {Fukami}, S. and {Fukazawa}, Y. and {Garc{\'\i}a L{\'o}pez}, R.~J. and {Gasparyan}, S. and {Gaug}, M. and {Giesbrecht Paiva}, J.~G. and {Giglietto}, N. and {Giordano}, F. and {Gliwny}, P. and {Grau}, R. and {Green}, J.~G. and {Hadasch}, D. and {Hahn}, A. and {Heckmann}, L. and {Herrera}, J. and {Hrupec}, D. and {H{\"u}tten}, M. and {Imazawa}, R. and {Inada}, T. and {Iotov}, R. and {Ishio}, K. and {Jim{\'e}nez Mart{\'\i}nez}, I. and {Jormanainen}, J. and {Kerszberg}, D. and {Kluge}, G.~W. and {Kobayashi}, Y. and {Kouch}, P.~M. and {Kubo}, H. and {Kushida}, J. and {L{\'a}inez Lez{\'a}un}, M. and {Lamastra}, A. and {Leone}, F. and {Lindfors}, E. and {Linhoff}, L. and {Lombardi}, S. and {Longo}, F. and {L{\'o}pez-Moya}, M. and {L{\'o}pez-Oramas}, A. and {Loporchio}, S. and {Lorini}, A. and {Machado de Oliveira Fraga}, B. and {Majumdar}, P. and {Makariev}, M. and {Maneva}, G. and {Mang}, N. and {Manganaro}, M. and {Mariotti}, M. and {Mart{\'\i}nez}, M. and {Mart{\'\i}nez-Chicharro}, M. and {Mas-Aguilar}, A. and {Mazin}, D. and {Menchiari}, S. and {Mender}, S. and {Miceli}, D. and {Miener}, T. and {Miranda}, J.~M. and {Mirzoyan}, R. and {Molero Gonz{\'a}lez}, M. and {Molina}, E. and {Mondal}, H.~A. and {Moralejo}, A. and {Morcuende}, D. and {Nakamori}, T. and {Nanci}, C. and {Neustroev}, V. and {Nigro}, C. and {Nikoli{\'c}}, L. and {Nishijima}, K. and {Njoh Ekoume}, T. and {Noda}, K. and {Nozaki}, S. and {Ohtani}, Y. and {Okumura}, A. and {Otero-Santos}, J. and {Paiano}, S. and {Palatiello}, M. and {Paneque}, D. and {Paoletti}, R. and {Paredes}, J.~M. and {Pavlovi{\'c}}, D. and {Persic}, M. and {Pihet}, M. and {Pirola}, G. and {Podobnik}, F. and {Prada Moroni}, P.~G. and {Prandini}, E. and {Principe}, G. and {Priyadarshi}, C. and {Rhode}, W. and {Rib{\'o}}, M. and {Rico}, J. and {Righi}, C. and {Sahakyan}, N. and {Saito}, T. and {Satalecka}, K. and {Saturni}, F.~G. and {Schleicher}, B. and {Schmidt}, K. and {Schmuckermaier}, F. and {Schubert}, J.~L. and {Schweizer}, T. and {Sciaccaluga}, A. and {Sitarek}, J. and {Spolon}, A. and {Stamerra}, A. and {Stri{\v{s}}kovi{\'c}}, J. and {Strom}, D. and {Suda}, Y. and {Tajima}, H. and {Takeishi}, R. and {Tavecchio}, F. and {Temnikov}, P. and {Terauchi}, K. and {Terzi{\'c}}, T. and {Teshima}, M. and {Tosti}, L. and {Truzzi}, S. and {Tutone}, A. and {Ubach}, S. and {van Scherpenberg}, J. and {Ventura}, S. and {Verguilov}, V. and {Viale}, I. and {Vigorito}, C.~F. and {Vitale}, V. and {Walter}, R. and {Wunderlich}, C. and {Yamamoto}, T. and {Multi-wavelength Collaborators} and {Perri}, M. and {Verrecchia}, F. and {Leto}, C. and {Das}, S. and {Chatterjee}, R. and {Raiteri}, C.~M. and {Villata}, M. and {Semkov}, E. and {Ibryamov}, S. and {Bachev}, R. and {Strigachev}, A. and {Damljanovic}, G. and {Vince}, O. and {Jovanovic}, M.~D. and {Stojanovic}, M. and {Larionov}, V.~M. and {Grishina}, T.~S. and {Kopatskaya}, E.~N. and {Larionova}, E.~G. and {Morozova}, D.~A. and {Savchenko}, S.~S. and {Troitskiy}, I.~S. and {Troitskaya}, Y.~V. and {Vasilyev}, A.~A. and {Chen}, W.~P. and {Hou}, W.~J. and {Lin}, C.~S. and {Tsai}, A. and {Jorstad}, S.~G. and {Weaver}, Z.~R. and {Acosta-Pulido}, J.~A. and {Carnerero}, M.~I. and {Carosati}, D. and {Kurtanidze}, S.~O. and {Kurtanidze}, O.~M. and {Jordan}, B. and {Ivanidze}, R.~Z. and {Gazeas}, K. and {Vrontaki}, K. and {Hovatta}, T. and {Liodakis}, I. and {Readhead}, A.~C.~S. and {Kiehlmann}, S. and {Zheng}, W. and {Filippenko}, A.~V. and {Fallah Ramazani}, V.},
        title = "{Multi-year characterisation of the broad-band emission from the intermittent extreme BL Lac 1ES 2344+514}",
      journal = {\aap},
         year = 2024,
        month = feb,
       volume = {682},
          eid = {A114},
        pages = {A114},
          doi = {10.1051/0004-6361/202347845},
archivePrefix = {arXiv},
       eprint = {2310.03922},
 primaryClass = {astro-ph.HE},
       adsurl = {https://ui.adsabs.harvard.edu/abs/2024A&A...682A.114M}
}

@ARTICLE{2000MNRAS.317..743G,
       author = {{Giommi}, P. and {Padovani}, P. and {Perlman}, E.},
        title = "{Detection of exceptional X-ray spectral variability in the TeV BL Lac 1ES 2344+514}",
      journal = {\mnras},
         year = 2000,
        month = oct,
       volume = {317},
       number = {4},
        pages = {743-749},
          doi = {10.1046/j.1365-8711.2000.03353.x},
archivePrefix = {arXiv},
       eprint = {astro-ph/9907377},
 primaryClass = {astro-ph},
       adsurl = {https://ui.adsabs.harvard.edu/abs/2000MNRAS.317..743G}
}

@ARTICLE{1991ApJS...76..813S,
       author = {{Stocke}, John T. and {Morris}, Simon L. and {Gioia}, I.~M. and {Maccacaro}, T. and {Schild}, R. and {Wolter}, A. and {Fleming}, Thomas A. and {Henry}, J. Patrick},
        title = "{The Einstein Observatory Extended Medium-Sensitivity Survey. II. The Optical Identifications}",
      journal = {\apjs},
         year = 1991,
        month = jul,
       volume = {76},
        pages = {813},
          doi = {10.1086/191582},
       adsurl = {https://ui.adsabs.harvard.edu/abs/1991ApJS...76..813S}
}

@ARTICLE{1991ApJ...374..431S,
       author = {{Stickel}, M. and {Padovani}, P. and {Urry}, C.~M. and {Fried}, J.~W. and {Kuehr}, H.},
        title = "{The Complete Sample of 1 Jansky BL Lacertae Objects. I. Summary Properties}",
      journal = {\apj},
         year = 1991,
        month = jun,
       volume = {374},
        pages = {431},
          doi = {10.1086/170133},
       adsurl = {https://ui.adsabs.harvard.edu/abs/1991ApJ...374..431S}
}

@ARTICLE{1993ApJ...412..541S,
       author = {{Schachter}, Jonathan F. and {Stocke}, John T. and {Perlman}, Eric and {Elvis}, Martin and {Remillard}, Ron and {Granados}, Arno and {Luu}, Jane and {Huchra}, John P. and {Humphreys}, Roberta and {Urry}, C. Megan and {Wallin}, John},
        title = "{Ten New BL Lacertae Objects Discovered by an Efficient X-Ray/Radio/Optical Technique}",
      journal = {\apj},
         year = 1993,
        month = aug,
       volume = {412},
        pages = {541},
          doi = {10.1086/172942},
       adsurl = {https://ui.adsabs.harvard.edu/abs/1993ApJ...412..541S}
}

@ARTICLE{2001A&A...367..809K,
       author = {{Katarzy{\'n}ski}, K. and {Sol}, H. and {Kus}, A.},
        title = "{The multifrequency emission of Mrk 501. From radio to TeV gamma-rays}",
      journal = {\aap},
         year = 2001,
        month = mar,
       volume = {367},
        pages = {809-825},
          doi = {10.1051/0004-6361:20000538},
       adsurl = {https://ui.adsabs.harvard.edu/abs/2001A&A...367..809K}
}

@BOOK{2004vhec.book.....A,
       author = {{Aharonian}, Felix A.},
        title = "{Very high energy cosmic gamma radiation : a crucial window on the extreme Universe}",
         year = 2004,
          doi = {10.1142/4657},
       adsurl = {https://ui.adsabs.harvard.edu/abs/2004vhec.book.....A}
}

@BOOK{2009herb.book.....D,
       author = {{Dermer}, Charles D. and {Menon}, Govind},
        title = "{High Energy Radiation from Black Holes: Gamma Rays, Cosmic Rays, and Neutrinos}",
         year = 2009,
       adsurl = {https://ui.adsabs.harvard.edu/abs/2009herb.book.....D}
}

@ARTICLE{2020MNRAS.496.3912M,
       author = {{MAGIC Collaboration} and {Acciari}, V.~A. and {Ansoldi}, S. and {Antonelli}, L.~A. and {Arbet Engels}, A. and {Babi{\'c}}, A. and {Banerjee}, B. and {Barres de Almeida}, U. and {Barrio}, J.~A. and {Becerra Gonz{\'a}lez}, J. and {Bednarek}, W. and {Bellizzi}, L. and {Bernardini}, E. and {Berti}, A. and {Besenrieder}, J. and {Bhattacharyya}, W. and {Bigongiari}, C. and {Blanch}, O. and {Bonnoli}, G. and {Bo{\v{s}}njak}, {\v{Z}}. and {Busetto}, G. and {Carosi}, R. and {Ceribella}, G. and {Cerruti}, M. and {Chai}, Y. and {Chilingaryan}, A. and {Cikota}, S. and {Colak}, S.~M. and {Colin}, U. and {Colombo}, E. and {Contreras}, J.~L. and {Cortina}, J. and {Covino}, S. and {D'Elia}, V. and {da Vela}, P. and {Dazzi}, F. and {de Angelis}, A. and {de Lotto}, B. and {Delfino}, M. and {Delgado}, J. and {Depaoli}, D. and {di Pierro}, F. and {di Venere}, L. and {Do Souto Espi{\~n}eira}, E. and {Dominis Prester}, D. and {Donini}, A. and {Doro}, M. and {Elsaesser}, D. and {Fallah Ramazani}, V. and {Fattorini}, A. and {Ferrara}, G. and {Foffano}, L. and {Fonseca}, M.~V. and {Font}, L. and {Fruck}, C. and {Fukami}, S. and {Garc{\'\i}a L{\'o}pez}, R.~J. and {Garczarczyk}, M. and {Gasparyan}, S. and {Gaug}, M. and {Giglietto}, N. and {Giordano}, F. and {Godinovi{\'c}}, N. and {Gliwny}, P. and {Green}, D. and {Hadasch}, D. and {Hahn}, A. and {Herrera}, J. and {Hoang}, J. and {Hrupec}, D. and {H{\"u}tten}, M. and {Inada}, T. and {Inoue}, S. and {Ishio}, K. and {Iwamura}, Y. and {Jouvin}, L. and {Kajiwara}, Y. and {Kerszberg}, D. and {Kobayashi}, Y. and {Kubo}, H. and {Kushida}, J. and {Lamastra}, A. and {Lelas}, D. and {Leone}, F. and {Lindfors}, E. and {Lombardi}, S. and {Longo}, F. and {L{\'o}pez}, M. and {L{\'o}pez-Coto}, R. and {L{\'o}pez-Oramas}, A. and {Loporchio}, S. and {Machado de Oliveira Fraga}, B. and {Maggio}, C. and {Majumdar}, P. and {Makariev}, M. and {Mallamaci}, M. and {Maneva}, G. and {Manganaro}, M. and {Maraschi}, L. and {Mariotti}, M. and {Mart{\'\i}nez}, M. and {Mazin}, D. and {Mender}, S. and {Mi{\'c}anovi{\'c}}, S. and {Miceli}, D. and {Miener}, T. and {Minev}, M. and {Miranda}, J.~M. and {Mirzoyan}, R. and {Molina}, E. and {Moralejo}, A. and {Morcuende}, D. and {Moreno}, V. and {Moretti}, E. and {Munar-Adrover}, P. and {Neustroev}, V. and {Nigro}, C. and {Nilsson}, K. and {Ninci}, D. and {Nishijima}, K. and {Noda}, K. and {Nogu{\'e}s}, L. and {Nozaki}, S. and {Ohtani}, Y. and {Oka}, T. and {Otero-Santos}, J. and {Paiano}, S. and {Palatiello}, M. and {Paneque}, D. and {Paoletti}, R. and {Paredes}, J.~M. and {Pavleti{\'c}}, L. and {Pe{\~n}il}, P. and {Peresano}, M. and {Persic}, M. and {Prada Moroni}, P.~G. and {Prandini}, E. and {Puljak}, I. and {Rib{\'o}}, M. and {Rico}, J. and {Righi}, C. and {Rugliancich}, A. and {Saha}, L. and {Sahakyan}, N. and {Saito}, T. and {Sakurai}, S. and {Satalecka}, K. and {Schleicher}, B. and {Schmidt}, K. and {Schweizer}, T. and {Sitarek}, J. and {{\v{S}}nidari{\'c}}, I. and {Sobczynska}, D. and {Spolon}, A. and {Stamerra}, A. and {Strom}, D. and {Strzys}, M. and {Suda}, Y. and {Suri{\'c}}, T. and {Takahashi}, M. and {Tavecchio}, F. and {Temnikov}, P. and {Terzi{\'c}}, T. and {Teshima}, M. and {Torres-Alb{\`a}}, N. and {Tosti}, L. and {van Scherpenberg}, J. and {Vanzo}, G. and {Vazquez Acosta}, M. and {Ventura}, S. and {Verguilov}, V. and {Vigorito}, C.~F. and {Vitale}, V. and {Vovk}, I. and {Will}, M. and {Zari{\'c}}, D. and {Baack}, D. and {Fact Collaboration} and {Balbo}, M. and {Beck}, M. and {Biederbeck}, N. and {Biland}, A. and {Blank}, M. and {Bretz}, T. and {Bruegge}, K. and {Bulinski}, M. and {Buss}, J. and {Doerr}, M. and {Dorner}, D. and {Hildebrand}, D. and {Iotov}, R. and {Klinger}, M. and {Mannheim}, K. and {Achim Mueller}, S. and {Neise}, D. and {Neronov}, A. and {N{\"o}the}, M. and {Paravac}, A. and {Rhode}, W. and {Schleicher}, B.},
        title = "{An intermittent extreme BL Lac: MWL study of 1ES 2344+514 in an enhanced state}",
      journal = {\mnras},
         year = 2020,
        month = aug,
       volume = {496},
       number = {3},
        pages = {3912-3928},
          doi = {10.1093/mnras/staa1702},
archivePrefix = {arXiv},
       eprint = {2006.06796},
 primaryClass = {astro-ph.HE},
       adsurl = {https://ui.adsabs.harvard.edu/abs/2020MNRAS.496.3912M}
}

@ARTICLE{1996ApJ...461..657B,
       author = {{Bloom}, Steven D. and {Marscher}, Alan P.},
        title = "{An Analysis of the Synchrotron Self-Compton Model for the Multi--Wave Band Spectra of Blazars}",
      journal = {\apj},
         year = 1996,
        month = apr,
       volume = {461},
        pages = {657},
          doi = {10.1086/177092},
       adsurl = {https://ui.adsabs.harvard.edu/abs/1996ApJ...461..657B}
}

@ARTICLE{1997A&A...320...19M,
       author = {{Mastichiadis}, A. and {Kirk}, J.~G.},
        title = "{Variability in the synchrotron self-Compton model of blazar emission.}",
      journal = {\aap},
         year = 1997,
        month = apr,
       volume = {320},
        pages = {19-25},
          doi = {10.48550/arXiv.astro-ph/9610058},
archivePrefix = {arXiv},
       eprint = {astro-ph/9610058},
 primaryClass = {astro-ph},
       adsurl = {https://ui.adsabs.harvard.edu/abs/1997A&A...320...19M}
}

@ARTICLE{2009RAA.....9..777Z,
       author = {{Zhang}, Jin},
        title = "{A synchrotron self-Compton scenario for the very high energy {\ensuremath{\gamma}}-ray emission of the intermediate BL Lacertae object W Comae}",
      journal = {Research in Astronomy and Astrophysics},
         year = 2009,
        month = jul,
       volume = {9},
       number = {7},
        pages = {777-782},
          doi = {10.1088/1674-4527/9/7/006},
archivePrefix = {arXiv},
       eprint = {0903.2344},
 primaryClass = {astro-ph.HE},
       adsurl = {https://ui.adsabs.harvard.edu/abs/2009RAA.....9..777Z}
}

@ARTICLE{2024ApJS..271...10W,
       author = {{Wang}, Ze-Rui and {Xue}, Rui and {Xiong}, Dingrong and {Wang}, Hai-Qin and {Sun}, Lu-Ming and {Peng}, Fang-Kun and {Mao}, Jirong},
        title = "{Broadband Multiwavelength Study of LHAASO-detected Active Galactic Nuclei}",
      journal = {\apjs},
         year = 2024,
        month = mar,
       volume = {271},
       number = {1},
          eid = {10},
        pages = {10},
          doi = {10.3847/1538-4365/ad168c},
archivePrefix = {arXiv},
       eprint = {2308.10200},
 primaryClass = {astro-ph.HE},
       adsurl = {https://ui.adsabs.harvard.edu/abs/2024ApJS..271...10W}
}

@ARTICLE{1992A&A...256L..27D,
       author = {{Dermer}, C.~D. and {Schlickeiser}, R. and {Mastichiadis}, A.},
        title = "{High-energy gamma radiation from extragalactic radio sources.}",
      journal = {\aap},
         year = 1992,
        month = mar,
       volume = {256},
        pages = {L27-L30},
       adsurl = {https://ui.adsabs.harvard.edu/abs/1992A&A...256L..27D}
}

@ARTICLE{1994ApJ...421..153S,
       author = {{Sikora}, Marek and {Begelman}, Mitchell C. and {Rees}, Martin J.},
        title = "{Comptonization of Diffuse Ambient Radiation by a Relativistic Jet: The Source of Gamma Rays from Blazars?}",
      journal = {\apj},
         year = 1994,
        month = jan,
       volume = {421},
        pages = {153},
          doi = {10.1086/173633},
       adsurl = {https://ui.adsabs.harvard.edu/abs/1994ApJ...421..153S}
}

@ARTICLE{2000ApJ...545..107B,
       author = {{B{\l}a{\.z}ejowski}, M. and {Sikora}, M. and {Moderski}, R. and {Madejski}, G.~M.},
        title = "{Comptonization of Infrared Radiation from Hot Dust by Relativistic Jets in Quasars}",
      journal = {\apj},
         year = 2000,
        month = dec,
       volume = {545},
       number = {1},
        pages = {107-116},
          doi = {10.1086/317791},
archivePrefix = {arXiv},
       eprint = {astro-ph/0008154},
 primaryClass = {astro-ph},
       adsurl = {https://ui.adsabs.harvard.edu/abs/2000ApJ...545..107B}
}

@ARTICLE{2009MNRAS.397..985G,
       author = {{Ghisellini}, G. and {Tavecchio}, F.},
        title = "{Canonical high-power blazars}",
      journal = {\mnras},
         year = 2009,
        month = aug,
       volume = {397},
       number = {2},
        pages = {985-1002},
          doi = {10.1111/j.1365-2966.2009.15007.x},
archivePrefix = {arXiv},
       eprint = {0902.0793},
 primaryClass = {astro-ph.CO},
       adsurl = {https://ui.adsabs.harvard.edu/abs/2009MNRAS.397..985G}
}

@ARTICLE{2009ApJ...704...38S,
       author = {{Sikora}, Marek and {Stawarz}, {\L}ukasz and {Moderski}, Rafa{\l} and {Nalewajko}, Krzysztof and {Madejski}, Greg M.},
        title = "{Constraining Emission Models of Luminous Blazar Sources}",
      journal = {\apj},
         year = 2009,
        month = oct,
       volume = {704},
       number = {1},
        pages = {38-50},
          doi = {10.1088/0004-637X/704/1/38},
archivePrefix = {arXiv},
       eprint = {0904.1414},
 primaryClass = {astro-ph.CO},
       adsurl = {https://ui.adsabs.harvard.edu/abs/2009ApJ...704...38S}
}

@ARTICLE{2017A&ARv..25....2P,
       author = {{Padovani}, P. and {Alexander}, D.~M. and {Assef}, R.~J. and {De Marco}, B. and {Giommi}, P. and {Hickox}, R.~C. and {Richards}, G.~T. and {Smol{\v{c}}i{\'c}}, V. and {Hatziminaoglou}, E. and {Mainieri}, V. and {Salvato}, M.},
        title = "{Active galactic nuclei: what's in a name?}",
      journal = {\aapr},
         year = 2017,
        month = aug,
       volume = {25},
       number = {1},
          eid = {2},
        pages = {2},
          doi = {10.1007/s00159-017-0102-9},
archivePrefix = {arXiv},
       eprint = {1707.07134},
 primaryClass = {astro-ph.GA},
       adsurl = {https://ui.adsabs.harvard.edu/abs/2017A&ARv..25....2P}
}

@ARTICLE{2012MNRAS.420.2899G,
       author = {{Giommi}, P. and {Padovani}, P. and {Polenta}, G. and {Turriziani}, S. and {D'Elia}, V. and {Piranomonte}, S.},
        title = "{A simplified view of blazars: clearing the fog around long-standing selection effects}",
      journal = {\mnras},
         year = 2012,
        month = mar,
       volume = {420},
       number = {4},
        pages = {2899-2911},
          doi = {10.1111/j.1365-2966.2011.20044.x},
archivePrefix = {arXiv},
       eprint = {1110.4706},
 primaryClass = {astro-ph.CO},
       adsurl = {https://ui.adsabs.harvard.edu/abs/2012MNRAS.420.2899G}
}

@ARTICLE{2008Natur.452..966M,
       author = {{Marscher}, Alan P. and {Jorstad}, Svetlana G. and {D'Arcangelo}, Francesca D. and {Smith}, Paul S. and {Williams}, G. Grant and {Larionov}, Valeri M. and {Oh}, Haruki and {Olmstead}, Alice R. and {Aller}, Margo F. and {Aller}, Hugh D. and {McHardy}, Ian M. and {L{\"a}hteenm{\"a}ki}, Anne and {Tornikoski}, Merja and {Valtaoja}, Esko and {Hagen-Thorn}, Vladimir A. and {Kopatskaya}, Eugenia N. and {Gear}, Walter K. and {Tosti}, Gino and {Kurtanidze}, Omar and {Nikolashvili}, Maria and {Sigua}, Lorand and {Miller}, H. Richard and {Ryle}, Wesley T.},
        title = "{The inner jet of an active galactic nucleus as revealed by a radio-to-{\ensuremath{\gamma}}-ray outburst}",
      journal = {\nat},
         year = 2008,
        month = apr,
       volume = {452},
       number = {7190},
        pages = {966-969},
          doi = {10.1038/nature06895},
       adsurl = {https://ui.adsabs.harvard.edu/abs/2008Natur.452..966M}
}

@ARTICLE{2011ApJ...743..171A,
       author = {{Ackermann}, M. and {Ajello}, M. and {Allafort}, A. and {Antolini}, E. and {Atwood}, W.~B. and {Axelsson}, M. and {Baldini}, L. and {Ballet}, J. and {Barbiellini}, G. and {Bastieri}, D. and {Bechtol}, K. and {Bellazzini}, R. and {Berenji}, B. and {Blandford}, R.~D. and {Bloom}, E.~D. and {Bonamente}, E. and {Borgland}, A.~W. and {Bottacini}, E. and {Bouvier}, A. and {Bregeon}, J. and {Brigida}, M. and {Bruel}, P. and {Buehler}, R. and {Burnett}, T.~H. and {Buson}, S. and {Caliandro}, G.~A. and {Cameron}, R.~A. and {Caraveo}, P.~A. and {Casandjian}, J.~M. and {Cavazzuti}, E. and {Cecchi}, C. and {Charles}, E. and {Cheung}, C.~C. and {Chiang}, J. and {Ciprini}, S. and {Claus}, R. and {Cohen-Tanugi}, J. and {Conrad}, J. and {Costamante}, L. and {Cutini}, S. and {de Angelis}, A. and {de Palma}, F. and {Dermer}, C.~D. and {Digel}, S.~W. and {Silva}, E. do Couto e. and {Drell}, P.~S. and {Dubois}, R. and {Escande}, L. and {Favuzzi}, C. and {Fegan}, S.~J. and {Ferrara}, E.~C. and {Finke}, J. and {Focke}, W.~B. and {Fortin}, P. and {Frailis}, M. and {Fukazawa}, Y. and {Funk}, S. and {Fusco}, P. and {Gargano}, F. and {Gasparrini}, D. and {Gehrels}, N. and {Germani}, S. and {Giebels}, B. and {Giglietto}, N. and {Giommi}, P. and {Giordano}, F. and {Giroletti}, M. and {Glanzman}, T. and {Godfrey}, G. and {Grenier}, I.~A. and {Grove}, J.~E. and {Guiriec}, S. and {Gustafsson}, M. and {Hadasch}, D. and {Hayashida}, M. and {Hays}, E. and {Healey}, S.~E. and {Horan}, D. and {Hou}, X. and {Hughes}, R.~E. and {Iafrate}, G. and {J{\'o}hannesson}, G. and {Johnson}, A.~S. and {Johnson}, W.~N. and {Kamae}, T. and {Katagiri}, H. and {Kataoka}, J. and {Kn{\"o}dlseder}, J. and {Kuss}, M. and {Lande}, J. and {Larsson}, S. and {Latronico}, L. and {Longo}, F. and {Loparco}, F. and {Lott}, B. and {Lovellette}, M.~N. and {Lubrano}, P. and {Madejski}, G.~M. and {Mazziotta}, M.~N. and {McConville}, W. and {McEnery}, J.~E. and {Michelson}, P.~F. and {Mitthumsiri}, W. and {Mizuno}, T. and {Moiseev}, A.~A. and {Monte}, C. and {Monzani}, M.~E. and {Moretti}, E. and {Morselli}, A. and {Moskalenko}, I.~V. and {Murgia}, S. and {Nakamori}, T. and {Naumann-Godo}, M. and {Nolan}, P.~L. and {Norris}, J.~P. and {Nuss}, E. and {Ohno}, M. and {Ohsugi}, T. and {Okumura}, A. and {Omodei}, N. and {Orienti}, M. and {Orlando}, E. and {Ormes}, J.~F. and {Ozaki}, M. and {Paneque}, D. and {Parent}, D. and {Pesce-Rollins}, M. and {Pierbattista}, M. and {Piranomonte}, S. and {Piron}, F. and {Pivato}, G. and {Porter}, T.~A. and {Rain{\`o}}, S. and {Rando}, R. and {Razzano}, M. and {Razzaque}, S. and {Reimer}, A. and {Reimer}, O. and {Ritz}, S. and {Rochester}, L.~S. and {Romani}, R.~W. and {Roth}, M. and {Sanchez}, D.~A. and {Sbarra}, C. and {Scargle}, J.~D. and {Schalk}, T.~L. and {Sgr{\`o}}, C. and {Shaw}, M.~S. and {Siskind}, E.~J. and {Spandre}, G. and {Spinelli}, P. and {Strong}, A.~W. and {Suson}, D.~J. and {Tajima}, H. and {Takahashi}, H. and {Takahashi}, T. and {Tanaka}, T. and {Thayer}, J.~G. and {Thayer}, J.~B. and {Thompson}, D.~J. and {Tibaldo}, L. and {Tinivella}, M. and {Torres}, D.~F. and {Tosti}, G. and {Troja}, E. and {Uchiyama}, Y. and {Vandenbroucke}, J. and {Vasileiou}, V. and {Vianello}, G. and {Vitale}, V. and {Waite}, A.~P. and {Wallace}, E. and {Wang}, P. and {Winer}, B.~L. and {Wood}, D.~L. and {Wood}, K.~S. and {Zimmer}, S.},
        title = "{The Second Catalog of Active Galactic Nuclei Detected by the Fermi Large Area Telescope}",
      journal = {\apj},
         year = 2011,
        month = dec,
       volume = {743},
       number = {2},
          eid = {171},
        pages = {171},
          doi = {10.1088/0004-637X/743/2/171},
archivePrefix = {arXiv},
       eprint = {1108.1420},
 primaryClass = {astro-ph.HE},
       adsurl = {https://ui.adsabs.harvard.edu/abs/2011ApJ...743..171A}
}

@ARTICLE{2007ApJ...669..862A,
       author = {{Albert}, J. and {Aliu}, E. and {Anderhub}, H. and {Antoranz}, P. and {Armada}, A. and {Baixeras}, C. and {Barrio}, J.~A. and {Bartko}, H. and {Bastieri}, D. and {Becker}, J.~K. and {Bednarek}, W. and {Berger}, K. and {Bigongiari}, C. and {Biland}, A. and {Bock}, R.~K. and {Bordas}, P. and {Bosch-Ramon}, V. and {Bretz}, T. and {Britvitch}, I. and {Camara}, M. and {Carmona}, E. and {Chilingarian}, A. and {Coarasa}, J.~A. and {Commichau}, S. and {Contreras}, J.~L. and {Cortina}, J. and {Costado}, M.~T. and {Curtef}, V. and {Danielyan}, V. and {Dazzi}, F. and {De Angelis}, A. and {Delgado}, C. and {de los Reyes}, R. and {De Lotto}, B. and {Domingo-Santamar{\'\i}a}, E. and {Dorner}, D. and {Doro}, M. and {Errando}, M. and {Fagiolini}, M. and {Ferenc}, D. and {Fern{\'a}ndez}, E. and {Firpo}, R. and {Flix}, J. and {Fonseca}, M.~V. and {Font}, L. and {Fuchs}, M. and {Galante}, N. and {Garc{\'\i}a-L{\'o}pez}, R.~J. and {Garczarczyk}, M. and {Gaug}, M. and {Giller}, M. and {Goebel}, F. and {Hakobyan}, D. and {Hayashida}, M. and {Hengstebeck}, T. and {Herrero}, A. and {H{\"o}hne}, D. and {Hose}, J. and {Hrupec}, D. and {Hsu}, C.~C. and {Jacon}, P. and {Jogler}, T. and {Kosyra}, R. and {Kranich}, D. and {Kritzer}, R. and {Laille}, A. and {Lindfors}, E. and {Lombardi}, S. and {Longo}, F. and {L{\'o}pez}, J. and {L{\'o}pez}, M. and {Lorenz}, E. and {Majumdar}, P. and {Maneva}, G. and {Mannheim}, K. and {Mansutti}, O. and {Mariotti}, M. and {Mart{\'\i}nez}, M. and {Mazin}, D. and {Merck}, C. and {Meucci}, M. and {Meyer}, M. and {Miranda}, J.~M. and {Mirzoyan}, R. and {Mizobuchi}, S. and {Moralejo}, A. and {Nieto}, D. and {Nilsson}, K. and {Ninkovic}, J. and {O{\~n}a-Wilhelmi}, E. and {Otte}, N. and {Oya}, I. and {Paneque}, D. and {Panniello}, M. and {Paoletti}, R. and {Paredes}, J.~M. and {Pasanen}, M. and {Pascoli}, D. and {Pauss}, F. and {Pegna}, R. and {Persic}, M. and {Peruzzo}, L. and {Piccioli}, A. and {Prandini}, E. and {Puchades}, N. and {Raymers}, A. and {Rhode}, W. and {Rib{\'o}}, M. and {Rico}, J. and {Rissi}, M. and {Robert}, A. and {R{\"u}gamer}, S. and {Saggion}, A. and {Saito}, T. and {S{\'a}nchez}, A. and {Sartori}, P. and {Scalzotto}, V. and {Scapin}, V. and {Schmitt}, R. and {Schweizer}, T. and {Shayduk}, M. and {Shinozaki}, K. and {Shore}, S.~N. and {Sidro}, N. and {Sillanp{\"a}{\"a}}, A. and {Sobczynska}, D. and {Stamerra}, A. and {Stark}, L.~S. and {Takalo}, L. and {Tavecchio}, F. and {Temnikov}, P. and {Tescaro}, D. and {Teshima}, M. and {Torres}, D.~F. and {Turini}, N. and {Vankov}, H. and {Vitale}, V. and {Wagner}, R.~M. and {Wibig}, T. and {Wittek}, W. and {Zandanel}, F. and {Zanin}, R. and {Zapatero}, J.},
        title = "{Variable Very High Energy {\ensuremath{\gamma}}-Ray Emission from Markarian 501}",
      journal = {\apj},
         year = 2007,
        month = nov,
       volume = {669},
       number = {2},
        pages = {862-883},
          doi = {10.1086/521382},
archivePrefix = {arXiv},
       eprint = {astro-ph/0702008},
 primaryClass = {astro-ph},
       adsurl = {https://ui.adsabs.harvard.edu/abs/2007ApJ...669..862A}
}

@ARTICLE{2010MNRAS.402..497G,
       author = {{Ghisellini}, G. and {Tavecchio}, F. and {Foschini}, L. and {Ghirlanda}, G. and {Maraschi}, L. and {Celotti}, A.},
        title = "{General physical properties of bright Fermi blazars}",
      journal = {\mnras},
         year = 2010,
        month = feb,
       volume = {402},
       number = {1},
        pages = {497-518},
          doi = {10.1111/j.1365-2966.2009.15898.x},
archivePrefix = {arXiv},
       eprint = {0909.0932},
 primaryClass = {astro-ph.CO},
       adsurl = {https://ui.adsabs.harvard.edu/abs/2010MNRAS.402..497G}
}

@ARTICLE{2024ApJ...967..104Z,
       author = {{Zhao}, X.~Z. and {Yang}, H.~Y. and {Zheng}, Y.~G. and {Kang}, S.~J.},
        title = "{The Energy Budget in the Jet of High-frequency Peaked BL Lacertae Objects}",
      journal = {\apj},
         year = 2024,
        month = jun,
       volume = {967},
       number = {2},
          eid = {104},
        pages = {104},
          doi = {10.3847/1538-4357/ad3ba9},
archivePrefix = {arXiv},
       eprint = {2406.01046},
 primaryClass = {astro-ph.HE},
       adsurl = {https://ui.adsabs.harvard.edu/abs/2024ApJ...967..104Z}
}

@ARTICLE{2014MNRAS.439.2933Y,
       author = {{Yan}, Dahai and {Zeng}, Houdun and {Zhang}, Li},
        title = "{The physical properties of Fermi BL Lac objects jets}",
      journal = {\mnras},
         year = 2014,
        month = apr,
       volume = {439},
       number = {3},
        pages = {2933-2942},
          doi = {10.1093/mnras/stu146},
archivePrefix = {arXiv},
       eprint = {1401.5552},
 primaryClass = {astro-ph.HE},
       adsurl = {https://ui.adsabs.harvard.edu/abs/2014MNRAS.439.2933Y}
}

\clearpage
\appendix
\section{Details of the sample}\label{appendix: Details}

\begin{itemize}
\item \textit{SHBL J001355.9--185406.}
The extreme blazar SHBL J001355.9--185406 is located at a redshift of z = 0.095 \citep{2009MNRAS.399..683J}, with equatorial coordinates (J2000) of R.A. = 3.483641$^\circ$ and Dec. = -18.901918$^\circ$. The equatorial coordinates of the sources in this appendix are all taken from the \citet{NED}\footnote{\url{https://ned.ipac.caltech.edu/}.}. SHBL J001355.9–-185406 was first detected in X-rays by \text{ROSAT} with flux in the 0.1–2.4 keV band is $1.26 \times 10^{-11}\rm\ erg\ s^{-1}\ cm^{-2}$ (1RXS J001356.6–185408, \citealt{1996IAUC.6420....2V}) and later identified as a BL Lac object by \citet{2000AN....321....1S}.  
The TeV $\gamma$-ray flux was measured at approximately 1\% of the Crab nebula flux above 300 GeV \citep{2013A&A...554A..72H}. 
\text{Fermi-LAT} did not detect a high-energy counterpart in 0.1-300 GeV for the initial two years of operation \citep{2012ApJS..199...31N}.
At 1.4 GHz, the radio flux density is $29.2 \pm 1.0$ mJy \citep{1998AJ....115.1693C}. The low-frequency radio spectrum is even flatter, with a spectral index $-0.1 \pm 0.1$ down to 156 MHz, suggesting that the jet exhibits a self-similar structure up to relatively large distances from the jet base \citep{2014A&A...563A.135Z}.

\item \textit{RGB J0152+017.}
The extreme blazar RGB J0152+017 is located at a redshift of z = 0.080 \citep{1999ApJ...525..127L}, with equatorial coordinates (J2000) of R.A. = 28.165046$^\circ$ and Dec. = 1.788162$^\circ$ in an elliptical galaxy with luminosity $M_R = -24.0$ \citep{2003A&A...400...95N}. Its TeV $\gamma$-ray emission was discovered by the Cherenkov telescope array H.E.S.S. in November 2007, showing no significant variation and a flux of $\rm 2.6 \times 10^{-12}\ cm^{-2}\ s^{-1}$ above 300 GeV; the X-ray and UV flux varies on time scales of $\sim 10$ days \citep{2008A&A...481L.103A}.

\item \textit{TXS 0210+515.}
The extreme blazar TXS 0210+515 is located at a redshift of z = 0.049 \citep{2011NewA...16..503M}, with equatorial coordinates (J2000) of R.A. = 33.574725$^\circ$ and Dec. = 51.747763$^\circ$.
According to the MAGIC observations, the intrinsic spectral index of TXS 0210+515 is $1.6\pm 0.3$ and it is classified as a hard-TeV EHBL \citep{2020ApJS..247...16A}.

\item \textit{1ES 0229+200.}
The extreme blazar 1ES 0229+200 resides in a faint elliptical host galaxy with luminosity $M_R = -24.53$ \citep{2000ApJ...542..731F} at a redshift of z = 0.139 \citep{2013ApJS..207...16M}. The position of 1ES 0229+200 in Equatorial (J2000) coordinates is R.A. = 38.202562$^\circ$ and Dec. = 20.288193$^\circ$. It was initially classified as HBL in 1995 after its discovery in 1992 by the Einstein Slew Survey \citep{1995A&AS..109..267G}. The TeV $\gamma$-ray emission from 1ES 0229+200 was first discovered by the H.E.S.S. observations during the period of 2005–2006 \citep{2007A&A...475L...9A}. As a prototypical EHBL, the spectrum of 1ES 0229+200 was measured to extend up to 10 TeV with a hard archival spectral index at TeV of $2.5\pm0.19$ \citep{2007A&A...475L...9A}.

\item \textit{1ES 0347--121.}
The extreme blazar 1ES 0347--121 was classified as a BL Lac object in 1993 after its discovery in 1992 by the Einstein Slew Survey \citep{1993ApJ...412..541S}. It resides in an elliptical host galaxy with luminosity $M_R = -23.2$ \citep{1999A&A...352...85F} at redshift z = 0.188 \citep{2013ApJS..207...16M}. The position of 1ES 0347--121 in Equatorial (J2000) coordinates is R.A.=57.346623$^\circ$ and Dec.=-11.990914$^\circ$. The first TeV $\gamma$-ray emission was discovered by H.E.S.S. telescopes in 2006 above an energy threshold of 250 GeV \citep{2007A&A...473L..25A}.

\item \textit{1ES 0414+009.}
The extreme blazar 1ES 0414+009 resides in an elliptical galaxy \citep{1991AJ....101..818H} at a redshift of z = 0.287 \citep{1991AJ....101..818H}, with equatorial coordinates (J2000) of R.A. = 64.218722$^\circ$ and Dec. = 1.089972$^\circ$. It is one of the most distant TeV BL Lacs detected to date. 1ES 0414+009 was first detected by the \text{HEAO-1} satellite \citep{1978ApJ...223..973G} in the energy range of 0.2 keV--10 MeV and was detected for the first time in the TeV band in 2005 by H.E.S.S. observations. The TeV $\gamma$-ray emission is observed both by the H.E.S.S. \citep{2012ApJ...755..118A} and VERITAS \citep{2012A&A...538A.103H} collaborations.

\item \textit{PKS 0548--322.}
The extreme blazar PKS 0548--322 resides in a giant elliptical galaxy with an absolute visual magnitude of $M_V = -23.4$ \citep{1995ApJ...438L...9F} at a redshift of z = 0.069 \citep{2013ApJS..207...16M}, with equatorial coordinates (J2000) of R.A. = 87.669037$^\circ$ and Dec. = -32.271197$^\circ$. 
PKS 0548--322 was not detected in the MeV-GeV range by the EGRET detector \citep{1999ApJS..123...79H}, but it is suggested as a candidate TeV emitter by \citet{2002A&A...384...56C}. 
The first discovery of TeV $\gamma$-rays from PKS 0548--322 was made using the H.E.S.S. Cherenkov telescopes, as presented in \citet{2010A&A...521A..69A}. 
The details about the X-ray history of this object can be found in \citet{2007A&A...462..889P}.

\item \textit{RGB J0710+591.}
The extreme blazar RGB J0710+591 was first discovered by \text{HEAO A-1} \citep{1984ApJS...56..507W}, hosted in an elliptical galaxy with a nuclear point source at redshift z = 0.125 \citep{1991ApJ...378...77G, 2000ApJ...532..740S}. The position of RGB J0710+591 in Equatorial (J2000) is R.A. = 107.629000$^\circ$, Dec. = 59.139000$^\circ$. 
Its first TeV emission was detected by VERITAS during the 2008-2009 observations with no evidence of variability in the $\gamma$-ray light curve above 300 GeV \citep{2010ApJ...715L..49A}. 

\item \textit{PGC 2402248.}
The extreme blazar PGC 2402248 is located at a redshift of z = 0.065 \citep{2023MNRAS.524..133M}, with equatorial coordinates (J2000) of R.A. = 113.361630$^\circ$, Dec. = 51.898871$^\circ$. 
The MAGIC telescopes observed the source from January 23 to April 19, 2018 (MJD 58141–58227). 
During this period, simultaneous broadband observations were also carried out by the KVA and the Swift/UVOT in the optical and ultraviolet bands, Swift/XRT in the X-ray band, and Fermi-LAT in the GeV $\gamma$-ray band, which showed no significant variability except for moderate variability in the Swift-UVOT/XRT data \citep{2023MNRAS.519..854M}.

\item \textit{RBS 0723.}
The extreme blazar RBS 0723 is a BL Lac object at z = 0.198 \citep{2012ApJS..203...21A}, identified as a bright X-ray source in the ROSAT All Sky Survey and characterized by a hard ($\alpha = 0.78$) X-ray continuum as observed by Swift/XRT \citep{2013ApJS..209...34A}. Its equatorial coordinates (J2000) are R.A. = 131.803899$^\circ$ and Dec. = 11.563962$^\circ$.  
RBS 0723 is included in the 1FHL and 2FHL catalogs, exhibiting a hard spectral index $\Gamma_{\rm 1FHL} = 1.4 \pm 0.4$ and a bright gamma-ray flux ($ 9.6 \times 10^{-11}\ \rm ph\ cm^{-2}\ s^{-1}$) above 10 GeV \citep{2013ApJS..209...34A}.  
The MAGIC telescopes detected TeV $\gamma$-rays from RBS 0723 during observations conducted in 2013 and 2014, with a flux of $2.6 \pm 0.5 \times 10^{-12}\ \rm ph\ cm^{-2}\ s^{-1}$ above 200 GeV \citep{2020ApJS..247...16A}.  

\item \textit{MRC 0910--208.}
The extreme blazar MRC 0910-208 is located at a redshift of z = 0.198 \citep{2013ApJS..207...16M}, with equatorial coordinates (J2000) of R.A. = 138.250930$^\circ$ and Dec. = -21.055861$^\circ$.  
Its TeV $\gamma$-rays were observed by H.E.S.S. in May 2018, with a spectral index of 3.63 \citep{2022icrc.confE.823B}.  

\item \textit{1ES 1101--232.}
The extreme blazar 1ES 1101--232 was first detected by the Ariel-5 X-ray satellite, hosted in an elliptical host galaxy at a redshift of z = 0.186 \citep{1994ApJS...93..125F}. The position of 1ES 1101--232 in Equatorial (J2000) is R.A. = 165.906738$^\circ$, Dec. = -23.491975$^\circ$.  
The discovery of TeV $\gamma$-ray emission from this source was reported in 2007 using H.E.S.S. observations performed during 2004-2005 in the energy range above 250 GeV, and no significant variation of the TeV $\gamma$-ray flux on any time scale was found \citep{2007A&A...470..475A}.

\item \textit{Mrk 421.}
The nearest extreme blazar Mrk 421 located at the redshift of z = 0.031 \citep{2021A&A...655A..89M} is one of the most extensively studied and brightest extragalactic $\gamma$-ray sources. The position of Mrk 421 in Equatorial (J2000) is R.A.=166.113808$^\circ$, Dec.=38.208833$^\circ$.
It is also one of the fastest varying $\gamma$-ray sources.
The observations about Mrk 421 are provided by various instruments that covering the energy range from radio to TeV $\gamma$-ray (e.g., \citealt{2003ApJ...598..242A, 2008ApJ...677..906F, 2013MNRAS.434.2684M, 2017ApJ...834....2A}. Its TeV $\gamma$-rays was firstly detected by the Whipple telescope in 1992 above 0.5 TeV \citep{1992Natur.358..477P}. 

\item \textit{1ES 1218+304.}
The extreme blazar 1ES 1218+304 is located at a redshift of z = 0.182 \citep{1998A&A...334..459B}, with equatorial coordinates (J2000) of R.A. = 185.341432$^\circ$ and Dec. = 30.176983$^\circ$.  
The TeV $\gamma$-rays were first observed by MAGIC \citep{2006ApJ...642L.119A} and subsequently by the VERITAS telescopes \citep{2009ApJ...695.1370A}. The $\gamma$-ray emission at $\sim160$ GeV and $\sim1.8$ TeV is characterized by a hard $\gamma$-ray photon index of $1.86 \pm 0.37$ after EBL correction \citep{2009ApJ...695.1370A}.  
The first evidence of variability in the TeV $\gamma$-ray emission from 1ES 1218+304 was observed by VERITAS during the active state in 2009, with a flux doubling time-scale of $\leqslant1$ day \citep{2010ApJ...709L.163A}.  
1ES 1218+304 exhibits a hard photon index of $1.72 \pm 0.02$ as observed by Fermi-LAT in the GeV $\gamma$-ray band \citep{2020ApJS..247...33A}.  

\item \textit{1ES 1312--423.}
The extreme blazar 1ES 1312--423, located at a redshift of z = 0.105 \citep{2000AJ....120.1626R}, was initially detected in X-rays by the Einstein observatory \citep{1990ApJ...356L..35G} and subsequently classified as a BL Lac object by \citet{1991ApJS...76..813S}. The position of 1ES 1312--423 in equatorial coordinates (J2000) is R.A. = 198.764134$^\circ$, Dec. = -42.613821$^\circ$.  
Between April 2004 and July 2010, it was observed by H.E.S.S. in the TeV $\gamma$-ray band above 280 GeV, with a spectral index of $\Gamma = 2.85 \pm 0.47$ and a differential flux at 1 TeV of $F = 1.89 \pm 0.58 \times 10^{-13}\ \rm cm^{-2}\ s^{-1}\ TeV^{-1}$ \citep{2013MNRAS.434.1889H}.  

\item \textit{1ES 1426+428.}
The extreme blazar 1ES 1426+428 is located at a redshift of z = 0.129 \citep{2024ApJS..270...22W}, with equatorial coordinates (J2000) of R.A. = 217.135851$^\circ$ and Dec. = 42.672493$^\circ$.  
It was first detected in X-rays by the Uhuru X-ray observatory \citep{1972ApJ...178..281G} and was classified as a BL Lac by \citet{1989ApJ...345..140R} based on the combined observations of featureless optical continua, compact radio emission, and images of the underlying host galaxies. 
1ES 1426+428 has a synchrotron peak at $10^{18.1}$ Hz \citep{2001AIPC..599..586C} and is characterized by being optically faint but very bright in the X-ray band.  

\item \textit{1ES 1440+122.}
The extreme blazar 1ES 1440+122, located at a redshift of z = 0.163 \citep{2006A&A...457...35S}, is identified as an extreme-high-synchrotron-peaked BL Lac object with a synchrotron peak frequency of $\nu_{\rm s,p} \sim 3 \times 10^{17}$ Hz \citep{2016MNRAS.461..202A}. The position of 1ES 1440+122 in equatorial coordinates (J2000) is R.A. = 220.700901$^\circ$, Dec. = 12.011167$^\circ$.  
1ES 1440+122 was included in the 3FGL catalogue (3FGL J1442.8+1200) with a spectral index of $1.80 \pm 0.12 $\citep{2015ApJS..218...23A}.  

\item \textit{Mrk 501.}
The extreme blazar Mrk 501, located at a redshift of z = 0.034 \citep{2024ApJS..270...22W}, is also a very well-studied TeV source, similar to Mrk 421. It has been intensively monitored by various astronomical detectors, from radio frequencies to TeV $\gamma$-rays (e.g., \citealt{2011ApJ...727..129A, 2012A&A...541A..31N, 2015A&A...573A..50A, 2015ApJ...798....2S, 2018A&A...620A.181A, 2022ApJ...929..125A}). The position of Mrk 501 in equatorial coordinates (J2000) is R.A. = 253.467569$^\circ$, Dec. = 39.760169$^\circ$.  
Its TeV $\gamma$-rays were first detected by the Whipple telescope in 1996 at energies above 0.3 TeV \citep{1996ApJ...456L..83Q}.  

\item \textit{1ES 1727+502.}
The extreme blazar 1ES 1727+502 is a nearby BL Lac located at a redshift of z = 0.055 \citep{1991rc3..book.....D}. The position of 1ES 1727+502 in equatorial coordinates (J2000) is R.A. = 262.077600$^\circ$, Dec. = 50.219575$^\circ$.  
Its TeV $\gamma$-rays were discovered by MAGIC, with an integral flux above 150 GeV estimated to be $(2.1 \pm 0.4)\%$ of the Crab nebula flux. The de-absorbed TeV spectrum has a photon index of $(2.7 \pm 0.5)$ \citep{2014A&A...563A..90A}.  
It was also detected by VERITAS, with an integral flux of $(1.1 \pm 0.2) \times 10^{-11}\ \rm cm^{-2}\ s^{-1}$ above 250 GeV \citep{2015ApJ...808..110A}.  
1ES 1727+502 shows little variability in the optical R-band, is bright in the X-ray band, and has a hard spectrum in the GeV $\gamma$-ray band \citep{2014A&A...563A..90A}.  

\item \textit{1ES 1741+196.}
The extreme blazar 1ES 1741+196 is a nearby BL Lac located at a redshift of z = 0.084 \citep{1996ApJS..104..251P}. The position of 1ES 1741+196 in equatorial coordinates (J2000) is R.A. = 265.990971$^\circ$, Dec. = 19.585839$^\circ$.  
TeV $\gamma$-ray emission was first detected by the MAGIC collaboration \citep{2017MNRAS.468.1534A}.  
It was included in the 3LAC catalogue as HBL 3FGL J1743.9+1934, with a spectral index of $\Gamma = 1.777 \pm 0.108$ \citep{2015ApJ...810...14A}.  

\item \textit{1RXS J195815.6--301119.}
The extreme blazar 1RXS J195815.6--301119 is located at a redshift of z = 0.119 \citep{2022icrc.confE.823B}, with equatorial coordinates (J2000) of R.A. = 299.562136$^\circ$ and Dec. = -30.186556$^\circ$.
Its TeV $\gamma$-rays were observed by H.E.S.S. in May 2018, with a spectral index of 2.78 \citep{2022icrc.confE.823B}. 

\item \textit{1ES 1959+650.}
The extreme blazar 1ES 1959+650, located at a redshift of z = 0.047 \citep{2024ApJS..270...22W}, is one of the earliest extragalactic TeV $\gamma$-ray emitters \citep{2003ApJ...583L...9H}, whose TeV emission was first detected in 1998 \citep{1999ICRC....3..370N}. The position of 1ES 1959+650 in equatorial coordinates (J2000) is R.A. = 299.999383$^\circ$, Dec. = 65.148514$^\circ$.  
Observations revealed that 1ES 1959+650 is very active across multiple wavebands \citep{2014ApJ...797...89A, 2014ChA&A..38..233Y, 2015AJ....150...67Y, 2016MNRAS.457..704K, 2018MNRAS.473.2542K, 2017ApJ...846..158K, 2017ApJ...847....8L, 2017MNRAS.469.1682Z, 2018A&A...611A..44P}.  

\item \textit{1ES 2037+521.}
The extreme blazar 1ES 2037+521 is located at a redshift of z = 0.053 \citep{2024ApJS..270...22W}, with equatorial coordinates (J2000) of R.A. = 309.847998$^\circ$ and Dec. = 52.330608$^\circ$.  
It was observed with the MAGIC telescopes in 2016, showing a flux of $\rm (1.8 \pm 0.4) \times 10^{-12}\ erg\ cm^{-2}\ s^{-1}$ and an EBL-corrected spectral index of $2.0 \pm 0.5$.  
It was also observed with Swift-XRT, showing a flux in the 2–10 keV band of $\rm (10.7 \pm 1.0) \times 10^{-12}\ erg\ cm^{-2}\ s^{-1}$ and a spectral index of $1.93 \pm 0.13$.  
It was observed with Fermi-LAT, showing a flux in the 1–300 GeV energy range of $\rm (4.6 \pm 1.5) \times 10^{-10}\ erg\ cm^{-2}\ s^{-1}$ and a spectral index of $1.7 \pm 0.2$ \citep{2020ApJS..247...16A}.

\item \textit{1ES 2344+514.}
The extreme blazar 1ES 2344+514 is a nearby BL Lac located at a redshift of z = 0.044 \citep{1996ApJS..104..251P}, with equatorial coordinates (J2000) of R.A. = 356.770154$^\circ$ and Dec. = 51.704967$^\circ$.  
The first TeV detection was achieved by the Whipple 10 m telescope during a bright flare in 1996, with a peak flux of approximately 60\% of the Crab Nebula flux above 350 GeV \citep{1998ApJ...501..616C}.  
Then, \citet{2017MNRAS.471.2117A} reported an average flux above 350 GeV of approximately 4\% of the Crab Nebula between 2008 and 2015, during a period without any flaring activity.  
\citet{2024A&A...682A.114M} reported that MAGIC observations from August 2019 to December 2021 revealed strong spectral variability occurring in the quiescent state. They also reported a strong X-ray flare, among the brightest observed from MAGIC.
The TeV flux can vary by a factor of approximately 2 on daily timescales \citep{2007ApJ...662..892A, 2017MNRAS.471.2117A, 2011ApJ...738..169A}, and it is highly variable in X-rays over timescales as short as 5000 seconds \citep{2000MNRAS.317..743G}.

\item \textit{H 2356--309.}
The extreme blazar H 2356--309 is one of the most distant BL Lacs, hosted by a normal giant elliptical galaxy located at a redshift of z = 0.165 \citep{1991AJ....101..821F}, with equatorial coordinates (J2000) of R.A. = 359.782919$^\circ$ and Dec. = -30.627960$^\circ$.  
It was first observed in X-rays by the UHURU satellite \citep{1978ApJS...38..357F}.  
The TeV $\gamma$-rays were discovered by H.E.S.S. during June to December 2004 \citep{2006A&A...455..461A}.  

\end{itemize}

\clearpage
\section{Model description}\label{section: model}
The model assumes a spherical, uniformly emitting region with a radius of $R$ and a magnetic field of $B$, which moves along the jet with the bulk Lorentz factor $\Gamma$. The angle between the jet and the observer's line of sight is $1/\Gamma$, and the Doppler beaming factor $\delta$ is considered to be equal to $\Gamma$.
\subsection{Electron Energy Distribution}\label{ped}
The steady electron energy distribution (EED) within this region is modeled as a broken power law without restricting the specific acceleration process.
\begin{equation}
N(\gamma)=N_0 \begin{cases}\gamma^{-p_1} & \gamma_{\min } \leqslant \gamma \leqslant \gamma_{\mathrm{b}}, \\ \gamma_{\mathrm{b}}^{p_2-p_1} \gamma^{-p_2} & \gamma_{\mathrm{b}}<\gamma<\gamma_{\max },\end{cases}
\end{equation}
where $p_1$ and $p_2$ are the electron spectral indices, $\gamma_{\rm min}$, $\gamma_b$, and $\gamma_{\rm max}$ are the electron Lorentz factors, and $N_0$ [$\rm cm^{-3}$] denotes the electron number density.

\subsection{Synchrotron}

Single particle synchrotron emissivity [$\rm erg\ s^{-1}\ Hz^{-1}$] \citep{2013LNP...873.....G} is
\begin{equation}
P_{\rm syn}(\nu, \gamma)=\frac{2 \pi \sqrt{3} e^2 \nu_{\mathrm{L}}}{c} \left[\frac{\nu}{\nu_{\rm c}} \int_{\nu / \nu_{\rm c}}^{\infty} K_{5 / 3}(t) d t\right],
\end{equation}
where Larmor frequency $\nu_{\rm L} = \frac{eB}{2 \pi m_e c} $, critical frequency $\nu_{\rm c}  = \frac{3}{2} \nu_{\rm s}$, synchrotron radiation frequency $\nu_{\rm s}= \gamma^2 \nu_{\rm L}$. $K_{5/3}(t)$ is the modified Bessel function of order 5/3.
The synchrotron emissivity coefficient from many electrons averaged over an isotropic distribution of pitch angles is given by [erg s$^{-1}$ cm$^{-3}$ Hz$^{-1}$ ster$^{-1}$] \citep{1979rpa..book.....R}
\begin{equation}
j_{\text {syn }}(\nu)=\frac{1}{4 \pi} \int P_{\text {syn}}( \nu, \gamma) N(\gamma) \mathrm{d} \gamma .
\end{equation}
Synchrotron self-absorption (SSA) coefficient $\kappa(\nu)$[$\mathrm{cm}^{-1}$] \citep{1979rpa..book.....R} is obtained by 
\begin{equation}\label{synchrotron absorb coeﬃcient}
\kappa(\nu)=- \frac{1}{8 \pi m_{\mathrm{e}} \nu^2} \int_{\gamma_{\rm min}}^{\gamma_{\rm max}} P_{\rm syn}(\nu, \gamma) \gamma^2 \frac{d}{d \gamma}\left[\frac{N(\gamma)}{\gamma^2}\right] d \gamma.
\end{equation}
The optical depth caused by SSA is $\tau = \kappa(\nu)R$, and when $\tau <10^{-4}$, the SSA is ignored.
The monochromatic luminosity [erg cm$^{-2}$ s$^{-1}$ Hz$^{-1}$] \citep{2017ApJ...842..129C} in the jet rest frame is 
\begin{equation}
    L_{\rm syn}\left(\nu\right)=
    \begin{cases}
        \frac{4}{3} \pi R^{3} \cdot 4 \pi j_{\rm syn}\left(\nu\right) \cdot\left(1-\frac{3}{4} \tau \right)& \tau <10^{-4}, \\
        2 \pi^2 R^{3} j_{\rm syn}\left(\nu\right) \frac{2 \tau^2-1+(2 \tau+1) e^{-2 \tau}}{\tau^3}  & \text {else}.
    \end{cases}
\end{equation}
Therefore, the synchrotron radiation flux density [$\rm erg\ s^{-1}cm^{-2}Hz^{-1}$] in the observer's reference frame is given by
\begin{equation}\label{flux_density}
    F_{\rm syn, obs}(\nu_{\rm obs}) = \frac{L_{\mathrm{syn}}(\nu)}{4 \pi d^2_{\rm L}} \delta^{3} (1+z),
\end{equation}
where $d_{\rm L}$ denotes the luminosity distance and the observed frequency is related to the source-frame frequency by $\nu_{\rm obs} = \nu \frac{\delta}{1+z}$.

\subsection{Synchrotron Self-Compton}

The SSC scattering power of a single electron moving in a seed radiation field is
\begin{equation}\label{IC_process}
P_{\mathrm{ssc}}(\nu, \gamma)=3 \sigma_{\rm T} \operatorname{hc} \int f\left(\gamma, \nu_i, \nu_{\rm s}\right) n_{\rm ph}\left(\nu_i\right) d \nu_i,
\end{equation}
where $\nu_{\rm i}$ and $\nu_{\rm s}$ are incident photon frequencies and scattered photon frequencies, respectively, and $\nu_{\rm i} \in [\frac{\nu_{\rm s}}{4\gamma}, \frac{3m_e c^2}{4h\gamma}]$. 
For the SSC process, the target photons required for inverse Compton scattering are provided by the synchrotron radiation itself, and the number density of the corresponding seed photon field is given by
\begin{equation}
n_{\rm ph} = \frac{9L_{\rm syn}}{16 \pi R^{2} c }\frac{1}{h\nu}.
\end{equation}
The inverse Compton scattering core function \citep{1968PhRv..167.1159J, 1970RvMP...42..237B} expressed as 
\begin{equation}
    f\left(\gamma, \nu_i, \nu_s\right)=
    \begin{cases}
        x\left(2 x \ln x+1+x-2 x^2\right) & 0<x<1, \\
        0 & \text {else},
    \end{cases}
\end{equation}
where $x \equiv \nu_{\rm s} / (4 \gamma^2 \nu_i)$.
The SSC emissivity [erg s$^{-1}$ cm$^{-3}$ Hz$^{-1}$ sr$^{-1}$], which arises from a large population of electrons, is given by
\begin{equation}
j_{\rm ssc}(\nu)=\frac{1}{4 \pi} \int P_{\mathrm{ssc}}(\nu, \gamma) N(\gamma) d \gamma.
\end{equation}
In addition, the approximation method for Klein-Nishina (KN) effect in \citet{1998MNRAS.301..451G} is applied to the SSC process,
\begin{equation}
    \sigma_{\rm KN} = 
    \begin{cases}
      \displaystyle \sigma_{\rm T} & \gamma\ \epsilon \leqslant 3/4,\\
      \displaystyle 0 & \rm else,\\
    \end{cases}
\end{equation}
where $\varepsilon$ is the dimensionless energy of the seed photon in units of $m_e c^2$.
Finally, the monochromatic SSC luminosity [$\rm erg\ s^{-1}\ Hz^{-1}$] in the jet rest frame is given by
\begin{equation}
    L_{\rm ssc} =\frac{16}{3} \pi^2 R^{3} j_{\rm ssc}(\nu).
\end{equation}
Therefore, the SSC radiation flux density [$\rm erg\ s^{-1}cm^{-2}\ Hz^{-1}$] in the observer's reference frame is given by
\begin{equation}\label{flux_density SSC}
    F_{\rm ssc, obs}(\nu_{\rm obs}) = \frac{L_{\mathrm{ssc}}(\nu)}{4 \pi d^2_{\rm L}} \delta^{3} (1+z).
\end{equation}

\subsection{Attenuated $\gamma$-rays by Pair Production.}
The $\gamma$-rays produced in the jet would be attenuated due to pair production ($\gamma\gamma\rightarrow \rm e^+e^-$) interactions with synchrotron pthones, the total cross-section \citep{2004vhec.book.....A} is given by 
\begin{equation}
\begin{gathered}
\sigma_{\gamma \gamma}\left(v, v_{\mathrm{i}}\right)=\frac{3 \sigma_{\mathrm{T}}}{2 s^2}\left[\left(s+\frac{1}{2} \ln s-\frac{1}{6}+\frac{1}{2 s}\right) \ln (\sqrt{s}+\sqrt{s-1})\right. \\
\left.-\left(s+\frac{4}{9}-\frac{1}{9 s}\right) \sqrt{1-\frac{1}{s}}\right],
\end{gathered}
\end{equation}
where $s \equiv h^2 \nu \nu_{\rm i}/m_{\rm e}^2 c^4$, $h\nu$ is the energy of the $\gamma$-rays and $h\nu_{\rm i}$ is the energy of the incident target photon. $\sigma_{\gamma\gamma}(\nu,\nu_{\rm i})$ reaches its maximum value $\sim 0.22\sigma_{\rm T}$ at $s = 3.5$.
The optical depth to pair production \citep{2025NatAs...9.1086L} is given by
\begin{equation}
\tau_{\gamma \gamma}(v)=\int_{v_{\min }}^{v_{\max }} n_{\mathrm{ph}}\left(v_{\mathrm{i}}\right) \sigma_{\gamma \gamma}\left(v, v_{\mathrm{i}}\right) R \mathrm{~d} v_{\mathrm{i}} .
\end{equation}
In the uniformity assumption of non-thermal electrons, the $\gamma$-ray photons remaining after attenuation are given by attenuation factor \citep{2009herb.book.....D}
\begin{equation}
A_{\rm syn} =\frac{3}{\tau_{\gamma \gamma}(v)}\left\{\frac{1}{2}+\frac{\exp \left[-\tau_{\gamma \gamma}(v)\right]}{\tau_{\gamma \gamma}(v)}-\frac{1-\exp \left[-\tau_{\gamma \gamma}(v)\right]}{\tau_{\gamma \gamma}(v)^2}\right\}.
\end{equation}

On the other hand, TeV $\gamma$-photons from the source are attenuated by photons of the EBL photons due to pair production.
EBL absorption optical depth $\tau_{\rm EBL}$ is  given by \citet{2022ApJ...941...33F}, and attenuation factor \citep{2022PhRvD.106j3021X} is given by
\begin{equation}
    A_{\rm EBL} = e^{-\tau_{\rm EBL}}.
\end{equation}
Finally, the observed flux [$\rm erg \ cm^{-2}\ s^{-1}$] is expressed as
\begin{equation}\label{flux_density SSC}
    F^{\prime}_{\rm ssc, obs} = \nu_{\mathrm{obs}} F_{\rm ssc, obs} A_{\rm syn} A_{\rm EBL}.
\end{equation}

\end{document}